\newif\iffigs\figstrue
\documentclass[Crown]{iopbk2e}
\usepackage{latexsym,amssymb}
\iffigs
 \input{epsf}
\else
\fi
\def\theequation{\arabic{section}.\arabic{equation}}

\renewcommand{\theequation}{\thesection.\arabic{equation}}
\newtheorem{definizione}{Definition}[section]
\newtheorem{teorema}{Theorem}[section]
\newtheorem{statement}{Statement}[section]

\newtheorem{lemma}{Lemma}[section]

\newtheorem{proposition}{Proposition}[section]
\newcommand{\blem}{\begin{lemma}}
\newcommand{\elem}{\end{lemma}}

\newcommand{\eqn}[1]{(\ref{#1})}
\newcommand{\ft}[2]{{\textstyle\frac{#1}{#2}}}

\def\Im{{\rm Im ~}}
\def\Re{{\rm Re ~}}
\def\IP{\relax{\rm I\kern-.18em P}}

\def\inbar{\vrule height1.5ex width.4pt depth0pt}
\def\IC{\relax\,\hbox{$\inbar\kern-.3em{\rm C}$}}
\def\IG{\relax\,\hbox{$\inbar\kern-.3em{\rm G}$}}
\def\IB{\relax{\rm I\kern-.18em B}}
\def\ID{\relax{\rm I\kern-.18em D}}
\def\IL{\relax{\rm I\kern-.18em L}}
\def\IF{\relax{\rm I\kern-.18em F}}
\def\IH{\relax{\rm I\kern-.18em H}}
\def\II{\relax{\rm I\kern-.17em I}}
\def\IN{\relax{\rm I\kern-.18em N}}
\def\IP{\relax{\rm I\kern-.18em P}}
\def\IQ{\relax\,\hbox{$\inbar\kern-.3em{\rm Q}$}}
\def\bfzero{\relax\,\hbox{$\inbar\kern-.3em{\rm 0}$}}
\def\IK{\relax{\rm I\kern-.18em K}}
\def\IG{\relax\,\hbox{$\inbar\kern-.3em{\rm G}$}}
\def\bfone{\relax{\rm 1\kern-.35em 1}}

\def\square{{\,\lower0.9pt\vbox{\hrule \hbox{\vrule height 0.2 cm
\hskip 0.2 cm \vrule height 0.2 cm}\hrule}\,}}
\def\bfone{\relax{\rm 1\kern-.35em 1}}
 
\def\bar{\overline}
\def\IGam{\relax{{\rm I}\kern-.18em \Gamma}}

\begin{document}
%%%%%%%%%%%%%%%%%%%%%%%%%%%%%%%%%%%%%%
\title{{\bf  GAUGINGS AND OTHER SUPERGRAVITY TOOLS of $p$--BRANE PHYSICS\\
 }}
\author{
\\
   PIETRO FRE' \\
\\
   {\small {\sl Dipartimento di Fisica
Teorica, Universit\'a di Torino,}}
\\
{\small {\sl  Via P. Giuria  1, I-10125 TORINO,   Italy }}\\
\\
{
%\small
{ Lectures at the School
  {\bf  Recent Advances in M--theory }}}\\
 {\small supported by EEC under RTN Contract HPRN-CT-2000-00131}\\
 {\small Paris, 1-8 Feb. 2001,
 IHP }
}
  \maketitle
%%%%%%%%%%%%%%%%%%%%%%%%%%%%%%%%%%
\tableofcontents
%%%%%%%%%%%%%%%%%%%%%%%%%%%%%%%%%%%%%
\chapter[Domain walls and the brane bestiary]{Domain walls and the brane
bestiary}
\label{bestia1}
\section[Introduction]{Introduction}
\setcounter{equation}{0}
\label{introbestia1}
I can best assess the purpose and the scope of this lecture series
starting with a summary of the second string revolution's moral \cite{huto,witten} as I
presently perceive it:
\begin{enumerate}
  \item There is just one non perturbative ten dimensional \textbf{string theory} that can
  also be identified as the mysterious \textbf{M--theory} having $D=11$ supergravity as its
  low energy limit.
  \item All \textbf{$p$--branes}, whether electric or magnetic, whether coupled
  to Neveu Schwarz or to Ramond $p+1$--forms encode noteworthy
  aspects of the unique M--theory.
  \item \textbf{Microscopically} the $p$--brane
  degrees of freedom are described by a suitable {\it gauge theory}
  $\mathcal{GT}_{p+1}$
  living on the  $p+1$ dimensional \emph{world volume} $\mathcal{WV}_{p+1}$
  that can be either conformal or not.
  \item\textbf{Macroscopically} each $p$-brane is a generalized soliton in the
  following sense. It  is a classical solution
  of  $D=10$ or $D=11$ supergravity interpolating between two
  asymptotic geometries that with some abuse of language we respectively name   the \emph{the
  geometry at infinity} $geo^\infty$ and the \emph{the near horizon geometry}
  $geo^H$.
  The latter which only occasionally corresponds to a true event horizon
  is  instead  universally characterized by the following property. It can be interpreted as a
  solution of some suitable $p+2$ dimensional supergravity $\mathcal{SG}_{p+2}$ times
  an appropriate \emph{internal space} $\Omega_{D-p-2}$.
  \item Because of the statement above, all space--time dimensions
  $11 \geq D \geq 3 $ are relevant and
  supergravities in these {\it diverse dimensions}  describe
  various perturbative and non--perturbative aspects of superstring
  theory. In particular we have a most intriguing $gauge/gravity$
  correspondence implying that classical supergravity $\mathcal{SG}_{p+2}$ expanded around
  the vacuum solution $geo^H$ is dual to the quantum gauge theory $\mathcal{GT}_{p+1}$
  in one lower dimension.
  \end{enumerate}
 In this general framework I will focus on the following issue that
 appears to be of great interest at the present time. The
 $gauge/gravity$ correspondence, by now largely tested at the level of
 the $AdS/CFT$ \cite{maldapasto,renatoine,review} duality,
 is presently under consideration in more
 general non conformal scenarios provided by the superstring world.
 One is the case of fractional branes
 \cite{Bertolini:2000dk,Herzog:2000rz}, the other is the issue of the
 $\mathrm{{DW/QFT}}$ correspondence between supergravity on domain
 wall space--times and quantum field theories living on the boundary wall of
 such space times. As a conspicuous aspect of such
 geometries appears the phenomenon of gravity trapping, suggested by
 Randall and Sundrum \cite{RS2,RS1} that can provide a
 phenomenologically very interesting alternative to compactification.
 This has initialized a world wide and so far unconclusive search for a proper
 embedding of this scenario within a well founded supersymmetric theory
 \cite{Cvetic:2000gj,Behrndt:2000tr,Behrndt:2000kz,Cvetic:1999pu,cveticdelta,renandrei,ceregatta}.
 Furthermore it has been clarified that the various domain wall
 geometries are the appropriate description of the \emph{near
 horizon} or \emph{near brane} regime of all $p$ and $Dp$--branes of
 string theory.
 The central unresolved question in all these interlaced issues where the
 domain wall appears is the systematic
 identification of the appropriate $p+2$ dimensional supergravity theory
 that accommodates the  wall as a classical solution. Clearly such a
 theory plays a fundamental role in the description of the $\mathrm{DW/QFT}$
 correspondence as the maximally compact gauged supergravities have
 played in the $\mathrm{AdS/CFT}$ duality.
 \par
 Hence after a survey of the supergravity $p$--branes and of their
 limiting case, the classical domain walls, I will devote the next two
 chapters to a  systematic  of supergravity gaugings. My main concern
 will be that of illustrating the geometric structures of
 supergravity, their meaning in relation with the parent string
 theory and their use in the gauging procedures.
%%%%%%%%%%%%%%%%%%%%%%%%%%%%%%%%%%%%%%%%%%%%%%%%%%%%%%%%%
\section[General aspects of supergravity $p$--branes]{General aspects
of supergravity $p$--branes}
\setcounter{equation}{0}
Supergravity $p$--branes can be obtained as classical solutions of
actions of the following form
\begin{eqnarray}
 \left \{\begin{array}{rcll}
 A_{p-brane}^{[D]}&=&
  \int \, d^Dx \, \sqrt{-g} \,\left[ 2 \, R[g]
  +\frac{1}{2}\partial^\mu \, \phi \partial_\mu \phi +
  \frac{(-1)^{p+1}}{2(p+2)!}\, e^{-a\phi} \vert F^{[p+2]} \vert^2
  \right]& \mbox{elec.}\\
 \null & \null & \null \\
  A_{\widetilde{p}-brane}^{[D]}&=&
\int \, d^Dx \, \sqrt{-g} \,\left[ 2 \, R[g]
  +\frac{1}{2}\partial^\mu \, \phi \partial_\mu \phi +
  \frac{(-1)^{D-\widetilde{p}-3}}{2(D-\widetilde{p}-2)!}\, e^{-a\phi} \vert F^{[D-\widetilde{p}-2]} \vert^2
  \right] & \mbox{magn.}\
  \end{array} \right.\nonumber\\
\label{pbranact}
\end{eqnarray}
where in both cases $F^{[n]}\equiv d A^{[n-1]}$ is the field strength of
an $n-1$--form gauge potential  and $a$ is some real number whose profound meaning
will become clear in my later discussion of the solutions.  As the
reader can notice the two formulae I have written for the $p$-brane
action are actually the same formula since $A_{p-brane}^{[D]}$ and
$A_{\widetilde{p}-brane}^{[D]}$ are mapped into each other by the replacement:
\begin{equation}
  \widetilde{p} = D-4-p \quad ; \quad p = D-4-\widetilde{p}
\label{exciangio}
\end{equation}
The reason why I doubled my writing  is that the essentially unique
action (\ref{pbranact}) admits two classical solutions each of which is interpreted
as describing a $p$--extended  and a $\widetilde{p}$--extended object
respectively. The first solution is driven by an electric $F^{[p+2]}$
form, while the second is driven by a magnetic $F^{[D-p+2]}$ form. The role of
electric and magnetic solutions of the action
$A_{p-brane}^{[D]}$ are interchanged as solutions of the dual action
$A_{\widetilde{p}-brane}^{[D]}$
For various values of
\begin{equation}
n=p+2  \quad \mbox{and } \quad  a
\end{equation}
the functional  $A_{p-brane}^{[D]}$ (or its dual) corresponds to
a consistent truncation of some supergravity bosonic action $S_D^{SUGRA}$
in dimension $D$. This is the reason why the classical configurations I am going to
describe are  generically named supergravity $p$--branes. Given
that supergravity is the low energy limit of superstring theory
supergravity $p$--branes are also solutions of superstring theory.
They can be approximate or exact solutions depending whether they
do or do not receive quantum corrections.  The second case is clearly
the most interesting one and occurs, in particular, when the
supergravity $p$--brane is a $BPS$--state that preserves some amount of
supersymmetry. This implies that it is part of a short supersymmetry multiplet
and for this reason cannot be renormalized.
By consistent truncation we mean that a
subset of the bosonic fields have been put equal to zero but
in such a way that all solutions of the truncated action are also
solutions of the complete one.
For instance if we choose:
\begin{equation}
a=1 \quad \quad n=\cases{3 \cr 7\cr}
\label{het}
\end{equation}
eq.(\ref{pbranact}) corresponds to the bosonic low energy action of $D=10$
heterotic superstring ($\mathcal{N}=1$, supergravity) where the $E_8\times
E_8$ gauge fields have been deleted. The two choices $3$ or $7$ in
eq.(\ref{het}) correspond to the two formulations (electric/magnetic)
of the theory. Other choices correspond to truncations of the type IIA or
type IIB action in the various intermediate dimensions $4\le D\le
10$. Since the $n-1$--form $A^{[n-1]}$ couples
to the world volume of an extended object of dimension:
\begin{equation}
p = n-2
\label{interpre}
\end{equation}
namely a $p$--brane, the choice of the truncated action
(\ref{pbranact})
is precisely motivated by the search for $p$--brane solutions of supergravity.
According with the interpretation (\ref{interpre}) we set:
\begin{equation}
  n=p+2  \qquad  d=p+1 \qquad
{\tilde d}= D-p-3
\label{wvol}
\end{equation}
where $d$ is the world--volume dimension of an electrically charged
{\it elementary} $p$--brane solution, while ${\tilde d}$ is
the world--volume dimension of a magnetically charged {\it solitonic}
${\tilde p}$--brane with ${\tilde p} = D-p-4$. The distinction between
elementary and solitonic is the following. In the elementary case
the field configuration we shall discuss is a true vacuum solution of
the field equations following from the action (\ref{pbranact}) everywhere in
$D$--dimensional space--time except for a singular locus of dimension
$d$. This locus can be interpreted as the location of an elementary
$p$--brane source that is coupled to supergravity via an electric charge
spread over its own world volume. In the solitonic case, the field
configuration I shall consider is instead a bona--fide solution of
the supergravity field equations everywhere in space--time without
the need to postulate external elementary sources. The field energy
is however concentrated around a locus of dimension ${\tilde p}$.
These  solutions  have  been  derived  and  discussed  thoroughly in the
literature \cite{stellebrane}. Good reviews of such results are
\cite{mbrastelle,mbratownsend}. Defining:
\begin{equation}
\Delta = a^2 +2\, \frac{d {\tilde d} }{ D-2}
\label{deltadef}
\end{equation}
it was shown in \cite{stellebrane} that the action (\ref{pbranact}) admits
the following elementary $p$--brane solution
\begin{eqnarray}
ds^2 & =& H(y)^{-  \frac {4\, { \tilde d}} {\Delta (D-2)}}
\, dx^\mu \otimes dx^\nu \, \eta_{\mu\nu}
- H(y)^{ \frac {4\, {d}} {\Delta (D-2)}}
\, dy^m \otimes dy^n \, \delta_{mn}\nonumber \\
F^{[p+2]} &= &\frac{2}{\sqrt{\Delta}}(-)^{p+1}\epsilon_{\mu_1\dots\mu_{p+1}} dx^{\mu_1}
\wedge \dots \wedge dx^{\mu_{p+1}}
\wedge \, d \left[ H(y)^{-1}\right] \nonumber\\
e^{\phi(r)} &=& H(y)^{-\frac {2a}{\Delta}}
\label{elem}
\end{eqnarray}
where the coordinates $X^M$ ($M=0,1\dots , D-1$) have been split into
two subsets:
\begin{itemize}
\item $x^\mu$, $(\mu=0,\dots ,p)$ are the coordinates on the
$p$--brane world--volume,
\item $y^m$, $(m=D-d+1,\dots ,D)$ are the coordinates  transverse to
the brane
\end{itemize}
and
\begin{equation}
H(y)=\left(1+\frac{k}{ r^{\tilde d} } \right)
\label{harmofun}
\end{equation}
is a harmonic function $\frac{\partial}{\partial y^m}
\frac{\partial}{\partial y^m} H(y)=0$ in the transverse space to the
brane--world volume. By  $r \equiv \sqrt{y^m y_m}$ we have denoted the radial distance from the
brane and by $k$   the value of its electric charge.
The same authors of \cite{stellebrane} show that that the
action (\ref{pbranact}) admits also
the following solitonic ${\tilde p}$--brane solution:
\begin{eqnarray}
ds^2 & =& H(y)^{- \frac {4\, {   d}}
{\Delta (D-2)}}
\, dx^\mu \otimes dx^\nu \, \eta_{\mu\nu}
- H(y)^{ \frac {4\, {\tilde d}} {\Delta (D-2)}}
\, dy^m \otimes dy^n \, \delta_{mn}\nonumber \\
{\tilde F}^{[D-p-2]} &= &\lambda  \epsilon_{\mu_1\dots\mu_{{\tilde d}}p} dx^{\mu_1}
\wedge \dots \wedge dx^{\mu_{\tilde d}} \, \frac{y^p}{r^{d+2}} \nonumber\\
e^{\phi(r)} &=& H(y)^{\frac {2a}{ \Delta}}
\label{solit}
\end{eqnarray}
where the $D-p-2$--form ${\tilde F}^{[D-p-2]}$
is the dual of $F^{[p+2]}$, $k$ is now the magnetic charge
and:
\begin{equation}
\lambda= - 2\, \frac{{ d} \, k}{\sqrt{\Delta}}
\label{constlam}
\end{equation}
The identification  (\ref{constlam}) of the constant $\lambda$ allows
to write the expression of the form ${\tilde F}^{[D-p-2]}$ in the
solitonic solution in the following more compact and inspiring way:
\begin{equation}
  {\tilde F}^{[D-p-2]}= \frac{2}{\sqrt{\Delta}} \, \star\,  d H(y)
\label{inspiro}
\end{equation}
These  $p$--brane configurations are solutions of the second order
field equations obtained by varying the action (\ref{pbranact}).
However, when (\ref{pbranact}) is the truncation of
a supergravity action it generically happens that
both (\ref{elem}) and (\ref{solit}) are also the
solutions of a {\it first order differential system of equations}
ensuring that they are BPS--extremal $p$--branes
preserving a fraction of the original supersymmetries.
The parameter (\ref{deltadef}) plays a particularly important role as
an intrinsic characterization of the brane solutions since it has the
very important property of being invariant under toroidal
compactifications. When we step down in dimensions compactifying
on a $\mathrm{T}^x$ torus  each $p$-brane
solution of the $D$-dimensional supergravity ends up in a $p^\prime$
brane of the $D-x$ supergravity that has the same value of
$\Delta$ its parent brane had in higher dimension. It also happens
that all elementary BPS branes of string or M--theory as the
various $Dp$--branes of the type II A or type II B theory,  the
M2 and M5 branes, the Neveu Schwarz $5$--brane and the elementary
type II or heterotic strings are characterized by the property that
$\Delta =4$. Namely we have:
\begin{equation}
  \Delta =4 \quad \Leftrightarrow \quad \mbox{elementary $p$--brane in $D=10$ or toroidal
  reduction thereof}
\label{delta=4}
\end{equation}
\section[The near brane geometry, the dual frame and the $\mathrm{DW/CFT}$ correspondence]{
The near brane geometry, the dual frame and the $\mathrm{DW/CFT}$ correspondence}
As I briefly recalled in the introduction the most exciting
new development  of the last three years has been, for the string
theory community the discovery of the $\mathrm{AdS/CFT}$ correspondence
\cite{maldapasto,renatoine,review},
between the superconformal quantum field theory describing the
microscopic degrees of freedom of certain $p$--branes and
classical supergravity compactified on $\mathrm{AdS}_{p+2} \times X^{D-p-2}$.
The origin of this correspondence is two-fold. On one side we have
the algebraic truth that the $\mathrm{AdS}_{p+2}$ isometry group, namely
$\mathrm{SO(2,p+1)}$ is also the conformal group in $p+1$ dimensions and, as
firstly noticed by the authors of \cite{renatoine}, this extends also
to the corresponding supersymmetric extensions appropriate to the
field theories leaving on the relevant brane volumes. On the other
hand we have the special behaviour of those $p$--branes that are
characterized by the conditions:
\begin{equation}
  \Delta = 4 \quad ; \quad a=0 \quad \Rightarrow \quad
  \frac{d \,\widetilde{d}}{D-2}=2
\label{adsbrane}
\end{equation}
In this case the $p$-brane metric takes the form:
\begin{equation}
  ds^2 =\left[  H(r)\right] ^{-\frac{\widetilde{d}}{D-2}}
  \, dx^\mu \otimes dx^\nu \, \eta_{\mu\nu} + \left[  H(r)\right]
  ^{\frac{d}{D-2}} \, \left( dr^2 +r^2 ds^2_{S^{D-p-2}} \right)
\label{polarpbran}
\end{equation}
where the flat metric $d^m \otimes dy^m $ in the $D-p-1$ dimensions
has been written in polar coordinates using the metric
$ds^2_{\mathrm{S}^{D-p-2}}$ on an $ \mathrm{S}^{D-p-2}$ sphere and
where the harmonic function is
\begin{equation}
  H(r)= \left( 1 + \frac{k}{r^{\widetilde{d}}}\right)
\label{harmospec}
\end{equation}
For large $r \to \infty$ the metric (\ref{polarpbran}) is
asymptotically flat, but for small values of the radial distance from the brane
$ r \mapsto 0$ the metric becomes a direct product metric:
\begin{equation}
  ds^2 \, \stackrel {r \to 0}{\Longrightarrow}  \, ds^2_H =
  \underbrace{(k)^{-\frac{\widetilde{d}}{D-2}} \, r^{\frac{\widetilde{d}^2}{D-2}}
  \, dx^\mu \otimes dx^\nu \, \eta_{\mu\nu} \, + \, (k)^{\frac{d}{D-2}} \,
  \frac{dr^2}{r^2} }_{\mbox{$AdS_{p+2}$ metric}}\, + \,
  (k)^{\frac{d}{D-2}} \, ds^2_{S^{D-p-2}}
\label{geoHor}
\end{equation}
We will see shortly from now why the underbraced metric is indeed that of an
anti de Sitter space. To this effect it suffices to set:
\begin{equation}
  r=\left( k\right) ^{\widetilde{d}/2(D-2)} \, \exp \,\left[ - \left(
  k\right) ^{-d/2(D-2)} \, {\bar r}\right]
\label{rbarvaria}
\end{equation}
and in the new variable ${\bar r}$ the underbraced metric of eq.(\ref{geoHor})
becomes identical to the metric (\ref{adsmetra}) with
\begin{equation}
  \lambda = \left(
  k\right) ^{-d/2(D-2)}  \, \frac{\widetilde{d}^2}{2(D-2)}
\label{lamdigoodbra}
\end{equation}
As we show in next section the metric (\ref{adsmetra}) is indeed the
$\mathrm{AdS}$ metric in horospherical coordinates. Hence the near
brane geometry of the special $p$--branes satisfying condition (\ref{adsbrane})
is $\mathrm{AdS}_{p+2} \times \mathrm{S}^{D-p-2}$ and this is the
very origin of the $\mathrm{AdS/CFT}$ correspondence. As it was shown
in \cite{Ghbrane} this mechanism can be extended to the case where
the sphere metric is replaced by the metric of  other coset manifolds
$\mathrm{G/H}$ of the same dimensions $D-p-2$ or even more generically
by the metric of some Einstein space $X^{D-p-2}$. This leads to the study of
many more non trivial examples of $\mathrm{AdS/CFT}$ correspondence, typically
characterized by a reduced non maximal supersymmetry.
\cite{m111,adscftcheckers,witkleb,noin0101,noin0102,gubser,
gubserkleb,poliv52,sergiotorino,sanssergio}.
\par
The relevant point of the $\mathrm{AdS/CFT}$ correspondence for the scope of
the present set of lectures is the following:
\begin{statement}
In all  cases of $a=0,\Delta=4$ $p-branes$ the low dimensional supergravity that one obtains
by compactifying the original $D$--dimensional supergravity on the
compact $X^{D-p-2}$ manifold is a \textbf{gauged supergravity}
$\mathcal{SG}_{p+2}^{gau}$ in $p+2$ space time dimensions that admits
$AdS_{p+2}$ as an exact solution. Furthermore the isometry group
$G_{iso}$ of $X^{D-p-2}$ is \textbf{the gauge group} of $\mathcal{SG}_{p+2}^{gau}$
and reappears in the dual conformal field theory as a global
$R$--symmetry or flavor symmetry.
\end{statement}
This shows the connection between supergravity gaugings and the
physics of superstring $p$--branes. Since $X^{D-p-2}$ is by definition
a compact manifold it follows that also its isometry group is compact
and that in the context of the $AdS/CFT$ correspondence one is lead
to consider compact gaugings. These do not exhaust the set of
supergravity gaugings. On the contrary as I explain in chapter
\ref{gaugchap} there is a wealth of non--compact and also of
non--semisimple gaugings that wait for interpretation in the context
of superstring theory. This is quite a good match since also the
branes satisfying condition (\ref{adsbrane}) are far from
exhausting the list of $p$-branes.
\par
The condition (\ref{adsbrane}) that leads to the wealth of interesting results
summarized above is the statement that the driving $p+2$--form $F^{[p+2]}$ does not
couple to the dilaton field $\phi$ which effectively drops out of the
game. This is the condition of conformal invariance and not too
surprisingly the $\mathrm{AdS/CFT}$ correspondence is a correspondence between
certain \emph{compact gauged supergravities} and certain
\emph{conformal field theories}. For all other $p$--branes $a \neq 0$
and we have a non trivial dilaton coupling. This forbids conformal
invariance and excludes a priori an $\mathrm{AdS/CFT}$
correspondence. Yet in a seminal and challenging paper Boonstra,
Skenderis and Townsend \cite{Boonstra:1999mp} have proposed the
following generalization.
\begin{statement}
In all $a \neq 0,\Delta=4$ $p$--branes, the low dimensional supergravity that one obtains
by compactifying the original $D$--dimensional supergravity on the
an $\mathrm{S}^{D-p-2}$ sphere or other compact manifold $X^{D-p-2}$
forming the base of the cone transverse to the brane is a \textbf{gauged supergravity}
$\mathcal{SG}_{p+2}^{gau}$ in $p+2$ space time dimensions that admits
an appropriate \textbf{domain wall} $\mathrm{DW}^{\overline{\Delta}}_{p+2}$ as an exact solution.
(${\overline{\Delta}}$ is a parameter labeling the type of domain wall). The isometry group
$G_{iso}$ of $X^{D-p-2}$ is \emph{part} of  \textbf{the gauge group} of
$\mathcal{SG}_{p+2}^{gau}$. The $p+1$--dimensional boundary
$\partial \mathrm{DW}^{\overline{\Delta}}$ of  the domain wall
space--time supports a quantum (non--conformal) field theory that is
dual to the $D$--dimensional supergravity compactified on $DW^{\overline{\Delta}}_{p+2}
\bigodot X^{D-p-2}$.
\end{statement}
The above statement is a challenging conjecture that has  so
far received many less checks than its conformal sister, yet there
are a lot of convincing hints that it should be right. Indeed it is just
a particularly circumstantial way of formulating the general
principle of the \textbf{gauge/gravity correspondence} which is
nowadays supported by the non trivial checks provided by fractional
D--branes \cite{Bertolini:2000dk,Herzog:2000rz,Billo:2000yb}. Domain
wall space--times $DW^\Delta_{p+2}$ will be described in the next section. They are
essentially the limiting case of a $p$--brane when $p=D-2$. From
another view point, as I explain in some detail in the next section, they are the
natural generalization of an anti de Sitter space--time $AdS_{p+2}$
since they are locally isometric to $AdS_{p+2}$. Globally $DW^{\overline{\Delta}}_{p+2}$ are
generically different from $AdS_{p+2}$ since they describe two
regions of an $AdS_{p+2}$ space separated by a thin $p+1$ dimensional
wall that is the location of a curvature singularity. Furthermore the
essential point is that in $DW^{\overline{\Delta}}_{p+2}$ solutions of
the gauged supergravity action there is a dynamical non constant
dilaton. Finally the notation $DW^{\overline{\Delta}}_{p+2}
\bigodot X^{D-p-2}$ recalls the fact that the direct product of the
$DW^{\overline{\Delta}}_{p+2}$ space--time with the compact space
$X^{D-p-2}$ is not a solution of higher dimensional supergravity but
a metric involving these two factors modulated by suitable warp
factors is.
\par
What is the main basis for this bold conjecture? It comes from an
observation made by the authors of \cite{Boonstra:1999mp} that
although in the Einstein frame  the metric of $p$--brane with $a\neq
0$ does not factorize in the limit $r \to 0$ as the conformal branes
do, yet one can always define a different frame, \textbf{the dual
frame} where this desired factorization occurs apart from an overall
warp factor.
\par
To this effect it is convenient to recall the general formula for the
Weyl transformation of the Einstein term in $D$-dimensions. Consider
the lagrangian density
\begin{equation}
\int \, 2 \,R[g] \, \sqrt{-g} \, d^Dx
\label{Einstein}
\end{equation}
where $R[g]$ denotes the curvature scalar and $g$ the determinant of
the metric and set the transformation:
\begin{equation}
  g_{\mu \nu } = \exp [2\alpha \,\phi] \, \overline{g}_{\mu \nu }
\label{Weyl}
\end{equation}
where $\alpha$ is a constant, $\phi$ a scalar field and $\overline{g}_{\mu \nu
}$ a new metric. In my conventions, after a partial integration one
finds:
\begin{equation}
  \int \, 2 \, R[g] \, \sqrt{-g} \, d^Dx = \int \,
  \exp \left[(D-2)\alpha \,\phi \right]
  \,  \left(2 R[\overline{g}] -\alpha ^2 \left(
  D-1\right) \left( D-2\right)  \partial ^\mu \phi  \partial _\mu \phi \right)
  \sqrt{-\overline{g}}\,d^Dx
\label{dopoweyl}
\end{equation}
Let us now apply this general formula to the $p$--brane action (\ref{pbranact})
and to its solutions (\ref{elem}) or (\ref{solit}). We introduce two
new metrics defined by a Weyl transformation and respectively named the string metric and
the dual metric:
\begin{eqnarray}
g^{(E)}_{\mu \nu } & = & g^{(string)}_{\mu \nu }  \, \exp[ \lambda _s \, \phi] \label{strfram}\\
g^{(E)}_{\mu \nu } & = & g^{(dual)}_{\mu \nu }  \, \exp[ \lambda _d \,
\phi]\label{dualfram}
\end{eqnarray}
where the parameters $\lambda _s$ and $\lambda _d$ are determined by
the following conditions. Considering the transformation of the
Einstein and $p+2$--form terms
\begin{eqnarray}
\int \, R[g] \, \sqrt{-g} \, & \mapsto & \exp \left[ (\ft D 2 -1 ) \lambda \phi\right]
 R[\overline{g}] \, \sqrt{-\overline{g}} + \mbox{$\partial \phi $ terms}
\nonumber\\
\exp[-a \phi ] \vert F^{[p+2]}  \vert^2 \sqrt{-g} & \mapsto & \exp \left[-a + \, \lambda \,\left( \ft D 2
 -D+p+2\right) \phi \right]
\vert F^{[p+2]}  \vert^2 \sqrt{-\overline{g}}
\label{trasfoentramb}
\end{eqnarray}
we determine the \emph{string frame} by requiring that after the
transformation the exponential of the dilaton field should not stand
in front of the $\vert F^{[p+2]}  \vert^2$ term. The \emph{dual
frame} is instead fixed by the request that after the transformation
the Einstein $ R[g]$ term and the $\vert F^{[p+2]}  \vert^2$ should
have the same power of the dilaton $e^\phi$  in front. With such a
definition we immediately get:
\begin{equation}
  \lambda _s = -\frac{2 \, a}{D-2p-4} \quad ; \quad \lambda _d = -\frac{a}{D-p-3}
\label{lammicoef}
\end{equation}
Choosing for definiteness the superstring critical dimension $D=10$
and the case of magnetic  $p$-branes, namely $\Delta =4$ we can
immediately write down the corresponding metric in the dual frame:
\begin{eqnarray}
  ds^2_{dual}&=& \left[ H(r)\right] ^{-\frac{5-p}{7-p}}
  \, dx^\mu \otimes dx^\nu \, \eta_{\mu\nu} + \left[ H(r)\right] ^{\frac{2}{7-p}}
 \left( dr^2 +r^2 ds^2_{S^{8-p}} \right)\nonumber\\
  H(r) & = & 1+ \frac{k}{r^{7-p}}
\label{pbrandualfram}
\end{eqnarray}
Now it happens that for $r\rightarrow 0$, independently from the
value of $p$ the dual metric has a \textbf{near brane factorized}
geometry. Indeed in this limit we find:
\begin{equation}
   ds^2_{dual} \simeq \underbrace{\left( k\right) ^{-\frac{5-p}{7-p}} r^{5-p}
   \, \, dx^\mu \otimes dx^\nu \, \eta_{\mu\nu} +
   \left( k\right) ^{\frac{2}{7-p}} \,   \frac{dr^2}{r^2}}_{\mbox{$\mathrm{AdS}$-metric}} + \,
   \left( k\right) ^{\frac{2}{7-p}} \, ds^2_{S^{8-p}}
\label{fattorpure}
\end{equation}
The underbraced metric is a locally an anti de Sitter metric by the same token
as the underbraced metric of eq.(\ref{geoHor}). So for all $p$ the
dual frame metric factorizes in the \emph{near brane regime} into the
product of an anti de Sitter metric times the metric of an $S^{8-p}$
sphere or other $8-p$--dimensional compact space. There is just one
noteworthy exception: that of the Neveu Schwarz five--brane. In this
case due to the exact cancelling of the $r$ powers the factorization
is even simpler. We get a flat $\mathbb{R}^{(1,6)}$ Minkowski space
times a three sphere $S^3$. This is related to the exact conformal
description of the Neveu Schwarz five--brane in terms of the conformal
field theory of an $SU(2)$ Wess Zumino model times a Feigin Fuchs
scalar plus the free conformal field theory of $6$ flat coordinates
\cite{Callan:1991dj,Billo:1993ei}.
\par
This near brane factorization suggests that we can perform a compactification of
the dual frame lagrangian on the internal compact manifold $S^{(8-p)}$
or $X^{(8-p)}$ obtaining an action in $p+2$ dimensions that we can subsequently
reduce to the Einstein frame. In the first step, namely in the
compactification,  what we do is the ordinary Kaluza-Klein reduction
of a scalar--Einstein theory where the background value of the
$F^{[8-p]}$ form is simply identified with the volume form of the
internal manifold. In the second step we simply apply the Weyl rule
transformation (\ref{dopoweyl}) in the reversed direction.
\par
The result of these operations is an action of the same form as
the action (\ref{Dwaction}) that we consider in the next section with
the following specific values of the constants \cite{Boonstra:1999mp}:
\begin{equation}
  a=-\frac{2 (p-3)}{\sqrt{2p(p-9)}} \quad ; \quad \Lambda = \ft 1 4
  (9-p)(7-p) \left( k\right) ^{\frac {4(p-3)}{p(p-7)4(p-3)}}
\label{skenderisul}
\end{equation}
As we explain in the next section the action (\ref{Dwaction}) is that
appropriate to discuss \textbf{domain wall} solutions, namely
$D-2$--branes. Hence it follows that when
we go to the near brane region, the  geometry of a non conformal $p$--brane
is well approximated   by a domain wall solution of some
suitable supergravity theory of which the action (\ref{Dwaction})
must be a consistent truncation. The fundamental question for which
we do not have a general ansatz yet is the following: of which
gauged supergravity the action (\ref{Dwaction}) with parameters
(\ref{skenderisul} is a truncation? One thing is certain: the gauge
group must contain the isometry group of $X^{[8-p]}$.
The authors of \cite{Boonstra:1999mp} have made a conjecture that I
entirely subscribe: it should be some non-compact, possibly non
semisimple gauging. In one case they could even make a prediction.
Take the $D2$--brane of type IIA theory. In that case $p+2=4$ so that
the candidate supergravity is a four--dimensional one and, since we
do not break any supersymmetry, it is also $\mathcal{N}=8$. Hence our
sought for theory must be one of the $\mathcal{N}=8$ gaugings that I
describe in section \ref{n8d4gaug}. The list is finite and presented
in table \ref{risulato}. Since the gauge group must contain the
compact subgroup $\mathrm{SO(7)}$ (the isometry group of
$\mathrm{S}^6$) it follows that there is a unique possibility, namely the
theory $\mathrm{CSO}(7,1)=\mathrm{ISO}(7)$ obtained by gauging the
Euclidean group in $7$ dimensions. Whether this conjecture is true or
not, so far  has not been verified but stands as a challenging
proposal.
\par
In view of the above discussion I turn, in the next section to a
survey of the notion of domain walls. The study of supergravity
gaugings presented in the next two chapters is mostly motivated by
the quest for domain wall solutions, their relation with higher
dimensional superstring $p$--branes and the testing of the
\textbf{gauge/gravity correspondence} in non conformal regimes.
\section[Domain walls in diverse space--time dimensions]{Domain walls in
diverse space--time dimensions}
\label{bobus}
The  generic coupling of a single scalar field to Einstein gravity is
described, in space--time dimensions $D$ by the following action
\begin{equation}
  A_{grav+scal}^{[D]}=\int \, d^Dx \, \sqrt{-g} \,\left[ 2 \, R[g]
  +\frac{1}{2}\partial^\mu \, \phi \partial_\mu \phi -
  \mathcal{V}(\phi)\right]
\label{scalfieldact}
\end{equation}
where $\mathcal{V}(\phi)$ is the scalar potential. If for this latter
we choose the very particular form:
\begin{equation}
  \mathcal{V}(\phi)=2 \, \Lambda \, e^{-a \, \phi} \quad ; \quad
  \cases {0 < \Lambda \in \mathbb{R}  \cr
a \in \mathbb{R} \cr}
\label{expopoto}
\end{equation}
then we have a limiting case of the general $p$--brane action (\ref{pbranact}) we have
considered above. Indeed if in the general formulae
(\ref{wvol}) we put
\begin{equation}
  p=D-2 \quad \Rightarrow \quad \widetilde{d} = -1 \quad ; \quad
  d=D-1
\label{pisDm2}
\end{equation}
we obtain that the \emph{electric} $D-2$--brane couples to a field strength which is a
top $D$--form $F^{[D]}$,
while the \emph{magnetic} solitonic brane couples to a $0$--form $F^{[0]}$,
namely to a cosmological constant. Indeed, we can formally set:
\begin{equation}
  F^{[0]} = 2 \, \sqrt{\Lambda} \quad \Rightarrow \quad
  \widetilde{F}^{[D]}=\mbox{Volume form on space--time}
\label{F0agnisco}
\end{equation}
and the action (\ref{scalfieldact}) with the potential (\ref{expopoto})
is reduced to the general form for an electric $D-2$--brane
(\ref{pbranact}). That $F^{[0]}$ should be constant and hence could be
identified as in eq.(\ref{F0agnisco}) follows from the
Bianchi identity that it is supposed to satisfy $dF^{[0]}=0$.
\par
Hence we can conclude that the action:
\begin{equation}
  A_{D-Wall}^{[D]}=\int \, d^Dx \, \sqrt{-g} \,\left[ 2 \, R[g]
  +\frac{1}{2}\partial^\mu \, \phi \partial_\mu \phi -
  2 \, \Lambda \, e^{-a \, \phi}\right]
\label{Dwaction}
\end{equation}
admits a distinguished class of solutions describing $D-2$--branes
that we name \textbf{domain walls} since at each instant of time
a brane of this type
separates the space manifold  into two adjacent non overlapping
regions.
\par
Specializing the general formulae (\ref{elem})
and (\ref{harmofun}) to our particular case we obtain the
domain wall solution of (\ref{Dwaction}) in the following form:
\begin{eqnarray}
ds^2_{DW} & = & H(y)^{2\alpha} \left( dx^\mu \otimes dx^\nu \eta_{\mu \nu } \right)
+ H(y)^{2\beta} \, dy^2 \label{dwmet1}\\
e^\phi & = & H(y)^{-\ft {2a}{\Delta}}\label{dwdila1}\\
H(y) &=&c  \, \pm \, Q \, y \label{harmdw}
\end{eqnarray}
where $y$ is the single coordinate transverse to the wall, $c$ is an arbitrary integration
constant and the other parameters appearing in the above formulae have the following values:
\begin{equation}
  \alpha = \frac{2}{\Delta (D-2)} \quad ; \quad \beta =2\,
  \frac{D-1}{\Delta(D-2)}\quad ; \quad Q= \sqrt{\Lambda \, \Delta}
\label{DWconstant}
\end{equation}
in terms of $\Delta$ whose expression (\ref{deltadef}) becomes:
\begin{equation}
  \Delta= a^2 -2 \frac{D-1}{D-2}
\label{Dwdelta}
\end{equation}
The form (\ref{harmdw}) of the function $H$ is easy to
understand because in one--dimension a harmonic function is just a
linear function. The arbitrariness of the sign in $H$ arises because
the equations of motion involve $m$ only quadratically
\cite{popino1}.
Since $a^2$ is a positive quantity, $\Delta$ is bounded from below by
the special value $\Delta_AdS$ that corresponds to the very simple
case of pure gravity with a negative cosmological constant (case
$a=0$ in eq.(\ref{Dwaction}):
\begin{equation}
  \Delta \geq \Delta_{AdS} \equiv -2 \frac{D-1}{D-2}
\label{Dads}
\end{equation}
The name given to $\Delta_{AdS}$ has an obvious explanation. As it was
originally shown by L\"u, Pope and Townsend in \cite{popino1}, for
$a=0$ the domain wall solution (\ref{dwmet1}) describes a region of
the anti de Sitter space $AdS_D$. To verify this statement it
suffices to insert the value (\ref{Dads}) into (\ref{DWconstant})
and (\ref{dwmet1}) to obtain:
\begin{equation}
  ds^2_{DW} = H^{-2/(D-1)}(y)\left( dx^\mu \otimes dx^\nu \eta_{\mu \nu } \right)
+ H(y)^{-2} \, dy^2
\label{steppo1}
\end{equation}
Performing the coordinate transformation:
\begin{equation}
  r = \frac{1}{Q} \, \ln \, (c\pm Q\, y)  \quad ; \quad
\label{trasfo}
\end{equation}
the metric becomes:
\begin{equation}ds^2_{DW} = e^{-2\lambda r}\, \eta_{\mu\nu}\, dx^\mu \, dx^\nu + dr^2\ ,
\label{adsmetra}
\end{equation}
where
\begin{equation}\lambda= \sqrt{\ft{2\Lambda}{(D-1)(D-2)}}= (D-1) Q\ .\label{lampara}
\end{equation}
In the same coordinates the solution for the dilaton field is:
\begin{equation}
  e^{\phi} = \exp \left[ -\frac{2\, a \,\lambda}{\Delta \, (D-1)} \,r\right]
\label{dilatoinr}
\end{equation}
Eq.(\ref{adsmetra}) is the metric of
$\mathrm{AdS}$ spacetime, in horospherical coordinates.  Following \cite{popino1}
we can verify this statement
by introducing the $(D+1)$ coordinates $(X,Y,Z^\mu)$
defined by
\begin{eqnarray}
X&=& \frac1{\lambda}\, \cosh\lambda r +\ft12\lambda\, \eta_{\mu\nu}\,
x^\mu x^\nu  \,
e^{-\lambda r} \ ,\nonumber\\
Y&=& -\frac1{\lambda}\, \sinh\lambda r -\ft12\lambda\, \eta_{\mu\nu}\,
x^\mu x^\nu \,
e^{-\lambda r} \ ,\label{embed}\\
Z^\mu&=&x^\mu\, e^{-\lambda r}\ .\nonumber
\end{eqnarray}
They satisfy
\begin{eqnarray}
\eta_{\mu\nu}\, Z^\mu Z^\nu +Y^2 -X^2 &=&-1/\lambda^2\ ,\label{emcon}\\
\eta_{\mu\nu}\, dZ^\mu dZ^\nu +dY^2 -dX^2&=&
e^{-2\lambda r} \eta_{\mu\nu}\, dx^\mu\, dx^\nu + dr^2\ ,
\end{eqnarray}
which shows that (\ref{adsmetra}) is the induced metric on the algebraic
locus (\ref{emcon}) which is the standard hyperboloid
corresponding to the $AdS$ space--time manifold. The signature of
embedding flat space is $(-,+,+,\cdots, +,-)$ and therefore
the  metric (\ref{adsmetra}) has the right $SO(2,D-1)$
isometry of the  $\mathrm{AdS}_D$ metric.
\par
Still following the discussion in \cite{popino1} we note that in horospherical coordinates
$X+Y=\lambda^{-1}\, e^{-\lambda r}$ is non-negative if $r$ is real. Hence
the region $X+Y<0$ of the full $\mathrm{AdS}$ spacetime is not accessible
in horospherical coordinates. Indeed this  coordinate patch
covers one half of the complete $\mathrm{AdS}$ space , and the metric describes
$\mathrm{AdS}_D /\mathbb{Z}_2$ where $\mathbb{Z}_2$ is the antipodal
involution $(X,Y, Z^\mu)\rightarrow
(-X, -Y, -Z^\mu)$. If $D$ is even, we can extend the metric
(\ref{steppo1}) to cover the whole anti de Sitter spacetime by setting the integration constant
$c=0$ which implies $H= Q\,y$.  So doing
the region with $y<0$ corresponds to the previously inaccessible region
$X+Y<0$.  If odd dimensions, we must restrict $H$ in
(\ref{steppo1}) to be non-negative in order to have a real metric and thus
in this case we have to choose $H=c + Q |y|$, with $c \ge 0$.  If the constant
$c$ is zero, the metric describes $\mathrm{AdS}_D/\mathbb{Z}_2$, while if $c$ is positive, the
metric describes a smaller portion of the complete $\mathrm{AdS}$ spacetime.  In any
dimension, if we set:
\begin{equation}
  H=c +Q |y|
\label{sofchoic}
\end{equation}
the solution can be interpreted as a domain wall
at $y=0$ that separates two regions of the anti de Sitter spacetime, with a delta function
curvature singularity at $y=0$ if the constant $c$ is positive.
\subsection{The Randall Sundrum  mechanism}
What we have just described is the \textbf{anti de Sitter domain
wall} that corresponds to $\Delta = \Delta_{AdS}$. The magic of this
solution is that, as shown by Randall and Sundrum in \cite{RS2},
it leads to the challenging phenomenon of \textbf{gravity trapping}.
These authors have found that because of the exponentially rapid
decrease of the factor
\begin{equation}
  \exp [-\lambda |r|] \quad \mbox{with} \quad \lambda >0
\label{decrease}
\end{equation}
away from the thin domain wall that separates the two asymptotic
anti de Sitter regions it happens that gravity in a certain sense is
localized near the brane wall. Instead of the $D$--dimensional
Newton's law that gives:
\begin{equation}
   \mbox{force} \sim \frac{1}{R^{D-2}}
\label{Dnewton}
\end{equation}
one finds the the $D-1$--dimensional Newton's law
\begin{equation}
   \mbox{force} \sim \frac{1}{R^{D-3}} + \mbox{small corrections $ \mathcal{O}\left(
   \frac{1}{R^{D-2}}\right) $}
\label{D-1newton}
\end{equation}
This can be seen by linearizing the Einstein equations for the metric fluctuations
around any domain wall background of the form:
\begin{equation}
  ds^2 = W(r)\, \eta_{\mu\nu}\, dx^\mu \, dx^\nu + dr^2\ ,
\label{warpfac}
\end{equation}
that includes in particular the $AdS$ case (\ref{adsmetra}). In a very sketchy
way if one sets:
\begin{equation}
  h_{\mu \nu }(x,y) = \exp \left[ \mbox{i} p\cdot x\right]  \,
  \psi_{\mu\nu}(y)
\label{factorization}
\end{equation}
one finds that the linearized Einstein equations translate into an
analog  Schroedinger equation  for the wave--function
$\psi(y)$. This problem has a potential that is determined by the warp
factor $W(y)$. If in the spectrum of this quantum mechanical problem
there is a normalizable zero mode then this is the wave function of
a $D-1$ dimensional graviton. This state is indeed a bound state and
falls off rapidly when leaving the brane. Since the extra dimension
is non compact the Kaluza Klein states form a continuous spectrum
without a gap. Yet $D-1$ dimensional physics is extremely well
approximated because the bound state mode reproduces conventional
gravity in $D-1$ dimensions while the massive states simply
contribute a small correction.
\par
It is clearly of utmost interest to establish which domain walls have
this magic trapping property besides the anti de Sitter one.
This has been recently done by Cvetic, L\"u and Pope in \cite{cveticdelta}
In order to summarize this and other related results I need first to
emphasize another aspect of domain walls that puts them into
distinguished special class among $p$--branes.
\subsection[The conformal gauge for domain
walls]{The conformal gauge for domain walls}
Going back to the general domain wall solution
(\ref{dwmet1}),(\ref{dwdila1}),(\ref{harmdw}),(\ref{DWconstant}),
classified by the value of $\Delta$ (eq.(\ref{Dwdelta})) we observe
that there is still an ambiguity in the powers of the harmonic
function (\ref{harmdw}) that appear as metric coefficients. This
ambiguity is due to coordinate transformations and it is a specific
property of $D-2$--branes not present in other $p$--branes, where the
harmonic function $H$ is not a linear function. Following a
discussion by Bergshoeff and van der Schaar \cite{Bergshoeff:1999bs}
we observe that in the range $y >0$ we can make the following linear
transformation:
$ y = -\frac{c}{Q}+y^\prime \quad \Rightarrow \quad H(y) = Q \,
  y^\prime $
that eliminates the integration constant $c$. Furthermore we can
redefine $y^\prime$ as some other fractional power of a third
coordinate ${\bar y}$, namely
$ y^\prime = -Q^{-\frac{1+\epsilon }{\epsilon }} \, {\bar y}^{-\frac{1}{\epsilon }}
$, then shifting it once again by a constant ${\bar
y}=z+\frac{c}{Q}$. Altogether this means that we introduce the
coordinate transformation:
\begin{equation}
  y=- \frac{c}{Q}- Q^{-\frac{1+\epsilon }{\epsilon }} \,\left( z+\frac{c}{Q}\right)
  ^{-\frac{1}{\epsilon }}
\label{coordo1}
\end{equation}
Under this transformation we have (for positive $y$):
\begin{equation}
  H(y) = - \left[ H(z)\right] ^{-1/\epsilon }
\label{trasfoharm}
\end{equation}
and the domain wall metric (\ref{dwmet1}) becomes:
\begin{equation}
  ds^2_{DW}  = H(z)^{-\frac{2\alpha}{\epsilon }} \left( dx^\mu \otimes
  dx^\nu \eta_{\mu \nu } \right)
+ H(z)^{-\frac{2\beta+\epsilon }{\epsilon }-2} \, \frac{dz^2}{\epsilon ^2} \label{dwmet2}
\end{equation}
This transformation allows for the remarkable possibility of choosing
a conformal gauge, namely a coordinate system where it becomes
manifest that the domain wall metric is conformally flat. Indeed it
suffices to impose that the two powers of the harmonic function
appearing in (\ref{dwmet2}) be equal:
\begin{equation}
  -\frac{2\alpha}{\epsilon }=-\frac{2\beta+\epsilon }{\epsilon }-2
\label{confocondo}
\end{equation}
Using eq.(\ref{DWconstant}) the solution of (\ref{confocondo}) for
$\epsilon$ is unique in all cases with the exception of $\Delta=-2$:
\begin{equation}
  \epsilon =-\frac{\Delta+2}{\Delta}
\label{epsDelta}
\end{equation}
Hence for $\Delta \neq -2$, redefining $z \mapsto \epsilon z$, $Q
\mapsto k\,|\epsilon|$ the domain wall solution (\ref{dwmet1}) can
always be rewritten in the following conformally flat way:
\begin{eqnarray}
  ds^2_{DW/conf} &=& \left[  H(z) \right]  ^{\frac {4}{(D-2)(\Delta+2)}}\,
  \left( \eta_{\mu \nu } \, dx^\mu
  \otimes dx^\nu  \, + \, dz^2 \right) \nonumber\\
  e^{\phi(z)}&=&H(z)^{-\frac{2 a}{\Delta+2}} \nonumber\\
  H(z) &=& 1+ k \,\vert z \vert \nonumber\\
  k & = & (\Delta+2) \, \sqrt{ \frac{\Lambda}{\Delta+2} }
\label{confDwall}
\end{eqnarray}
Obviously the solution (\ref{confDwall}) could have been obtained by directly
solving the Einstein equations associated with the action (\ref{Dwaction})
starting from a conformal ansatz of the type:
\begin{equation}
  ds^2_{DW/conf} = \exp [A(z)] \left(  \eta_{\mu\nu}\, dx^\mu \, dx^\nu + dz^2 \right)  ,
\label{confans}
\end{equation}
Yet I preferred to obtain it from the general solution (\ref{elem}) for
supergravity $p$--branes in order to emphasize its interpretation as
a domain wall, namely a $D-2$--brane.
The direct method of solution can be used to find the conformal representation
of the domain wall metric in the exceptional case $\Delta=-2$. As
shown in \cite{cveticdelta} one obtains:
\begin{eqnarray}
ds^2 &=& e^{-\frac {2k}{d-2} \, |z|}\, (\eta_{\mu\nu}\, dx^\mu\, dx^\nu +
dz^2)\,,\nonumber\\
\phi &=& \frac {\sqrt2\, k}{ \sqrt{d-2}}\, |z|\,,\label{d2sol}
\end{eqnarray}
where $k$ is now given by
\begin{equation}
k^2 = -2\Lambda\, (d-2)\,,
\end{equation}
which is real for negative $\Lambda$.
There is another important
point that we should note. Our starting point, prior to all the
subsequent manipulations, has been the form
(\ref{dwmet1}),(\ref{DWconstant}) which is that of an electric
$p$--brane and not that of a solitonic one (see eq.\ref{solit})).
This implies that our domain wall solutions are not exactly bona fide
solutions of the action (\ref{Dwaction}) but require also the
coupling to a source term that is  the world-volume action of
the domain wall, localized at $z=0$ in the last  coordinate frame we
have used. Namely the true action is
\begin{equation}
 A= \int_{M_D} \, d^Dx \, \sqrt{-g} \,\left[ 2 \, R[g]
  +\frac{1}{2}\partial^\mu \, \phi \partial_\mu \phi -
  2 \, \Lambda \, e^{-a \, \phi}\right] + \, \mathcal{T}\, \int_{WV_{D-1}} d^{D-1}\xi
  \, \mathcal{L}_{source}
\label{actsource}
\end{equation}
where $\mathcal{L}_{source}$ is world--volume lagrangian of the
$D-2$--brane and the parameter $\mathcal{T}$ denotes its tension.
An important issue is to relate the wall-tension to the parameters
appearing in the classical domain wall solution. This was done
in \cite{cveticdelta} following a standard analysis developed in
previous papers \cite{cs1,cs2}.
%%%%%%%%%%%%%%%%%%%%%%%%
The matching conditions across the singular domain wall source imply
that that the energy density (tension) of the wall is related to the
values of the cosmological constant parameters on either side of the
wall, namely the authors of \cite{cveticdelta} find:
%%%%%
\begin{equation}
\sigma = {\cal T}=   2 (A'_{z=0^-}-A'_{z=0^+})\,, \end{equation}
%%%%%
where the prime denotes a derivative with respect to $z$.  This leads
to
%%%%%
\begin{eqnarray} \Delta\ne -2:&&  {\cal T}= -8\, {\rm sign}[k\,
(\Delta+2)]\sqrt{\Lambda\over \Delta}\,,\nonumber\\
\Delta=-2: && {\cal T} = \frac {8k}{d-2}\,.
\label{tension}\end{eqnarray}
%%%%%
Thus positive-tension domain-wall solutions exist for $\Delta \le -2$
with $k>0$ and for $\Delta>-2$ with $k<0$.  Conversely,
negative-tension domain walls arise for $\Delta \le -2$ with $k<0$
and for $\Delta>-2$ with $k>0$.  So for our domain walls with
$\Delta\le -2$, we assume the lower bound (\ref{Dads}).
To avoid  { {\textbf{naked singularites}}} we also need {
{$k>0$}}.
\par
Using the simple conformal gauge (\ref{confDwall}) the authors of \cite{cveticdelta}
have analyzed the fluctuations of the metric around such a background
and have found that the graviton wave function obeys, as predicted by
Randall-Sundrum \cite{RS2,RS1} a Schr\"odinger equation with a
potential that is completely fixed by the value of $\Delta$. More
precisely one finds that in the conformal gauge the fluctuations of
the $D$--dimensional graviton satisfy the Klein Gordon equation of a
scalar field in the gravitational background
namely $\partial_M\,(\sqrt{-g}\, g^{MN}\, \partial_N\, \Phi) =0$.
Parametrizing:
\begin{equation}
  \Phi = \phi(z)\, e^{{\rm i}\, p\cdot x} = e^{-k\,z}\,
\psi(z)\, e^{{\rm i}\ p\cdot x}\,,
\label{pametro}
\end{equation}
where $p$ is the $D-1$--dimensional momentum
the Klein Gordon equation becomes the following  Schr\"odinger-type equation,
%%%%%
\begin{equation}
-\ft 1 2 \psi'' + U\, \psi = -\ft12 p^2\, \psi\,,\label{schrod}
\end{equation}
%%%%%
where the  potential, calculated in \cite{cveticdelta} is given by
%%%%%
\begin{eqnarray}
 \Delta\ne -2:&& U = -\frac {(\Delta+1)\, k^2}{2(\Delta+2)^2\,
H(z)^2}\, + \frac  {k}{\Delta+2}\, \delta(z)\,,\nonumber\\
\Delta=-2:&& U = \ft 18 k^2\, - \ft 12\, k\, \delta(z)\,.\label{upot}
\end{eqnarray}
Such an
equation has a normalizable zero--mode wave function if the following
condition is satisfied $\Delta \leq -2$. Indeed it is
evident from these expressions that  for  $\Delta \le -2$, $U$ has a volcano
shape as in fig.\ref{volcano} since the delta
function has a negative coefficient, and the ``bulk'' term is
non-negative for all $z$.
\iffigs
\begin{figure}
\begin{center}
\caption{\label{volcano} The volcano potential}
\epsfxsize = 10cm
\epsffile{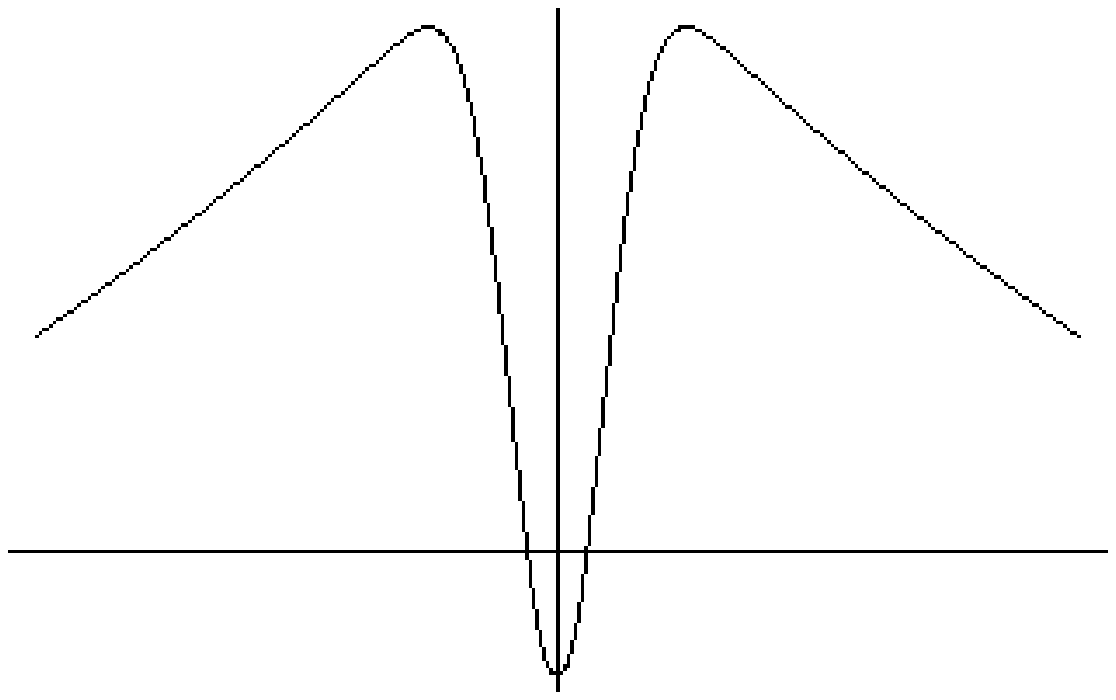}
\vskip -0.1cm
\unitlength=1mm
\end{center}
\end{figure}
\fi
 Hence the trapping of
gravity occurs for positive tension $D-2$--branes in the following
window:
\begin{equation}
  \Delta_{AdS} \leq \Delta \leq -2
\label{delwindo}
\end{equation}
\section{Conclusion of this first bestiary}
The brief survey of the $p$--brane bestiary I have presented in the
present chapter was meant to illustrate and single out one main
issue. It now appears that the near brane geometry of all the
superstring $p$--brane is a domain wall, anti de Sitter space being
just a particular case that corresponds to conformal invariance.
There is a challenging proposal of a $\mathrm{DW/QFT}$ correspondence
that should relate non--conformal gauge theories on the wall world
volume to supergravity compactified on the domain wall space--time.
The unresolved question is how to identify the appropriate gauged
supergravities that corresponds to each choice of brane
configuration. For this reason I devote the next two chapters to
describe the basic geometric structures of supergravity (the
supergravity bestiary) and how these latter are used to construct the
gaugings.
\par
Let me stress that a complete and unambiguous pairing between
supergravity gaugings and  the $Dp$--brane spectrum cannot fail to
contribute a new profound insight in superstring theory. The quest
for supersymmetric realizations of the Randall-Sundrum scenario that
I will shortly touch in the last chapter has to be viewed as part of
this more general problem
%%%%%%%%%%%%%%%%%%%%%%%%%%%%%%%%%%%%%%%%%%%%%%%%%%%%%%%%%
% Nuovo capitolo %%%%%%%%%%%%%%%%%%%%%%%%%%%%%%%%%%%%%%%
%%%%%%%%%%%%%%%%%%%%%%%%%%%%%%%%%%%%%%%%%%%%%%%%%%%%%%%%%
\chapter[Supergravity bestiary and the diverse
dimensions of superstring theory]{Supergravity bestiary and the diverse
dimensions of superstring theory}
\label{bestia2}
\section[Introduction]{Introduction}
\label{intro2}
\setcounter{equation}{0}
In the previous chapter we discussed the $p$--brane motivations to
consider supergravity theories in diverse dimensions. From this
viewpoint the basic information one would like to master is the
following:
\begin{itemize}
  \item The scalar field dependence of the kinetic terms of
  $p$--forms $ \mathcal{\mathcal{N}}_{\Lambda\Sigma} (\phi) \,
  F^\Lambda \, \wedge \star \, F^\Sigma$ since this latter eventually
  decides the values of the coefficients $a$ in the exponential
  factors of the $p$--brane  actions (\ref{pbranact}).
  \item The scalar field potential $\mathcal{V}(\phi)$ which
  eventually decides the form of the cosmological term in the domain wall
  actions (\ref{Dwaction})
  \item The  metric $g_{ij}(\phi)$ appearing in the kinetic term $g_{ij}(\phi)\partial
  _\mu  \phi^i  \, \partial ^\mu \phi ^j$ of the scalar fields since
  it is needed as much as the matrix $\mathcal{\mathcal{N}}_{\Lambda\Sigma}
  (\phi)$ to determine the values of $a$ and eventually of $\Delta$
\end{itemize}
It turns out that each of the above items involves a wealth of
surprisingly sophisticated geometric structures that are skillfully
utilized by supergravity, first to stand on its feet at the ungauged
level and, secondly, to be gauged producing non abelian symmetries and
the scalar potential. In the present chapter I survey all these
structures and I try to illustrate their meaning in relation with the
parent string theory. Obviously the cause that imposes on the theory
all such structures is supersymmetry and the presence of the
fermions. Yet since the fermions are ugly objects to deal with
while their product, namely the geometric structure of the theory is
beautiful, I will only stick to the latter and mention the fermions as
seldom as possible. This implies that my presentation is mostly
descriptive. I nowhere pretend to give the proof that the various
supergravities are as they are but I do my best to illustrate their
miraculous geometric functioning that eventually governs the
$p$--brane classical physics we are interested in. In view of the
advocated correspondences such classical physics is also the quantum physics
of the underlying world volume theories.
\section[Supergravity and homogeneous scalar manifolds $\mathrm{G/H}$]{Supergravity
and homogeneous scalar manifolds
$\mathrm{G/H}$}
\label{SugraGsuH}
\setcounter{equation}{0}
If we consider the whole set of supergravity theories in diverse
dimensions we discover an important general property. With the
caveat of three noteworthy exceptions in all the other cases the
constraints imposed by supersymmetry imply that the scalar manifold
$\mathcal{M}_{scalar}$ is necessarily a \emph{homogeneous coset manifold}
$ \mathcal{G}/\mathcal{H}$
of the  \emph{non--compact type}, namely a suitable non compact Lie group
$ \mathcal{G}$ modded by the action of a maximal compact subgroup
$ \mathcal{H}\subset \mathcal{G}$.  By $\mathcal{M}_{scalar}$ we mean the
manifold parametrized by the scalar fields $\phi^I$ present in the theory. The
metric $g_{IJ}(\phi)$ defining the Riemannian structure of the scalar
manifold appears in the supergravity lagrangian through
the scalar kinetic term which is of the $\sigma$--model type:
\begin{equation}
  \mathcal{L}_{scalar}^{kin} =\frac{1}{2} \,
g_{IJ}(\phi) \, \partial_\mu \phi^I \, \partial^\mu \phi^J
\label{scalkinter}
\end{equation}
The three noteworthy exceptions where the scalar manifold is allowed to be
something more general than a coset $ \mathcal{G}/\mathcal{H}$ are
the following
\begin{enumerate}
  \item $\mathcal{N}=1$ supergravity in $D=4$ where
  $\mathcal{M}_{scalar}$ is simply requested to be a \emph{Hodge K\"ahler
  manifold}
  \item $\mathcal{N}=2$ supergravity in $D=4$ where
  $\mathcal{M}_{scalar}$ is simply requested to be the product of a \emph{special K\"ahler
  manifold} $ \mathcal{SK}_n$\footnote{Special K\"ahler geometry was
  introduced in a coordinate dependent way in the first papers on
  the vector multiplet coupling to supergravity in the middle
  eighties \cite{CFGVP},\cite{dWLVP}. Then it was formulated in a coordinate--free way at the beginning of the
  nineties from a Calabi Yau standpoint by Strominger \cite{Strominger:1990pd} and from a supergravity standpoint by
  Castellani, D'Auria and Ferrara \cite{Castellani:1990zd,Strominger:1990pd}.
  The properties of holomorphic isometries of special K\"ahler manifolds, namely the geometric structures of
  special geometry involved in the gauging were clarified by D'Auria, Fr\'e and Ferrara in \cite{D'Auria:1991fj}.
  For a review of special K\"ahler geometry in the setup and notations of the present lectures see
  \cite{mylecture}}
   containing the $n$ complex scalars of
  the $n$ vector multiplets with a \emph{quaternionic manifold}
  $\mathcal{QM}_m$ containing the $4m$ real scalars of the $m$
  hypermultiplets \footnote{The notion of quaternionic geometry, as it enters
  the formulation of hypermultiplet coupling was introduced by
  Bagger and Witten in \cite{bagwit} and formalized by Galicki in \cite{gal} who also explored the relation
  with the notion of HyperK\"ahler quotient, whose use in the construction of supersymmetric $N=2$ lagrangians
  had already been emphasized in \cite{hklr}. The general problem of classifying quaternionic homogeneous spaces
  had been addressed in the mathematical literature by Alekseevski
  \cite{alex}.}
  \item $\mathcal{N}=2$ supergravity in $D=5$ where
  $\mathcal{M}_{scalar}$ is simply requested to be the product of a \emph{very
  special manifold} $ \mathcal{VS}_n$ \footnote{The notion of very special geometry
  is essentially due to the work of G\"unaydin Sierra and Townsend who discovered it
  their work on coupling $D=5$ supergravity to vector multiplets
  \cite{GST1,GST2}. The notion was subsequently refined and properly related to special
  K\"ahler geometry in four dimensions through the work by de Wit and Van Proeyen
  \cite{deWit:1992nm,deWit:1993wf,deWit:1995tf}} containing the $n$ real scalars of
  the $n$ vector multiplets with a \emph{quaternionic manifold}
  $\mathcal{QM}_m$ containing the $4m$ real scalars of the $m$
  hypermultiplets.
\end{enumerate}
I shall come back to the case of $\mathcal{N}=2$ supergravity in
five dimensions because of its relevance in the quest of domain walls
and supersymmetric realizations  of the Randall Sundrum scenario and
there I shall briefly discuss  both  very special geometry and
quaternionic geometry. Instead for special K\"ahler geometry and the
structure of $\mathcal{N}=2$ supergravity in four dimensions I refer
the reader to \cite{mylecture} and \cite{bertolo} where they
are extensively discussed. Probably the most relevant aspect of special K\"ahler manifolds is
their interpretation as moduli spaces of Calabi Yau three--folds which connects
the structures of $ \mathcal{N}=2$  supergravity to superstring theory
via the algebraic geometry of compactifications on such three--folds. Here
I do not address these topics and I rather focus on the case of
homogeneous scalar manifolds which covers all the other types of
supergravity lagrangians and also specific instances of $\mathcal{N}=2$
theories since there exist subclasses of special K\"ahler and
very special manifolds that are homogeneous spaces $ {\mathcal{G}/\mathcal{H}}$.
\par
By means of my choice I aim at illustrating some of the very ample collection of
supergravity features that encode quite non trivial aspects of
superstring theory and that can be understood in terms of Lie algebra
theory and differential geometry of homogeneous coset spaces.
\subsection[How to determine the scalar cosets $\mathrm{G/H}$ from supersymmetry]{
How to determine the scalar cosets $\mathrm{G/H}$ of supergravities from supersymmetry}
\label{howGsuH}
The best starting point of our discussion is provided by presenting
the table of coset structures in four--dimensional supergravities.
This is done in the next subsection in table \ref{topotable} where supergravities are
classified according to the number $\mathcal{N}$ of the preserved supersymmetries.
Recalling that a Majorana spinor in $D=4$ has four real components
the total number of supercharges preserved by each theory is
\begin{equation}
  \# \mbox{\,of supercharges}\, = \, 4 \, \mathcal{N}
\label{supchargnum}
\end{equation}
and becomes maximal for the $\mathcal{N}=8$ theory where it is $32$.
\par
Here I present a short general discussion that applies to all the
diverse dimensions.
\par
There are two ways to determine the scalar manifold structure of
a supergravity theory:
\begin{itemize}
  \item By compactification from higher dimensions. In this case the
  scalar manifold is identified as the \emph{moduli space} of the internal compact space
  \item By direct construction of each supergravity theory in the
  chosen space--time dimension. In this case one uses all the \emph{a
  priori} constraints provided by supersymmetry,
  namely the field content of the various multiplets, the global and
  local symmetries that the action must have and, most prominently,
  as I am going to explain, the \emph{duality symmetries}.
\end{itemize}
The first method makes direct contact with important aspects of
superstring theory but provides answers that are specific to the
chosen compact internal space $\Omega_{10-D}$ and not fully general.
The second method gives instead fully general answers. Obviously the
specific answers obtained by compactification must fit into the
general scheme provided by the second method. In the next section I
highlight the basic arguments that lead to the construction of table
\ref{topotable}. Obviously the table relies on the fact that each of
the listed lagrangians has been explicitly constructed and  shown to
be supersymmetric\footnote{For a review of supergravity theories both in
$D=4$ and in diverse dimensions the reader is referred to the book\cite{castdauriafre}.
Furthermore for a  review of the geometric structure of all supergravity theories
in a modern perspective I refer to \cite{Andrianopoli:1998ve} } but it is quite instructive to see
how the scalar manifold, which is the very \emph{hard core} of the theory
determining its interaction structure, can be predicted a priori with
simple group theoretical arguments.
\par
The first thing to clarify is this:  what is classified
in table \ref{topotable} are the
\emph{ungauged supergravity theories} where all vector fields are abelian
and the isometry group of the scalar manifold is a global symmetry.
\emph{Gauged supergravities} which are the main
concern of these lectures are constructed only in a second time
starting from the ungauged ones and by means of a \textbf{gauging}
procedure that I will describe in further chapters. Each ungauged
supergravity admits a finite number of different gaugings where
suitable subgroups of the isometry group of the scalar manifold
are promoted to local symmetries using some or all of the available
vector fields of the theory. It is clear that which gaugings are
possible is once again  determined by the structure of the
scalar manifold plus additional constraints that I will explain
later.
\par
In every space--time dimension $D$
the reasoning that leads to single out the scalar coset manifolds
$\mathcal{G}/\mathcal{H}$ is based on the following elements:
\begin{description}
  \item[A] Knowledge of the field content of the various
  supermultiplets $\mu_i$
  that constitute irreducible representations of the $ \mathcal{N}$--extended
  supersymmetry algebra in $D$--dimensions. In particular this means
  that we know the total number of scalar fields. The scalars
  pertaining to the various types of multiplets must fill separate
  submanifolds $ \mathcal{M}_i$ of the total scalar manifold which is the direct
  product of all such subspaces: $ \mathcal{M}_{scalar} = \bigotimes_{i} \, \mathcal{M}_i$.
  \item[B] Knowledge of the automorphism group $H_{\mathrm{Aut}}$ of the
  relevant supersymmetry algebra. This latter acts on the gravitinos
  and on the other fermion fields as a local symmetry group. The
  gauge connection for this gauge symmetry is not elementary, rather
  it is a composite connection derived from the $\sigma$--model of the scalar
  fields:
\begin{equation}
  \Omega^{\mathrm{Aut}}_\mu = \mathbb{P}_{\mathrm{Au}t} \,\left[ \,  g^{-1} \left( \phi\right)
  \frac{\partial }{\partial \phi^I} \, g\left( \phi \right) \right]
  \, \partial _\mu  \phi^I
\label{compcongen}
\end{equation}
where $\mathbb{P}_{\mathrm{Aut}}$ denotes the projection onto the automorphism
subalgebra $\mathrm{Aut} \subset \mathcal{H}$ of the isotropy algebra $ \mathcal{H}
$ of the scalar coset manifold $ \mathcal{G}/ \mathcal{H}$. This is
consistent  only if the isotropy group has the following direct product structure:
\begin{equation}
  \mathcal{H} =\mathrm{Aut} \, \bigotimes \, \mathcal{H}^\prime
\label{HHprime}
\end{equation}
$\mathcal{H}^\prime$ being some other closed Lie group.
 \item[C]{Existence of appropriate irreducible representations of $
 \mathcal{G}$ in which we can accommodate each type of $p+1$--forms
 $A^{[p+1]\Lambda}$ appearing in the various supermultiplets. Indeed each
$p+1$--form sits in some supermultiplet together with fermion fields
and with scalars. The transformations of $\mathcal{G}$ commute with
supersymmetry and must rotate an entire supermultiplet into another
one of the same sort. Since the action of $\mathcal{G}$ is
well defined on scalars we must be able to lift it also to the
$p+1$--form partners of the scalars. Here we have a bifurcation:
\begin{itemize}
  \item When the magnetic dual of the $p_i+1$--forms, that are
  $D-p_i-3$--forms have a different degree, namely $D-p_i-3 \neq p_i+1$, then
  the group $\mathcal{G}$ must have irreducible representations $D_i$ of
  dimensions:
\begin{equation}
  \mbox{dim}\left( D_i\right) = { n}_i
\label{dimni}
\end{equation}
where ${ n}_i$ is the number of $p_i+1$--forms present in the
theory
  \item When there are  $\overline{p}+1$--forms, whose magnetic duals have the
  same degree, namely $D-\overline{p}-3 = \overline{p}+1$, then
  the group $\mathcal{G}$ must have, in addition to the  irreducible representations $D_i$
  that accommodate the other $p_i+1$-forms as in eq.(\ref{dimni})
  also a representation $\overline{D}$ of dimension
\begin{equation}
  \mbox{dim}\left( \overline{D}\right) = 2 {\bar n}
\label{dimbarn}
\end{equation}
which accommodates the ${\bar n}$ forms of degree $\overline{p} $ and
has the following additional property. In $D=6,10$ it is realized by
pseudorthogonal matrices in the fundamental of $\mathrm{SO}\left(
{\bar n}\, , \, {\bar n} \right) $ while in $D=4,8$ it is realized by
symplectic matrices in the fundamental of $\mathrm{Sp}\left( 2\,
{\bar n},\mathrm{R}\right) $. The reason for this apparently
extravagant request is that in the case of $\overline{p}+1$--forms
the lifting of the action of the group $ \mathcal{G}$ is realized by
means of electric/magnetic duality rotations as I explain in section \ref{dualsym}.
Furthermore the reason why, in this discussion,  I consider only the even dimensional cases
 is that self--dual $\overline{p}+1$--forms can exist only
when $D=2 r$ is even.
\end{itemize}
}
\end{description}
\subsection[The scalar cosets of $D=4$ supergravities]{The scalar cosets of $D=4$
supergravities}
\label{d4coset}
%%%%%%%%%%%%%%%%%%%%%%%%%%%%%%%%%%%%%%%%%%%%%%%%%%%%%%%%%%%%%%%%%%%
In four dimensions the only relevant $p+1$--forms are the $1$--forms
that correspond to ordinary gauge vector fields. Indeed $3$--forms do
not have degrees of freedom and  $2$--forms can be dualized to
scalars. On the contrary $D=4$ is an even number and $1$--forms are
self--dual in the sense described in section \ref{dualsym} and alluded
above in section \ref{howGsuH}. Furthermore the automorphism
group of the $\mathcal{N}$ extended supersymmetry algebra in $D=4$ is
\footnote{The role of the $\mathrm{SU}(\mathcal{N})$ symmetry in
$\mathcal{N}$--extended supergravity  was firstly emphasized in
\cite{Cremmer:1977zt,Cremmer:1978tt}.}:
\begin{equation}
\begin{array}{rclcl}
  \mathcal{H}_{\mathrm{Aut}}&=& \mathrm{SU}\left(  \mathcal{N}\right)
  \times \mathrm{U}\left( 1\right) & ; &
  \mathcal{N}=1,2,3,4,5,6  \\
\mathcal{H}_{\mathrm{Aut}}&=& \mathrm{SU}\left(  \mathrm{8}\right)
    & ; &
  \mathcal{N}=7,8\
  \end{array}
\label{Haut}
\end{equation}
Hence applying the strategy outlined in section \ref{howGsuH} the
requests to be imposed on the coset $\mathcal{G}/\mathcal{H}$ in
four--dimensional supergravities are:
\begin{enumerate}
  \item The total number of spin zero fields must be equal to the
  dimension of the coset:
\begin{equation}
  \# \mbox{\, of spin zero fields}\,  \equiv \, m \, = \, \mbox{dim} \,
  \mathcal{G}\, - \, \mbox{dim} \,  \mathcal{H}
\label{dimG-dimH}
\end{equation}
  \item The total number of vector fields in the theory ${\bar n}$
  must be equal to one half the dimension of a symplectic irreducible
  representation $D_{Sp}$ of the group $\mathcal{G}$:
\begin{equation}
   \# \mbox{\, of spin 1 fields}\,  \equiv \, {\bar n} \, = \,\frac{1}{2}\, \mbox{dim} \,
   D_{Sp} \left( \mathcal{G}\right)
\label{numbspin1}
\end{equation}
 \item  The isotropy group $\mathcal{H}$ must be of the form\footnote{
 The difference between the $\mathcal{N}=7,8$ cases and the others is properly explained
 in the following way. As far as superalgebras are concerned the automorphism group is always
 $\mathrm{U}(\mathcal{N})$ for all $\mathcal{N}$, which can extend, at this level also beyond
 $\mathcal{N}=8$. Yet for the $\mathcal{N}=8$ graviton multiplet, which is
 identical to the $\mathcal{N}=7$ multiplet, it happens that the $\mathrm{U(1)}$
 factor in $\mathrm{U(8)}$ has vanishing action on all physical states
 since the multiplet is self--conjugate under $CPT$--symmetries. From here it follows that the
 isotropy group of the scalar manifold must be $\mathrm{SU(8)}$ rather than $\mathrm{U(8)}$.
 A similar situation occurs for the $\mathcal{N}=4$ vector multiplets that are also $CPT$
 self--conjugate. From this fact follows that the isotropy group of the scalar submanifold
 associated with the vector multiplet scalars is $\mathrm{SU(4)} \times \mathcal{H}^\prime$
 rather than $\mathrm{U(4)} \times \mathcal{H}^\prime$. In $N=4$
 supergravity, however, the $\mathrm{U(1)}$ factor of the automorphism
 group appears in the scalar manifold as isotropy group of the submanifold associated
 with the graviton multiplet scalars. This is so because the $\mathcal{N}=4$ graviton
 multiplet is not CPT self conjugate. I warmly thank A. Van Proeyen for pointing out to me the
 need to explain this point more explicitly than in the first draft of these notes.}:
\begin{equation}
\begin{array}{rclcl}
\mathcal{H} & = & \mathrm{SU}\left (\mathcal{N}\right )  \times \mathrm{U(1)}  \times
 \mathcal{H}^\prime & ; &
\mathcal{N}= 3,4  \\
\mathcal{H} & = & \mathrm{SU}\left (\mathcal{N}\right)  \times \mathrm{U(1)}   & ;
&
\mathcal{N}= 5,6 \\
\mathcal{H} & = & \mathrm{SU}\left (8\right)    & ; &
\mathcal{N}= 7,8\
\end{array}
\label{Hgroups}
\end{equation}
The distinction between the cases $\mathcal{N}=3,4$ and the cases
$\mathcal{N}= 5,6$ comes from the fact that in the former we have
both the graviton multiplet plus vector multiplets, while in the
latter there is only the graviton multiplet. The vector multiplets
can transform non trivially under the additional group
$\mathcal{H}^\prime$ for which there is no room in the latter cases.
Finally the $\mathcal{N}= 7,8$ supergravities that contain only the
graviton multiplet are indistinguishable theories since their field
content and interactions are the same.
\par
Using the above rules and the known list of Lie groups one arrives at
the unique solution provided in table \ref{topotable}
\end{enumerate}
\begin{table*}
\begin{center}
\caption{\sl Scalar Manifolds of Extended Supergravities in $D=4$}
\label{topotable}
\begin{tabular}{|c||c|c|c||c|c||c||c| }
\hline \hline ~ & $\#$ scal. & $\#$ scal. & $\#$ scal. & $\#$ vect. &
 $\#$ vect. &~ & $~ $ \\
$\mathcal{N}$ & in & in & in & in  &
 in  &$\Gamma_{cont}$ & ${\cal M}_{scalar}$   \\
 ~ & scal.m. & vec. m. & grav. m. & vec. m. & grav. m. & ~ &~
\\
\hline \hline ~    &~    &~   &~   &~  &~  & ~ & ~ \\ $1$  & 2 m &~ &
~  & n &~  &  ${\cal I}$  & ~   \\ ~    &~    &~   &~   &~  &~  &
$\subset \mathrm{Sp}(2n,\mathbb{R})$ & K\"ahler \\ ~    &~    &~   &~   &~  &~
& ~ & ~ \\ \hline ~    &~    &~   &~   &~  &~  & ~ & ~ \\ $2$  & 4 m
& 2 n& ~  & n & 1 &  ${\cal I}$ & Quaternionic $\otimes$
\\
~    &~    &~   &~   &~  &~  &  $\subset \mathrm{Sp}(2n+2,\mathbb{R})$ &
Special K\"ahler \\ ~    &~    &~   &~   &~  &~  & ~ & ~ \\ \hline ~
&~ &~   &~   &~  &~  & ~ & ~ \\ $3$  & ~   & 6 n& ~  & n & 3 &
$S\mathrm{U}(3,n)$ &~  \\ ~    &~    &~   &~   &~  &~  & $\subset
\mathrm{Sp}(2n+6,\mathbb{R})$ & $\frac{\mathrm{SU}(3,n)} {S(\mathrm{U}(3)\times \mathrm{U}(n))}$ \\ ~
&~ & ~ &~   &~  &~  & ~ & ~ \\ \hline ~    &~    &~   &~   &~  &~  &
~ & ~
\\ $4$  & ~   & 6 n& 2  & n & 6 &  $\mathrm{SU}(1,1)\otimes \mathrm{SO}(6,n)$ &
$\frac{\mathrm{SU}(1,1)}{\mathrm{U}(1)} \otimes $ \\ ~    &~    &~   &~   &~  &~  &
$\subset \mathrm{Sp}(2n+12,\mathbb{R})$ & $\frac{\mathrm{SO}(6,n)}{\mathrm{SO}(6)\times
\mathrm{SO}(n)}$
\\ ~    &~ &~   &~   &~  &~  & ~ & ~ \\ \hline ~    &~    &~   &~
&~ &~  & ~ & ~ \\ $5$  & ~   & ~  & 10 & ~ & 10 & $\mathrm{SU}(1,5)$ & ~  \\ ~
&~ &~   &~   &~  &~  & $\subset \mathrm{Sp}(20,\mathbb{R})$ & $\frac{\mathrm{SU}(1,5)}
{S(\mathrm{U}(1)\times \mathrm{U}(5))}$ \\ ~    &~    &~   &~   &~  &~  & ~ & ~ \\
\hline ~    &~ &~   &~   &~  &~  & ~ & ~ \\ $6$  & ~   & ~  & 30 & ~
& 16 & $SO^\star(12)$ & ~ \\ ~    &~    &~   &~   &~  &~  & $\subset
\mathrm{Sp}(32,\mathbb{R})$ & $\frac{SO^\star(12)}{\mathrm{U}(1)\times \mathrm{SU}(6)}$ \\ ~
&~ &~ &~   &~  &~  & ~ & ~ \\ \hline ~    &~    &~   &~   &~  &~  & ~
& ~
\\ $7,8$& ~   & ~  & 70 & ~ & 56 & $E_{7(-7)}$  & ~ \\ ~    &~    &~
&~   &~  &~  & $\subset \mathrm{Sp}(128,\mathbb{R})$ & $\frac{ E_{7(-7)}
}{\mathrm{SU}(8)}$ \\ ~ &~    &~   &~   &~  &~  & ~ & ~ \\ \hline
 \hline
\end{tabular}
\end{center}
\end{table*}
%%%%%%%%%%%%%%%%%%%%%%%%%%%%%%%%%%%%%%%%%%%%%%%%%%%%%%%%%%%%
\section[Maximal supergravities in diverse dimensions and their scalar manifolds]
{Maximal supergravities in diverse dimensions and their scalar manifolds}
\label{maxsugra}
\setcounter{equation}{0}
In table \ref{topotable} we have classified
supergravities at fixed space--time dimension according to the number
of supersymmetries. Another possible classification is according to
space time dimensions $D$ at fixed number of supercharges $N_Q$.  In
particular one can consider maximal supergravities where $N_Q=32$ and
discuss their structure in the diverse dimensions $3\leq D \leq 10$.
Such a study is very much rewarding since  we can  relate it to the
alternative way of deriving the scalar manifold of supergravity,
namely via compactification. There is indeed a class of hierarchical
compactifications that have the distinguished property of preserving
the number of supersymmetries at each step of the hierarchy. These
are the toroidal compactifications where $D$-dimensional space--time
$ \mathcal{M}_D$ is replaced by:
\begin{equation}
  \mathcal{M}_D \mapsto \mathcal{M}_{D-x} \, \times \, T^{x}
\label{toroidco}
\end{equation}
$T^{x}$ denoting an $x$--dimensional torus and $\mathcal{M}_{D-x}$
being a new space--time in $D-x$--dimensions. By means of sequential toroidal
compactifications we can reach all maximally extended supergravities
in lower dimensions starting from either type IIA or type IIB
supergravity in $D=10$. The result is always the same since
supersymmetry allows for unique maximal theories in $D\leq 9$ and
there is just one scalar coset manifold, that listed in table \ref{maxsugrat}.
Yet this result can be interpreted in
two ways depending on whether we look at it from the type IIA or from the type
IIB viewpoint.
%%%%%%%%%%%%%%%%%%%%%%%%%%%%%%%%%%%%%%%%%%%%%%%%%%%%%
\begin{table}[ht]
\caption{\label{maxsugrat} Scalar geometries in maximal
supergravities}
\begin{center}
\begin{tabular}{|c|c|c|l|}
\hline
$D=9$ &      $\mathrm{E_{2(2)}}  \equiv \mathrm{SL}(2,\mathbb{R})\otimes O(1,1)$ & $H =
O(2) $ &
$\mbox{dim}_{\mathbb{R}}\,({\mathcal{G}}/{\mathcal{H}}) \, =\, 3$ \\
 \hline
$D=8$ &      $\mathrm{E_{3(3)}}  \equiv \mathrm{SL}(3,\mathbb{R})\otimes \mathrm{SL}(2,
\mathbb{R})$ & $
\mathcal{H} =
\mathrm{O(2)}\otimes \mathrm{O(3)} $ &
$\mbox{dim}_{\mathbb{R}}\,({\mathcal{G}}/{\mathcal{H}}) \, =\, 7$ \\
\hline
$D=7$ &      $\mathrm{E_{4(4)}}  \equiv \mathrm{SL}(5,\mathbb{R}) $ & $\mathcal{H} = O(5) $ &
$\mbox{dim}_{\mathbb{R}}\,({\mathcal{G}}/{\mathcal{H}}) \, =\, 14$ \\
\hline
$D=6$ &      $\mathrm{E_{5(5)}}  \equiv \mathrm{O(5,5)} $ & $H = \mathrm{O(5)\otimes O(5)} $ &
$\mbox{dim}_{\mathbb{R}}\,({\mathcal{G}}/{\mathcal{H}}) \, =\, 25$ \\
\hline
$D=5$ &      $\mathrm{E_{6(6)}}$   & $\mathcal{H} = \mathrm{Usp(8)} $ &
$\mbox{dim}_{\mathbb{R}}\,({\mathcal{G}}/{\mathcal{H}}) \, =\, 42$ \\
\hline
$D=4$ &      $\mathrm{E_{7(7)}}$   & $\mathcal{H} = \mathrm{SU(8)} $ &
$\mbox{dim}_{\mathbb{R}}\,({\mathcal{G}}/{\mathcal{H}}) \, =\, 70$ \\
\hline
$D=3$ &      $\mathrm{E_{8(8)}}$   & $\mathcal{H} =\mathrm{O(16)} $ &
$\mbox{dim}_{\mathbb{R}}\,({\mathcal{G}}/{\mathcal{H}}) \, =\, 128$ \\
\hline
\end{tabular}
\end{center}
\end{table}
%%%%%%%%%%%%%%%%%%%%%%%%%%%%%%%%%%%%%%%%%%%%%%%%%%%%%
There is indeed a challenging problem that corresponds
to retrieving the steps of the two possible chains of sequential
toroidal compactifications within the algebraic structure of the
isometry groups $\mathcal{G}_x$ and identifying which scalar field
appears at which step of the sequential chain. Such a problem has a
very elegant and instructive solution in terms of a rather simple and classical
mathematical theory, namely the \emph{solvable Lie algebra parametrization} of
non--compact cosets. This mathematical theory that makes a perfect match with
the string theory origin of supergravities plays an important role in
the discussion of $p$--brane solutions. I will review it in chapter
\ref{solvchap}. As we are going to see there, in the
Solvable Lie algebra parametrizations the scalar fields are divided
into two groups, those that are associated with Cartan generators of
the solvable algebra and those that are associated with nilpotent
generators. The \emph{Cartan scalars} are those that play the role of
generalized \emph{dilatons} and couple to the field strength $p+2$--forms as
in eq.(\ref{pbranact}). Within the algebraic approach the  $a$ parameters
appearing in the couplings of type $\exp \left[ -a \phi\right] \, \vert F^{[p+2]} \vert^2$
have an interpretation in terms of  \emph{roots} and \emph{weights} of the $
\mathcal{G}_x$ Lie algebras which provides a very important insight
into the whole matter. The solvable Lie algebra approach that in maximal supergravities
helps so clearly to master the string theory origin of the
cosets $\mathcal{G}/\mathcal{H} $ can be
extended also to the scalar manifolds of theories with a lesser number of supercharges.
Indeed, from a mathematical point of view it works for all
non--compact cosets. We refer the reader to chapter \ref{solvchap}
for a review of these ideas and of this geometrical setup.
\par

%%%%%%%%%%%%%%%%%%%%%%%%%%%%%%%%%%%%%%%%%%%%%%%%%%%%%%%%
\section[Duality symmetries in even dimensions]{Duality symmetries
in even dimensions and the coupling of self--dual
 forms}
\label{dualsym}
\setcounter{equation}{0}
Generically a $p$--brane in $D$--dimensions either carries an electric charge
with respect a $(p+1)$-form gauge field $A^{[p+1]}$ or a magnetic charge with
respect to the dual $D-p-3$--form $A^{[D-p-3]}_{dual}$. In the
general case it cannot be dyonic with respect to the same gauge field
since
\begin{equation}
  p+1 \neq D-p-3
\label{nodualp}
\end{equation}
However, in even dimension $D=2r$, the Diophantine eq.(\ref{nodualp})
admits one solution $p=\frac{D-4}{2}$, so that we always
have, in this case, a special instance of branes which can be dyonic: they are
particles or $0$--branes in $D=4$, strings or $1$--branes in $D=6$
and $2$--branes in $D=8$. The possible presence of such dyonic
objects has profound implications on the structure of the
even dimensional supergravity lagrangians.
Indeed most of the  dualities, $T$, $S$ and $U$ that relate the five
perturbative superstrings have a non trivial action on $p$-branes and generically
transform them as electric--magnetic duality rotations.
Hence, when self--dual $r-1$--forms are available, string dualities
reflect into \emph{duality symmetries} of the supergravity
lagrangians which constitute an essential ingredient
in their construction. By duality symmetry we mean the following: a
certain group of transformations $ \mathcal{G}_{dual}$ acts on the set of
\emph{field equations} of supergravity plus the \emph{Bianchi identities} of the $r-1$--forms
mapping this set into itself. Clearly $ \mathcal{G}_{dual}$ acts also on
the scalar fields $\phi^I$ and in order to be a symmetry it must respect their
kinetic term $g_{IJ} (\phi) \partial_\mu \phi^I \, \partial^\mu
\phi^J$. This happens if and only if $ \mathcal{G}_{dual}$ is a group
of isometries for the scalar metric $g_{IJ} (\phi)$. In other words
string dualities are encoded in the isometry group of the scalar
manifold  of supergravity which is lifted to act as a
group of electric--magnetic duality rotations on the $r-1$--forms.
\par
The request that these duality symmetries do exist determines the
general form  of the supergravity lagrangian and is a key
ingredient in its construction. For this reason in the present
section  I consider the case of even
dimensions $D=2r$  and I review the general structure of an abelian
theory  containing $\overline{n}$ differential $(r-1)$--forms:
\begin{eqnarray}
A^\Lambda & \equiv & A^\Lambda_{\mu_1 \dots\ \mu_{r-1}} \, dx^{\mu_1}
\, \wedge \, \dots \, \wedge dx^{\mu_{r-1}}  \quad ; \quad
\left ( \Lambda = 1, \dots , {\bar n} \right )
\end{eqnarray}
and $\overline{m}$ real scalar fields $\phi^I$. The  field strengths of the
$r-1$--forms  and their Hodge duals are defined as follows:
\begin{equation}
\begin{array}{lclclcl}
{ F}^\Lambda  &\equiv&  d \, A^\Lambda \,  \equiv  \,
{{1}\over{r!}} \, {\cal F}^\Lambda_{\mu_1 \dots\ \mu_{r}} \,
dx^{\mu_1} \, \wedge \, \dots \, \wedge dx^{\mu_{r}} & ;  &
{\cal F}^\Lambda_{\mu_1 \dots\ \mu_{r}}  & \equiv &  \partial_{\mu_1}
A^\Lambda_{\mu_2 \dots\ \mu_{r}} \, + \, \mbox{r-2 terms}  \\
{ F}^{\Lambda\star}  & \equiv &  {{1}\over{r!}} \, {\tilde {\cal
F}}^\Lambda_{\mu_1 \dots\ \mu_{r}} \, dx^{\mu_1} \, \wedge \, \dots
\, \wedge dx^{\mu_{r}} & ; & {\tilde {\cal F}}^\Lambda_{\mu_1
\dots\ \mu_{r}}  & \equiv & {{1}\over{r!}} \varepsilon_{\mu_1\dots
\mu_r \nu_1\dots \nu_r}\, {\cal F}^{\Lambda \vert \nu_1 \dots \nu_r}\
\end{array}
\label{campfort}
\end{equation}
Defining the space--time integration volume as :
\begin{equation}
\mbox{d}^D x \, \equiv \, {{1}\over{D!}} \, \varepsilon_{\mu_1\dots
\mu_D} \, dx^{\mu_1} \, \wedge \, \dots \, \wedge dx^{\mu_{D}}
\label{volume}
\end{equation}
we obtain:
\begin{eqnarray}
 F^\Lambda \, \wedge \, F^\Sigma \,& = &
{{1}\over{(r!)^2}} \, \varepsilon^{\mu_1\dots \mu_r \nu_1\dots
\nu_r}\, {\cal F}^\Lambda_{\mu_1 \dots\ \mu_{r}} \, {\cal
F}^\Sigma_{\nu_1 \dots\ \nu_{r}} \nonumber\\  F^\Lambda \, \wedge
\, F^{\Sigma\star} & = & (-)^r \, {{1}\over{(r!)}} \,
{\cal F}^\Lambda_{\mu_1 \dots\ \mu_{r}} \, {\cal F}^{\Sigma \vert
\mu_1 \dots\ \mu_{r}}  \label{cinetici}
\end{eqnarray}
The real scalar
fields $\phi^I$ ( $I=1,\dots , {\bar m}$) span  an ${\bar
m}$--dimensional manifold ${\cal M}_{scalar}$ \footnotemark
\footnotetext{whether the $\phi^I$ can be arranged into complex
fields is not relevant at this level of the discussion } endowed with
a metric $g_{IJ}(\phi)$. Utilizing the above field content
we can write the following action functional:
\begin{eqnarray}
{\cal S}&=&{\cal S}_{tens} \, + \, {\cal S}_{scal}\nonumber\\
{\cal S}_{tens}&=& \int \, \left [ \, \frac{(-)^r}{2} \,
\gamma_{\Lambda\Sigma}(\phi) \, F^\Lambda \, \wedge \,
F^{\Sigma\star} \, +  \frac{1}{2} \,
\theta_{\Lambda\Sigma}(\phi) \, F^\Lambda \, \wedge \, F^{\Sigma}
\right ]\nonumber\\ {\cal S}_{scal}&=& \int \, \left [ \frac{1}{2} \,
g_{IJ}(\phi) \, \partial_\mu \phi^I \, \partial^\mu \phi^J \right] \,
\mbox{d}^D x \label{gaiazuma}
\end{eqnarray}
where the scalar field dependent ${\bar n} \times {\bar n}$ matrix
$\gamma_{\Lambda\Sigma}(\phi)$ generalizes the inverse of the squared
coupling constant $\frac{1}{g^2}$ appearing in ordinary 4D--gauge
theories. The field dependent matrix $\theta_{\Lambda\Sigma}(\phi)$
is instead a generalization of the $theta$--angle of quantum
chromodynamics. The matrix $\gamma$ is symmetric in every space--time
dimension, while $\theta$ is symmetric or antisymmetric depending on
whether $r=D/2$ is an even or odd number. In view of this fact it is
convenient to distinguish the two cases, setting:
\begin{equation}
D\, = \, \cases{4 \nu \quad \qquad\nu \in \mathbb{Z} \, \vert \quad
r=2\nu \cr 4 \nu + 2\quad \, \nu \in \mathbb{Z} \, \vert \quad r=2\nu
+ 1 \cr} \label{duecasi}
\end{equation}
Introducing a formal operator $j$ that maps a field strength into its
Hodge dual:
\begin{equation} \left ( j \, {\cal
F}^\Lambda \right )_{\mu_1 \dots \mu_r} \, \equiv \, {{1}\over{(r!)}}
\, \epsilon_{\mu_1 \dots\mu_r \nu_1 \dots \nu_r} \, {\cal F}^{\Lambda
\vert\nu_1 \dots \nu_r} \label{opjei}
\end{equation} and a formal scalar product:
\begin{equation} \left (
G \, , \, K \right ) \equiv G^T K \, \equiv \, {{1}\over{(r!)}}
\sum_{\Lambda=1}^{\bar n} G^{\Lambda}_{\mu_1\dots\mu_r} K^{\Lambda
\vert \mu_1\dots\mu_r } \label{formprod}
\end{equation} the total
lagrangian of eq. (\ref{gaiazuma}) can be rewritten as
\begin{eqnarray}
\mathcal{L}^{(tot)}\, & =&  {\cal F}^T \, \left ( \gamma
\otimes \bfone + \theta \otimes j \right ) {\cal F}   +
\frac{1}{2} \, g_{IJ}(\phi) \, \partial_\mu \phi^I \,
\partial^\mu \phi^J
\label{gaiazumadue}
\end{eqnarray}
and the essential distinction between the two cases of
eq.(\ref{duecasi}) is given, besides the symmetry of $\theta$, by the
involutive property of $j$, namely we have:
\begin{equation} \matrix{ D=4 \nu & \vert &
\theta = \theta^T & j^2 = - \bfone \cr D=4 \nu + 2 & \vert & \theta =
- \theta^T & j^2 = \bfone \cr } \label{splitta}
\end{equation}
Introducing dual and antiself--dual combinations:
\begin{equation}
\matrix{ D=4 \nu & \cases{ {\cal F}^{\pm} = {\cal F}\, \mp i \, j
{\cal F} \cr j \, {\cal F}^{\pm} = \pm \mbox{i} {\cal F}^{\pm} \cr}
\cr D=4 \nu+2 & \cases{ {\cal F}^{\pm} = {\cal F} \, \pm \, j {\cal
F}\cr j \, {\cal F}^{\pm} = \pm {\cal F}^{\pm} \cr} \cr}
\label{selfduals}
\end{equation}
and the field--dependent matrices:
\begin{equation} \matrix{D=4 \nu & \cases{ {\cal N} = \theta -
\mbox{i} \gamma \cr {\bar {\cal N}} = \theta + \mbox{i} \gamma \cr
}\cr D=4 \nu+2 & \cases{ {\cal N} = \theta + \gamma \cr -{\cal N}^T =
\theta - \gamma \cr }\cr } \label{scripten}
\end{equation} the
tensor part of the lagrangian (\ref{gaiazumadue}) can be rewritten in
the following way in the two cases:
\begin{eqnarray} D=4\nu &
: &   {\cal L}_{tens}   =   {{\mbox{i}}\over{8}} \, \left [
{\cal F}^{+T} {\cal N} {\cal F}^{+} - {\cal F}^{-T} {\bar {\cal N}}
{\cal F}^{-} \right ] \nonumber\\
D=4\nu+2 & : &   {\cal
L}_{tens}   =   {{\mbox{1}}\over{8}} \, \left [ {\cal F}^{+T} {\cal N}
{\cal F}^{+} + {\cal F}^{-T} { {\cal N}^T} {\cal F}^{-} \right
]  \label{lagrapm}
\end{eqnarray}
Introducing the new tensor:
\begin{equation} \matrix { {\tilde
G}^\Lambda_{\mu_1\dots\mu_r} &\equiv & -(r!)  { {\partial {\cal
L}}\over{\partial {\cal F}^\Lambda_{\mu_1\dots\mu_r}}} & D=4\nu \cr
{\tilde G}^\Lambda_{\mu_1\dots\mu_r} &\equiv & (r!)  { {\partial
{\cal L}}\over{\partial {\cal F}^\Lambda_{\mu_1\dots\mu_r}}} &
D=4\nu+2 \cr} \label{gtensor}
\end{equation}
which, in matrix notation, corresponds to:
\begin{eqnarray} j \, G \, \equiv \, a \, {
{\partial {\cal L}}\over{\partial {\cal F}^T}} &  = & \,
{\frac{a}{r!}}\, \left ( \gamma\otimes\bfone + \theta\otimes j \right
) \, {\cal F} \label{ggmatnot}
\end{eqnarray}
where $a=\mp$ depending on whether $D=4\nu$ or $D=4\nu+2$, the
Bianchi identities and field equations associated with the lagrangian
(\ref{gaiazuma}) can be written as follows:
\begin{equation}  \partial^{\mu_1}
{\tilde {\cal F}}^{\Lambda}_{\mu_1\dots\mu_r}  =  0   \quad ; \quad
\partial^{\mu_1} {\tilde {\cal G}}^{\Lambda}_{\mu_1\dots\mu_r} =
0   \label{biafieq}
\end{equation}
This suggests that we introduce the $2{\bar n}$ column vector :
\begin{equation}
{\bf V} \, \equiv \, \left ( \matrix { j \, {\cal F}\cr j \, {\cal
G}\cr}\right ) \label{sympvec}
\end{equation}
and that we consider general linear transformations on such a vector:
\begin{equation}
\left ( \matrix { j \, {\cal F}\cr j \, {\cal G}\cr}\right )^\prime
\, =\, \left (\matrix{ A & B \cr C & D \cr} \right ) \left ( \matrix
{ j \, {\cal F}\cr j \, {\cal G}\cr}\right ) \label{dualrot}
\end{equation}
For any matrix $\left (\matrix{ A & B \cr C & D \cr} \right ) \, \in
\, \mathrm{GL}(2{\bar n},\mathbb{R} )$ the new vector ${\bf V}^\prime$ of {\it
magnetic and electric} field--strengths satisfies the same equations
~\ref{biafieq} as the old one. In a condensed notation we can write:
\begin{equation}
\partial \, {\bf V}^\prime \, = \, 0 \quad \Longleftrightarrow \quad
\partial \, {\bf V}^\prime \, = \, 0
\label{dualdue}
\end{equation}
Separating the self--dual and anti--self--dual parts
\begin{eqnarray}
{\cal F}={\frac{1}{2}}\left ({\cal F}^+ +{\cal F}^- \right )
 \quad ; \quad {\cal G}={\frac{1}{2}}\left ({\cal G}^+ +{\cal G}^-
\right ) \label{divorzio}
\end{eqnarray}
and taking into account that for $D=4\nu$ we have:
\begin{equation}
{\cal G}^+ \, = \, {\cal N}{\cal F}^+  \quad {\cal G}^- \, = \, {\bar
{\cal N}}{\cal F}^- \label{gigiuno}
\end{equation}
while for $D=4\nu +2$ the same equation reads:
\begin{equation}
{\cal G}^+ \, = \, {\cal N}{\cal F}^+  \quad {\cal G}^- \, = \, {-
{\cal N}^T}{\cal F}^- \label{gigidue}
\end{equation}
the duality rotation of eq.(\ref{dualrot}) can be rewritten as:
\begin{equation}
\begin{array}{lcrcl}
D=4\nu    &: &  \left ( \matrix {   {\cal F}^+
\cr
 {\cal G}^+\cr}\right )^\prime & = &
\left (\matrix{ A & B \cr C & D \cr} \right ) \left ( \matrix {
{\cal F}^+\cr {\cal N} {\cal F}^+\cr}\right )  \\
\null & \null &
\left (
\matrix {   {\cal F}^- \cr
 {\cal G}^-\cr}\right )^\prime & = &
\left (\matrix{ A & B \cr C & D \cr} \right ) \left ( \matrix {
{\cal F}^-\cr {\bar {\cal N}} {\cal F}^-\cr}\right )  \\
D=4\nu+2 & : &  \left ( \matrix {   {\cal F}^+ \cr
 {\cal G}^+\cr}\right )^\prime & = &
\left (\matrix{ A & B \cr C & D \cr} \right ) \left ( \matrix {
{\cal F}^+\cr {\cal N} {\cal F}^+\cr}\right )  \\
\null & \null & \left (
\matrix {   {\cal F}^- \cr
 {\cal G}^-\cr}\right )^\prime & = &
\left (\matrix{ A & B \cr C & D \cr} \right ) \left ( \matrix {
{\cal F}^-\cr { - {\cal N}^T} {\cal F}^-\cr}\right ) \
\end{array}
\label{trasform}
\end{equation}
In both cases the problem is that the transformation rule
(\ref{trasform}) of ${\cal G}^\pm$ must be consistent with the
definition of the latter as variation of the Lagrangian with respect
to ${\cal F}^\pm$ (see eq.(\ref{gtensor})). This request restricts the
form of the matrix $\Lambda =\left (\matrix{ A & B \cr C & D \cr}
\right )$. As we are just going to show, in the $D=4\nu$ case
$\Lambda$ must belong to the symplectic subgroup $\mathrm{Sp}(2\bar
n,\mathbb{R})$ of the special linear group, while in the $D=4\nu +2$
case it must be in the pseudorthogonal subgroup $\mathrm{SO}(\bar n , \bar
n)$:
\begin{equation}
\begin{array}{lcrcl}
D=4\nu \quad &:&  \left (\matrix{ A & B \cr C & D
\cr} \right ) & \in &\mathrm{Sp}(2\bar n,\mathbb{R}) \,\subset \, \mathrm{GL}(2\bar n
,\mathbb{R} ) \\
D=4\nu +2 &:& \left
(\matrix{ A & B \cr C & D \cr} \right ) & \in & \mathrm{SO}(\bar n , \bar n)
\,\subset \, \mathrm{GL}(2\bar n ,\mathbb{R} ) \
\end{array}
\label{distinguo}
\end{equation}
the above subgroups being defined as the set of $2\bar n \times 2\bar
n$ matrices satisfying, respectively, the following conditions:
\begin{equation}
\begin{array}{rcccl}
  \Lambda \in \mathrm{Sp}(2\bar n,\mathbb{R}) & \to &
   \Lambda^T  \,
\left (\matrix{ {\bf 0}_{} & \bfone_{} \cr -\bfone_{} & {\bf 0}_{}
\cr }\right )
 \, \Lambda  & = &
   \left (\matrix{ {\bf 0}_{} &
\bfone_{} \cr -\bfone_{} & {\bf 0}_{} \cr }\right )   \\
 \Lambda \in \mathrm{SO}(\bar n , \bar n)  &\to &    \Lambda^T
\, \left (\matrix{ {\bf 0}_{} & \bfone_{} \cr \bfone_{} & {\bf 0}_{}
\cr }\right ) \, \Lambda   & = &   \left (\matrix{ {\bf 0}_{} &
\bfone_{} \cr \bfone_{} & {\bf 0}_{} \cr }\right )
\end{array}
\label{ortosymp}
\end{equation}
To prove the statement we just made, we calculate the transformed
lagrangian ${\cal L}^\prime$ and then we compare its variation
${\frac{\partial {\cal L}^\prime}{\partial {\cal F}^{\prime T}}}$
with ${\cal G}^{\pm\prime}$ as it follows from the postulated
transformation rule (\ref{trasform}). To perform such a calculation we
rely on the following basic idea. While the duality
rotation (\ref{trasform}) is performed on the field strengths and on
their duals, also the scalar fields are transformed by the action of
some diffeomorphism ${\xi }\,  \in \, {\rm Diff}\left ( {\cal
M}_{scalar}\right )$ of the scalar manifold and, as a consequence of
that, also the matrix ${\cal N}$ changes. In other words given the
scalar manifold ${\cal M}_{scalar}$ we assume that in the two cases
of interest there exists a surjective homomorphism of the following
form :
\begin{equation}
\iota _{\delta} : \,  {\rm Diff}\left ( {\cal M}_{scalar}\right ) \,
\longrightarrow \, \mathrm{GL}(2\bar n,\mathbb{R}) \label{immersione}
\end{equation}
so that:
\begin{eqnarray}
\forall &  \xi   &\in \, {\rm Diff}\left ( {\cal M}_{scalar}\right )
\, : \, \phi^I \, \stackrel{\xi}{\longrightarrow} \,  \phi^{I\prime}
\nonumber\\ \exists  & \iota _{\delta}(\xi) & = \left (\matrix{ A_\xi
& B_\xi \cr C_\xi & D_\xi \cr }\right ) \, \in \,  \mathrm{GL}(2\bar
n,\mathbb{R}) \label{apnea}
\end{eqnarray}
Using such a homomorphism we can define the simultaneous action of
$\xi$ on all the fields of our theory by setting:
\begin{equation}
\xi \, : \, \cases{   \phi \, \longrightarrow \, \xi (\phi) \cr {\bf
V} \, \longrightarrow \, \iota _{\delta}(\xi) \, {\bf V} \cr {\cal
N}(\phi) \, \longrightarrow \, {\cal N}(\xi (\phi)) \cr }
\end{equation}
where the notation (\ref{sympvec}) has been utilized. In the tensor
sector the transformed lagrangian, is
\begin{eqnarray}
  {\cal L}^{\prime}_{tens} & =  &  {\frac{{\rm i}}{8}} \,
\Bigl [ {\cal F}^{+T} \, \bigl ( A + B {\cal N} \bigr )^T {\cal
N}^\prime ( A + B {\cal N} \bigr ) {\cal F}^{+}    - \,
{\cal F}^{-T} \, \bigl ( A + B {\bar {\cal N}} \bigr )^T {\bar {\cal
N}}^\prime ( A + B {\bar {\cal N}} \bigr ) {\cal F}^{-} \Bigr ]
\nonumber\\
\label{elleprima}
\end{eqnarray}
for the $D=4\nu$ case and
\begin{eqnarray}
 {\cal L}^{\prime}_{tens}  & =  &  {\frac{{\rm i}}{8}} \,
\Bigl [ {\cal F}^{+T} \, \bigl ( A + B {\cal N} \bigr )^T {\cal
N}^\prime ( A + B {\cal N} \bigr ) {\cal F}^{+}     -
\, {\cal F}^{-T} \, \bigl ( A - B {  {\cal N}^T} \bigr )^T {  {\cal
N}^T}^\prime ( A - B {  {\cal N}^T} \bigr ) {\cal F}^{-} \Bigr ]\nonumber\\
\label{elleprimap}
\end{eqnarray}
Consistency with the definition of ${\cal G}^+$ requires, in both
cases that
\begin{eqnarray}
{\cal N}^\prime \, \equiv \, {\cal N}(\xi(\phi)) & =&
\left ( C_\xi + D_\xi  {\cal N} \right )   \left ( A_\xi + B_\xi
{\cal N}\right )^{-1}
 \label{Ntrasform}
\end{eqnarray}
while consistency with the definition of ${\cal G}^-$ imposes, in the
$D=4\nu$ case the transformation rule:
\begin{eqnarray}
{\bar {\cal N}}^\prime  \, \equiv \, {\bar {\cal N}}(\xi(\phi)) \,
& = &  \left ( C_\xi + D_\xi  {\bar {\cal N}} \right )
\left ( A_\xi + B_\xi {\bar {\cal N}}\right )^{-1}
\label{Nbtrasform}
\end{eqnarray}
and in the case $D=4\nu+2$ the other transformation rule:
\begin{eqnarray}
{- {\cal N}^T}^\prime  \, \equiv \, {-{\cal N}^T}(\xi(\phi)) \, & =&
\left ( C_\xi - D_\xi  {  {\cal N}^T} \right )   \left (
A_\xi - B_\xi {  {\cal N}^T}\right )^{-1}
\label{Nttrasform}
\end{eqnarray}
It is from the transformation rules
(\ref{Ntrasform}),(\ref{Nbtrasform})
and (\ref{Nttrasform}) that we derive a restriction on the form of the
duality rotation matrix $\Lambda_\xi \equiv \iota_\delta(\xi)$.
Indeed, in the $D=4\nu$ case we have that by means of the fractional
linear transformation (\ref{Ntrasform}) $\Lambda_\xi$ must map an
arbitrary {\it complex symmetric} matrix into another matrix of the
{\it same sort}. It is straightforward to verify that this condition
is the same as the first of conditions (\ref{ortosymp}), namely the
definition of the symplectic group $\mathrm{Sp}(2\bar n ,\mathbb{R})$.
Similarly in the $D=4\nu +2$ case the matrix $\Lambda_\xi$ must obey
the property that taking the {\it negative of the transpose} of an
arbitrary real matrix ${\cal N}$ {\it before or after} the fractional
linear transformation induced by $\Lambda_\xi$ {\it is immaterial}.
Once again, it is easy to verify that this condition is the same as
the second property in eq.(\ref{ortosymp}), namely the definition of
the pseudorthogonal group $\mathrm{SO}(\bar n, \bar n)$. Consequently the
surjective homomorphism of eq.(\ref{immersione}) specializes as
follows in the two relevant cases
 \begin{equation}
\iota _{\delta} : \, \cases{ {\rm Diff}\left ( {\cal
M}_{scalar}\right ) \, \longrightarrow \, \mathrm{Sp}(2\bar n,\mathbb{R}) \cr
{\rm Diff}\left ( {\cal M}_{scalar}\right ) \, \longrightarrow \,
\mathrm{SO}(\bar n , \bar n) \cr} \label{spaccoindue}
\end{equation}
Clearly, since both $\mathrm{Sp}(2\bar n,\mathbb{R})$ and $\mathrm{SO}(\bar n , \bar
n)$ are finite dimensional Lie groups, while ${\rm Diff}\left ( {\cal
M}_{scalar}\right )$ is infinite--dimensional, the homomorphism
$\iota _{\delta}$ can never be an isomorphism. Defining the Torelli
group of the scalar manifold as:
\begin{equation}
{\rm Diff}\left ( {\cal M}_{scalar}\right ) \, \supset \, \mbox{Tor}
\left ({\cal M}_{scalar} \right ) \, \equiv \, \mbox{ker} \,
\iota_\delta \label{torellus}
\end{equation}
we always have:
\begin{equation}
\mbox{dim} \, \mbox{Tor} \left ({\cal M}_{scalar} \right ) \, = \,
\infty \label{infitor}
\end{equation}
The reason why have given the name of Torelli to the group defined by
eq.~\ref{torellus} is because of its similarity with the Torelli
group that occurs in algebraic geometry. There one deals with the
{\it moduli space} ${\it M}_{moduli}$ of complex structures of a
$(p+1)$--fold ${\cal M}_{p+1}$ and considers the action of the
diffeomorphism group $\mbox{Diff}\left ( {\it M}_{moduli} \right )$
on canonical homology bases of $(p+1)$--cycles.  Since this action
must be linear and must respect the intersection matrix, which is
either  symmetric or antisymmetric depending on the odd or even
parity of $p$, it follows that one obtains a homomorphism similar to
that in eq.~\ref{spaccoindue}:
\begin{equation}
\iota _h : \, \cases{ {\rm Diff}\left ( {\cal M}_{moduli}\right ) \,
\longrightarrow \, \mathrm{Sp}(2\bar n,\mathbb{R}) \cr {\rm Diff}\left ( {\cal
M}_{moduli}\right ) \, \longrightarrow \, \mathrm{SO}(\bar n , \bar n) \cr}
\label{spaccointre}
\end{equation}
The Torelli group is usually defined as the kernel of such a
homomorphism. When cohomology with real coefficients is replaced by
cohomology with integer coefficients the homomorphism of
eq.~\ref{spaccointre} reduces to
\begin{equation}
\iota _h : \, \cases{ {\rm Diff}\left ( {\cal M}_{moduli}\right ) \,
\longrightarrow \, \mathrm{Sp}(2\bar n,\mathbb{Z}) \cr {\rm Diff}\left ( {\cal
M}_{moduli}\right ) \, \longrightarrow \, \mathrm{SO}(\bar n , \bar
n,\mathbb{Z}) \cr} \label{spaccointrez}
\end{equation}
and the Torelli group becomes even larger.
\par
This similarity between two problems that are, at first sight,
totally disconnected is by no means accidental. When the supergravity
lagrangian that we consider emerges from some compactification of string
theory, the scalar manifold ${\it
M}_{scalar}$ is identified with the moduli--space of  complex
structures for suitable complex $(p+1)$--folds or tori and the duality
rotations are related with changes of integer homology basis. From
the physical point of view what requires the restriction from the
continuous duality groups $\mathrm{Sp}(2\bar n,\mathbb{R})$, $\mathrm{SO}(\bar n , \bar
n,\mathbb{Z})$ to their discrete counterparts $\mathrm{Sp}(2\bar
n,\mathbb{Z})$, $\mathrm{SO}(\bar n , \bar n,\mathbb{Z})$ is the Dirac
quantization condition of electric and magnetic charges
\begin{equation}
\frac{q_e \, q_m}{4 \pi \, \hbar} \, = \, \frac{n}{2} \qquad n \, \in
\,\mathbb{Z}
\label{dirquant}
\end{equation}
which obviously occurs when electric and magnetic
currents are introduced. Indeed the lattice spanned by electric and
magnetic charges is eventually identified with the integer homology
lattice of the corresponding $(p+1)$--fold.
\par
In view of this analogy, the natural question which arises is the
following: what is the counterpart in algebraic geometry of the
matrix ${\cal N}$ that appears in the kinetic terms of the gauge
fields? In view of its transformation property eq.(\ref{Ntrasform})
the answer is very simple: it is the {\it period matrix}. Consider
for instance the situation, occurring in Calabi--Yau three--folds,
where the middle cohomology group $H^{3}_{DR}\left ({\cal
M}_{3}\right )$ admits a Hodge--decomposition of the type:
\begin{eqnarray}
 H^{(3)}_{DR}({\cal M}_3)& =& H^{(3,0)}\, \oplus
 \, H^{(2,1)}   \oplus \, H^{(1,2)} \, \oplus \, H^{(0,3) }
\label{residue20}
\end{eqnarray}
and where the canonical bundle is trivial:
\begin{equation}
c_1\left( T{\cal M}\right) \, =\, 0 \, \longleftrightarrow \,
\mbox{dim} H^{(3,0)} \, = \, 1
\end{equation}
naming $\Omega^{(3,0)}$ the unique (up to a multiplicative constant)
holomorphic $3$--form, and choosing a canonical homology basis of
$3$--cycles $(A^\Lambda , B_ \Sigma)$ satisfying :
\begin{equation}
\matrix{
 A^\Lambda \, \cap \, A^\Sigma \, = 0 \, & A^\Lambda \, \cap \, B_ \Delta \,
 = \, \delta^\Lambda_\Delta \cr
 B_ \Gamma \, \cap \, A^\Sigma  \,
 = - \, \delta^\Lambda_\Sigma  &  B_\Gamma \, \cap \, B_\Delta \, = \,
 0 \cr }
 \label{interseca}
\end{equation}
where
\begin{equation}
\Lambda , \Sigma \, \dots \, = \, 1,\dots\, {\bar n}=1+h^{(2,1)}
\end{equation}
we can define the periods:
\begin{eqnarray}
X^\Lambda (\phi) & = &  \int_{A^\Lambda} \, \Omega^{(3,0)}(\phi)
\quad ; \quad  F_ \Sigma (\phi)  =  \int_{A_\Sigma} \,
\Omega^{(3,0)}(\phi) \label{periodando}
\end{eqnarray}
where $\phi^i$ ($i=1,\dots\, h^{(2,1)}$) are the moduli of the
complex structures and we can implicitly define the {\it period
matrix} by the relation:
\begin{equation}
F_ \Lambda \,  =  \, {\bar {\cal N}}_{\Lambda \Sigma} \, X^\Sigma
\label{matper}
\end{equation}
Under a diffeomorphism $\xi$ of the manifold of complex structures
the period vector
 \begin{equation}
   V (\phi) \, = \, \left ( \matrix{ X^\Lambda (\phi) \cr
 F_ \Sigma (\phi) \cr } \right )
 \label{pervec}
\end{equation}
will transform linearly through the $\mathrm{Sp}(2\bar n, \mathbb{R})$ matrix
$\iota_h (\xi)$ defined by the homomorphism in
eq.(\ref{spaccointre});
at the same time the period matrix ${\cal N}$ will obey the linear fractional
transformation rule of eq.(\ref{Ntrasform}). Indeed the intersection
relations in eq.(\ref{interseca}) define the symplectic invariant
metric
$\left (\matrix { {\bf 0} & \bfone \cr -\bfone & {\bf 0} \cr }
\right )$
%\label{sympinvmet}
%\end{equation}
\par
 What should be
clear from the above discussion is that a family of Lagrangians as in
eq. (\ref{gaiazuma}) will admit a group of
duality--rotations/field--redefinitions that will map one into the
other member of the family, as long as a {\it kinetic matrix} ${\cal
N}_{\Lambda\Sigma}$ can be constructed  that transforms as in
eq.(\ref{Ntrasform}). A way to obtain such an object is to identify it
with the {\it period matrix} occurring in problems of algebraic
geometry. At the level of the present discussion, however, this
identification is by no means essential: any construction of ${\cal
N}_{\Lambda\Sigma}$ with the appropriate transformation properties is
acceptable.
\par
Note also that so far we have used the words {\it
duality--rotations/field--redefinitions} and not the word duality
symmetry. Indeed the diffeomorphisms of the scalar manifold we have
considered were quite general and, as such had no claim to be
symmetries of the action, or of the theory. Indeed the question we
have answered is the following: what are the appropriate
transformation properties of the tensor gauge fields and of the
generalized coupling constants under diffeomorphisms of the scalar
manifold? The next question is obviously that of duality symmetries.
Suppose that a certain diffeomorphism $\xi \in \mbox{Diff}\left (
{\cal M}_{scalar} \right )$ is actually an {\it isometry} of the
scalar metric $g_{IJ}$. Naming $\xi^\star : \, T{\cal M}_{scalar} \,
\rightarrow \, T{\cal M}_{scalar}$ the push--forward of $\xi$, this
means that
\begin{eqnarray}
&\forall\,  X,Y \, \in \, T{\cal M}_{scalar}& \nonumber\\ & g\left (
X, Y \right )\, = \, g \left ( \xi^\star X, \xi^\star Y \right )&
\label{isom}
\end{eqnarray}
and $\xi$ is an exact global symmetry of the scalar  part of the
Lagrangian in eq~(\ref{gaiazuma}). The obvious question is: {\it " can
this symmetry be extended to a symmetry of the complete action?}
Clearly the answer  is that, in general, this is not possible. The
best we can do is to extend it to a symmetry of the field equations
plus Bianchi identities letting it act as a duality rotation on the
field--strengths plus their duals. This requires that the group of
isometries of the scalar metric ${\cal G}_{iso}({\cal M}_{scalar})$
be suitably embedded into the duality group (either $\mathrm{Sp}(2\bar
n,\mathbb{R})$ or $\mathrm{SO}(\bar n , \bar n)$ depending on the case) and
that the kinetic matrix ${\cal N}_{\Lambda\Sigma}$ satisfies the
covariance law:
\begin{equation}
{\cal N}\left ( \xi (\phi)\right ) \, = \, \left ( C_\xi + D_\xi
{\cal N}(\phi) \right ) \left ( A_\xi + B_\xi {\cal N}( \phi )\right
) \label{covarianza}
\end{equation}
A general class of solutions to this programme can be derived in the
case where the scalar manifold is taken to be a homogeneous space
${\cal G}/{\cal H}$. This is the subject of next section.
%%%%%%%%%%%%%%%%%%%%%%%%%%%%%%%%%
% FILE TRST 3 %%%%%%%%%%%%%%%%%%%
%%%%%%%%%%%%%%%%%%%%%%%%%%%%%%%%%
\section[The kinetic matrix $\mathcal{N}$ and symplectic embeddings]{The kinetic
matrix $\mathcal{N}$ and symplectic embeddings}
In our survey of the geometric features  of bosonic
supergravity  lagrangians that are specifically relevant for $p$--brane solutions
the next important item we have to consider is the kinetic term of the
$p+1$--form gauge fields. Generically it is of the form:
\begin{equation}
  \mathcal{L}^{Kin}_{forms}=\mathcal{N}_{\Lambda\Sigma} \left( \phi
  \right)\, F^\Lambda_{\mu_1\dots\mu_{p+2}} \, F^{\Sigma\vert
  \mu_1\dots\mu_{p+2}}
\label{formkinter}
\end{equation}
where  $\mathcal{N}_{\Lambda\Sigma}$ is a suitable scalar
field dependent symmetric matrix. In the case of self--dual
$p+1$--forms, that occurs only in even dimensions, the matrix
$\mathcal{N}$ is completely fixed by the requirement that the
\emph{ungauged supergravity theory} should admit duality symmetries.
Furthermore as remarked in the previous section, the
problem of constructing duality--symmetric lagrangians of the
type (\ref{gaiazuma}) admits general solutions  when the scalar
manifold is a homogeneous space ${\cal G}/{\cal H}$. Hence I devote the
present section to  review  the construction of the {\it kinetic
period matrix} ${\cal N}$ in the case of homogeneous spaces.
The case of odd space dimensions where there are no dualities will be
addressed in a subsequent section.
\par The relevant cases of even dimensional supergravities are:
\begin{enumerate}
  \item In $D=4$ the self--dual forms are  ordinary gauge vectors and
  the duality rotations are symplectic. There are several theories
  depending on the number of supersymmetries. They are summarized in table
  \ref{topotable}. Each theory involves a different
  number ${\bar n}$ of vectors $A^\Lambda$ and different cosets ${{\cal G}\over{\cal H}} $
  but the relevant homomorphism $\iota_\delta$ (see eq.~\ref{spaccoindue}) is always of
  the same type:
\begin{equation}
\iota_\delta : \, \mbox{Diff}\left ({{\cal G}\over{\cal H}} \right )
\, \longrightarrow \, \mathrm{Sp}(2\bar n, \mathbb{R}) \label{embeddif}
\end{equation}
having denoted by ${\bar n}$ the total number of vector fields that
is displayed in table \ref{topotable}
  \item In $D=6$ we have self-dual $2$--forms. Also here
  we have a few different possibilities depending on the number
  $(\mathcal{N}_+,\mathcal{N}_-)$ of left and right handed
  supersymmetries with a variable number  ${\bar n}$
  of $2$--forms. In particular for the $(2,2)$ theory that originates
  from type IIA compactifications the scalar manifold is:
\begin{equation}
  {\cal G}/{\cal H}= \frac{\mathrm{O(4,n)}}{\mathrm{O(4)}\times
  \mathrm{O(n)}}\times \mathrm{O(1,1)}
\label{IIAd6}
\end{equation}
while for the $(4,0)$ theory that originates from type IIB
compactifications the scalar manifold is the following:
\begin{equation}
  {\cal G}/{\cal H}= \frac{\mathrm{O(5,n)}}{\mathrm{O(5)}\times
  \mathrm{O(n)}}
\label{IIBd6}
\end{equation}
Finally in the case of ($\mathcal{N}_+=2,\mathcal{N}_-=0$)
supergravity, the scalar manifold is
\begin{equation}
  \mathcal{M}_{scalar}= \frac{\mathrm{O(1,n)}}{\mathrm{O(n)}} \times \mathcal{QM}
\label{20d6}
\end{equation}
the first homogeneous factor $\frac{\mathrm{O(1,n)}}{\mathrm{O(n)}}$
containing the scalars of the tensor multiplets, while the second
factor denotes a generic \emph{quaternionic manifold} that contains
the scalars of the hypermultiplets.
In all cases the relevant embedding is
\begin{equation}
\iota_\delta : \, \mbox{Diff}\left ({{\cal G}\over{\cal H}} \right )
\, \longrightarrow \, \mathrm{SO}({\bar n}, {\bar n}) \label{embed6}
\end{equation}
where ${\bar n}$ is the total number of $2$--forms, namely:
\begin{equation}
  \cases{ {\bar n} = 4+n \quad \mbox{for the $(2,2)$ theory} \cr
  {\bar n} = 5+n \quad \mbox{for the $(4,0)$ theory} \cr
  {\bar n} = 1+n \quad \mbox{for the $(2,0)$ theory} \cr}
\label{nbarro}
\end{equation}
 \item In $D=8$ we have self--dual three--forms. There are two
 theories. The first is maximally extended $N=2$ supergravity
 where the number of three--forms is ${\bar n}=3$ and the scalar coset
 manifold is:
\begin{equation}
  {\cal G}/{\cal H}= \frac{\mathrm{SL}(3,\mathbb{R})}{\mathrm{O(3)}}
  \times\frac{\mathrm{SL}(2,\mathbb{R})}{\mathrm{O(2)}}
\label{d8max}
\end{equation}
The second theory is $N=1$ supergravity that contains ${\bar n}=1$
three--forms and where the scalar coset is:
\begin{equation}
  {\cal G}/{\cal H}=\frac{\mathrm{SO(2,n)}}{\mathrm{SO(2)}\times
  \mathrm{SO(n)}}\times \mathrm{O(1,1)}
\label{d8n1}
\end{equation}
having denoted $n=\# \mbox{of vector multiplets}$.
In the two cases the relevant embedding is symplectic and
specifically it is:
\begin{equation}
\iota_\delta : \, \mbox{Diff}\left ({{\cal G}\over{\cal H}} \right )
\, \longrightarrow \, \cases {\mathrm{Sp}(6,  \mathbb{R}) \quad
\mbox{maximal supergravity}\cr
\mathrm{Sp}(2,  \mathbb{R}) \quad\mbox{$N=1$ supergravity}\cr}
 \label{embed8}
\end{equation}
\end{enumerate}
\subsection{Symplectic embeddings in general}
Let us begin with the case of symplectic embeddings relevant to $D=4$
and $D=8$ theories.
\par
Focusing on the isometry group of the canonical metric
defined on ${{\cal G}\over{\cal H}}$\footnotemark
\footnotetext{Actually, in order to be true, eq.(\ref{isogroup})
requires that that the normaliser of ${\cal H}$ in ${\cal G}$ be the
identity group, a condition that is verified in all the relevant
examples}:
\begin{equation}
 {\cal G}_{iso}\left ({{\cal G}\over{\cal H}}\right ) \, = \, {\cal G}
 \label{isogroup}
\end{equation}
we must consider the embedding:
\begin{equation}
\iota_\delta : \,  {\cal G}  \, \longrightarrow \, \mathrm{Sp}(2\bar n,
\mathbb{R}) \label{embediso}
\end{equation}
That in eq.(\ref{embeddif}) is a homomorphism of finite dimensional
Lie groups and as such it constitutes a problem that can be solved in
explicit form. What we just need to know is the dimension of the
symplectic group, namely the number $\bar n$ of
$\frac{D-4}{2}$--forms
appearing in the theory. Without supersymmetry the dimension $m$ of
the scalar manifold (namely the possible choices of ${{\cal
G}\over{\cal H}}$) and the number of vectors $\bar n$ are unrelated
so that the possibilities covered by eq.(\ref{embediso}) are
infinitely many. In supersymmetric theories, instead, the two numbers
$m$ and $\bar n$ are related, so that there are finitely many cases
to be studied corresponding to the possible  embeddings of given
groups ${\cal G}$ into a symplectic group $\mathrm{Sp}(2\bar n, \mathbb{R})$
of fixed dimension $\bar n$. Actually taking into account further
conditions on the holonomy of the scalar manifold that are also
imposed by supersymmetry, the solution for the symplectic embedding
problem is unique for all extended supergravities  as
we have already remarked. In $D=4$ this yields the unique scalar manifold
choice displayed in table~\ref{topotable}, while in the other
dimensions gives the  results recalled above.
\par
Apart from the details of the specific case considered once a
symplectic embedding is given there is a general formula one can
write down for the {\it period matrix} ${\cal N}$ that guarantees
symmetry (${\cal N}^T = {\cal N}$) and the required transformation
property (\ref{covarianza}). This is the first result I want to
present.
\par
The real symplectic group $\mathrm{Sp}(2\bar n ,\mathbb{R})$ is defined as the
set of all {\it real} $2\bar n \times 2\bar n$ matrices
\begin{equation}
\Lambda \, = \, \left ( \matrix{ A & B \cr C & D \cr } \right )
\label{matriciana}
\end{equation}
satisfying the first of equations (\ref{ortosymp}), namely
\begin{equation}
\Lambda^T \, \IC \, \Lambda \, = \, \IC \label{condiziona}
\end{equation}
 where
\begin{equation}
\IC  \, \equiv \, \left ( \matrix{ {\bf 0} & \bfone \cr -\bfone &
{\bf 0} \cr } \right ) \label{definizia}
\end{equation}
If we relax the condition that the matrix should be real but we still
impose eq.(\ref{condiziona}) we obtain the definition of the complex
symplectic group $\mathrm{Sp}(2\bar n, \IC)$. It is a well known fact that the
following isomorphism is true:
\begin{equation}
\mathrm{Sp}(2\bar n, \mathbb{R})  \sim  \mathrm{\mathrm{USp}}(\bar n , \bar n)   \equiv
\mathrm{Sp}(2\bar n, \IC)   \cap   \mathrm{U}(\bar n , \bar n) \label{usplet}
\end{equation}
By definition an element ${\cal S}\,\in \, \mathrm{\mathrm{USp}}(\bar n , \bar
n)$ is a complex matrix that satisfies simultaneously
eq.(\ref{condiziona}) and a pseudounitarity condition, that is:
\begin{eqnarray}
{\cal S}^T \, \IC \, {\cal S} &=& \IC \quad ; \quad {\cal S}^\dagger \,
\IH \, {\cal S} = \IH \quad ; \quad  \IH  \equiv   \left ( \matrix{
\bfone & {\bf 0} \cr {\bf 0} & -\bfone
 \cr } \right )
\label{uspcondo}
\end{eqnarray}
The general block form of the matrix ${\cal S}$ is:
\begin{equation}
{\cal S}\, = \, \left ( \matrix{ T & V^\star \cr V & T^\star \cr }
\right ) \label{blocusplet}
\end{equation}
and eq.s (\ref{uspcondo}) are equivalent to:
\begin{eqnarray}
T^\dagger \, T \, - \, V^\dagger \, V &=& \bfone  \quad ; \quad
T^\dagger \, V^\star  \, - \,  V^\dagger \, T^\dagger = {\bf 0}
\label{relazie}
\end{eqnarray}
The isomorphism of eq.(\ref{usplet}) is explicitly realized by the so
called Cayley matrix:
\begin{equation}
{\cal C} \, \equiv \, {\frac{1}{\sqrt{2}}} \, \left ( \matrix{ \bfone
& {\rm i}\bfone \cr \bfone & -{\rm i}\bfone
 \cr } \right )
\label{cayley}
\end{equation}
via the relation:
\begin{equation}
{\cal S}\, = \, {\cal C} \, \Lambda \, {\cal C}^{-1} \label{isomorfo}
\end{equation}
which yields:
\begin{eqnarray}
T &=& {\frac{1}{2}}\, \left ( A - {\rm i} B \right ) +
{\frac{1}{2}}\, \left ( D + {\rm i} C \right ) \quad ; \quad V =
{\frac{1}{2}}\, \left ( A - {\rm i} B \right ) - {\frac{1}{2}}\,
\left ( D + {\rm i} C \right )  \label{mappetta}
\end{eqnarray}
When we set $V=0$ we obtain the subgroup $\mathrm{U}(\bar n) \subset \mathrm{USp} (\bar
n , \bar n)$, that in the real basis is given by the subset of
symplectic matrices of the form $\left ( \matrix{ A & B \cr -B & A
 \cr } \right )$. The basic idea, to obtain the
general formula for the period matrix, is that the symplectic
embedding of the isometry group ${\cal G}$ will be such that the
isotropy subgroup ${\cal H}\subset {\cal G}$ gets embedded into the
maximal compact subgroup $\mathrm{U}(\bar n)$, namely:
\begin{eqnarray}
{\cal G} & {\stackrel{\iota_\delta}{\longrightarrow}} & \mathrm{USp} (\bar n ,
\bar n) \qquad ; \qquad {\cal G} \supset {\cal H}
{\stackrel{\iota_\delta}{\longrightarrow}}  \mathrm{U}(\bar n) \subset \mathrm{USp}
(\bar n , \bar n) \label{gruppino}
\end{eqnarray}
If this condition is realized let $\mathbb{L}(\phi)$ be a
parametrization of the coset ${\cal G}/{\cal H}$ by means of coset
representatives. By this we mean the following. Let $\phi^I$ be local
coordinates on the manifold ${\cal G}/{\cal H}$: to each point $\phi
\in {\cal G}/{\cal H}$ we assign an element $\mathbb{L}(\phi) \in
{\cal G}$ in such a way that if $\phi^\prime \ne \phi$, then no $h
\in {\cal H}$ can exist such that
$\mathbb{L}(\phi^\prime)=\mathbb{L}(\phi)\cdot h$. In other words for
each equivalence class of the coset (labelled by the coordinate
$\phi$) we choose one representative element $\mathbb{L}(\phi)$ of
the class. Relying on the symplectic embedding of eq.(\ref{gruppino})
we obtain a map:
\begin{eqnarray}
& \mathbb{L}(\phi)  \, \longrightarrow  {\cal O}(\phi)\, =  \,
\left ( \matrix{ U_0(\phi) & U^\star_1(\phi) \cr
U_1(\phi) & U^\star_0(\phi) \cr } \right )\,  \in  \,
\mathrm{USp}(\bar n , \bar n) &\label{darstel}
\end{eqnarray}
that associates to $\mathbb{L}(\phi)$ a coset representative of
$\mathrm{USp}(\bar n , \bar n)/\mathrm{U}(\bar n)$. By construction if
$\phi^\prime \ne \phi$ {\it no} unitary $\bar n \times \bar n$ matrix
$W$ {\it can exist} such that:
\begin{equation}
 {\cal O}(\phi^\prime)  =  {\cal O}(\phi) \,
 \left ( \matrix{ W & {\bf 0} \cr {\bf 0}
& W^\star \cr } \right )
\end{equation}
On the other hand let $\xi \in {\cal G}$ be an element of the
isometry group of ${{\cal G}/{\cal H}}$. Via the symplectic embedding
of eq.(\ref{gruppino}) we obtain a $\mathrm{USp}(\bar n, \bar n)$
matrix
\begin{equation}
{\cal S}_ \xi \, = \, \left ( \matrix{ T_\xi & V^\star_\xi \cr V_\xi
& T^\star_\xi \cr } \right ) \label{uspimag}
\end{equation}
such that
\begin{equation}
{\cal S}_ \xi \,{\cal O}(\phi) \, = \, {\cal O}(\xi(\phi)) \, \left (
\matrix{ W(\xi,\phi) & {\bf 0} \cr {\bf 0} & W^\star(\xi,\phi) \cr }
\right ) \label{cosettone}
\end{equation}
where $\xi(\phi)$ denotes the image of the point $\phi \in  {{\cal
G}/{\cal H}}$ through $\xi$ and $W(\xi,\phi)$ is a suitable $\mathrm{U}(\bar
n)$ compensator depending both on $\xi$ and $\phi$. Combining
eq.s (\ref{cosettone}),(\ref{darstel}), with eq.s (\ref{mappetta}) we
immediately obtain:
\begin{eqnarray}
U_0^\dagger \left( \xi(\phi) \right ) + U^\dagger_1 \left (\xi(\phi)
\right)  & = &   W^\dagger   \left [ U_0^\dagger \left(
\phi \right )   \left ( A^T + {\rm i}B^T \right ) + U_1^\dagger
\left( \phi \right )   \left ( A^T - {\rm i}B^T \right ) \right ]
\nonumber\\ U_0^\dagger \left( \xi(\phi) \right ) - U^\dagger_1
\left (\xi(\phi) \right)  & = &  W^\dagger \, \left [
U_0^\dagger \left( \phi \right )   \left ( D^T - {\rm i}C^T \right )
- U_1^\dagger \left( \phi \right )   \left ( D^T + {\rm i}C^T \right
) \right ]   \label{semitrasform}
\end{eqnarray}
Setting:
\begin{equation}
{\cal N} \, \equiv \, {\rm i} \left [ U_0^\dagger + U_1^\dagger
\right ]^{-1} \, \left [ U_0^\dagger - U_1^\dagger \right ]
\label{masterformula}
\end{equation}
and using the result of eq.(\ref{semitrasform}) one verifies that the
transformation rule (\ref{covarianza}) is verified. It is also an
immediate consequence of the analogue of eq.s (\ref{relazie}) satisfied
by $U_0$ and $U_1$ that the matrix in eq.(\ref{masterformula}) is symmetric
\begin{equation}
{\cal N}^T \, = \, {\cal N} \label{massi}
\end{equation}
Eq. (\ref{masterformula}) is the masterformula derived in 1981 by
Gaillard and Zumino \cite{gaizum}. It explains the structure of the
gauge field kinetic terms in all $N\ge 3$ extended supergravity
theories and also in those $N=2$ theories where,
the {\it special K\"ahler manifold} ${\cal SM}$
is a homogeneous manifold ${\cal G}/{\cal H}$. Similarly it applies to the kinetic terms of
the three--forms in $D=8$. In particular, using
eq. (\ref{masterformula}) we can easily retrieve the structure of $\mathcal{N}=4$
supergravity, on which I have more to say in the sequel.
Actually, given the information (following from $\mathcal{N}=4$ supersymmetry)
that the scalar manifold is the following coset manifold (see
table~\ref{topotable}):
\begin{eqnarray}
{\cal M}^{N=4}_{scalar} & = & {\cal ST}\left [ 6,n \right ]
\nonumber\\ {\cal ST}\left [ m,n \right ] & \equiv &
{\frac{\mathrm{SU}(1,1)}{\mathrm{U}(1)}} \, \otimes \, {\frac{\mathrm{SO}(m,n)}
{\mathrm{SO}(m)\otimes
\mathrm{SO}(n)}} \nonumber\\ \null & \null & \null \label{stmanif}
\end{eqnarray}
what we just need to study is the symplectic embedding of the coset
manifolds ${\cal ST}\left [ 6,n \right ]$ where $n$ is the number
vector multiplets in the theory. This is what I do   next
  considering the general case of  ${\cal
ST}\left [ m,n \right ]$ manifolds.
\subsection{Symplectic embedding of the ${\cal ST}\left [ m,n \right ]$
homogeneous manifolds} The first thing I should do is to justify the
name I have given to the particular class of coset manifolds I
propose to study. The letters ${\cal ST}$ stand for space--time and
target space duality. Indeed, the isometry group of the ${\cal
ST}\left [ m,n \right ]$ manifolds defined in eq.(\ref{stmanif})
contains a factor ($\mathrm{SU}(1,1)$) whose transformations act as
non--perturbative $S$--dualities and another factor $(\mathrm{SO}(m,n)$ whose
transformations act as $T$--dualities, holding true at each order in
string perturbation theory. Furthermore $S$ is the traditional name
given, in superstring theory, to the complex field obtained by
combining together the {\it dilaton} $D$ and {\it axion} ${\cal A}$:
\begin{eqnarray}
{\cal S} & = & {\cal A} - {\rm i} \mbox{exp}[D]  \qquad ; \qquad
\partial^\mu {\cal A}  \equiv  \varepsilon^{\mu\nu\rho\sigma} \,
\partial_\nu \, B_{\rho\sigma}
\label{scampo}
\end{eqnarray}
while $t^i$ is the name usually given to the moduli--fields of the
compactified target space. Now in string and supergravity
applications $S$ is identified with the complex coordinate on
the manifold ${\frac{\mathrm{SU}(1,1)}{\mathrm{U}(1)}}$, while  $t^i$ are the
coordinates of the coset space  ${\frac{\mathrm{SO}(m,n)}{\mathrm{SO}(m)\otimes
\mathrm{SO}(n)}}$. Although as differentiable and metric manifolds the spaces
${\cal ST}\left [ m,n \right ]$ are just direct products of two
factors (corresponding to the above mentioned different physical
interpretation of the coordinates $S$ and $t^i$), from the point of
view of the symplectic embedding and duality rotations they have to
be regarded as a single entity. This is even more evident in the case
$m=2,n=\mbox{arbitrary}$, where the following theorem has been proven
by Ferrara and Van Proeyen \cite{ferratoine}: ${\cal ST}\left [ 2,n
\right ]$ are the only special K\"ahler manifolds with a direct
product structure. For the definition of special K\"ahler manifolds I
refer to \cite{mylecture}.
\paragraph{Moduli spaces of string compactifications and discrete duality groups}
In
the $\mathcal{N}=4$ case to make direct contact with string theory compactifications,
I can recall that the tree--level moduli space of the heterotic
string compactified on a $T^6$ torus is
\begin{equation}
{\cal M}^{N=4}_{moduli} \, = \, {  {\cal ST}}\left [ 6,22 \right ]
\label{n4moduli}
\end{equation}
the number of abelian gauge fields being $22= 6\, \mbox{(moduli of
$T^6$) } \oplus    16 \, \mbox{ (rank of $E_8 \times E_8$ )}$.
Because of the uniqueness of $\mathcal{N}=4$ supergravity the quantum
moduli--space ${\hat  {\cal ST}}\left [ 6,22 \right ]$ cannot be
anything else but a manifold with the same covering space as ${
{\cal ST}}\left [ 6,22 \right ]$, namely a manifold with the same
local structure. Indeed the only thing which is not fixed by $\mathcal{N}=4$
supersymmetry is the global structure of the scalar manifold.  What
actually comes out is the following result
\begin{equation}
{\hat  {\cal ST}}\left [ 6,22 \right ] \, = \, {\frac{{  {\cal
ST}}\left [ 6,22 \right ] }{\mathrm{SL}(2,\mathbb{Z})\otimes
\mathrm{SO}(6,22,\mathbb{Z})}} \label{qn4mod}
\end{equation}
The homotopy group of the quantum moduli space:
\begin{equation}
 \pi_1 \left ( {\hat  {\cal ST}}\left [ 6,22 \right ] \right ) \,= \,
 {\mathrm{SL}(2,\mathbb{Z})\otimes \mathrm{SO}(6,22,\mathbb{Z})}
\label{homotop}
\end{equation}
is just the restriction to the integers $\mathbb{Z}$ of the original
continuous duality group $\mathrm{SL}(2,\mathbb{R})\otimes
\mathrm{SO}(6,22,\mathbb{R})$ associated with the manifold ${  {\cal ST}}\left
[ 6,22 \right ]$. After modding by this discrete group the
duality--rotations that survive as exact duality symmetries of the
quantum theory are those contained in $\pi_1 \left ( {\hat  {\cal
ST}}\left [ 6,22 \right ] \right )$ itself. This happens because of
the Dirac quantization condition ~\ref{dirquant} of electric and
magnetic charges, the lattice spanned by these charges being
invariant under the discrete group~\ref{homotop}. At this junction
the relevance, in the quantum theory, of the symplectic embedding
should appear. What does restriction to the integers exactly, mean?
It means that the image in $\mathrm{Sp}(56,\mathbb{R})$ of those matrices of
$\mathrm{SL}(2,\mathbb{R}) \times \mathrm{SO}(6,22,\mathbb{R})$ that are
retained as elements of $\pi_1 \left ( {\hat  {\cal ST}}\left [ 6,22
\right ] \right )$ should be integer valued. In other words we
define:
\begin{eqnarray}
  \mathrm{SL}(2,\mathbb{Z}) \times \mathrm{SO}(6,22,\mathbb{Z}) &
  \equiv &
 \iota_\delta \left ( \mathrm{SL}(2,\mathbb{R}) \times \mathrm{SO}(6,22,\mathbb{R}) \right )
  \cap
 \mathrm{Sp}(56,\mathbb{Z})
\label{astrazeta}
\end{eqnarray}
As we see the statement in eq. (\ref{astrazeta}) is dependent on the
symplectic embedding. What is integer valued in one embedding is not
integer valued in another embedding. This raises the question of the
correct symplectic embedding. Such a question has two aspects:
\begin{enumerate}
\item{Intrinsically inequivalent embeddings}
\item{Symplectically equivalent embeddings that become inequivalent
after gauging}
\end{enumerate}
The first issue in the above list is group--theoretical in nature.
When we say that the group ${\cal G}$ is embedded into $\mathrm{Sp}(2\bar
n,\mathbb{R})$ we must specify how this is done from the point of
view of irreducible representations. Group--theoretically the matter
is settled by specifying how the fundamental representation of
$\mathrm{Sp}(2\bar n)$ splits into irreducible representations of ${\cal G}$:
\begin{eqnarray}
& {\bf {2 \bar n}} \, {\stackrel{{\cal G}}{\longrightarrow}}
\oplus_{i=1}^{\ell} \, {\bf D}_i & \label{splitsplit}
\end{eqnarray}
Once eq. (\ref{splitsplit}) is given (in supersymmetric theories such
information is provided by supersymmetry ) the only arbitrariness
which is left is that of conjugation by arbitrary $\mathrm{Sp}(2\bar
n,\mathbb{R})$ matrices. Suppose we have determined an embedding
$\iota_delta$ that obeys law (\ref{splitsplit}), then:
\begin{equation}
\forall \, {\cal S} \, \in \, \mathrm{Sp}(2\bar n,\mathbb{R}) \, : \,
\iota_\delta^\prime \, \equiv \, {\cal S} \circ  \iota_\delta \circ
{\cal S}^{-1} \label{matrim}
\end{equation}
will obey the same law. That in eq. (\ref{matrim}) is a symplectic
transformation that corresponds to an allowed
duality--rotation/field--redefinition in the abelian theory of
type (\ref{gaiazuma}) discussed in the previous subsection. Therefore
all abelian lagrangians related by such transformations are
physically equivalent.
\paragraph{Gaugings and Embeddings}
The matter changes in presence of {\it gauging}. When we switch on
the gauge coupling constant and the electric charges, symplectic
transformations cease to yield physically equivalent theories. This
is the second issue in the above list. The choice of a symplectic
gauge becomes physically significant. As I have emphasized in the
introduction, the construction of supergravity theories proceeds in
two steps. In the first step, which is the most extensive and
complicated, one constructs the abelian theory: at that level the
only relevant constraint is that encoded in eq.(\ref{splitsplit}) and
the choice of a symplectic gauge is immaterial. Actually one can
write the entire theory in such a way that {\it symplectic
covariance} is manifest. In the second step one {\it gauges} the
theory. This {\it breaks symplectic covariance} and the choice of the
correct symplectic gauge becomes a physical issue.
\par
These facts being cleared I proceed to discuss the symplectic
embedding of the ${\cal ST}\left [ m,n \right ]$ manifolds.
\par
Let $\eta$ be the symmetric flat metric with signature $(m,n)$ that
defines the $\mathrm{SO}(m,n)$ group, via the relation
\begin{equation}
L \, \in \, \mathrm{SO}(m,n) \, \Longleftrightarrow \, L^T \, \eta L \, = \,
\eta \label{ortogruppo}
\end{equation}
Both in the $\mathcal{N}=4$ and in the $\mathcal{N}=2$ theory, the number of gauge fields
is given by:
\begin{equation}
\# \mbox{vector fields} \, = \, m \oplus n \label{vectornum}
\end{equation}
$m$ being the number of {\it graviphotons}, namely of vectors contained in the graviton multiplet
and $n$ being the number of {\it
vector multiplets}. Hence we have to embed $\mathrm{SO}(m,n)$ into
$\mathrm{Sp}(2m+2n,\mathbb{R})$ and the explicit form of the decomposition in
eq.(\ref{splitsplit}) required by supersymmetry is:
\begin{equation}
{\bf {2m+2n}} \, {\stackrel{\mathrm{SO}(m,n)}{\longrightarrow}} \, {\bf {
m+n}} \oplus  {\bf { m+n}} \label{ortosplitsplit}
\end{equation}
where ${\bf { m+n}}$ denotes the fundamental representation of
$\mathrm{SO}(m,n)$. Eq.(\ref{ortosplitsplit}) is easily understood in physical
terms. $\mathrm{SO}(m,n)$ must be a T--duality group, namely a symmetry
 holding true order by order in perturbation theory. As such it must
 rotate electric  field strengths into electric field strengths and
 magnetic field strengths into magnetic field  strengths. The
 two irreducible representations into which the  fundamental
 representation of the symplectic group decomposes when reduced to
 $\mathrm{SO}(m,n)$ correspond precisely to the electric and magnetic sectors,
 respectively.
In the {\it simplest  gauge} the symplectic embedding satisfying
eq.(\ref{ortosplitsplit}) is block--diagonal and takes the form:
\begin{eqnarray}
\forall \,  L \, \in \, \mathrm{SO}(m,n) & {\stackrel{\iota_\delta}
{\hookrightarrow}} &  \left ( \matrix{ L & {\bf 0}\cr
{\bf 0} & (L^T)^{-1} \cr } \right ) \, \in \,
\mathrm{Sp}(2m+2n,\mathbb{R})
\label{ortoletto}
\end{eqnarray}
Consider instead the group $\mathrm{SU}(1,1) \sim \mathrm{SL}(2,\mathbb{R})$.
This is the factor in the isometry group of ${\cal ST}[m,n]$ that is
going to act by means of S--duality non perturbative rotations.
Typically it will rotate each electric field strength into its
homologous magnetic one. Correspondingly supersymmetry implies that
its embedding into the symplectic group must satisfy the following
condition:
\begin{equation}
{\bf {2m+2n}} \,
{\stackrel{\mathrm{SL}(2,\mathbb{R})}{\longrightarrow}} \,
\oplus_{i=1}^{m+n} \, {\bf 2} \label{simposplisplit}
\end{equation}
where  ${\bf 2}$ denotes the fundamental representation of
$\mathrm{SL}(2,\mathbb{R})$. In addition it must commute with the
embedding (\ref{ortoletto}) of $\mathrm{SO}(m,n)$. Both conditions are
fulfilled by setting:
\begin{eqnarray}
 \forall \,   \left ( \matrix{a & b \cr  c &d \cr }\right ) \, \in
\, \mathrm{SL}(2,\mathbb{R}) & {\stackrel{\iota_\delta}
{\hookrightarrow}} &  \left ( \matrix{ a \, \bfone & b \,
\eta \cr c \, \eta & d \, \bfone \cr } \right ) \, \in \,
\mathrm{Sp}(2m+2n,\mathbb{R})  \label{ortolettodue}
\end{eqnarray}
Utilizing eq.s (\ref{isomorfo}) the corresponding  embeddings into the
group $\mathrm{USp}(m+n,m+n)$ are immediately derived:
\begin{eqnarray}
 \forall \,  L \, \in \, \mathrm{SO}(m,n) & {\stackrel{\iota_\delta}
{\hookrightarrow}} &  \left ( \matrix{ {\frac{1}{2}}
\left ( L+ \eta L \eta \right ) & {\frac{1}{2}}  \left ( L- \eta L
\eta \right )\cr {\frac{1}{2}}  \left ( L - \eta L \eta \right ) &
{\frac{1}{2}} \left ( L+ \eta L \eta \right ) \cr } \right ) \, \in \,
\mathrm{USp}(m+n,m+n)  \nonumber\\   \forall
\,   \left ( \matrix{t & v^\star \cr  v &t^\star \cr }\right ) \, \in
\, \mathrm{SU}(1,1) & {\stackrel{\iota_\delta} {\hookrightarrow}} &
 \left ( \matrix{ {\rm Re}t \bfone +{\rm i}{\rm Im}t\eta
& {\rm Re}v \bfone -{\rm i}{\rm Im}v \eta   \cr {\rm Re}v \bfone
+{\rm i}{\rm Im}v\eta & {\rm Re}t \bfone - {\rm i}{\rm Im}t\eta \cr }
\right ) \, \in \, \mathrm{USp}(m+n,m+n)\nonumber\\
\label{uspembed}
\end{eqnarray}
where the relation between the entries of the $\mathrm{SU}(1,1)$ matrix and
those of the corresponding $\mathrm{SL}(2,\mathbb{R})$ matrix are
provided by the relation (\ref{mappetta}).
\par
Equipped with these relations we can proceed to derive the explicit
form of the {\it period matrix} ${\cal N}$.
\par
The homogeneous manifold $\mathrm{SU}(1,1)/\mathrm{U}(1)$ can be conveniently
parametrized in terms of a single complex coordinate $S$, whose
physical interpretation will be that of {\it axion--dilaton},
according to eq. (\ref{scampo}). The coset parametrization appropriate
for comparison with other constructions (Dimensional reduction (see~\cite{senlecture}))
is given by the matrices:
\begin{eqnarray}
 M({\cal S}) & \equiv & {\frac{1}{n({\cal S})} } \, \left (
\matrix{ \bfone & { \frac{{\rm i} -{\cal S} }{ {\rm i} + {\cal S} }
}\cr {\frac{ {\rm i} + {\bar {\cal S}} }{ {\rm i} -{\bar {\cal S}} }
} & \bfone \cr} \right ) \quad ; \quad  n({\cal S}) \, \equiv \,
\sqrt{ {\frac{4 {\rm Im}{\cal S} } {
 1+\vert {\cal S} \vert^2 +2 {\rm Im}{\cal S} } } }
 \label{su11coset}
\end{eqnarray}
To parametrize the coset $\mathrm{SO}(m,n)/\mathrm{SO}(m)\times \mathrm{SO}(n)$ we can instead
take the usual coset representatives (see for
instance~\cite{castdauriafre}):
\begin{equation}
\mathbb{L}(X) \, \equiv \, \left (\matrix{ \left ( \bfone + XX^T
\right )^{1/2} & X \cr X^T & \left ( \bfone + X^T X \right )^{1/2}\cr
} \right ) \label{somncoset}
\end{equation}
where the $m \times n $ real matrix $X$ provides a set of independent
coordinates. Inserting these matrices into the embedding formulae of
eq.s (\ref{uspembed}) we obtain a matrix:
\begin{eqnarray}
 \mathrm{USp}(n+m , n+m) \, \ni  \, \iota_\delta \left ( M (S)
\right ) \circ \iota_\delta \left ( \mathbb{L}(X) \right )  \,
 = \, \left ( \matrix{ U_0({\cal S},X) & U^\star_1({\cal
S},X) \cr U_1({\cal S},X) & U^\star_0({\cal S},X) \cr } \right )
\label{uspuspusp}
\end{eqnarray}
that inserted into the master formula (\ref{masterformula}) yields the
following result:
\begin{equation}
{\cal N}\, = \, {\rm i} {\rm Im}{\cal S} \, \eta \mathbb{L}(X) L^T(X)
\eta + {\rm Re}{\cal S} \, \eta \label{maestrina}
\end{equation}
Alternatively, remarking that if $\mathbb{L}(X)$ is an $\mathrm{SO}(m,n)$
matrix also $\mathbb{L}(X)^\prime =\eta \mathbb{L}(X) \eta$ is such a
matrix and represents the same equivalence class, we can rewrite
(\ref{maestrina}) in the simpler form:
\begin{equation}
{\cal N}\, = \, {\rm i} {\rm Im}{\cal S} \,   \mathbb{L}(X)^\prime
\mathbb{L}^{T\prime} (X) + {\rm Re}{\cal S} \, \eta \label{maestrino}
\end{equation}
%%%%%%%%%%%%%%%%%%%%%%%%%%%%%%%%%%
% Insert d5gen %%%%%%%%%%%%%%%%%%%
%%%%%%%%%%%%%%%%%%%%%%%%%%%%%%%%%%%
\section[Supergravities in five dimension and more scalar geometries]
{Supergravities in five dimension and more scalar geometries}
\label{minid5geo}
\setcounter{equation}{0}
The recent renewed interest in five--dimensional gauged supergravities
stems from two developments. On one hand we have the $AdS_5/CFT_4$
correspondence between
\begin{enumerate}
  \item [a] superconformal gauge theories in $D=4$, viewed as
the world volume description of a stack of $\mathrm{D3}$--branes
  \item [b] type IIB supergravity compactified on  $AdS_5$ times a five--dimensional
internal manifold $X^5$ which yields a \textbf{gauged} supergravity
model in $D=5$
\end{enumerate}
On the other hand we have the quest for supersymmetric realizations
of the Randall-Sundrum scenarios which also correspond to \textbf{domain
wall} solutions of appropriate $D=5$ gauged supergravities.
It is, however, noteworthy that five dimensional supergravity has a long and
interesting history. The minimal theory ($\mathcal{N}=2$ ), whose field content is given
by the metric $g_{\mu \nu }$, a doublet of pseudo Majorana gravitinos
$\psi_{A\mu}$ ($A=1,2$) and a vector boson $A_\mu$ was constructed
twenty years ago \cite{D'Auria:1981kq}  as the first non--trivial
example of a rheonomic construction\footnote{We leave aside pure
$\mathcal{N}=1,D=4$ supergravity that from the rheonomic viewpoint
is a completely trivial case.}. This simple model remains to the
present day the unique example of a perfectly geometric theory where,
notwithstanding the presence of a gauge boson $A_\mu$, the action
can be written solely in terms of differential forms and wedge
products without introducing Hodge duals. This feature puts pure
$D=5$ supergravity into a selective club of few  ideal theories whose
other members are just pure gravity in arbitrary dimension and pure
$\mathcal{N}=1$ supergravity in four dimensions. The miracle that
allows the boson $A_\mu$ to propagate without introducing its kinetic
term is due to the conspiracy of first order formalism for the spin
connection $\omega^{ab}$ together with the presence of two
Chern--Simons terms. The first Chern Simons is the standard gauge
one:
\begin{equation}
   {CS}_{gauge} = F\wedge F\wedge A
\label{purchsi}
\end{equation}
while the second  is a mixed, gravitational-gauge  Chern Simons that
reads as follows
\begin{equation}
  {CS}_{mixed} = T^a \wedge F\wedge V_a
\label{mixchsi}
\end{equation}
where $V^a$ is the vielbein and $T^a=\mathcal{D}V^a$ is its
\emph{curvature}, namely the torsion.
\par
The possible matter multiplets for $\mathcal{N}=2,D=5$ are the
\textbf{vector/tensor} multiplets and the \textbf{hypermultiplets}.
The field content of the first type of multiplets is the following
one:
\begin{equation}
    \left\{ \begin{array}{cccll}
    A^{I}_\mu & \null & \null & (\mathbf{I}=1,\dots,n_V) & \mbox{vectors} \\
    \null & \lambda^i_A & \phi^i & (i=1,\dots,n_V+n_T \equiv n) & (A=1,2) \\
    B^\mathcal{M}_{\mu \nu } & \null & \null &(\mathcal{M}=1,\dots,n_T)&  \mbox{tensors} \
  \end{array} \right\}
\label{tensvect}
\end{equation}
where  by $n_V$ I have denoted the number of  vectors or gauge
$1$--forms $A^{I}_\mu$,  $n_T$ being instead the number of tensors
or gauge $2$--forms $B^\mathcal{M}_{\mu \nu }=-B^\mathcal{M}_{\nu \mu
}$. In ungauged supergravity, where everything is abelian, vectors and
tensors are equivalent since they can be dualised into each other by
means of the transformation:
\begin{equation}
  \partial _{[\mu} \, A _{\nu ]}=\epsilon _{\mu \nu }^{\phantom{\mu \nu
  }\lambda \rho \sigma } \, \partial _\lambda B_{\rho \sigma }
\label{dual2to3}
\end{equation}
but in gauged supergravity it is only the $1$--forms that can be
promoted to non--abelian gauge vectors while the $2$--forms  describe
massive degrees of freedom. The other members of each vector/tensor
multiplet are a doublet of pseudo Majorana spin 1/2 fields:
\begin{equation}
\lambda^i_A=\epsilon^{AB} \, \mathcal{C} \, \left(
\overline{\lambda}^{iB}\right) ^T \, \quad ; \quad
\overline{\lambda}^{iB}= \left( \lambda_{B}^i\right) ^\dagger
\gamma_0 \quad ; \quad
 ~~~A,B=1,\dots,2\,.
\end{equation}
and a real scalar $\phi ^i$.
The field content of hypermultiplets is the following:
\begin{equation}
  \mbox{hypermultiplets}=\left \{ q^u \,(u=1,\dots,4 \,m )\, ,
  \zeta^\alpha \, (\alpha = 1, \dots \, 2m) \right \}
\label{hypmulcont}
\end{equation}
where, having denoted $m$ the number of hypermultiplets, $q^u$ are $m$
quadruplets of real scalar fields and $\zeta^\alpha$ are $m$ doublets
of pseudo Majorana spin 1/2 fields:
\begin{equation}
\zeta^\alpha=\mathbb{C}^{\alpha \beta } \, \mathcal{C} \, \left(
\overline{\zeta}_{\beta }\right) ^T \, \quad ; \quad
\overline{\zeta}_{\beta }= \left( \zeta^{\beta} \right) ^\dagger
\gamma_0 \quad ; \quad
 ~~~\alpha ,\beta =1,\dots,2\,m
\end{equation}
the matrix $\mathbb{C}^T=-\mathbb{C}$, $\mathbb{C}^2 =-{\bf 1}$ being
the symplectic invariant metric of $\mathrm{Sp}(2m,\mathbb{R})$.
\par
In the middle of the eighties Gunaydin Sierra and Townsend \cite{GST1,GST2} considered
the general structure of $ \mathcal{N}=2,D=5$ supergravity coupled to an
arbitrary number $n=n_V+n_T$ of  vector/tensor multiplets and discovered
that this is dictated by a peculiar geometric structure imposed by
supersymmetry on the scalar manifold $\mathcal{SV}_n$ that contains
the $\phi^i$ scalars. In  modern nomenclature this
peculiar geometry  is named \textbf{very special geometry} and $\mathcal{SV}_n$
are referred to as real \textbf{very special manifolds}.  The
characterizing property of very special geometry arises from the need
to reconcile the transformations of the scalar members of each
multiplet with those of the vectors in presence of the Chern-Simons
term (\ref{purchsi}) which generalizes to:
\begin{equation}
  \mathcal{L}^{CS} = \frac{1}{8} \, d_{\Lambda\Sigma\Gamma}
  F^\Lambda_{\mu \nu } \, F^\Sigma_{\rho \sigma } \, A^\Gamma_\tau \,
  \epsilon ^{\mu \nu \rho \sigma \tau }
\label{Csmult}
\end{equation}
the symbol $ d_{\Lambda\Sigma\Gamma}$ denoting some appropriate
constant symmetric tensor and, having dualised all $2$--forms to
vectors, the range of the indices $\Lambda,\Sigma,\Gamma$ being:
\begin{equation}
  \Lambda= 1,\dots,n+1 \, = \left \{\underbrace{\, 0\, , \, I }_{ \mathbf{I}}\,,
   , \, \mathcal{M}\right\}
\label{Lamrang}
\end{equation}
Indeed the total number of vector fields, including the graviphoton
that belongs to the graviton multiplet, is always $n+1$, $n$ being
the number of vector multiplets. It turns out that  very special geometry
is completely defined in terms of the constant tensors $ d_{\Lambda\Sigma\Gamma}$
that are further restricted by a condition ensuring positivity of
the energy. At the beginning of the nineties special manifolds were
classified and thoroughly studied by de Wit, Van Proeyen and some
other collaborators \cite{deWit:1992nm,deWit:1993wf,deWit:1995tf}
who also explored the dimensional reduction from $D=5$ to $D=4$,
clarifying the way \emph{real very special geometry} is mapped into the
\emph{special K\"ahler geometry} featured by vector multiplets in $D=4$ and
generically related to Calabi--Yau moduli spaces.
\par
The $4m$ scalars  of the hypermultiplet sector have instead exactly
the same geometry in $D=4$ as in $D=5$ dimensions, namely they fill
a quaternionic manifold $\mathcal{QM}$.  These scalar geometries are
a crucial ingredient in the construction of both the ungauged and the
gauged supergravity lagrangians. Indeed the  basic operations
involved by the gauging procedure are based on the specific
geometric structures of very special and quaternionic manifolds, in
particular the crucial existence of a moment--map (see sect.\ref{momentm1}).
For this reason the present section is devoted to a summary of these
geometries and to an illustration of the general form of the bosonic
$D=5$ lagrangians. Yet, before entering these mathematical topics, I
want to recall the structure of maximally extended ($\mathcal{N}=8$)
supergravity in the same dimensions. Indeed  it
will be fruitful, in the next chapter, to see the general structure
of the gauged theories and compare the $\mathcal{N}=8$ with the $\mathcal{N}=2 $
case within a unified framework.
\par
As explained in section \ref{maxsugra} (see in particular table
\ref{maxsugrat}) the scalar manifold of maximal supergravity in
five-dimensions is the $42$--dimensional  homogeneous space:
\begin{equation}
  M_{scalar}^{max} = \frac{\mathrm{E_{6(6)}}}{\mathrm{USp(8)}}
\label{maxd5scalma}
\end{equation}
The holonomy subgroup $ \mathcal{H}=\mathrm{USp(8)}$ is the
largest invariance group of complex linear transformations that
respects the pseudo--Majorana condition on the $8$ gravitino
$1$--forms:
\begin{equation}
\psi^A=\Omega^{AB} \, \mathcal{C} \, \left( \overline{\psi}_A\right) ^T \,
 ~~~A=1,\dots,8\,.
\end{equation}
where $\Omega^{AB}=-\Omega^{BA}$ is an antisymmetric $8 \times 8$
matrix such that $\Omega^2=-{\bf 1}$. Using these notations where
the capital Latin indices transform in the fundamental $8$--representation
of $\mathrm{USp(8)}$ we can summarize the field content of the
$\mathcal{N}=8$  graviton multiplet as:
\begin{enumerate}
  \item the graviton field, namely the f\"unfbein 1--form $V^a$
  \item eight gravitinos $\psi^A \equiv \psi^A_\mu \, dx^\mu$ in the
$\bf 8$ representation of $\mathrm{USp(8)}$
  \item $27$ vector fields
$A^{\Lambda} \equiv A^{\Lambda} _\mu \, dx^\mu$ in the $\bf 27$ of
 $\mathrm{E_{6(6)}}$\footnote{In the ungauged theory all two--forms
 have been dualized to vectors}
 \item $48$ dilatinos $\chi^{ABC}$ in the $\bf 48$ of $\mathrm{USp(8)}$
 \item $42$ scalars $\phi$ that
parametrize the coset manifold $E_{\left(6\right)6}/\mathrm{USp(8)}$.
They appear in the theory through the coset representative
$\mathbb{L}_{\Lambda}^{~AB}(\phi)$, which is regarded as covariant
in the $\bf (27,\overline{{27}})$ of $\mathrm{USp(8)}\times
\mathrm{E_{6(6)}}$. This means the following. Since the fundamental
${\bf 27}$ (real) representation of $\mathrm{E_{6(6)}}$ remains irreducible
under reduction to the subgroup
$\mathrm{USp(8)}\subset\mathrm{E_{6(6)}}$ it follows that there exists a
constant intertwining $27 \times 27$ matrix $\mathcal{I}_\Sigma^{AB}$
that transforms the index $\null^\Sigma$ running in the fundamental of
$\mathrm{E_{6(6)}}$ into an antisymmetric pair of indices $\null^{AB}$ with
the additional property that ${\null}^{AB} \, \Omega_{AB}=0$ which is
the definition of the $27$ of $\mathrm{USp(8)}$. The
coset representative we use is to be interpreted as
$\mathbb{L}_{\Lambda}^{~AB}(\phi)=\mathbb{L}_{\Lambda}^\Sigma
\,\mathcal{I}_\Sigma^{AB}$.
\end{enumerate}
The construction of the ungauged theory proceeds then through well
established general steps and the basic ingredients, namely the
$\mathrm{USp(8)}$ connection in the $36$ adjoint representation
$\mathcal{Q}_A^{\phantom{A}B}$ and the scalar vielbein
$\mathcal{P}^{ABCD}$ (fully antisymmetric in $ABCD$) are extracted
from the left--invariant $1$--form on the scalar coset according to:
\begin{eqnarray}
\mathbb{L}^{-1~\Lambda}_{AB}d\mathbb{L}_{\Lambda}^{~CD} &=&
\mathcal{Q}_{AB}^{~~~CD}+\mathcal{P}_{AB}^{~~~CD}\, \nonumber\\
\mathcal{Q}_{AB}^{~~~CD}&=& 2\, \delta_{[A}^{[C} \,
\mathcal{Q}_{B]}^{\phantom{C} D]} \nonumber \\
\mathcal{P}_{AB}^{~~~CD}&=&\Omega_{AE} \Omega_{BF} \,\mathcal{P}^{EFCD}
\label{defungaugedconnection}
\end{eqnarray}
which is fully analogous to eq.(\ref{uspYconnec}) that applies to the
$D=4$ case and to the $\mathrm{E_{7(7)}/SU(8)}$ coset.
\par
Independently from the number of supersymmetries we can write a
general form for the bosonic action of any $D=5$ \emph{ungauged
supergravity}, namely the following one:
\begin{eqnarray}
  \mathcal{L}^{(ungauged)}_{(D=5)}&=&\sqrt{-g} \, \left( \frac{1}{2}
  \,R \, - \, \frac{1}{4}\,\mathcal{N}_{\Lambda\Sigma} F^\Lambda_{\mu \nu
  } \, F^{\Sigma \vert\mu \nu} + \frac{1}{2} \, g_{ij} \, \partial
  _\mu\phi^i \, \partial ^\mu \, \phi^j \right ) \nonumber\\
  &&+ \frac{1}{8} d_{\Lambda\Sigma\Gamma} \, \epsilon ^{\mu \nu \rho \sigma \tau
  } \, F^\Lambda_{\mu \nu } \, F^\Sigma_{\rho \sigma } \, A^\Gamma_\tau
\label{genford5bos}
\end{eqnarray}
where $g_{ij}$ is the metric of the scalar manifold $\mathcal{M}_{scalar}$ ,
$\mathcal{N}_{\Lambda\Sigma}(\phi)$ is a positive definite symmetric function of the
scalars that under the isometry group $\mathcal{G}_{iso}$ of $\mathcal{M}_{scalar}$ transforms
in $\bigotimes^2_{sym} \mathbf{R}$, having denoted by $ \mathbf{R}$ a
linear representation of $\mathcal{G}_{iso}$ to which the vector fields $A^\Gamma$
are assigned. Finally $d_{\Lambda\Sigma\Gamma}$ is a three--index
symmetric tensor invariant with respect to the representation
$\mathbf{R}$.
At this point we invite the reader to compare the above statements
with the general discussion of section \ref{howGsuH}, in particular
points \textbf{B} and \textbf{C}. As stated in eq. (\ref{HHprime})
the automorphism group of $\mathcal{N}$--extended supersymmetry (which in $D=5$
is $\mathrm{USp}(\mathcal{N})$ due to pseudo Majorana fermions) must
be contained as a factor in the holonomy group of the scalar
manifold. On the other hand the $p_i+1$--forms must be assigned to linear
representations $D_i$ of the isometry group for
$\mathcal{M}_{scalar}$. In our case having dualised the two forms we
just have vectors, namely $p+1=1$--forms and the representation $\mathbf{R}$
is the only $D_i$ we need to discuss. In the four--dimensional case
the construction of the lagrangian was mainly dictated by the symplectic embedding
of eq.(\ref{embediso}). Indeed, since  the $1$--forms
are self--dual in $D=4$, then the isometries of the scalar manifolds must
be realized on the vectors as symplectic duality symmetries, according to the general
discussion of section \ref{dualsym}. In five dimensions, where no such
duality symmetry can be realized, the isometry of the scalar manifold
has to be linearly realized on the vectors in such a way as to make it
an \textbf{exact symmetry} of the lagrangian (\ref{genford5bos}).
This explains while the kinetic matrix $\mathcal{N}$ must transform in the
representation $\bigotimes^2_{sym} \mathbf{R}$
\par
In maximal $\mathcal{N}=8$ supergravity the items involved in the construction of the bosonic
lagrangian have the following values:
\begin{enumerate}
  \item  The scalar metric is the $E_{6(6)}$ invariant metric on
  the coset (\ref{maxd5scalma}), namely:
\begin{equation}
  g_{ij} =\frac{1}{6} P^{ABCD}_i \, P_{ABCD\vert j}
\label{scalmete6}
\end{equation}
  \item The vector kinetic metric is given by the following quadratic
  form in terms of the coset representative:
\begin{equation}
  \mathcal{N}_{\Lambda\Sigma} = 4\, \left(
  \mathbb{L}_\Lambda^{\phantom{\Lambda} AB} \, \mathbb{L}_\Sigma^{\phantom{\Lambda} CD}
  \, \Omega_{AC} \, \Omega_{BD} \right)
\label{veckinmet}
\end{equation}
  \item The representation $\mathbf{R}$ is the fundamental $27$ of $E_{6(6)}$
  \item The tensor $d_{\Lambda\Sigma\Gamma}$ is the coefficient of
  the cubic invariant of $E_{6(6)}$ in the $27$ representation.
\end{enumerate}
To see how the same items are realized in the case of an $\mathcal{N}=2$ theory
we have to introduce very special and quaternionic geometry. Just
before entering this it is worth nothing that also the supersymmetry
transformation rule of the gravitino field admits a general form
(once restricted to the purely bosonic terms), namely:
\begin{equation}
  \delta \psi_{A\mu} = \mathcal{D}_\mu \, \epsilon_A - \frac{1}{3} \,
  \mathcal{T}_{AB}^{\rho \sigma } \left( g_{\mu \rho } \, \gamma _\sigma
  - \frac{1}{8}\, \epsilon _{\mu \rho \sigma \lambda \nu } \, \gamma
  ^{\lambda \nu }   \right) \, \epsilon ^B
\label{gensusrul}
\end{equation}
where the indices $A,B$ run in the fundamental representation of the
automorphism (R-symmetry) group $\mathrm{USp}(\mathcal{N})$ and the tensor
$\mathcal{T}_{AB}^{\rho \sigma }$, antisymmetric both in $AB$ and in
$\rho \sigma $ and named the graviphoton field strength, is given by:
\begin{equation}
  \mathcal{T}_{AB}^{\rho \sigma } = \Phi^\Lambda_{AB}(\phi) \,
  \mathcal{N}_{\Lambda\Sigma} F^{\Sigma\vert \rho \sigma }
\label{gravfotft}
\end{equation}
the scalar field dependent tensor $X^\Lambda_{AB}(\phi)$ being
intrinsically defined as the coefficient of the term ${\bar \epsilon
}^A \, \psi^B_\mu$ in the supersymmetry transformation rule of the
vector field $A^\Lambda_\mu$, namely:
\begin{equation}
  \delta A^\Lambda_\mu = \dots + 2 \, \mbox{i}\,
  \Phi^\Lambda_{AB}(\phi) \, {\bar \epsilon
}^A \, \psi^B_\mu
\label{CapFi}
\end{equation}
From its own definition it follows that under isometries of the scalar manifold
$\Phi^\Lambda_{AB}(\phi)$ must transform in the representation
$\mathbf{R}$ of $\mathcal{G}_{iso}$ times $\bigwedge^2 \mathcal{N}$
of the R-symmetry $\mathrm{USp}(\mathcal{N})$. In the case of $\mathcal{N}=8$
supergravity the tensor $\Phi^\Lambda_{AB}(\phi)$ is simply the
inverse coset representative:
\begin{equation}
  \Phi^\Lambda_{AB}(\phi)= \left(\mathbb{L}^{-1}
  \right)_{AB}^{\phantom{AB} \Lambda}
\label{PhiN8}
\end{equation}
We see in the next subsection how the same object is generally
realized in an $\mathcal{N}=2$ theory via very special geometry.
%%%%%%%%%%%%%%%%%%%%%%%%%%%%%%
\subsection[Very special geometry]{Very special geometry}
\emph{Very special geometry} is the peculiar metric and associated
Riemannian  structure that can be constructed on a very special
manifold. By definition a \emph{very special manifold}
$\mathcal{VS}_n$ is a real manifold of dimension $n$ that can be
represented as the following algebraic locus in $\mathbb{R}^{n+1}$:
\begin{equation}
  1= \mathrm{N}(X) \, \equiv \, \sqrt{ \, d_{\Lambda\Sigma\Delta} \,
  X^\Lambda \, X^\Sigma \, X^\Delta }
\label{defispec}
\end{equation}
where $X^\Lambda$ ($\Lambda=1,\dots,n+1$) are the coordinates of $\mathbb{R}^{n+1}$
while
\begin{equation}
d_{\Lambda\Sigma\Delta} \label{symtens}
\end{equation}
is a \textbf{constant symmetric tensor} fulfilling some additional
defining properties that I will recall later on.
\par
A coordinate system $\phi^i$ on $\mathcal{VS}_n$ is provided by any
parametric solution of eq.(\ref{defispec}) such that:
\begin{equation}
  X^\Lambda = X^\Lambda (\phi) \quad ; \quad \phi^i = \mbox{free}
  \quad ; \quad i=1,\dots , n
\label{coordspec}
\end{equation}
The \emph{very special metric} on the very special manifold is
nothing else but the pull--back
on the algebraic surface (\ref{defispec}) of the following $\mathbb{R}^{n+1}$ metric:
\begin{eqnarray}
  ds^2_{\mathbb{R}^{n+1}}&=&
  \mathcal{N}_{\Lambda\Sigma} \, dX^\Lambda \otimes dX^\Sigma
  \label{rn+1met}\\
\mathcal{N}_{\Lambda\Sigma} & \equiv & - \partial _\Lambda \partial _\Sigma \, \ln \,
\mathrm{N}(X)
\label{scriptNsp}
\end{eqnarray}
In other words in any coordinate frame the coefficients of the very
special metric are the following ones:
\begin{equation}
  g_{ij}(\phi) = \mathcal{N}_{\Lambda\Sigma} \, f_i^\Lambda \, f_j^\Sigma
\label{specmet}
\end{equation}
where we have introduced the new objects:
\begin{equation}
  f_i^\Lambda \, \equiv \, \partial _i X^\Lambda = \frac{\partial }{\partial
  \phi^i} X^\Lambda
\label{fidefi}
\end{equation}
If we also define
\begin{equation}
  F_\Lambda = \frac{\partial }{\partial X^\Lambda} \, \ln \,
  \mathrm{N(X)} \quad ; \quad h_{\Lambda i}\equiv \, \partial _i F_\Lambda
\label{Flam}
\end{equation}
and introduce the $2(n+1)$-vectors:
\begin{equation}
  U = \left( \begin{array}{c}
    X^\Lambda \\
    F_\Sigma \
  \end{array} \right)  \quad; \quad U_i = \partial_i U =  \left( \begin{array}{c}
    f^\Lambda_i \\
    h_{\Sigma i} \
  \end{array} \right)
\label{UiUvec}
\end{equation}
taking a second covariant derivative it can be shown that the
following identity is true:
\begin{equation}
  \nabla_i U_j = \frac{2}{3} \, g_{ij} U + \sqrt{\frac{2}{3}} \,
  T_{ijk} \, g^{k\ell} \, U_\ell
\label{dUide}
\end{equation}
where the world--index symmetric coordinate dependent tensor
$T_{ijk}$ is related to the constant tensor $d_{\Lambda\Gamma\Sigma}$
by:
\begin{equation}
  d_{\Lambda\Gamma\Sigma}=\frac{20}{27} \, F_\Lambda \, F_\Gamma
  \,F_\Sigma \, -\, \frac{2}{3} \, \mathcal{N}_{(\Lambda\Gamma} \,
  F_{\Sigma)} + \frac{8}{27} T_{ijk} \, g^{ip} g^{jq} g^{kr} \,
  h_{\Lambda p} \, h_{\Gamma q} \, h_{\Sigma r}
\label{Tidedide}
\end{equation}
The identity (\ref{dUide}) is the real counterpart of a completely
similar identity that holds true in special K\"ahler geometry and
also defines a symmetric $3$--index tensor.  In the use  of very
special geometry to construct a supersymmetric field theory the
essential property is the existence of the \emph{section}
$X^\Lambda(\phi)$. Indeed it is this object that allows the
writing of the tensor $\Phi^\Lambda_{AB}(\phi)$ appearing in the
vector transformation rule (\ref{CapFi}). It suffice to set:
\begin{equation}
  \Phi^\Lambda_{AB}(\phi) = X^\Lambda(\phi) \, \epsilon _{AB}
\label{Phiforn2}
\end{equation}
Why do we call it a section? Since it is just a section of a
\textbf{flat vector bundle} of rank $n+1$
\begin{equation}
  \mathrm{FB} \stackrel{\pi}{\rightarrow} \mathcal{SV}_n
\label{flatbundvsg}
\end{equation}
with base manifold the very special manifold and structural group
some subgroup of the $n+1$ dimensional linear group:
$\mathcal{G}_{iso}\subset \mathrm{GL}(n+1,\mathbb{R})$. The bundle is flat because
the transition functions from one local trivialization to another one
are constant matrices:
\begin{equation}
  \forall g \in \mathcal{G}_{iso} \quad : \quad X^\Lambda (g \, \phi) = \left(
  M[g]\right )^\Lambda_{\phantom{\Lambda}\Sigma} \,X^\Sigma (\phi) \quad ;
  \quad M[g]=\mbox{constant matrix}
\label{urbano}
\end{equation}
The structural group $\mathcal{G}_{iso}$ is implicitly defined as the set of
matrices $M$ that leave the $d_{\Lambda\Gamma\Sigma}$ tensor
invariant:
\begin{equation}
M\in \mathcal{G}_{iso} \quad \Leftrightarrow \quad  M_{\Lambda_1}^{\phantom{\Lambda}\Sigma_1} \,
  M_{\Lambda_2}^{\phantom{\Lambda}\Sigma_2}\, M_{\Lambda_3}^{\phantom{\Lambda}\Sigma_3}
  \, d_{\Lambda_1 \Lambda_2 \Lambda_3}\,=\, d_{\Sigma_1 \Sigma_2
  \Sigma_3}
\label{defiGs}
\end{equation}
Since the very special metric is defined by eq.(\ref{specmet}) it immediately
follows that $\mathcal{G}_{iso}$ is also the isometry group of  such a metric, its
action in any coordinate patch (\ref{coordspec}) being defined by the
action (\ref{urbano}) on the section $X^\Lambda$. This fact explains
the name given to this group.
\par
By means of this reasoning I have shown that the classification of
very special manifolds is fully reduced to the classification of the
constant tensors $d_{\Lambda\Gamma\Sigma}$ such that their group of invariances
acts transitively on the manifold $\mathcal{SV}_n$ defined by  eq.
(\ref{defispec}) and that the special metric (\ref{specmet}) is positive definite.
This is the algebraic problem that was completely solved by de Wit
and Van Proeyen in \cite{deWit:1992nm}. They found all such tensors
and the corresponding manifolds. There is a large subclass of very
special manifolds that are homogeneous spaces but there are also infinite families
of manifolds that are not $\mathcal{G}/\mathcal{H}$ cosets.
%%%%%%%%%%%%%%%%%%%%%%%%%%
\subsection{The very special geometry of the $\mathrm{SO(1,1)} \times
\mathrm{SO(1,n)/SO(n)}$ manifold}
 As an  example of very special manifold we consider the following class
 of homogeneous spaces:
\begin{equation}
  \mathcal{RT}[n] \equiv \mathrm{SO(1,1)} \times \frac{\mathrm{SO(1,n)}}{\mathrm{SO(n)}}
\label{RTn}
\end{equation}
This example is particularly simple and relevant
to string theory because reducing it on a circle $S^1$  from five to
four dimensions one finds a supergravity  model where the special K\"ahler
geometry is that of the $\mathcal{ST}[2,n]$ manifolds described in the previous
sections.\par
To see that the $\mathcal{RT}[n]$ are indeed very special manifolds
we consider the following instance of cubic norm:
\begin{eqnarray}
  \mathrm{N(X)}&=& \sqrt{ \mathrm{C(X)}} \label{norm1}\\
  \mathrm{C(X)}&=& X^0 \left( X^+X^- - \mathbf{X}^2 \right) \quad ; \quad \mathbf{X}^2=
  \sum_{\ell=1}^{r} \,\left(  X^\ell\right) ^2
\label{norm2}
\end{eqnarray}
It is immediately verified that the infinitesimal linear transformations
$X^\Lambda \rightarrow X^\Lambda + \delta X^\Lambda$ that
leave the cubic polynomial $\mathrm{C(X)}$ invariant are the following ones:
\begin{eqnarray}
 \delta_0 \left( \begin{array}{c}
    X^0 \\
    X^+ \\
    X^- \\
    \mathbf{X} \
  \end{array} \right) &=&  \left(
  \begin{array}{c|ccc}
    -4 & 0 & 0 & 0 \\
    \hline
    0 & 2 & 0 & 0 \\
    0 & 0 & 2 & 0 \\
    0 & 0 & 0 & 2 \\
  \end{array}
  \right)   \left( \begin{array}{c}
    X^0 \\
    X^+ \\
    X^- \\
    \mathbf{X} \
  \end{array} \right)
\label{delDelt} \\
\delta_L \left( \begin{array}{c}
    X^0 \\
    X^+ \\
    X^- \\
    \mathbf{X} \
  \end{array} \right) &=&  \left(
  \begin{array}{c|ccc}
    0 & 0 & 0 & 0 \\
    \hline
    0 & 0 & -4 & 0 \\
    0 & 4 & 0 & 0 \\
    0 & 0 & 0 & 0 \\
  \end{array}
  \right)   \left( \begin{array}{c}
    X^0 \\
    X^+ \\
    X^- \\
    \mathbf{X} \
  \end{array} \right)
\label{delL} \\
\delta_\mathbf{v} \left( \begin{array}{c}
    X^0 \\
    X^+ \\
    X^- \\
    \mathbf{X} \
  \end{array} \right) &=&  \left(
  \begin{array}{c|ccc}
    0 & 0 & 0 & 0 \\
    \hline
    0 & 0 & 0 & \mathbf{v}^T \\
    0 & 0 & 0 & 0 \\
    0 & 0 & \mathbf{v} & 0 \\
  \end{array}
  \right)   \left( \begin{array}{c}
    X^0 \\
    X^+ \\
    X^- \\
    \mathbf{X} \
  \end{array} \right)
\label{delv} \\
\delta_\mathbf{u} \left( \begin{array}{c}
    X^0 \\
    X^+ \\
    X^- \\
    \mathbf{X} \
  \end{array} \right) &=&  \left(
  \begin{array}{c|ccc}
    0 & 0 & 0 & 0 \\
    \hline
    0 & 0 & 0 &  0 \\
    0 & 0 & 0 & \mathbf{v}^T \\
    0 & \mathbf{v} & 0 & 0 \\
  \end{array}
  \right)   \left( \begin{array}{c}
    X^0 \\
    X^+ \\
    X^- \\
    \mathbf{X} \
  \end{array} \right)
\label{delu} \\
\delta_\mathbf{A} \left( \begin{array}{c}
    X^0 \\
    X^+ \\
    X^- \\
    \mathbf{X} \
  \end{array} \right) &=&  \left(
  \begin{array}{c|ccc}
    0 & 0 & 0 & 0 \\
    \hline
    0 & 0 & 0 &  0 \\
    0 & 0 & 0 & 0 \\
    0 & 0 & 0 & \mathbf{A} \\
  \end{array}
  \right)   \left( \begin{array}{c}
    X^0 \\
    X^+ \\
    X^- \\
    \mathbf{X} \
  \end{array} \right) \quad ; \quad \mathbf{A}^T = - \mathbf{A} \,
  \in  \,  \mathrm{SO(r)}
\label{delA}
\end{eqnarray}
The transformation $\delta_\Delta$ generates an $\mathrm{SO(1,1)}$ group
that commutes with the $\mathrm{SO(1,r+1)}$ group generated by the
transformations $\delta_L,\delta _\mathbf{u},\delta _\mathbf{v}$ and
$\delta _\mathbf{A}$, hence the symmetry group of the symmetric
tensor:
\begin{equation}
  d_{\Lambda\Sigma\Gamma}=\cases{d_{0+-}=1\cr
d_{0\ell m}=-\delta_{\ell m} \cr
0 \qquad \mbox{otherwise}\cr}
\label{symtensd5}
\end{equation}
defined by the cubic polynomial $\mathrm{C(X)}$ is indeed the group
$\mathrm{SO(1,1)\times SO(1,n)}$. This is quite simple and evident.
What is important is that the same group has also a transitive action
on the manifold defined by the equation $\mathrm{C(X)}=1$ that can be
identified with the product $\mathrm{SO(1,1)} \times
\mathrm{SO(1,n)/SO(2)}$. To verify this statement it suffices to
consider that the quadratic equation
\begin{equation}
  H^+ H^- - \mathbf{H}^2 =1
\label{H+H-}
\end{equation}
defines the homogeneous manifold $\mathrm{SO(1,n)/SO(2)}$ on which $\mathrm{SO(1,n)}$
has a transitive action. For instance we can use as independent $r+1$
coordinates the following ones:
\begin{equation}
  \phi^0 =H^+ \quad; \quad \phi^\ell = H^\ell \,\, (\ell=1,\dots,r)
  \quad \Rightarrow \quad H^- =\frac{1+\mathbf{\phi}^2}{\phi^0}
\label{phicoord}
\end{equation}
and then it suffices to set:
\begin{equation}
  X^0[\sigma,\phi] = e^{-2\sigma} \quad ; \quad \left( X^+,X^-,\mathbf{X}\right)
  = e^{\sigma} \, \left( H^+[\phi], \, H^-[\phi],\, \mathbf{H}[\phi] \right )
\label{lillo}
\end{equation}
to obtain a parametrization of the section $X$ in terms of
coordinates $\sigma ,\phi$ of the manifold $\mathrm{SO(1,1)} \times
\mathrm{SO(1,n)/SO(2)}$. This achieves the desired proof that the
group $\mathcal{G}_{iso}$ has a transitive action on the special
manifold and consequently that the cubic norm
(\ref{norm1}), (\ref{norm2}) is admissible as a definition of a very
special manifold
%%%%%%%%%%%%%%%%%%%%%%%%%%%%
% Hypergeometry %%%%%%%%%%%%
%%%%%%%%%%%%%%%%%%%%%%%%%%%%
\subsection{Quaternionic Geometry}
\label{hypgeosec}
Next I turn my attention to the hypermultiplet sector of an
$\mathcal{N}=2$ supergravity. For these multiplets no distinction arises
between the $D=4$ and $D=5$. Each hypermultiplet contains $4$ real scalar fields
and, at least locally, they can be regarded as the
four components of a quaternion. The locality caveat is, in this
case, very substantial because global quaternionic coordinates can be
constructed only occasionally even on those manifolds that are
denominated quaternionic in the mathematical literature
\cite{alex}, \cite{gal}. Anyhow, what is important  is that, in
the hypermultiplet sector, the scalar manifold $\mathcal{QM}$ has
dimension multiple of four:
\begin{equation}
\mbox{dim}_{\bf R} \, \mathcal{QM} \, = \, 4 \, m \,\equiv \, 4 \, \# \,
\mbox{of hypermultiplets}
\label{quatdim}
\end{equation}
and, in some appropriate sense, it has a quaternionic structure.
\par
We name {\it Hypergeometry} that pertaining to the
hypermultiplet sector, irrespectively whether we deal with global or
local $\mathcal{N}$=2 theories. Yet there are two kinds of hypergeometries.
Supersymmetry requires the existence
of a principal $\mathrm{SU}(2)$--bundle
\begin{equation}
{\cal SU} \, \longrightarrow \, \mathcal{QM} \label{su2bundle}
\end{equation}
The bundle ${\cal SU}$ is
{\bf flat} in the {\it rigid supersymmetry case} while its curvature is
proportional to the K\"ahler forms in the {\it local case}.
\par
These two versions of hypergeometry were already known in mathematics prior to
their use \cite{dWLVP}, \cite{skgsugra_13}, \cite{D'Auria:1991fj},
\cite{sabarwhal}, \cite{vanderseypen}  in the context of $\mathcal{N}=2$
supersymmetry and are identified as:
\begin{eqnarray}
\mbox{rigid hypergeometry} & \equiv & \mbox{HyperK\"ahler geometry.}
\nonumber\\ \mbox{local hypergeometry} & \equiv & \mbox{quaternionic
geometry} \label{picchio}
\end{eqnarray}
\subsubsection{Quaternionic, versus HyperK\"ahler manifolds}
Both a quaternionic or a HyperK\"ahler manifold $\mathcal{QM}$
is a $4 m$-dimensional real manifold endowed with a metric $h$:
\begin{equation}
d s^2 = h_{u v} (q) d q^u \otimes d q^v   \quad ; \quad u,v=1,\dots,
4  m \label{qmetrica}
\end{equation}
and three complex structures
\begin{equation}
(J^x) \,:~~ T(\mathcal{QM}) \, \longrightarrow \, T(\mathcal{QM}) \qquad
\quad (x=1,2,3)
\end{equation}
that satisfy the quaternionic algebra
\begin{equation}
J^x J^y = - \delta^{xy} \, \bfone \,  +  \, \epsilon^{xyz} J^z
\label{quatalgebra}
\end{equation}
and respect to which the metric is hermitian:
\begin{equation}
\forall   \mbox{\bf X} ,\mbox{\bf Y}  \in   T\mathcal{QM}   \,: \quad h
\left( J^x \mbox{\bf X}, J^x \mbox{\bf Y} \right )   = h \left(
\mbox{\bf X}, \mbox{\bf Y} \right ) \quad \quad
  (x=1,2,3)
\label{hermit}
\end{equation}
From eq. (\ref{hermit}) it follows that one can introduce a triplet
of 2-forms
\begin{equation}
\begin{array}{ccccccc}
K^x& = &K^x_{u v} d q^u \wedge d q^v & ; & K^x_{uv} &=&   h_{uw}
(J^x)^w_v \cr
\end{array}
\label{iperforme}
\end{equation}
that provides the generalization of the concept of K\"ahler form
occurring in  the complex case. The triplet $K^x$ is named the {\it
HyperK\"ahler} form. It is an $\mathrm{SU}(2)$ Lie--algebra valued
2--form  in the same way as the K\"ahler form is a $U(1)$
Lie--algebra valued 2--form. In the complex case the definition of
K\"ahler manifold involves the statement that the K\"ahler 2--form is
closed. At the same time in Hodge--K\"ahler manifolds (those
appropriate to local supersymmetry in $D=4$ ) the K\"ahler 2--form can be
identified with the curvature of a line--bundle which in the case of
rigid supersymmetry is flat. Similar steps can be taken also here and
lead to two possibilities: either HyperK\"ahler or quaternionic
manifolds.
\par
Let us  introduce a principal $\mathrm{SU}(2)$--bundle ${\cal SU}$ as
defined in eq. (\ref{su2bundle}). Let $\omega^x$ denote a connection
on such a bundle. To obtain either a HyperK\"ahler or a quaternionic
manifold we must impose the condition that the HyperK\"ahler 2--form
is covariantly closed with respect to the connection $\omega^x$:
\begin{equation}
\nabla K^x \equiv d K^x + \epsilon^{x y z} \omega^y \wedge K^z    \,
= \, 0 \label{closkform}
\end{equation}
The only difference between the two kinds of geometries resides in
the structure of the ${\cal SU}$--bundle.
\begin{definizione} A
HyperK\"ahler manifold is a $4 m$--dimensional manifold with the
structure described above and such that the ${\cal SU}$--bundle is
{\bf flat}
\end{definizione}
 Defining the ${\cal SU}$--curvature by:
\begin{equation}
\Omega^x \, \equiv \, d \omega^x + {1\over 2} \epsilon^{x y z}
\omega^y \wedge \omega^z \label{su2curv}
\end{equation}
in the HyperK\"ahler case we have:
\begin{equation}
\Omega^x \, = \, 0 \label{piattello}
\end{equation}
Viceversa \begin{definizione} A quaternionic manifold is a $4
m$--dimensional manifold with the structure described above and such
that the curvature of the ${\cal SU}$--bundle is proportional to the
HyperK\"ahler 2--form \end{definizione} Hence, in the quaternionic
case we can write:
\begin{equation}
\Omega^x \, = \, { {\lambda}}\, K^x \label{piegatello}
\end{equation}
where $\lambda$ is a non vanishing real number.
\par
As a consequence of the above structure the manifold $\mathcal{QM}$ has
a holonomy group of the following type:
\begin{eqnarray}
{\rm Hol}(\mathcal{QM})&=& \mathrm{SU}(2)\otimes {\cal H} \quad
(\mbox{quaternionic}) \nonumber \\ {\rm Hol}(\mathcal{QM})&=& \bfone
\otimes {\cal H} \quad (\mbox{HyperK\"ahler}) \nonumber \\ {\cal H} &
\subset & Sp (2m,\mathbb{R}) \label{olonomia}
\end{eqnarray}
In both cases, introducing flat indices $\{A,B,C= 1,2\}
\{\alpha,\beta,\gamma = 1,.., 2m\}$  that run, respectively, in the
fundamental representations of $\mathrm{SU}(2)$ and
$\mathrm{Sp}(2m,\mathbb{R})$, we can find a vielbein 1-form
\begin{equation}
{\cal U}^{A\alpha} = {\cal U}^{A\alpha}_u (q) d q^u
\label{quatvielbein}
\end{equation}
such that
\begin{equation}
h_{uv} = {\cal U}^{A\alpha}_u {\cal U}^{B\beta}_v
\mathbb{C}_{\alpha\beta}\epsilon_{AB} \label{quatmet}
\end{equation}
where $\mathbb{C}_{\alpha \beta} = - \mathbb{C}_{\beta \alpha}$ and $\epsilon_{AB}
= - \epsilon_{BA}$ are, respectively, the flat $\mathrm{Sp}(2m)$ and
$\mathrm{Sp}(2) \sim \mathrm{SU}(2)$ invariant metrics. The vielbein
${\cal U}^{A\alpha}$ is covariantly closed with respect to the
$\mathrm{SU}(2)$-connection $\omega^z$ and to some
$\mathrm{Sp}(2m,\mathbb{R})$-Lie Algebra valued connection
$\Delta^{\alpha\beta} = \Delta^{\beta \alpha}$:
\begin{eqnarray}
\nabla {\cal U}^{A\alpha}& \equiv & d{\cal U}^{A\alpha} +{i\over 2}
\omega^x (\epsilon \sigma_x\epsilon^{-1})^A_{\phantom{A}B}
\wedge{\cal U}^{B\alpha} \nonumber\\ &+& \Delta^{\alpha\beta} \wedge
{\cal U}^{A\gamma} \mathbb{C}_{\beta\gamma} =0 \label{quattorsion}
\end{eqnarray}
\noindent where $(\sigma^x)_A^{\phantom{A}B}$ are the standard Pauli
matrices. Furthermore ${ \cal U}^{A\alpha}$ satisfies  the reality
condition:
\begin{equation}
{\cal U}_{A\alpha} \equiv ({\cal U}^{A\alpha})^* = \epsilon_{AB}
\mathbb{C}_{\alpha\beta} {\cal U}^{B\beta} \label{quatreality}
\end{equation}
Eq.(\ref{quatreality})  defines  the  rule to lower the symplectic
indices by means   of  the  flat  symplectic   metrics
$\epsilon_{AB}$   and $\mathbb{C}_{\alpha \beta}$. More specifically we can
write a stronger version of eq. (\ref{quatmet}) \cite{sugkgeom_3}:
\begin{eqnarray}
({\cal U}^{A\alpha}_u {\cal U}^{B\beta}_v + {\cal U}^{A\alpha}_v
{\cal
 U}^{B\beta}_u)\mathbb{C}_{\alpha\beta}&=& h_{uv} \epsilon^{AB}\nonumber\\
 \label{piuforte}
\end{eqnarray}
\noindent We have also the inverse vielbein ${\cal U}^u_{A\alpha}$
defined by the equation
\begin{equation}
{\cal U}^u_{A\alpha} {\cal U}^{A\alpha}_v = \delta^u_v \label{2.64}
\end{equation}
Flattening a pair of indices of the Riemann tensor ${\cal
R}^{uv}_{\phantom{uv}{ts}}$ we obtain
\begin{equation}
{\cal R}^{uv}_{\phantom{uv}{ts}} {\cal U}^{\alpha A}_u {\cal
U}^{\beta B}_v = -\,{{\rm i}\over 2} \Omega^x_{ts} \epsilon^{AC}
 (\sigma_x)_C^{\phantom {C}B} \mathbb{C}^{\alpha \beta}+
 \mathbb{R}^{\alpha\beta}_{ts}\epsilon^{AB}
\label{2.65}
\end{equation}
\noindent where $\mathbb{R}^{\alpha\beta}_{ts}$ is the field strength
of the $\mathrm{Sp}(2m) $ connection:
\begin{equation}
d \Delta^{\alpha\beta} + \Delta^{\alpha \gamma} \wedge \Delta^{\delta
\beta} \mathbb{C}_{\gamma \delta} \equiv \mathbb{R}^{\alpha\beta} =
\mathbb{R}^{\alpha \beta}_{ts} dq^t \wedge dq^s \label{2.66}
\end{equation}
Eq. (\ref{2.65}) is the explicit statement that the Levi Civita
connection associated with the metric $h$ has a holonomy group
contained in $\mathrm{SU}(2) \otimes \mathrm{Sp}(2m)$. Consider now
eq.s (\ref{quatalgebra}), (\ref{iperforme}) and (\ref{piegatello}).
We easily deduce the following relation:
\begin{equation}
h^{st} K^x_{us} K^y_{tw} = -   \delta^{xy} h_{uw} +
  \epsilon^{xyz} K^z_{uw}
\label{universala}
\end{equation}
that holds true both in the HyperK\"ahler and in the quaternionic
case. In the latter case, using eq. (\ref{piegatello}), eq.
(\ref{universala}) can be rewritten as follows:
\begin{equation}
h^{st} \Omega^x_{us} \Omega^y_{tw} = - \lambda^2 \delta^{xy} h_{uw} +
\lambda \epsilon^{xyz} \Omega^z_{uw} \label{2.67}
\end{equation}
Eq.(\ref{2.67}) implies that the intrinsic components of the curvature
 2-form $\Omega^x$ yield a representation of the quaternion algebra.
In the HyperK\"ahler case such a representation is provided only by
the HyperK\"ahler form. In the quaternionic case we can write:
\begin{equation}
\Omega^x_{A\alpha, B \beta} \equiv \Omega^x_{uv} {\cal U}^u_{A\alpha}
{\cal U}^v_{B\beta} = - i \lambda \mathbb{C}_{\alpha\beta}
(\sigma_x)_A^{\phantom {A}C}\epsilon _{CB} \label{2.68}
\end{equation}
\noindent Alternatively eq.(\ref{2.68}) can be rewritten in an
intrinsic form as
\begin{equation}
\Omega^x =\,-{\rm i}\, \lambda \mathbb{C}_{\alpha\beta} (\sigma
_x)_A^{\phantom {A}C}\epsilon _{CB} {\cal U}^{\alpha A} \wedge {\cal
U}^{\beta B} \label{2.69}
\end{equation}
\noindent whence we also get:
\begin{equation}
{i\over 2} \Omega^x (\sigma_x)_A^{\phantom{A}B} = \lambda{\cal
U}_{A\alpha} \wedge {\cal U}^{B\alpha} \label{2.70}
\end{equation}
\par
The quaternionic manifolds are not requested to be homogeneous
spaces, however there exists a subclass of quaternionic homogeneous
spaces that are displayed in Table~\ref{quatotable}.
\par
%%%%%%%%%%%%%%%%%%%%%%%%%%%%%%%%
%%%%%%%%%%INIZIO TABELLA 2 %%%%%
%%%%%%%%%%%%%%%%%%%%%%%%%%%%%%%%
\begin{table}
\begin{center}
\caption{\sl Homogeneous symmetric quaternionic manifolds}
\label{quatotable}
\begin{tabular}{|c||c|}
\hline m  & $G/H$
\\
\hline ~~&~~\\ $m$ & $\frac {\mathrm{Sp}(2m+2)}{\mathrm{Sp}(2)\times
\mathrm{Sp}(2m)}$ \\ ~~&~~\\ \hline ~~&~~\\ $m$ & $\frac
{\mathrm{SU}(m,2)}{SU(m)\times \mathrm{SU}(2)\times U(1)}$  \\
~~&~~\\ \hline ~~&~~\\ $m$ & $\frac
{\mathrm{SO}(4,m)}{\mathrm{SO}(4)\times \mathrm{SO}(m)}$  \\ ~~&~~\\
\hline \hline ~~&~~\\ $2$ & $\frac {G_2}{\mathrm{SO}(4)}$ \\ ~~&~~\\
\hline ~~&~~\\ $7$ & $\frac {F_4}{\mathrm{Sp}(6)\times
\mathrm{Sp}(2)}$\\ ~~&~~~\\ \hline ~~&~~\\ $10$ & $\frac
{E_6}{\mathrm{SU}(6)\times U(1)}$\\ ~~&~~\\ \hline ~~&~~\\ $16$ &
$\frac {E_7}{S0(12)\times \mathrm{SU}(2)}$\\ ~~&~~\\ \hline ~~&~~\\
$28$ & $\frac {E_8}{E_7\times \mathrm{SU}(2)}$\\ ~~&~~\\ \hline
\end{tabular}
\end{center}
\end{table}
\subsection{$\mathcal{N}=2$, $D=5$ supergravity before gauging}
\label{beforgau}
Relying on the geometric lore developed in the previous sections it
is now easy to state what is the bosonic Lagrangian of a general $\mathcal{N}=2$ theory
in five--dimensions. We just have to choose an $n$--dimensional very
special manifold and some quaternionic manifold $\mathcal{QM}$ of
quaternionic dimension $m$. Then recalling eq.(\ref{genford5bos}) we
can specialize it to:
\begin{eqnarray}
  \mathcal{L}^{(ungauged)}_{(D=5,\mathcal{N}=2)}&=&\sqrt{-g} \, \left( \frac{1}{2}
  \,R \, - \, \frac{1}{4}\,\mathcal{N}_{\Lambda\Sigma} (\phi)  F^\Lambda_{\mu \nu
  } \, F^{\Sigma \vert\mu \nu} + \frac{1}{2} \, g_{ij}(\phi) \, \partial
  _\mu\phi^i \, \partial ^\mu \, \phi^j
  +\frac{1}{2} \, h_{uv}(q) \, \partial
  _\mu q^u \, \partial ^\mu \, q^v\right ) \nonumber\\
  &&+  \frac{1}{8} d_{\Lambda\Sigma\Gamma} \, \epsilon ^{\mu \nu \rho \sigma \tau
  } \, F^\Lambda_{\mu \nu } \, F^\Sigma_{\rho \sigma } \, A^\Gamma_\tau
\label{genn2d5ung}
\end{eqnarray}
where $h_{uv}(q)$ is the quaternionic metric on the quaternionic
manifold $\mathcal{QM}$, while $g_{ij}(\phi)$ is the very special metric
on the very special manifold. At the same time the constant tensor
$d_{\Lambda\Sigma\Gamma}$ is that defining the cubic norm
(\ref{defispec}) while the kinetic metric $\mathcal{N}$ is that
defined in eq.(\ref{scriptNsp}). The transformation rule of the
gravitino field takes the general form (\ref{gensusrul}) with the
graviphoton defined as in eq.(\ref{gravfotft}) and the tensor
$\Phi^\Lambda_{AB}$ given by eq.(\ref{Phiforn2}). In this respect it
is noteworthy that gravitino supersymmetry transformation rule does
depend only on the vector multiplet scalars and it is independent
from the hypermultiplets. Such a situation will be changed by the
gauging that introduces a gravitino mass-matrix depending also on the
hypermultiplets.
%%%%%%%%%%%%%%%%%%%%%%%%%%%%%%%%%%
\chapter[Supergravity Gaugings]{Supergravity Gaugings}
\label{gaugchap}
\section[Gaugings and Vacua]{Gaugings and Vacua}
\setcounter{equation}{0}
The conventional lore is that a vacuum of gravity or supergravity is
a configuration with maximal symmetry, namely with Lorentz invariance
$\mathrm{SO(1,D-1)}$ in $D$--dimensions. Adding translation
invariance one ends up with either Poincar\'e or de Sitter or anti de
Sitter symmetry which forces the vacuum expectation values of all
scalar fields to be constant. The new insight provided by the role of
the domain wall solutions and by the developments discussed in chapter
\ref{bestia1} suggests that we might also consider vacua where
there is Poincar\'e invariance in one dimension less $\mathrm{ISO(1,D-2)}$
and where  the vacuum value of the scalar fields depends on the last $Dth$
coordinate. These are precisely the domain wall vacua which are
expected to be a distinguished property of gauged supergravities.
Yet, as I explained in section \ref{bobus}, these wall geometries are
like solitons or kinks that interpolate between conventional vacua.
Conventional vacua are also effectively characterized by their
properties with respect to supersymmetry breaking or preservation.
Hence I begin my analysis of the supergravity gaugings by recalling
the general properties of conventional vacua and of the possible
supersymmetry breaking patterns, that, as it will immediately appear,
encode fundamental information about the  basic new ingredients produced
by the gaugings, namely the fermion shifts.
\section[General aspects of supergravity gaugings and susy breaking]{General
aspects of supergravity gaugings and susy breaking}
\label{genaspect}
\setcounter{equation}{0}
Let me begin by
recalling some very general aspects of the super-Higgs mechanism in
extended supergravity that were codified in the literature of the
early  and middle eighties \cite{susyb1,deRoo:1985np,susyb2,susyb3}
(for a review see chapter II.8 of \cite{castdauriafre}) and were
further analyzed and extended in the middle nineties
\cite{Zinovev:1992mw,Ferrara:1996gu,Fre:1997js,Girardello:1997hf}.
Because of the fundamental property of extended supergravity that the
scalar potential is generated by the gauging procedure, the
discussion of spontaneous supersymmetry breaking goes hand in hand
with the discussion of possible gaugings.
\subsection[Supersymmetry breaking in conventional vacua]{Supersymmetry breaking in
conventional vacua}
A conventional vacuum of $p+2$--dimensional supergravity corresponds
to a space--time geometry with a maximally extended group of isometries, namely with
$\frac{1}{2}(p+2)(p+3)$ Killing vectors. This means that the metric
$ds^2 = g_{\mu\nu} dx^\mu \otimes dx^\nu$ necessarily has constant
curvature in $p+2$--dimensions and is one of the following three:
\begin{equation}
  \mathcal{M}_{space--time}=\left \{ \begin{array}{lcl}
  AdS_{p+2} & ;& \mbox{negative
  curvature}\\
  \mathrm{Minkowski_{p+2}}&;& \mbox{zero curvature}\\
\mathrm{dS_{p+2}} &;& \mbox{positive curvature}\ \end{array}\right.
\label{trecass}
\end{equation}
At the same time, in order to be consistent with this maximal
symmetry, the v.e.v.s of the scalar  fields, $< \phi^i >=\phi^i_0$
must be constant and be extrema  of the scalar potential:
\begin{equation}
  \left.\frac{\partial \mathcal{V}}{\partial \phi ^i}\right|_{\phi=\phi_0} =  0~,
\label{extrepot}
\end{equation}
Minkowski space occurs when $\mathcal{V}(\phi_0)=0$,  anti de Sitter space AdS$_{p+2}$ occurs
when $\mathcal{V}(\phi_0)< 0$ and finally de Sitter space dS$_{p+2}$ is generated by
$\mathcal{V}(\phi_0) > 0$.
To be definite I  focus on the $4$--dimensional case, but all
the mechanisms and properties I describe below have straightforward counterparts in higher
dimensions.  In particular I will apply them to the $5$--dimensional case in section
\ref{gaugd5theo}. So let me state that in relation
with the super-Higgs mechanism, there are just three
relevant items of the entire $D=4$ supergravity construction that  have
to be considered.
\begin{enumerate}
\item The \emph{gravitino mass matrix} $S_{AB}(\phi)$ , namely the
non-derivative scalar field dependent term that appears in the
gravitino supersymmetry transformation rule:
\begin{equation}
  \delta \psi _{A\vert \mu} = \mathcal{D}_\mu \, \epsilon_A \, + \,
  S_{AB} \left( \phi \right) \, \gamma_\mu \, \epsilon^B~ +\dots,
\label{gravsusy}
\end{equation}
and reappears as a mass term in the Lagrangian:
\begin{equation}
  \mathcal{L}^{\rm SUGRA} \, = \, \dots \, + \, \mbox{const} \,\left( \,
  S_{AB} (\phi ) \, \psi^A_\mu \, \gamma^{\mu\nu} \, \psi^B_\nu \, +
  \,  S^{AB} (\phi ) \, \psi_{A\vert\mu} \, \gamma^{\mu\nu} \,
  \psi_{B\vert\nu} \,\right)
\label{gravmasmat}
\end{equation}
\item The \emph{fermion shifts}, namely the non-derivative scalar field
  dependent terms in the supersymmetry transformation rule of the spin $\ft 12$
  fields :
\begin{eqnarray}
  \delta \, \lambda^i_R & =& \mbox{derivative terms} \, + \,
  \Sigma_{A}^{\phantom{A} i} \left( \phi \right) \,  \epsilon^A~,
  \nonumber\\
  \delta \, \lambda^i_L & =& \mbox{derivative terms} \, + \,
  \Sigma^{A\vert i} \left( \phi \right) \,  \epsilon_A~.
\label{fermioshif}
\end{eqnarray}
\item The scalar potential itself, $\mathcal{V}(\phi)$.
\end{enumerate}
These three items are related by a general supersymmetry Ward
identity, firstly discovered in the context of gauged $\mathcal{N}=8$
supergravity \cite{dewit1} and later extended to all supergravities
\cite{susyb1,susyb2,susyb3}, that, in the conventions of
\cite{deRoo:1985np,wageman,deroowag,wagthesis} reads as follows:
\begin{equation}
  24\,S_{AC} \, S^{CB} - 4 \, K_{i,j} \Sigma_{A}^{\phantom{A} i}
  \Sigma^{B \vert j} \, = \, -\delta_{A}^{B} \,\mathcal{V}~,
\label{wardide}
\end{equation}
where $K_{i,j}$ is the kinetic matrix of the spin--1/2 fermions. The
numerical coefficients appearing in (\ref{wardide}) depend on the
normalization of the kinetic terms of the fermions, while $A,B,\dots
= 1,\dots,\mathcal{N}$ are $\mathrm{SU}(\mathcal{N})$ indices that
enumerate the supersymmetry charges. We also follow the standard
convention that the upper or lower position of such indices denotes
definite chiral  projections of Majorana spinors, right or left
depending on the species of fermions considered\footnote{For
instance, we have $\gamma_5 \, \epsilon_A= \epsilon_A$ and $\gamma_5
\, \epsilon^A = - \epsilon^A$.}. The position denotes also the way of
transforming of the fermion with respect to
$\mathrm{SU}(\mathcal{N})$, with lower indices in the fundamental and
upper indices in the fundamental bar. In this way we have $S^{AB} =
\left(S_{AB} \right) ^\star$ and $\Sigma_A^{\phantom{A} i}
=\left(\Sigma^{B \vert i} \right) ^\star$. Finally, the index $i$ is
a collective index that enumerates all spin--1/2  fermions $ \lambda^i$ present in
the theory\footnote{We denote by $ \lambda^i$ the right
handed chiral projection while $\lambda_i$ are the left handed ones}.
\par
  The corresponding  fermion shifts are defined by
\begin{equation}
  \delta \, \lambda^i \, = \,\mbox{derivative terms} \, + \,
  \Sigma_{A}^{\phantom{A}i} \left( \phi \right) \,  \epsilon^A~.
\label{gaushif}
\end{equation}
\par
A vacuum configuration $\phi_0$ that preserves $\mathcal{N}_0$
supersymmetries is characterized by the existence of $\mathcal{N}_0$
vectors $\rho^A_{(\ell)}$ ($\ell=1,\ldots,\mathcal{N}_0$) of
$\mathrm{SU}(\mathcal{N})$, such that
\begin{eqnarray}
\label{breakpat} S_{AB} \left( \phi_0 \right) \, \rho^A_{(\ell)} & =
& e^{\rm i\theta}\, \sqrt{\ft{-\mathcal{V}(\phi_0)}{24}} \, \rho_{A(\ell)}
%\sqrt{-V(\phi_0)/3}\, \rho_{A(\ell)}
~,\nonumber\\ \Sigma_A^{\phantom{A}i} \left( \phi_0 \right) \,
\rho^A_{(\ell)} & = & 0~,
\end{eqnarray}
where $\theta$ is an irrelevant phase. Indeed, consider the spinor
\begin{equation}
  \epsilon ^A(x) = \sum_{\ell =1}^{\mathcal{N}_0} \, \rho^A_{(\ell)}
  \epsilon^{(\ell)}(x)~,
\label{Eadix}
\end{equation}
where $\epsilon^{(\ell)}(x)$ are $\mathcal{N}_0$ independent
solutions of the equation for covariantly
constant spinors in $\mathrm{AdS}_4$ (or Minkowski space) with
$2\,e= \sqrt{-\mathcal{V}(\phi_0)/{24}}$:
\begin{equation}
\label{susygravads4}
D_a^{(\mathrm{AdS})}\epsilon(x)\equiv
(\partial_a - {1\over 4}\omega^{bc}_{~~a} \gamma_{bc} -
2\, e\, \gamma_5\gamma_a)\epsilon(x)=0~,
\end{equation}
The integrability of eq.(\ref{susygravads4})
is guaranteed by the expression of the AdS$_4$ curvature,
$R^{ab}_{~~cd}=-16\,e^2\,\delta^{ab}_{cd}$, that
corresponds to the Ricci tensor:
\begin{equation}
\label{cosmol4e7}
\mathcal{R}_{ab} = -24\, e^2\, \eta_{ab}~=~\frac{1}{4}\,\mathcal{V}(\phi_0) \, \eta_{ab},
\end{equation}
 Then it follows that under supersymmetry
transformations of parameter (\ref{Eadix}) the chosen vacuum
configuration $\phi=\phi_0$ is invariant\footnote{As already
stressed, the v.e.v.s of all the fermions are zero and equation
(\ref{breakpat}) guarantees that they remain zero under supersymmetry
transformations of parameters (\ref{Eadix}).}. That such a
configuration is a true vacuum follows from another property proved,
for instance, in  \cite{susyb3}: all vacua that admit at least one
vector $\rho^A$ satisfying eq. (\ref{breakpat}) are automatically
extrema  of the potential, namely they satisfy eq. (\ref{extrepot}).
Furthermore, as one can immediately check, the action
(\ref{scalfieldact}), for constant scalar field configurations implies
that the Ricci tensor must be $\mathcal{R}_{\mu \nu
}=\frac{1}{4}\,\mathcal{V}(\phi_0)\, g_{\mu \nu }$ as in equation
(\ref{cosmol4e7}).
\par
The above integrability argument can be easily generalized to all
dimensions and to all numbers of supersymmetries $\mathcal{N}$.
%%% marca %%%%%%%%
Consider a supergravity action in $D$ dimensions that, once reduced to the
gravitational plus scalar field sector, has the same normalization as
the action (\ref{scalfieldact}) considered in section \ref{bobus},
namely:
\begin{equation}
  A_{grav+scal}^{[D]}=\int \, d^Dx \, \sqrt{-g} \,\left[ 2 \, R[g]
  +\alpha \, \frac{1}{2} g_{ij} (\phi) \, \partial^\mu \, \phi^i \partial_\mu \phi^j -
  \mathcal{V}(\phi)\right]
\label{scalfielsect}
\end{equation}
where $\alpha$ is a normalization constant that can vary from case to
case since it can always be reabsorbed into the definition of the
scalar metric but the scalar potential  $\mathcal{V}$ has an
unambiguous and unique normalization with respect to the Einstein
term. For constant field configurations $\phi_0$ the Einstein equations
derived from (\ref{scalfielsect}) imply that:
\begin{equation}
  R_{\mu \nu  }= \frac{1}{2(D-2)} \, \mathcal{V}(\phi_0) \, g_{\mu \nu }
\label{Ricciustens}
\end{equation}
Then the Riemann tensor of an anti de Sitter space $AdS_D$ consistent
with eq. (\ref{Ricciustens}) is necessarily the following:
\begin{equation}
  R^{\rho \sigma }_{\mu \nu } = \frac{1}{(D-1)(D-2)}\, \mathcal{V}(\phi_0) \,
  \delta^{[\rho}_{[\mu} \, \delta ^{\sigma]}_{\nu]}
\label{Riemantens}
\end{equation}
Consider next the equation for a covariantly constant  spinor in $AdS_D$. Its
general form is:
\begin{eqnarray}
D_\mu^{(\mathrm{AdS})}\epsilon &\equiv& \mathcal{D}_\mu \epsilon(x)-
\mu \, \gamma_\mu \, \epsilon
=(\partial_\mu - {1\over 4}\omega^{bc}_{~~\mu} \gamma_{bc} -
\mu \, \gamma_\mu)=0~,
\label{gencovconstspi}
\end{eqnarray}
where the parameter $\mu$ is fixed by integrability in terms of the
vacuum value of potential $\mathcal{V}(\phi_0) $. Indeed from the
condition $D_{[\mu}^{(\mathrm{AdS})} D_{\nu]}^{(\mathrm{AdS})} =0$ we
immediately get:
\begin{equation}
  |\mu |^2 = \frac{1}{4} \, \frac{|\mathcal{V}(\phi_0)|}{(D-1) (D-2)}
\label{modulmu}
\end{equation}
On the other hand the general form of the gravitino transformation
rule is, independently from the number of space--time dimensions,
that given in eq.(\ref{gravsusy}), so that, in a conventional vacuum
with an unbroken supersymmetry $\mu$ is to be interpreted as
\textbf{eigenvalue} of the gravitino mass--matrix. So the general
conditions for the preservation of $\mathcal{N}_0$ supersymmetries
in $D$ dimensions are fully analogous to those in eq.(\ref{breakpat})
and correspond to the existence of $\mathcal{N}_0$ independent
vectors $\rho^A_{(\ell)}$ ($\ell=1,\dots,\mathcal{N}_0$), such that:
%%%%%%%%%%%%%%%%%%%%%%%%%%%%
\begin{eqnarray}
\label{genbreakpat} S_{AB} \left( \phi_0 \right) \, \rho^A_{(\ell)} & =
& e^{\rm i\theta}\, \sqrt{\ft{|\mathcal{V}(\phi_0)|}{4(D-2)(D-1)}} \, \rho_{A(\ell)}
%\sqrt{-V(\phi_0)/3}\, \rho_{A(\ell)}
~,\nonumber\\ \Sigma_A^{\phantom{A}i} \left( \phi_0 \right) \,
\rho^A_{(\ell)} & = & 0~,
\end{eqnarray}
%%%%%%%%%%%%%%%%%%%%%%%%%%%%
By extension of language the vectors $\rho^A_{(\ell)}$ are named
\textbf{Killing spinors}
\subsection{Gaugings and fermion shifts}
As we have already recalled few lines above the most important
general feature  of extended $\mathcal{N}\ge 2$ supergravities in
$D=4$, is that the fermion shifts and the gravitino
mass--matrix are uniquely determined by the \emph{gauging} of the
theory and are proportional to the gauge group coupling constants
$g_i$. Indeed there are very general formulae for these objects
expressing them in terms of geometrical data of the scalar manifold
and of the structure constants of the gauge group (or of
representation  matrices if, in addition to vector multiplets, also
hyper-multiplets are  present).  Hyper-multiplets are present only
for the case $\mathcal{N}=2$, whose most general  form and gauging is
discussed in \cite{bertolo} and whose partial breaking is discussed
in \cite{Fre:1997js,Girardello:1997hf,Ferrara:1996gu,Zinovev:1992mw}.
\par
For $\mathcal{N}\ge 5$  hyper and vector multiplets are absent and
the scalar manifold is a uniquely fixed non--compact
coset space as we have already stressed in chapter \ref{bestia2}.
For $\mathcal{N}=3,4 $ there are no hyper-multiplets
and, in addition to the graviton multiplet, there are at most vector
multiplets. Also in this case, the geometry of the scalar manifold is
fixed to be  that of a  non--compact coset space (see table
\ref{topotable}).  Similar considerations apply to higher dimensions.
Indeed we can conclude that for all theories with a number of supercharges $N_Q \geq 8$
which kind of supersymmetry breaking patterns and which kind of domain wall
solutions can be obtained depends uniquely on the choice of a
gauging. Indeed both the above aspects, besides being intimately related,
are controlled by the scalar potential and the latter is uniquely
determined by the gauging procedure.
\par
So I discuss the general properties of gauging beginning with maximal
theories where the number of gauge fields is fixed and just the gauge
algebra has to be chosen.  Non--maximal theories
where the matter multiplets come into play will be addressed later
choosing the physically relevant $5$--dimensional case.
%%%%%%%%%%%%%%%%%%%%%%%%%%%%%
%file{fromfin8}
%%%%%%%%%%%%%%%%%%%%%%%%%%%%%%%%%
\section[Gaugings of $\mathcal{N}=8$ Supergravity in $D=4$]{Gaugings of
$\mathcal{N}=8$ Supergravity in $D=4$}
\label{n8d4gaug}
\setcounter{equation}{0}
To illustrate the general ideas in a case of maximal supersymmetry,
I consider the possible \emph{gaugings}
of the $\mathcal{N}=8$ theory in four dimensions. The complete
classification that can be reached in this case constitutes an
inspiring model for the analogous problem in maximal $5$--dimensional
supergravity. Relying on the comparison with the $D=4$ case we shall
be better equipped to appreciate the additional subtleties occurring
in five--dimensions. In turn the comparison at fixed space--time
dimension $D=5$ between the maximal and non maximal matter coupled theory
will be of special value.
\par
So coming to the present $N=8$,$D=4$ case we recall that here
there is no other multiplet besides the graviton
multiplet which contains the graviton $g_{\mu \nu }$, $8$ gravitinos
$\psi_{A\vert \mu} \, dx^\mu$,
$28$ one--form gauge fields $A^{\Lambda\Sigma}_\mu \, dx^\mu = -
A^{\Sigma\Lambda}_\mu \, dx^\mu $ transforming in the $28$
antisymmetric representation of the electric subgroup
$\mathrm{SL}(8,\mathbb{R}) \,\subset \,  \mathrm{E_{7(7)}}$ (see eq.(\ref{elecgrou})),
$56$ spin $1/2$ dilatinos $\chi_{ABC} =\gamma_5 \chi_{ABC}$ (anti-symmetric in $ABC$)
and $70$ scalars parametrizing the
$\mathrm{E_{7(7)}/SU(8)}$ coset manifold. I have labeled the
vector fields with a pair of antisymmetric indices
\footnote{For later convenience we have slightly
changed the conventions used in the general discussion of section \ref{dualsym}}, each of them
ranging on $8$ values $\Lambda,\Sigma,\Delta,\Pi,=1, \dots, 8$
and transforming in the fundamental representation of $
\mathrm{SL}(8,\mathbb{R})$.
The capital latin indices  carried by the fermionic fields
range also on eight values $A,B,C,=1,\dots, 8$ but they are covariant
under the maximal compact subgroup $\mathrm{SU(8)}\subset \mathrm{E_{7(7)}} $
rather than the non compact  $\mathrm{SL}(8,\mathbb{R})\subset
 \mathrm{E_{7(7)}}$.
As in previous sections, also here I use the convention that
upper and lower $\mathrm{SU(8)}$ indices  denote different chirality
projections of the fermion fields: $\psi^A=-\gamma_5\psi^A$
and $\chi^{ABC} =-\gamma_5 \chi^{ABC}$.
\par
\subsection{The bosonic action}
In these notations, the bosonic action of the theory is
%%%%%%%%%%%%%%%%%%%%%%%%%%%%%%%%%%%%%%%%%%%%%%%%
\begin{eqnarray}
\label{lN=8}
{\cal A}_{bosonic}&=&\int\sqrt{-g}\, d^4x\left(2R+\Im{\cal
N}_{\Lambda\Sigma|\Gamma\Delta}F_{\mu\nu}^{~~\Lambda\Sigma}F^{\Gamma\Delta|\mu\nu}+
{1 \over 2}g_{ij}\partial_{\mu}
\phi^{i}\partial^{\mu}\phi^{j}+\right.\nonumber\\
&+&\left.{1 \over 2}\Re{\cal N}_{\Lambda\Sigma|\Gamma\Delta}
{\epsilon^{\mu\nu\rho\sigma}\over\sqrt{-g}}
F_{\mu\nu}^{~~\Lambda\Sigma}F^{\Gamma\Delta}_{~~\rho\sigma} + {\cal V} (\phi)\right)
\label{lagrared}
\end{eqnarray}
%%%%%%%%%%%%%%%%%%%%%%%%%%%%%%%%%%%%%%%%%%%%%%%%
where the scalar metric $ g_{ij}\left( \phi \right)$ and the vector kinetic metric
$ {\cal N}_{\Lambda\Sigma|\Gamma\Delta} $ are uniquely
determined  in terms of coset representatives and
left--invariant  $1$--forms of $\mathrm{E_{7(7)}/SU(8)}$.
Following the general set up of chapter \ref{bestia2} we  name $\mathbb{L}(\phi)$ the coset
representative  parametrizing the equivalence classes
of $\mathrm{E_{7(7)}/SU(8)}$.
\par
Since the gauging procedure is completely \emph{coordinate free} the
choice of a coset parametrization is completely immaterial in the
following discussion. Just to fix ideas and to avoide the technicalities of
the solvable decomposition we can think of $\mathbb{L}(\phi)$  as
the exponential of the $70$--dimensional coset $\mathbb{K}$ in the orthogonal
decomposition:
\begin{equation}
\mathrm{E_{7(7)}} = \mathrm{SU(8)} \, \oplus \, \mathbb{K}
\label{ortodeco}
\end{equation}
In practice this means that we can write:
\begin{equation}
\mathbb{L}(\phi)=\exp \left( \matrix{  0               &  \phi^{EFGH} \cr
                        \phi_{ABCD}   & 0              } \right)
 =\left( \matrix{  u^{\Lambda\Xi}{}_{AB}  &  v^{\Lambda\Xi CD} \cr
                   v_{\Delta\Gamma AB}
&  u_{\Delta\Gamma}{}^{CD}  } \right)  \,
\label{cosetrep}
\end{equation}
where the $70$ parameters $\phi_{ABCD}$ satisfy the self--duality
condition \footnote{
Here we have used the notation,
$\phi^{ABCD} \equiv (\phi_{ABCD})^*$} :
\begin{equation}
 \phi_{ABCD}=\frac{1}{4!}\varepsilon_{ABCDEFGH}\phi^{EFGH}
 \label{selfdual1}
\end{equation}
%%%%%%%%%%%%%%%%%%%%%%%%%%%%%%%%%%%%%%%
% Si inserisce il nuovo qui
%%%%%%%%%%%%%%%%%%%%%%%%%%%%%%%%%%%%%%%
According to eq.(\ref{masterformula}) the period matrix ${\cal N}_{\Lambda\Sigma,\Delta\Pi}$
has the following general expression:
\begin{equation}
{\cal N}_{\Lambda\Sigma,\Delta \Pi}=h \cdot f^{-1}
\label{gaiazumap}
\end{equation}
where the complex $ 28 \times 28 $ matrices $ f,h$ are
defined by the realization $\mathbb{L}_{Usp} \left(\phi\right)$ of the
coset representative in the $\mathrm{Usp(28,28)}$ complex basis.  This latter is related to its
counterpart $\mathbb{L}_{Sp}(\phi)$ in the real $\mathrm{Sp(56},\mathbb{R})$
basis, by a Cayley transformation, as displayed in the
following formula (see eq.s (\ref{cayley}) and (\ref{isomorfo})):
\begin{eqnarray}
\mathbb{L}_{Usp} \left(\phi\right) &=& \frac{1}{\sqrt{2}}\left(\matrix{ f+ {\rm i}h &
\bar f+ {\rm i}\bar h \cr f- {\rm i}h
& \bar f - {\rm i}\bar h  \cr } \right)
\equiv {\cal C} \, \mathbb{L}_{Sp}\left(\phi\right) {\cal C}^{-1} \nonumber\\
\mathbb{L}_{Sp}(\phi) & \equiv & \exp \left[ \phi^i \, T_i \right ] \, = \,
\left(\matrix{ A(\phi) & B(\phi) \cr C(\phi) & D(\phi) \cr } \right)
\label{cayleytra}
\end{eqnarray}
As explained in \cite{noi3}
there are actually four bases where the $ 56 \times 56 $ matrix
$\mathbb{L}(\phi)$ can be written:
\begin{enumerate}
\item {The $\mathrm{SpD(56)}$--basis}
\item {The $\mathrm{UspD(28,28)}$--basis}
\item {The $\mathrm{SpY(56)}$--basis}
\item {The $\mathrm{UspY(28,28)}$--basis}
\end{enumerate}
corresponding to two cases where $\mathbb{L}$ is symplectic
real ($\mathrm{SpD(56)}$,$\mathrm{SpY(56)}$) and two cases where it is
pseudo--unitary symplectic ($\mathrm{UspD}(56)$,$\mathrm{UspY}(56)$). This further
distinction in a pair of subcases corresponds
to choosing either a basis composed of weights or of Young tableaux.
By relying on (\ref{cosetrep}) I have chosen to utilize the
$\mathrm{UspY}(28,28)$--basis which is directly related to the $\mathrm{SU(8)}$ indices
carried by the fundamental fields of supergravity.
%%%%%%%%%%%%%%%%%%%%%%%%%%%%%%%%%%%%%%
However, for the description
of the gauge generators the Dynkin basis is more convenient.
We can optimize the advantages of both bases introducing a mixed one
where  the coset representative $\mathbb{L} $ is
multiplied on the left by the constant matrix ${\cal S}$ performing the
transition from the pseudo--unitary Young basis to the real symplectic
Dynkin basis. Explicitly we have:
\begin{eqnarray}
\left( \matrix { u^{AB} \cr  v_{AB} \cr }\right) & = & {\cal S} \,
\left( \matrix { W^{i} \cr  W^{i+28} \cr }\right) (i,1,\dots\, 28) \nonumber\\
\end{eqnarray}
where
\begin{eqnarray}
{\cal S}&=& \left( \matrix{ {\bf S} & {\bf 0} \cr {\bf 0} & {\bf S^\star}
\cr } \right) \, {\cal C} = \frac{1}{\sqrt{2}}\left( \matrix{ {\bf S} & {\rm i}\,{\bf S} \cr
{\bf S^\star} & -{\rm i}\,{\bf S^\star} \cr } \right) \nonumber\\
\end{eqnarray}
the $ 28 \times 28 $ matrix ${\bf S}$ being unitary:
\begin{eqnarray}
&& \null \nonumber\\
 {\bf S}^\dagger {\bf S} &=& \bfone
\end{eqnarray}
The explicit form of the $\mathrm{U(28)}$ matrix ${\bf S}$ was obtained in
section 5.4 of \cite{noi3} and has the following explicit expression:
\begin{eqnarray}
& \sqrt{2} \,{\bf S}= & \nonumber
\end{eqnarray}
{\tiny
\begin{eqnarray}
&\left(\matrix{ 0 & 0 & 0 & 0 & 0 & 0 & 0 & 0 & 0 & 0 & 0 & 0 & 0 & 0 & 0 & 0 &
  {i\over {{\sqrt{2}}}} & {1\over {{\sqrt{2}}}} & 0 & 0 & 0 & 0 &
  {1\over {{\sqrt{2}}}} & {{-i}\over {{\sqrt{2}}}} & 0 & 0 & 0 & 0 \cr 0 & 0
   & 0 & 0 & 0 & 0 & 0 & 0 & 0 & 0 & 0 & 0 & 0 & 0 & 0 & 0 & -i & 0 & 0 & 0 &
  0 & 0 & 0 & -i & 0 & 0 & 0 & 0 \cr 0 & 0 & 0 & 0 & 0 & 0 & 0 & 0 & 0 & 0 & 0
   & 0 & 0 & 0 & 0 & 0 & -{1\over {{\sqrt{3}}}} & 0 & 0 & 0 & 0 & 0 &
  {{-2\,i}\over {{\sqrt{3}}}} & {1\over {{\sqrt{3}}}} & 0 & 0 & 0 & 0 \cr 0 &
  0 & 0 & 0 & 0 & 0 & 0 & 0 & 0 & 0 & 0 & 0 & 0 & 0 & 0 & 0 &
  -{1\over {{\sqrt{6}}}} & -i\,{\sqrt{{3\over 2}}} & 0 & 0 & 0 & 0 &
  {i\over {{\sqrt{6}}}} & {1\over {{\sqrt{6}}}} & 0 & 0 & 0 & 0 \cr 1 & 0 & 0
   & 0 & 0 & 0 & -1 & 0 & 0 & 0 & 0 & 0 & 0 & 0 & 0 & 0 & 0 & 0 & 0 & 0 & 0 &
  0 & 0 & 0 & 0 & 0 & 0 & 0 \cr 0 & 1 & i & 0 & 0 & 0 & 0 & 0 & 0 & 0 & 0 & 0
   & 0 & 0 & 0 & 0 & 0 & 0 & 0 & 0 & 0 & 0 & 0 & 0 & 0 & 0 & 0 & 0 \cr 0 & 0
   & 0 & 1 & 0 & 0 & 0 & 0 & 0 & 0 & 0 & 0 & 1 & 0 & 0 & 0 & 0 & 0 & 0 & 0 & 0
   & 0 & 0 & 0 & 0 & 0 & 0 & 0 \cr 0 & 0 & 0 & 0 & 1 & 0 & 0 & 0 & 0 & 0 & 0
   & -1 & 0 & 0 & 0 & 0 & 0 & 0 & 0 & 0 & 0 & 0 & 0 & 0 & 0 & 0 & 0 & 0 \cr 0
   & 0 & 0 & 0 & 0 & 1 & 0 & 0 & 0 & 0 & 0 & 0 & 0 & 0 & 0 & 1 & 0 & 0 & 0 & 0
   & 0 & 0 & 0 & 0 & 0 & 0 & 0 & 0 \cr 0 & 0 & 0 & 0 & 0 & 0 & 0 & 1 & 0 & 0
   & 0 & 0 & 0 & 0 & i & 0 & 0 & 0 & 0 & 0 & 0 & 0 & 0 & 0 & 0 & 0 & 0 & 0
   \cr 0 & 0 & 0 & 0 & 0 & 0 & 0 & 0 & 1 & 0 & 0 & 0 & 0 & i & 0 & 0 & 0 & 0
   & 0 & 0 & 0 & 0 & 0 & 0 & 0 & 0 & 0 & 0 \cr 0 & 0 & 0 & 0 & 0 & 0 & 0 & 0
   & 0 & 1 & i & 0 & 0 & 0 & 0 & 0 & 0 & 0 & 0 & 0 & 0 & 0 & 0 & 0 & 0 & 0 & 0
   & 0 \cr 0 & 0 & 0 & 0 & 0 & 0 & 0 & 0 & 0 & 0 & 0 & 0 & 0 & 0 & 0 & 0 & 0
   & 0 & 1 & 0 & 0 & 0 & 0 & 0 & i & 0 & 0 & 0 \cr 0 & 0 & 0 & 0 & 0 & 0 & 0
   & 0 & 0 & 0 & 0 & 0 & 0 & 0 & 0 & 0 & 0 & 0 & 0 & 1 & 0 & 0 & 0 & 0 & 0 & i
   & 0 & 0 \cr 0 & 0 & 0 & 0 & 0 & 0 & 0 & 0 & 0 & 0 & 0 & 0 & 0 & 0 & 0 & 0
   & 0 & 0 & 0 & 0 & 1 & 0 & 0 & 0 & 0 & 0 & i & 0 \cr 0 & 0 & 0 & 0 & 0 & 0
   & 0 & 0 & 0 & 0 & 0 & 0 & 0 & 0 & 0 & 0 & 0 & 0 & 0 & 0 & 0 & 1 & 0 & 0 & 0
   & 0 & 0 & i \cr 1 & 0 & 0 & 0 & 0 & 0 & 1 & 0 & 0 & 0 & 0 & 0 & 0 & 0 & 0
   & 0 & 0 & 0 & 0 & 0 & 0 & 0 & 0 & 0 & 0 & 0 & 0 & 0 \cr 0 & 1 & -i & 0 & 0
   & 0 & 0 & 0 & 0 & 0 & 0 & 0 & 0 & 0 & 0 & 0 & 0 & 0 & 0 & 0 & 0 & 0 & 0 & 0
   & 0 & 0 & 0 & 0 \cr 0 & 0 & 0 & 1 & 0 & 0 & 0 & 0 & 0 & 0 & 0 & 0 & -1 & 0
   & 0 & 0 & 0 & 0 & 0 & 0 & 0 & 0 & 0 & 0 & 0 & 0 & 0 & 0 \cr 0 & 0 & 0 & 0
   & 1 & 0 & 0 & 0 & 0 & 0 & 0 & 1 & 0 & 0 & 0 & 0 & 0 & 0 & 0 & 0 & 0 & 0 & 0
   & 0 & 0 & 0 & 0 & 0 \cr 0 & 0 & 0 & 0 & 0 & 1 & 0 & 0 & 0 & 0 & 0 & 0 & 0
   & 0 & 0 & -1 & 0 & 0 & 0 & 0 & 0 & 0 & 0 & 0 & 0 & 0 & 0 & 0 \cr 0 & 0 & 0
   & 0 & 0 & 0 & 0 & 1 & 0 & 0 & 0 & 0 & 0 & 0 & -i & 0 & 0 & 0 & 0 & 0 & 0 &
  0 & 0 & 0 & 0 & 0 & 0 & 0 \cr 0 & 0 & 0 & 0 & 0 & 0 & 0 & 0 & 1 & 0 & 0 & 0
   & 0 & -i & 0 & 0 & 0 & 0 & 0 & 0 & 0 & 0 & 0 & 0 & 0 & 0 & 0 & 0 \cr 0 & 0
   & 0 & 0 & 0 & 0 & 0 & 0 & 0 & 1 & -i & 0 & 0 & 0 & 0 & 0 & 0 & 0 & 0 & 0 &
  0 & 0 & 0 & 0 & 0 & 0 & 0 & 0 \cr 0 & 0 & 0 & 0 & 0 & 0 & 0 & 0 & 0 & 0 & 0
   & 0 & 0 & 0 & 0 & 0 & 0 & 0 & 1 & 0 & 0 & 0 & 0 & 0 & -i & 0 & 0 & 0 \cr 0
   & 0 & 0 & 0 & 0 & 0 & 0 & 0 & 0 & 0 & 0 & 0 & 0 & 0 & 0 & 0 & 0 & 0 & 0 & 1
   & 0 & 0 & 0 & 0 & 0 & -i & 0 & 0 \cr 0 & 0 & 0 & 0 & 0 & 0 & 0 & 0 & 0 & 0
   & 0 & 0 & 0 & 0 & 0 & 0 & 0 & 0 & 0 & 0 & 1 & 0 & 0 & 0 & 0 & 0 & -i & 0
   \cr 0 & 0 & 0 & 0 & 0 & 0 & 0 & 0 & 0 & 0 & 0 & 0 & 0 & 0 & 0 & 0 & 0 & 0
   & 0 & 0 & 0 & 1 & 0 & 0 & 0 & 0 & 0 & -i \cr  }\right ) &
   \nonumber\\
\end{eqnarray}
}
while the weights of the
$\mathrm{E_{7(7)}}$ ${\bf 56}$ representation were constructed in section
\ref{pesanti} and are listed in table \ref{e7weight}. In the \textbf{Dynkin
basis} the basis vectors of the real
symplectic representation  are eigenstates
of the Cartan generators with eigenvalue one of the $56$ weight vectors
($\pm {\vec \Lambda } =\{ \Lambda_1 , \dots , \Lambda_7 \}$ pertaining
to the representation:
\begin{eqnarray}
\label{dynkbas}
(W=1,\dots\, 56 )& :&  \vert \, {W} \, \rangle \, = \, \cases
{ \vert {\vec \Lambda} \rangle ~~~~\, : \quad H_i \vert \, {\vec \Lambda} \, \rangle
~~~~ = ~~~\Lambda_i \,
\vert \, {\vec \Lambda} \, \rangle   ~~~\quad (\Lambda= 1,\dots \, 28)
\cr
\vert \, -{\vec \Lambda} \rangle \, :
\quad H_i \vert \, -{\vec \Lambda} \, \rangle = \, -\Lambda_i \,
\vert \, -{\vec \Lambda} \, \rangle   \quad (\Lambda= 1,\dots \, 28)
\cr}\nonumber\\
\vert \, V  \, \rangle & = & f^{\Lambda} \,   \vert {\vec \Lambda} \rangle
\, \,  \oplus \,\,  g_{\Lambda} \,   \vert {- \vec \Lambda} \rangle
\nonumber\\
\mbox{or in matrix notation} && \nonumber\\
\nonumber\\
{\vec V}_{SpD} & = & \left(\matrix{ f^\Lambda \cr g_\Sigma \cr }\right )
\end{eqnarray}
In the \textbf{Young basis}, instead,
the basis vectors of the complex pseudounitary
representation correspond to the natural basis of the
${\bf 28}$ + ${\bar {\bf 28}}$ antisymmetric representation
of the maximal compact subgroup $\mathrm{SU(8)}$. In other
words, in this  realization of the fundamental $\mathrm{E_{7(7)}}$
representation a generic vector is of the following form:
\begin{eqnarray}
\label{youngbas}
\vert {V} \rangle &=& u^{AB} \,
\mbox{$\begin{array}{|c|}
\hline
\stackrel{ }{A}\cr
\hline
\stackrel{}{B}\cr
\hline
\end{array}
$} \,\,    \oplus \, \,   {v}_{AB} \,  \mbox{$\begin{array}{|c|}
\hline
\stackrel{ }{\bar A}\cr
\hline
\stackrel{ }
{\bar B}\cr
\hline
\end{array} $} \quad ; \quad (A,B=1,\dots,8) \nonumber\\
&&\null\nonumber\\
 \mbox{ or in matrix notation}&& \nonumber\\
&&\null\nonumber\\
{\vec V}_{UspY} &=&  \left( \matrix { u^{AB} \cr  v_{AB} \cr }\right)
\end{eqnarray}
%%%%%%%%%%%%%%%%%%%%%%%%%%%%%%%%%%%%%%%%%%%%%%%%%%%%%%%%%%%%%%%%%%%%%%%
This discussion suffices to make the implicit form of the vector
kinetic matrix (\ref{gaiazumap}) explicitly calculable given any
explicit parametrization of the coset representative. As for the
kinetic matrix of the scalars we have:
\begin{equation}
  g_{ij}\left( \phi \right)  = {1 \over 3}P_{ABCD,i}\bar{P}^{ABCD}_{j}
\label{n8scalmet}
\end{equation}
where $P_{ABCD,i} \, d \phi^i$ is the scalar vielbein obtained from
the \textbf{gauged} left--invariant $1$--form of the scalar coset
that we discuss below (see eq.(\ref{uspYconnec}). Because of that the
lagrangian (\ref{lagrared}) contains also the minimal coupling of the
scalar fields to the gauge bosons of the chosen gauge group.
\par
To complete the  illustration of the bosonic lagrangian we need to
discuss the scalar potential ${\cal V}(\phi)$. This cannot be done
without referring to the supersymmetry transformation rules since, as we have explained
in section \ref{genaspect}, the potential is determined by the
fundamental relation (\ref{wardide}) that gives it as a quadratic
form in terms of the {\it fermion shifts}. These latter appear
in the supersymmetry transformation
rules of the fermionic fields and  are the primary objects determined by the
choice of the gauge algebra.
%%%%%%%%%%%%%%%%%%%%%%%%%%%%%%%%%%%%%%%%
% Punto di sutura  nuovo vecchio
%%%%%%%%%%%%%%%%%%%%%%%%%%%%%%%%%%%%%%%
\subsection{Supersymmetry transformation rules of the Fermi fields}
Since the $\mathcal{N}=8$ theory has no matter multiplets the
fermions are just, as already pointed out,
the ${\bf 8}$ spin $3/2$ gravitinos and the
${\bf 56}$ spin $1/2$ dilatinos. The two numbers ${\bf 8}$ and ${\bf
56}$ have been written boldfaced since they also single out the
dimensions of the two irreducible $\mathrm{SU(8)}$ representations to which
the two kind of fermions are respectively assigned, namely the
fundamental and the three times antisymmetric:
\begin{equation}
 \psi_{\mu\vert A} \, \leftrightarrow \,
 \mbox{$ \begin{array}{|c|}
\hline
\stackrel{ }{A}\cr
\hline
\end{array}$} \, \equiv \,  {\bf 8}\quad ; \quad \chi _{ABC}  \,
\leftrightarrow \,
\mbox{$ \begin{array}{|c|}
\hline
\stackrel{ }{A}\cr
\hline
\stackrel{}{B}\cr
\hline
\stackrel{}{C}\cr
\hline
\end{array}
$}   \, \equiv \,  {\bf 56}
\end{equation}
Following the conventions of \cite{unquart} the fermionic
supersymmetry transformation rules of are written as follows:
\begin{eqnarray}
\delta\psi_{A\mu}&=&\nabla_{\mu}\epsilon_A~
~-\frac{\rm 1}{4}\, T^{(-)}_{AB|\rho\sigma}\gamma^{\rho\sigma}\gamma_{\mu}\epsilon^{B}+
S_{AB} \, \gamma_\mu  \, \epsilon^B \, +
\cdots \nonumber\\
\delta\chi_{ABC}&=&~4 \mbox{\rm i} ~P_{ABCD|i}\partial_{\mu}\Phi^i\gamma^{\mu}
\epsilon^D-3T^{(-)}_{[AB|\rho\sigma}\gamma^{\rho\sigma}
\epsilon_{C]}+  \Sigma^{D}_{ABC} \, \epsilon_D \cdots
\label{trasforma}
\end{eqnarray}
where
$T^{-}_{AB\vert \mu\nu}$ is the antiselfdual part of the
graviphoton field strength, $P_{ABCD\vert i }$ is the already mentioned vielbein
of the scalar
coset manifold completely antisymmetric in $ABCD$ and satisfying the
same pseudoreality condition as our choice of the scalars $\phi_{ABCD}$:
\begin{equation}
P_{ABCD}=\frac{1}{4!}\epsilon_{ABCDEFGH}\bar P^{EFGH}.
\end{equation}
By comparison with eq.s(\ref{gravsusy}) and (\ref{fermioshif}) we see that
$S_{AB}$,$\Sigma^{D}_{ABC}$ are the appropriate {\it gravitino mass matrix}
and {\it fermion shifts}. Recalling also the normalization of the
fermion kinetic terms:
\begin{equation}
  \mathcal{L}^{kin}_{fermion}=\int d^4x\left [ 2\left( {\bar \psi}^A_\mu \, \gamma_\nu
   \nabla_\rho \psi_{A \vert \mu} +\mbox{h.c} \right) -
   \mbox{i}\,\sqrt{-g}\,\frac{1}{24}\left(
{\bar \chi}^{ABC} \gamma^\mu \, \nabla_\mu \chi_{ABC} -\mbox{h.c.} \right) \right]
\label{kinfermact}
\end{equation}
the general Ward identity (\ref{wardide}) takes, in this theory, the
following explicit form:
\begin{equation}
-{\cal V} \,\delta^{A}_{B} \,  = 24\, S_{AM} \, S^{BM} \, - \,\frac{1}{6}
\Sigma_{A}^{PQR} \, \Sigma^{B}_{PQR}
\label{specwardid}
\end{equation}
\par
What we need is the explicit expression of the two items appearing   in
the supersymmetry transformations (\ref{trasforma}) in terms of the coset representatives.
For the graviphoton such an expression is \emph{independent of the gauging} and
coincides with that appearing in the case of ungauged supergravity.
On the contrary, the  expression of the scalar vielbein and of the fermion shifts,
involves the choice of the gauge group and can be given
only upon introducing the {\it gauged  Maurer Cartan equations}.
Hence we first   recall  the structure of the graviphoton and
then we turn our attention to the second kind of items entering the
transformation rules that are the most relevant ones in our
discussion.
\subsubsection{The graviphoton field strength}
We  introduce the multiplet of electric and magnetic field strengths
according to the general discussion of section \ref{dualsym}:
\begin{equation}
{\vec V}_{\mu\nu} \equiv \left(\matrix {
F^{\Lambda\Sigma}_{\mu\nu} \cr G_{\Delta\Pi\vert\mu\nu} \cr}\right)
\label{symvecft}
\end{equation}
where
\begin{eqnarray}
G_{\Delta\Pi\vert\mu\nu} &=& -\mbox{Im}{\cal N}_{\Delta\Pi,\Lambda\Sigma} \,
{\widetilde F}^{\Lambda\Sigma}_{\mu\nu} -
\mbox{Re}{\cal N}_{\Delta\Pi,\Lambda\Sigma} \,
{ F}^{\Lambda\Sigma}_{\mu\nu}\nonumber\\
{\widetilde F}^{\Lambda\Sigma}_{\mu\nu}&=&\frac{1}{2}\, \epsilon_{\mu
\nu\rho\sigma} \, F^{\Lambda\Sigma\vert \rho\sigma}
\end{eqnarray}
The $56$--component field strength vector ${\vec V}_{\mu\nu}$
transforms in the real symplectic representation of the U--duality
group $\mathrm{E_{7(7)}}$. We can also write  a column vector containing
the $ 28 $ components of the graviphoton field strengths and their
complex conjugate:
\begin{equation}
{\vec T}_{\mu\nu} \equiv \left(\matrix{
T^{\phantom{\mu \nu}\vert AB}_{\mu \nu}  \cr
T_{\mu \nu \vert AB} \cr }\right) \quad T^{\vert AB}_{\mu \nu} =
\left(T_{\mu \nu \vert AB}\right)^\star
\label{gravphotvec}
\end{equation}
in which the upper and lower components  transform in the canonical
{\it Young basis} of $\mathrm{SU(8)}$ for the ${\bar {\bf 28}}$ and  ${\bf 28}$
representation respectively.\par
The relation between the graviphoton field strength vectors and the
electric magnetic field strength vectors involves the coset
representative in the $\mathrm{SpY(56)}$ representation and it is the following one:
\begin{equation}
{\vec T}_{\mu \nu} = - {\cal C} \, \mathbb{C} \, \mathbb{L}_{SpY}^{-1}(\phi) \, {\vec
V}_{\mu \nu}
\label{T=SCLV}
\end{equation}
The matrix $\mathbb{C}$  being symplectic invariant matrix
(\ref{definizia}). Eq.(\ref{T=SCLV}) reveals that the
graviphotons transform under the $\mathrm{SU(8)}$ compensators  associated
with the $\mathrm{E_{7(7)}}$ rotations.
It is appropriate to express the upper and lower components of $\vec{T}$
in terms of the self--dual and antiself--dual parts of the graviphoton
field strengths, since only the latter enter the transformation rules (\ref{trasforma}).\\
These components are defined as follows:
\begin{eqnarray}
T^{+\vert AB}_{ \mu \nu}&=& \frac{1}{2}\left(T^{ \vert AB}_{\mu \nu}+
\frac{{\rm i}}{2}
\, \epsilon_{\mu\nu\rho\sigma}g^{\rho\lambda} g^{\sigma\pi}
\, T^{ \vert AB}_{\lambda \pi}\right) \nonumber\\
T^-_{ AB \vert \mu \nu}&=& \frac{1}{2}\left(T_{  AB \vert\mu \nu}-\frac{{\rm i}}{2}
\, \epsilon_{\mu\nu\rho\sigma}g^{\rho\lambda} g^{\sigma\pi}
\, T_{ AB \vert\lambda \pi} \right)
\label{selfdual}
\end{eqnarray}
As shown in \cite{noi3} the following equalities hold true:
\begin{eqnarray}
 T_{\mu  \nu}^{\phantom{\mu  \nu}\vert AB}&=& T^{+ \vert AB}_{\mu
 \nu} \quad; \quad
 T_{ \mu  \nu\vert AB} =  T^{-}_{ \mu  \nu\vert AB}
\label{symprop}
\end{eqnarray}
and we can simply write:
\begin{equation}
{\vec T}_{\mu \nu} \equiv \left(\matrix{T^{+\vert AB}_{\mu \nu}  \cr
T^{-}_{\mu \nu \vert AB} \cr }\right)
\end{equation}
\subsection{The gauged Maurer Cartan equations and the fermion shifts}
\label{3.3.3}
As  stressed in section \ref{genaspect} the key ingredient
in constructing the gauged version of any extended supergravity is
provided by the \emph{gauged} left-invariant 1--forms on the coset manifold.
We illustrate this notion in the present example.
\par
First note that in the $\mathrm{UspY(28,28)}$ basis we have chosen the coset
representative (\ref{cosetrep}) satisfies the following identities:
\begin{eqnarray}
u^{\Pi\Delta}{}_{AB}u^{AB}{}_{\Lambda\Sigma}-
v^{\Pi\Delta AB}v_{AB\Lambda\Sigma}&=&
\delta^{\Pi\Delta}_{\Lambda\Sigma}\,,\nonumber\\
u^{\Pi\Delta}{}_{AB}v^{AB\Lambda\Sigma}-
v^{\Pi\Delta AB}u_{AB}{}^{\Lambda\Sigma}&=&0\,,
\label{identuv}
\end{eqnarray}
\begin{eqnarray}
u^{AB}{}_{\Lambda\Sigma}u^{\Lambda\Sigma}{}_{CD}-
v^{AB \Lambda\Sigma}v_{\Lambda\Sigma CD}&=&
\delta^{AB}_{CD}\,,\nonumber\\
u^{AB}{}_{\Lambda\Sigma}v^{\Lambda\Sigma CD}-
v^{AB\Lambda\Sigma}u_{\Lambda\Sigma}{}^{CD}&=&0\,,
\label{2identuv}
\end{eqnarray}
and the inverse coset representative is given by:
\begin{equation}
\mathbb{L}^{-1} =\left( \matrix{  u^{AB}{}_{\Lambda\Sigma}  &
-v^{AB\Delta\Gamma} \cr
                   -v_{CD\Lambda\Sigma}    &  u_{CD}{}^{\Delta\Gamma}  }
                   \right)  \,.
\label{Linvers}
\end{equation}
where, by raising and lowering the indices, complex conjugation
is understood.
\par
Secondly recall that in our basis the generators of the electric
subalgebra $\mathrm{SL}(8,\mathbb{R}) \, \subset \, \mathrm{E_{7(7)}}$ have the following
form
\begin{equation}
G_\alpha  =
\left( \matrix{    q^{\Lambda\Sigma}{}_{\Pi\Delta}(\alpha )  &
                    p^{\Lambda\Sigma\Psi\Xi} (\alpha ) \cr
                   p_{\Delta\Gamma\Pi\Delta}(\alpha )  &
                   q_{\Lambda\Sigma}{}^{\Psi\Xi}(\alpha )  } \right)
\label{sl8gener}
\end{equation}
where the matrices $q$ and $p$ are real and have the following form
\begin{eqnarray}
q^{\Lambda\Sigma}{}_{\Pi\Delta}&=&2\delta^{[\Lambda}{}_{[\Pi}q^{\Sigma]}
{}_{\Delta]}={2\over 3}
\delta^{[\Lambda}{}_{[\Pi}q^{\Sigma]\Gamma}{}_{\Delta]\Gamma}\,,\nonumber\\
p_{\Delta\Gamma\Pi\Omega}&=&{1\over 24}
\varepsilon_{\Delta\Gamma\Pi\Omega\Lambda\Sigma\Psi\Xi}
p^{\Lambda\Sigma\Psi\Xi}\,.\label{uspconstraints}
\end{eqnarray}
The index $\alpha=1,\dots \, 63$  in (\ref{sl8gener}) spans the adjoint
representation of
$\mathrm{SL}(8,\mathbb{R})$ according to some chosen basis and we can freely raise
and lower the Greek indices $ \Lambda, \Sigma, ...$ because of the
reality of the representation.
\par
Next let us introduce the fundamental item in the gauging
construction. It is the $28 \times 63 $ constant embedding matrix:
\begin{equation}
{\bf {\cal E}} \equiv  e_{\Lambda\Sigma}^\alpha
\label{alettomat}
\end{equation}
transforming under $\mathrm{SL}(8,\mathbb{R})$ as its indices specify, namely in the
tensor product of the adjoint with the antisymmetric ${\bf 28}$ and
that specifies which
generators of $\mathrm{SL}(8,\mathbb{R})$ are gauged and by means of which vector
fields in the $28$--dimensional stock. In particular, using this matrix
${\cal E}$, one writes the gauge 1--form as:
\begin{equation}
 A \equiv A^{\Lambda\Sigma} e_{\Lambda\Sigma}^\alpha G_\alpha
 \label{gaugeform}
\end{equation}
In \cite{Cordaro:1998tx}   the
most general form of the embedding matrix
$e_{\Lambda\Sigma}^\alpha$ was determined that is consistent with
supersymmetry. Analyzing its structure and modding out all the irrelevant gauge
transformations one could determine all the different gauge groups.
I review that result and summarize its derivation. Here
In terms of the gauge 1--form $A$ and of the coset representative
$\mathbb{L}(\phi)$ we can write the {\it gauged left--invariant 1--form}:
\begin{equation}
\Omega=\mathbb{L}^{-1} d \mathbb{L} + g \mathbb{L}^{-1} A \mathbb{L} \,
\label{gaul1f}
\end{equation}
which   belongs to the $\mathrm{E_{7(7)}}$ Lie algebra in the
$\mathrm{UspY(28,28)}$ representation and defines the {\it gauged} scalar
vielbein $P^{ABEF}$ and the $\mathrm{SU(8)}$ connection $Q^{\phantom{D} B}_{D}$:
\begin{equation}
\Omega = \, \left( \matrix{
2 \delta^{[A }_{[C} \, Q^{\phantom{D} B]}_{D]}
& P^{ABEF} \cr P_{CDGH} &
- 2 \delta^{[E}_{[G} \, Q^{ F]}_{\phantom{F}H]}\cr }\right)
\label{uspYconnec}
\end{equation}
Because of its definition  the 1--form $\Omega$ satisfies
{\it gauged Maurer Cartan equations}:
\begin{equation}
d\Omega + \Omega\wedge\Omega =
g \big[F^{\Lambda\Sigma}
-\big( \sqrt{2}(u^{\Lambda\Sigma}{}_{AB}+v_{\Lambda\Sigma AB})
\bar\psi^A \wedge \psi^B + {\rm h.c.} \big)\big]
e_{\Lambda\Sigma}^\alpha \mathbb{L}^{-1} G_\alpha  \mathbb{L}\,,
\label{MaurerCartan}
\end{equation}
with $F^{\Lambda\Sigma} $ the supercovariant field strength of the vectors
$A^{\Lambda\Sigma}$. Let us focus on the last factor in
eq.(\ref{MaurerCartan}):
\begin{equation}
{\bf U}_\alpha  \, \equiv \, \mathbb{L}^{-1} G_\alpha  \mathbb{L}  =
\left( \matrix{    \cal A (\alpha ) &  \cal B (\alpha )\cr
                   \bar{\cal B} (\alpha )  &  \bar{\cal A} (\alpha ) } \right)\,
\label{boosted}
\end{equation}
Since ${\bf U}_\alpha$ is an $\mathrm{E_{7(7)}}$ Lie algebra element,
for each gauge generator $G_\alpha $ we necessarily have:
\begin{eqnarray}
{\cal A}^{AB}_{\phantom{AB}CD}(\alpha ) &=& \frac{2}{3} \,
\delta^{[A }_{[C} \, {\cal A}^{B]M}_{\phantom{B]M}D]M} \nonumber\\
 {\cal B}^{ABFG}(\alpha ) &=& {\cal B}^{[ABFG]}(\alpha )
\label{proprieta}
\end{eqnarray}
Comparing with eq.(\ref{MaurerCartan}) we see that the scalar field
dependent $\mathrm{SU(8)}$ tensors multiplying the gravitino bilinear terms
are the following ones:
\begin{eqnarray}
T^A{}_{BCD}& \equiv & (u^{\Omega\Sigma}{}_{CD} + v_{\Omega\Sigma CD}) \,
              e^\alpha_{\Omega\Sigma} \, {\cal A}^{AM}{}_{BM}(\alpha)
\label{Ttensor} \nonumber\\
Z_{CD}^{ABEF}&  \equiv &
(u^{\Omega\Sigma}{}_{CD} + v_{\Omega\Sigma CD}) \,
              e^\alpha_{\Omega\Sigma} \, {\cal B}^{ABEF}(\alpha )
\label{uvetta}
\end{eqnarray}
As shown in the original papers by de Wit and Nicolai \cite{dewit1}
(or Hull \cite{hull}) and reviewed in \cite{castdauriafre},
closure of the supersymmetry algebra
\footnote{ in the rheonomy approach closure of
the Bianchi identities} and hence existence of the corresponding
gauged supergravity models is obtained {\it if and only if}
the following $T$--identities are satisfied:
\begin{eqnarray}
T^A{}_{BCD} &=& T^A{}_{[BCD]} + \frac{2}{7} \delta^A_{[C} T^M{}_{D]MB}
\label{T-id} \\
Z^{CD}_{ABEF}&=&\frac{4}{3} \delta^{[C}{}_{[A} T^{D]}{}_{BEF]}
\label{T-idbis}
\end{eqnarray}
Eq.s \eqn{T-id} and \eqn{T-idbis} have a clear group theoretical
meaning. Namely, they state that both the $T^A{}_{BCD}$ tensor and
the $Z_{CD}^{ABEF}$ tensor can be expressed in a basis spanned by two
irreducible $\mathrm{SU(8)}$ tensors corresponding to the {\bf
420} and {\bf 36} representations  respectively:
\begin{equation}
{\stackrel {\circ}{^{\phantom{A}}T^A}}_{BCD}
\,  \equiv \,  \epsilon^{AI_1\dots I_7} \,
\mbox{$ \begin{array}{l}
{\begin{array}{|c|c|}
\hline
\stackrel{ }{I_1} & \stackrel { }{B}\cr
\hline
\stackrel{ }{I_2} & \stackrel { }{C}\cr
\hline
\stackrel{ }{I_3} & \stackrel { }{D}\cr
\hline
\end{array} } \cr
{\begin{array}{|c|}
\hline
\stackrel{ }{I_4} \cr
\hline
\end{array} }  \cr
{\begin{array}{|c|}
\hline
\stackrel{ }{I_5} \cr
\hline
\end{array} } \cr
{\begin{array}{|c|}
\hline
\stackrel{ }{I_6} \cr
\hline
\end{array} }   \cr
{\begin{array}{|c|}
\hline
\stackrel{ }{I_7} \cr
\hline
\end{array} } \cr
\end{array}$ }  \, \equiv \, {\bf 420} \nonumber  \, \qquad ; \qquad
 {\stackrel {\circ}{T}}_{DB}  \,  \equiv \,  \mbox{$
 \begin{array}{|c|c|}
\hline
\stackrel{ }{D} & \stackrel { }{B}\cr
\hline
\end{array} $} \, \equiv \, {\bf 36}
\label{representazie}
\end{equation}
 To see this let us consider first eq.(\ref{T-id}). In general a
 tensor of type $T^A{}_{B[CD]}$  would have $8 \times 8 \times 28$
 components and contain several irreducible representations of $\mathrm{SU(8)}$. However, as a
 consequence of eq. (\ref{T-id}) only the representations {\bf 420},
 {\bf 28} and {\bf 36} can appear. (see fig. \ref{young1}).
In addition, since the $\cal A$ tensor, being in the adjoint of $\mathrm{SU(8)}$, is traceless
also the $T$-tensor appearing in (\ref{T-id}) is traceless: $T^A{}_{ABC}=0$.
Combining this information   with eq.(\ref{T-id})  we obtain
\begin{equation}
T^M{}_{[AB]M}=0,
\label{no28}
\end{equation}
Eq. (\ref{no28}) is the statement that the {\bf 28} representation
appearing in fig. $1$ vanishes so that the $T^A{}_{B[CD]}$ tensor is
indeed expressed solely in terms of the irreducible tensors
\eqn{representazie}.
%%%%%%%%%%%%%%%%%%%%%
\iffigs
\begin{figure}
\begin{center}
\caption{\label{young1} Decomposition
of a tensor of type $T^A{}_{BCD}$ into irreducible representations } \epsfxsize = 10cm
\epsffile{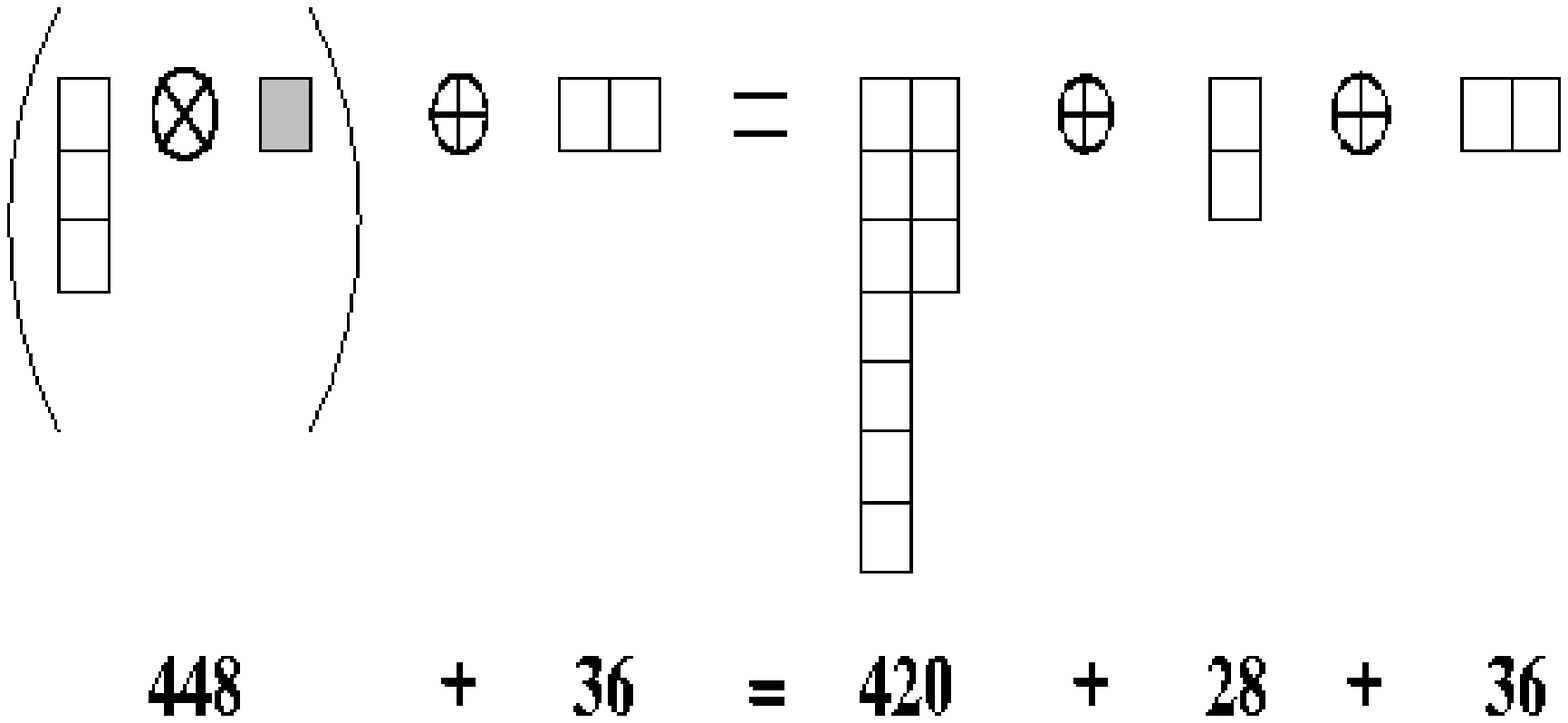}
\vskip -0.1cm
\unitlength=1mm
\end{center}
\end{figure}
\fi
%%%%%%%%%%%%%%%%%%%%%
A similar argument can be given to interpret the second $T$--identity (\ref{T-idbis}).
A tensor of type $Z_{[CD]}^{[ABEF]}$ contains, a priori, $70 \times 28$
components and contains the irreducible representations  {\bf 1512},
{\bf 420} and {\bf 28} (see fig. \ref{young2}). Using eq.(\ref{T-idbis}) one immediately
sees that the representations {\bf 1512} and {\bf 28} must vanish and
that the surviving  {\bf 420} is proportional through a fixed coefficient to the
{\bf 420} representations appearing in the decomposition of the $T^A{}_{B[CD]}$ tensor.
%%%%%%%%%%%%%%%%%%%%%
\iffigs
\begin{figure}
\begin{center}
\caption{\label{young2} Decomposition
of a tensor of type $Z^{CD}_{ABEF}$ into irreducible representations }
\epsfxsize = 10cm
\epsffile{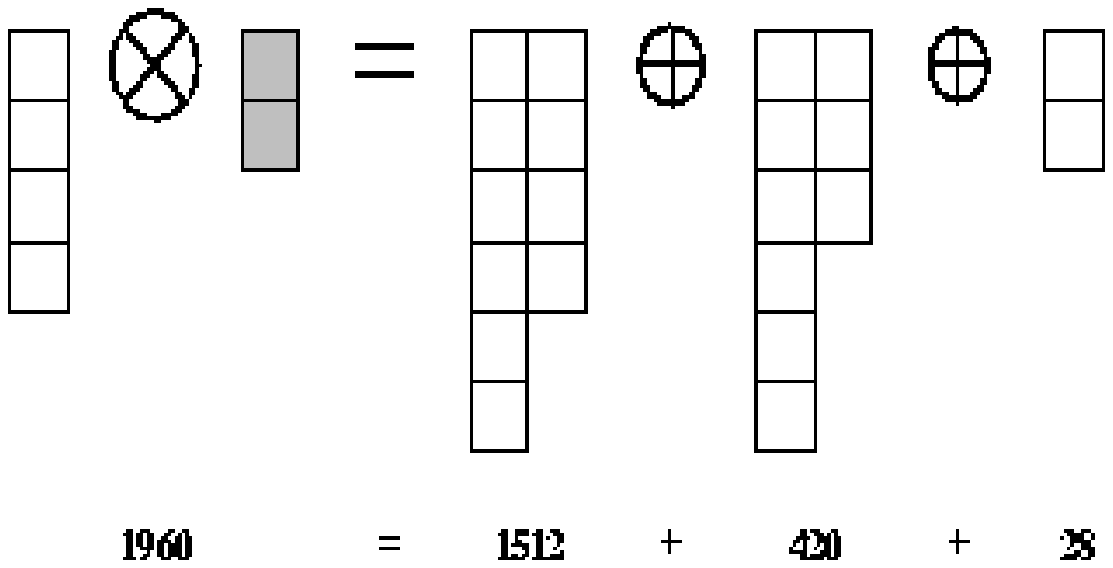}
\vskip -0.1cm
\unitlength=1mm
\end{center}
\end{figure}
\fi
%%%%%%%%%%%%%%%%%%%%%
\noindent
In view of this discussion, the $T$--identities can be
rewritten as follows in the basis of the independent irreducible
tensors
\begin{eqnarray}
{\stackrel {\circ}{^{\phantom{A}}T^A}}{}_{ BCD } &=& T^A{}_{[BCD]}
  \quad ; \quad {\stackrel {\circ}{T}}_{AB} = T^M{}_{AMB}
\label{panettone}
\end{eqnarray}
%%%%%%%%%%%%%%%%%%%%%%%%%%%%%%%%%%%%%%%%%%%%%%%%%%%%%%%%%%%%%%%%
The irreducible tensors {\bf 420} and {\bf 36}   can be identified, through a
suitable coefficient fixed by Bianchi identities, with the fermion shifts appearing
in the supersymmetry transformation rules (\ref{trasforma}):
\begin{equation}
\Sigma^{A}_{BCD} \, = \, \sigma  {\stackrel {\circ}{^{\phantom{A}}T^A}}_{ BCD } \quad ; \quad
S_{DB} \, = \, s \, {\stackrel {\circ}{T}}_{DB}
\label{agnizio}
\end{equation}
%%%%%%%%%%%%%%%%%%%%%%%%%%%%%%%%%%%%%%%%%%%%%%%%%%%%%%%%%%%%%%%%
Finally, as shown by de Wit and Nicolai \cite{dewit1} the crucial Ward identity
(\ref{specwardid}) is satisfied if
and only if the ratio between the two constants in eq. (\ref{agnizio}) is:
\begin{equation}
 \frac{s^2}{\sigma^2} \, = \, \frac{1}{392}
 \label{razione}
\end{equation}
\subsection{Algebraic characterization of the gauge group embedding
$G_{gauge} \longrightarrow \mathrm{SL}(8,\mathbb{R})$}
As we have seen in the previous section the existence of gauged
supergravity models relies on a peculiar pair of identities to
be satisfied by the $T$--tensors. Therefore a classification of all
possible gaugings involves a parametrization of all $\mathrm{SL}(8,\mathbb{R})$
subalgebras that lead to satisfied $T$--identities. Since the
$T$--tensors are scalar field dependent objects it is not immediately
obvious how such a program can be carried through. On the other
hand since the problem is algebraic in nature (one looks for all
Lie subalgebras of $\mathrm{SL}(8,\mathbb{R})$ fulfilling a certain property) it is
clear that it should admit a completely algebraic formulation.
It turns out that such an algebraic formulation is possible
and actually very simple. Indeed the $T$--identities
imposed on the $T$--tensors are nothing else but a single algebraic
equation imposed on the embedding matrix ${\cal E}$
introduced in eq.(\ref{alettomat}). This is what we outline next.
\par
To begin with we recall a general and obvious constraint to be satisfied
by ${\cal E}$ which embeds a subalgebra of the $\mathrm{SL}(8,\mathbb{R})$ Lie
algebra into its ${\bf 28}$ irreducible representation:
the \textbf{vectors} should be in the
\textbf{coadjoint representation} of the gauge group. Hence under the
reduction to $G_{gauge}\subset \mathrm{SL}(8,\mathbb{R})$ we must obtain the following
decomposition of the entire set of the electric vectors:
\begin{equation}
{\bf 28}\stackrel{G_{gauge}}{\rightarrow }{\bf coadj} {G_{gauge}}
\oplus {\cal R}
\label{adjdec}
\end{equation}
where ${\cal R}$ denotes the subspace of vectors not entering the
adjoint representation of $G_{gauge}$ which is not necessarily a
representation of $G_{gauge}$ itself.
\par
Next in order to reduce the field dependent $T$--identities to an
algebraic equation on ${\cal E}$ we introduce the following constant
tensors:\footnote{
For example, in the de Wit--Nicolai theory, where one gauges
$G_{gauge} =SO(8)$ we have:
\[
t^{(1)}_{\Omega\Sigma}{}^{\Pi\Gamma}{}_{\Delta\Lambda}=
\delta_{[\Delta}^{[\Pi}\delta_{\Lambda][\Omega}
\delta_{\Sigma]}^{\Gamma]}\,,\qquad t^{(2)}_{\Omega\Sigma}{}^{\Pi\Gamma\Delta\Lambda}=0  \,.
\]
}
\begin{eqnarray}
 t^{(1)}_{\Omega\Sigma}{}^{\Pi\Gamma}{}_{\Delta\Lambda}
&\equiv& \sum_{\alpha } e^\alpha_{\Omega\Sigma} \,
q^{\Pi\Gamma}{}_{\Delta\Lambda}(\alpha ) \, \quad ,
\, \quad
t^{(2)}_{\Omega\Sigma}{}^{\Pi\Gamma\Delta\Lambda}
\equiv  \sum_{\alpha } e^\alpha_{\Omega\Sigma} \,
p^{\Pi\Gamma\Delta\Lambda}(\alpha )
\,.
\end{eqnarray}
In terms of $t^{(1)}$ and  $t^{(2)}$  the field dependent $T$-tensor is
rewritten as
\begin{eqnarray}
T^A{}_{BCD} = (u^{\Omega\Sigma}{}_{CD} + v_{\Omega\Sigma CD})
             \big[\,
             t^{(1)}_{\Omega\Sigma}{}^{\Pi\Gamma}{}_{\Delta\Lambda}\,
                  (u^{AM}{}_{\Pi\Gamma} \, u^{\Delta\Lambda}{}_{BM}
                   - v^{AM\Phi\Gamma} \, v_{\Delta\Lambda BM} ) \,\,
                   \nonumber\\
                  + t^{(2)}_{\Omega\Sigma}{}^{\Pi\Gamma\Delta\Lambda}\,
                  (u^{AM}{}_{\Pi\Gamma} \, v^{\Delta\Lambda}{}_{BM}
                   - v^{AM\Phi\Gamma} \, u_{\Delta\Lambda BM} ) \, \big]\,.
\end{eqnarray}
By means of lengthy algebraic manipulations in \cite{Cordaro:1998tx} the
following  statement was shown to be true
\begin{teorema}
{\large The field dependent $T$-identities are fully equivalent to the
following algebraic equation:}
\begin{eqnarray}
t^{(1)}_{\Omega\Sigma}{}^{\Pi\Gamma}{}_{\Delta\Lambda} + t^{(1)}_{\Delta\Lambda}{}^{\Pi\Gamma}{}_{\Omega\Sigma}
+ t^{(2)}_{\Pi\Gamma}{}^{\Delta\Lambda}{}^{\Omega\Sigma} &=& 0
\,,  \nonumber
\\
\label{lt-id}
\end{eqnarray}
\end{teorema}
Here we omit the proof but we stress the relevance of the result. All
possible gauged supergravities have been put into one-to-one
correspondence with the inequivalent solutions of an algebraic
equation to be satisfied by the embedding matrix.
\par
The algebraic $t$--identity (\ref{lt-id}) is a linear equation imposed
on the embedding matrix ${\cal E}$. In \cite{Cordaro:1998tx} it was solved    by
means of a computer program yielding a 36--parameter
solution. It was then shown that all the $36$ parameters could be absorbed
by means of conjugations with elements of the electric subgroup leaving only a
finite discrete set of inequivalent solutions corresponding to as many
inequivalent compact and non compact subalgebras of $\mathrm{SL}(8,\mathbb{R})$.
In order to describe this result more explicitly we need to discuss
the embedding of the electric group in some detail
\subsubsection{Embedding of the electric subalgebra $\mathrm{SL}(8,\mathbb{R})$ }
The Electric $Sl(8,\mathbb{R})$ subalgebra is identified in $\mathrm{E_{7(7)}}$ by
specifying its simple roots $\beta_i$ spanning the standard $A_7$
Dynkin diagram. The Cartan generators are the same for the
$\mathrm{E_{7(7)}}$ Lie algebra as for the $\mathrm{SL}(8,\mathbb{R})$ subalgebra and if we
give $\beta_i$ every other generator is defined.
The basis we have chosen is the following one:
\begin{equation}
\begin{array}{ccccccc}
\beta _1\,&=&\,\alpha _2+2\alpha _3+3\alpha _4+2\alpha _5+2 \alpha _6+
\alpha _7
&;&
\beta _2 &=& \alpha_1 \cr
\beta _3\,&=&\,\alpha _2 & ; & \beta _4 & =&  \alpha _3 \cr
\beta _5\,&=&\,\alpha _4 &; & \beta _6 & =& \alpha _6 \cr
\beta _7\,&=&\,\alpha_7  & \null & \null & \null & \null \cr
\end{array}
\label{simrot}
\end{equation}
The complete set of positive roots of $\mathrm{SL}(8\mathbb{R})$ is then composed
of $28$ elements that we name $\rho_i$ ($i=1,\dots,28$) and that are
enumerated according to our chosen order in table \ref{h_igauging}.
%%%%%%%%%%%%%%%%%%%%%%%%%%%%%%%%%%%%%%%%%%%%%%%%%%%%%%%%%%%%%%%%%%%
%%%%%%%%%%%%%%%%%%%%%%%%%%%%%%%%%%%%%%%%%%%%%%%%%%%%%%%%%%%
\begin{table}[ht]\caption{{\bf}
The choice of the order of the $\mathrm{SL}(8,\mathbb{R})$ roots:}
\label{h_igauging}
\begin{center}
\begin{tabular}{||cl||}
\hline
\hline
\null & \null \\
$ \rho_1 \, \equiv\, $ &$ \beta_2  $ \\
$ \rho_2 \, \equiv\, $ &$ \beta_2+\beta_3  $   \\
$ \rho_3 \, \equiv\, $ &$ \beta_2+\beta_3+\beta_4 $ \\
$ \rho_4 \, \equiv\, $ &$ \beta_2+\beta_3+\beta_4+\beta_5  $    \\
$ \rho_5 \, \equiv\, $ &$ \beta_2+\beta_3+\beta_4+\beta_5+\beta_6  $ \\
$ \rho_6 \, \equiv\, $ &$ \beta_3 $    \\
$ \rho_7 \, \equiv\, $ &$ \beta_3+\beta_4 $  \\
$ \rho_8 \, \equiv\, $ &$ \beta_3+\beta_4+\beta_5 $    \\
$ \rho_9 \, \equiv\, $ &$ \beta_3+\beta_4+\beta_5+\beta_6  $  \\
$ \rho_{10} \, \equiv\, $ &$ \beta_4 $   \\
$ \rho_{11} \, \equiv\, $ &$ \beta_4+\beta_5 $  \\
$ \rho_{12} \, \equiv\, $ &$ \beta_4+\beta_5+\beta_6   $    \\
$ \rho_{13} \, \equiv\, $ &$ \beta_5 $ \\
$ \rho_{14} \, \equiv\, $ &$ \beta_5+\beta_6  $    \\
$ \rho_{15} \, \equiv\, $ &$ \beta_6   $  \\
$ \rho_{16} \, \equiv\, $ &$ \beta_1+\beta_2+\beta_3+\beta_4+\beta_5+\beta_6+\beta_7  $    \\
$ \rho_{17} \, \equiv\, $ &$ \beta_2+\beta_3+\beta_4+\beta_5+\beta_6+\beta_7  $  \\
$ \rho_{18} \, \equiv\, $ &$ \beta_3+\beta_4+\beta_5+\beta_6+\beta_7   $    \\
$ \rho_{19} \, \equiv\, $ &$ \beta_4+\beta_5+\beta_6+\beta_7  $ \\
$ \rho_{20} \, \equiv\, $ &$ \beta_5+\beta_6+\beta_7  $    \\
$ \rho_{21} \, \equiv\, $ &$ \beta_6+\beta_7  $\\
$ \rho_{22} \, \equiv\, $ &$ \beta_1  $    \\
$ \rho_{23} \, \equiv\, $ &$ \beta_1+\beta_2  $ \\
$ \rho_{24} \, \equiv\, $ &$ \beta_1+\beta_2+\beta_3 $    \\
$ \rho_{25} \, \equiv\, $ &$ \beta_1+\beta_2+\beta_3+\beta_4  $  \\
$ \rho_{26} \, \equiv\, $ &$ \beta_1+\beta_2+\beta_3+\beta_4+\beta_5   $    \\
$ \rho_{27} \, \equiv\, $ &$ \beta_1+\beta_2+\beta_3+\beta_4+\beta_5+\beta_6  $ \\
$ \rho_{28} \, \equiv\, $ &$ \beta_7 $ \\
\null & \null         \\
\hline
\end{tabular}
\end{center}
\end{table}
\par
Hence the $63$ generators of   $\mathrm{SL}(8,\mathbb{R})$   are:
\begin{equation}
\begin{array}{ccc}
\mbox{The 7 Cartan generators} & C_i   =   H_{\alpha_i} & i=1,\dots,7\cr
\mbox{The 28 positive root generators} & E^{\rho_i} & i=1,\dots,28
\cr
\mbox{The 28 negative root generators} & E^{-\rho_i} & i=1,\dots,28
\cr
\end{array}
\label{genenume}
\end{equation}
and since the $56 \times 56$ matrix representation of each $\mathrm{E_{7(7)}}$
Cartan generator or step operator was constructed in section
\ref{pesanti} it is obvious that it is in particular given for the subset
of those that belong to the $\mathrm{SL}(8,\mathbb{R})$ subalgebra. The basis of this
matrix representation is provided by the weights enumerated in table
\ref{e7weight}. \par
In this way we have concluded our illustration of the basis in which
we have solved the algebraic $t$--identity.
The result found in \cite{Cordaro:1998tx} by means of a computer programme
is a $28 \times 63 $ matrix:
\begin{equation}
{\cal E}(h,\ell) \, \longrightarrow \, e^{\alpha}_{W}(h,\ell)
 \label{solvomat}
\end{equation}
where the index $W$ runs on the $28$ negative weights of table
\ref{e7weight}, while the index $\alpha$ runs on all the $\mathrm{SL}(8,\mathbb{R})$
generators according to eq. (\ref{genenume}). The matrix ${\cal E}(h,p)$
depends on $36$ parameters that were  named $h_i   i=1,\dots, 8$ and
$\ell_i  i=1, \dots ,28 $, respectively. The distinction between the two sets
of parameters is drawn in the following way: the $8$ parameters  $h_i$ are those that
never multiply a Cartan generator, while the $28$ parameters $\ell_i$ are those
that multiply at least one Cartan generator.
In other words if we set all the $\ell_i=0$ the gauge subalgebra
$G_{gauge} \subset \mathrm{SL}(8,\mathbb{R})$ is composed solely of step operators
while if you switch on also the $\ell_i$.s then some Cartan generators
appear in the Lie algebra. This distinction is very useful
in classifying the independent solutions.
\subsection[Classification of gauge algebras]{Classification of gauge algebras}
First we observe that the solution of $t$--identities encoded in the matrix ${\cal
E}(h,\ell)$ is certainly overcomplete since we are still free to
conjugate any gauge algebra $G_{gauge}$ with an arbitrary finite
element of the electric group $g \, \in \, \mathrm{SL}(8,\mathbb{R})$:
$G^\prime_{gauge}=g \,G_{gauge} \, g^{-1} $ yields a completely
physically equivalent gauging as $G_{gauge}$. This means that we
need to consider the $\mathrm{SL}(8,\mathbb{R})$ transformations of the matrix
${\cal E}(h,\ell)$ defined as:
\begin{equation}
\forall \, g \,\in \, \mathrm{SL}(8,\mathbb{R}) \, : \,  g  \cdot {\cal E}(h,\ell)
\equiv D_{28}(g^{-1}) \, {\cal E}(h,\ell) \, D_{63}(g)
\label{conjug}
\end{equation}
where $D_{28}(g)$ and $D_{63}(g)$ denote the matrices of the ${\bf 28}$
and the ${\bf 63}$ representation respectively.
If two set of parameters $\{h,\ell\}$ and
$\{ h^\prime,\ell^\prime \}$ are related by an $\mathrm{SL}(8,\mathbb{R})$
conjugation, in the sense that:
$\exists g \, \in \,\mathrm{SL}(8,\mathbb{R}) $ : $
 {\cal E}(h^\prime,\ell^\prime)=g \cdot {\cal E}(h,\ell)
$ then the theories described by  $\{h,\ell\}$  and
$\{ h^\prime,\ell^\prime \}$ are the same theory. In other words what
we need is the space of orbits of $\mathrm{SL}(8,\mathbb{R})$ inequivalent embedding
matrices. Possible theories obtained by choosing a set of $\{h,\ell\}$
parameters are further restricted by the constraints that:
the selected generators of $\mathrm{SL}(8,\mathbb{R})$ should close a Lie
subalgebra $G_{gauge}$
and that the selected vectors (=weights) should
transform in the coadjoint representation $Coadj
\left(G_{gauge}\right)$. Hence the $28$ linear combinations of $\mathrm{SL}(8,\mathbb{R})$
generators:
\begin{equation}
  T_W \equiv e_{W}^{\phantom{W}\alpha} \left(h,\ell\right) \, G_\alpha
\label{wgene}
\end{equation}
must span the adjoint representation of a $28$--dimensional
subalgebra $G_{gauge}(h,\ell) \subset \mathrm{SL}(8,\mathbb{R})$ algebra.
Naming ${\cal G}_{gauge}(h,\ell) $ the corresponding Lie subgroup, because of its very
definition we have that the matrix ${\cal E}(h,\ell)$ is invariant under
transformations of ${\cal G}_{gauge}(h,\ell)$, i.e.
$ \forall   \gamma \, \in\, {\cal G}_{gauge}(h,\ell )$ we have
$ \gamma \, \cdot \, {\cal E}(h,\ell) = {\cal E}(h,\ell)$.
\footnote{Note that some of the 28 generators of ${\cal G}_{gauge}(h,\ell)  \,
\subset \, \mathrm{SL}(8,\mathbb{R})$
may be trivially represented  in the adjoint representation, but in this case also the
corresponding group transformations leave  the embedding matrix invariant.}:
Hence, fixing a matrix  $\mathcal{E}(h,\ell)$ and hence an algebra $\mathcal{G}_{gauge}(h,\ell)$,
by acting on it with $\mathrm{SL}(8,\mathbb{R})$ we obtain a $35$--dimensional orbit of
equivalent embedding matrices, parametrized by the elements of the
coset manifold $\frac{\mathrm{SL}(8,\mathbb{R})}{{\cal G}_{gauge}(h,\ell)}$,
whose dimension is precisely $63 -28=35$. \par
Since the explicit solution
of the algebraic $t$--identities has produced an embedding matrix ${\cal E}(h,\ell)$
depending on no more than $36$--parameters, then the only
continuous parameter which is physically relevant is an overall
constant, the remaining $35$--parameters being
reabsorbed by $\mathrm{SL}(8,\mathbb{R})$ conjugations.
An essential and a priori unexpected conclusion follows from this discussion.
\par
\begin{proposition}
{ The gauged $\mathcal{N}=8$ supergravity models cannot depend on more than a
single continuous parameter  (=coupling constant), even if they
correspond to gauging a multidimensional abelian algebra}
\end{proposition}
In other words what we have found is that the space of orbits we were looking for is a discrete
space. The classifications of gauged supergravity models is just a
classification of gauge algebras a single coupling constant being
assigned to each case. This is considerably different from other supergravities
with less supersymmetries, like the $\mathcal{N}=2$ case.
There gauging a group $G_{gauge}$ involves
as many coupling constants as there are simple factors in
$G_{gauge}$. This difference
is an yield of supersymmetry and not of Lie algebra theory.
\par
To classify the orbits and hence the gauged supergravity theories
we consider invariants.
The natural invariant associated with the embedding matrix ${\cal E}(h,\ell)$
is the {\it signature} of the
{\it Killing--Cartan 2--form} for the resulting gauge algebra ${\cal G}_{gauge}(h,\ell)$.
Consider the $28$ generators (\ref{wgene}) and define:
\begin{equation}
\begin{array}{rcl}
\eta_{W_1W_2}\left( h,\ell\right ) & \equiv & \mbox{Tr}  \, \left( T_{W_1}
\, T_{W_1} \right)   \cr
\null & = & e_{W_1}^{\phantom{W_1}\alpha }\left(h,\ell \right ) \,
e_{W_2}^{\phantom{W_1}\beta }\left(h,\ell\right) \,
\mbox{Tr}  \,
\left( G_ \alpha \, G_ \beta \right) \cr
\null & =  & e_{W_1}^{\phantom{W_1}\alpha }\left(h,\ell\right) \,
 e_{W_2}^{\phantom{W_1}\beta }\left(h,\ell\right)
\, B_{\alpha\beta}\cr
\end{array}
\label{assas1}
\end{equation}
where the trace $Tr$ is taken over any representation and the
constant matrix $ B_{\alpha\beta} \equiv \mbox{Tr} \,\left( G_ \alpha \, G_
\beta \right)$ is the Killing--Cartan
2--form of the $\mathrm{SL}(2,\mathbb{R})$ Lie algebra. The $ 28 \times 28 $
matrix
is the Killing--Cartan 2--form of the gauge algebra $G_{gauge}$. As
it is well known from general Lie algebra theory, by means of
suitable changes of bases inside the same Lie algebra the matrix
$\eta_{W_1W_2}\left( h,\ell\right )$
can be diagonalized and its eigenvalues can be reduced to be either of
modulus one or zero. What cannot be done since it corresponds to an
intrinsic characterization of the Lie algebra is to change the
signature of $\eta_{W_1W_2}\left( h,\ell\right )$, namely the ordered set
of $28$ signs (or zeros) appearing on the principal diagonal when
$\eta_{W_1W_2}\left( h,\ell\right )$
is reduced to diagonal form. Hence what is constant throughout   an
$\mathrm{SL}(8,\mathbb{R})$ orbit is the signature. Let us   name
$\Sigma \left( \mbox{orbit} \right)$ the
$28$ dimensional vector characterizing the signature of an orbit.
From our discussion we conclude that
the classification of gauged $\mathcal{N}=8$ models has been reduced to the
classification of the signature vectors $\Sigma\left( \mbox{orbit}
\right)$. In \cite{Cordaro:1998tx} it was shown
that for generic   values of $h_i$
and $\ell_i$ the  matrix
$\eta_{W_1W_2}\left( h,\ell \right )\,$   has $28 \times 28 $ non vanishing entries,
while setting $\ell_i=0$ it becomes automatically diagonal and of the form:
\begin{eqnarray}
&\eta\left( h,\ell = 0\right ) \,= & \nonumber \\
& \begin{array}{rcccccccl}
\mbox{diag}
\Bigl \{
 & - {h_7}\,{h_8}  , &{h_1}\,{h_6}, &{h_2}\,{h_6},&
   - {h_3}\,{h_6} & ,{h_4}\,{h_6},& - {h_5}\,{h_6}  ,  &
   {h_1}\,{h_2}, & \null \\
      \null & - {h_1}\,{h_3}  , & {h_1}\,{h_4},   &
   - {h_1}\,{h_5}  , & - {h_2}\,{h_5}  , & {h_3}\,{h_5},
 &   - {h_4}\,{h_5}  , & {h_2}\,{h_4}, & \null \\
   \null &  - {h_2}\,{h_3}  ,  &
   - {h_3}\,{h_4}  , & {h_1}\,{h_7}, & {h_2}\,{h_7},    &
   - {h_3}\,{h_7}  ,& {h_4}\,{h_7}, & - {h_5}\,{h_7}, & \null  \\
   \null &   {h_6}\,{h_7}, & - {h_1}\,{h_8}  ,& - {h_2}\,{h_8}  ,  &
   {h_3}\,{h_8}, &- {h_4}\,{h_8}  , & {h_5}\,{h_8},
 &  - {h_6}\,{h_8} & \Bigr \} \
 \end{array} & \nonumber\\
\end{eqnarray}
Hence all possible signatures $\Sigma \left(orbit \right)$ are
obtained by assigning to the parameters $h_i$ the values $1,-1,0$ in all possible ways.
An explicit computer evaluation shows that for each signature
there is a corresponding acceptable $h$ vector such that the $28$
generators (\ref{wgene}) it singles out do close a Lie subalgebra.
In this way we have obtained an exhaustive
classification of supergravity gaugings that is displayed
in eq.(\ref{risulato})
 This classification is a list of $\mathrm{SL}(8,\mathbb{R})$ Lie subalgebras
identified by an acceptable $h$--vector
and a corresponding signature of the Killing--Cartan form. This is denoted by
writing the numbers $n_+$,$n_-$,$n_0$ of its positive, negative and zero
eigenvalues. In addition we have also written the actual dimension of
the gauge algebra namely the number of generators that have a
non--vanishing representations or correspondingly the number of
gauged vectors that are gauged (=paired to a non vanishing
generator):
\begin{table}[ht]
\caption{\label{risulato}{\bf Classification of $\mathcal{N}=8, D=4$ gaugings}}
$$
\begin{array}{|c|c|c|c|c|c|}
\hline
\mbox{Algebra} & {n_+} & {n_-} & {n_0}& \{h_1,h_2,h_3,h_4,h_5,h_6,h_7\} &
\mbox{dimension}
\cr
\hline
\hline
  \mathrm{SO}( 8 ) & 28 & 0 & 0 & \{1,1,-1,1,-1,1,1,-1\} & 28 \cr
  \mathrm{SO}( 1,7 ) & 21 & 7 & 0 & \{1,1,-1,1,-1,1,1,1\}& 28 \cr
  \mathrm{SO}( 2,6 ) & 16 & 12 & 0 & \{-1,1,-1,1,-1,1,1,1\}& 28\cr
  \mathrm{SO}( 3,5 ) & 13 & 15 & 0 & \{-1,-1,-1,1,-1,1,1,1\}& 28 \cr
  \mathrm{SO}( 4,4 ) & 12 & 16 & 0 & \{-1,-1,1,1,-1,1,1,1\} & 28\cr
  \mathrm{SO}(5,3  ) & 13 & 15 & 0 & \{-1,-1,1,-1,-1,1,1,1\}& 28\cr
  \mathrm{SO}( 6,2 ) & 16 & 12 & 0 & \{-1,-1,1,-1,1,1,1,1\} & 28\cr
  \mathrm{SO}( 7,1 ) & 21 & 7 & 0 & \{-1,-1,1,-1,1,-1,1,1\} & 28\cr
  \hline
  \mathrm{CSO}( 1 , 7) & 0 & 0 & 28 & \{0,0,0,0,0,0,0,1\}& 7 \cr
  \mathrm{CSO}(  2 , 6) & 1 & 0 & 27 & \{-1,0,0,0,0,0,0,1\} & 13 \cr
  \mathrm{CSO}(  3 ,5 ) & 3 & 0 & 25 & \{-1,-1,0,0,0,0,0,1\}& 18 \cr
  \mathrm{CSO}(  4 ,4 ) & 6 & 0 & 22 & \{-1,-1,1,0,0,0,0,1\}& 22 \cr
  \mathrm{CSO}( 5  , 3) & 10 & 0 & 18 & \{-1,-1,1,-1,0,0,0,1\}& 25 \cr
  \mathrm{CSO}( 6  , 2) & 15 & 0 & 13 & \{-1,-1,1,-1,1,0,0,1\}& 27\cr
  \mathrm{CSO}(  7 , 1) & 21 & 0 & 7 & \{-1,-1,1,-1,1,-1,0,1\}& 28\cr
  \hline
  \mathrm{CSO}(  1 ,1 , 6) & 0 & 1 & 27 &\{1,0,0,0,0,0,0,1\}& 13 \cr
  \mathrm{CSO}(  1 ,2 , 5) & 1 & 2 & 25 & \{1,-1,0,0,0,0,0,1\} & 18\cr
  \mathrm{CSO}(  2 ,1 , 5) & 1 & 2 & 25 & \{1,1,0,0,0,0,0,1\} & 18\cr
  \mathrm{CSO}( 1  ,3, 4) & 3 & 3 & 22 & \{1,-1,1,0,0,0,0,1\}& 22\cr
  \mathrm{CSO}( 2  ,2 , 4) & 2 & 4 & 22 & \{1,1,1,0,0,0,0,1\}  &  22  \cr
  \mathrm{CSO}(  3 ,1, 4) & 3 & 3 & 22 & \{1,1,-1,0,0,0,0,1\} & 22\cr
  \mathrm{CSO}(  1 ,4,3 ) & 6 & 4 & 18 & \{1,-1,1,-1,0,0,0,1\} & 25\cr
  \mathrm{CSO}(  2 ,3 , 3) & 4 & 6 & 18 &\{1,1,1,-1,0,0,0,1\} & 25\cr
  \mathrm{CSO}(  3 ,2, 3) & 4 & 6 & 18 & \{1,1,-1,-1,0,0,0,1\} & 25 \cr
  \mathrm{CSO}(  4 ,1, 3) & 6 & 4 & 18 & \{1,1,-1,1,0,0,0,1\} & 25\cr
  \mathrm{CSO}(  1 ,5 ,2 ) & 10 & 5 & 13 & \{1,-1,1,-1,1,0,0,1\} & 27\cr
  \mathrm{CSO}(  2 ,4, 2) & 7 & 8 & 13 & \{1,1,1,-1,1,0,0,1\} & 27 \cr
  \mathrm{CSO}( 3  ,3,2 ) & 6 & 9 & 13 & \{1,1,-1,-1,1,0,0,1\} & 27 \cr
  \mathrm{CSO}(  4 ,2,2 ) & 7 & 8 & 13 & \{1,1,-1,1,1,0,0,1\} & 27\cr
  \mathrm{CSO}(  5 ,1, 2) & 10 & 5 & 13 & \{1,1,-1,1,-1,0,0,1\} & 27\cr
  \mathrm{CSO}(  1 ,6, 1) & 15 & 6 & 7 &\{1,-1,1,-1,1,-1,0,1\} & 28 \cr
  \mathrm{CSO}(  2 ,5 ,1 ) & 11 & 10 & 7 &\{1,1,1,-1,1,-1,0,1\} & 28 \cr
  \mathrm{CSO}( 3  ,4, 1) & 9 & 12 & 7 & \{1,1,-1,-1,1,-1,0,1\}& 28 \cr
  \mathrm{CSO}( 4  ,3,1 ) & 9 & 12 & 7 & \{1,1,-1,1,1,-1,0,1\} & 28 \cr
  \mathrm{CSO}(  5 ,2, 1) & 11 & 10 & 7 & \{1,1,-1,1,-1,-1,0,1\} & 28 \cr
  \mathrm{CSO}( 6  ,1, 1) & 15 & 6 &  7   &\{1,1,-1,1,-1,1,0,1\}& 28 \cr
\hline
\hline
\end{array}
$$
\end{table}
By restricting the matrix  $e_{W}^{\phantom{W} \alpha}$ to the parameters
$h_i$ we can immediately write the correspondence between the vectors ${\vec W}^{(28+i)}$
and the generators of the gauge algebra that applies to all the gaugings we have classified above.
This correspondence is summarized in  table \ref{piripic},
where it suffices to substitute the corresponding values of $h_i$ to
obtain the generators of each gauge algebra expressed as linear
combinations of the $56$ positive and negative root step operators
of $\mathrm{SL}(8,\mathbb{R})$.
 %\par
%\vskip 0.3cm
\begin{table}[ht]
\begin{center}
\caption{{\bf
Gauge group generators in $\mathcal{N}=8$ gaugings:}}
\label{piripic}
\begin{tabular}{||cl||}
\hline
\hline
    Electric & \qquad \qquad\qquad Gauge \\
      vector &  \qquad\qquad\qquad generator\\
\hline
\null & \null \\
$ {\vec W}^{(35)} \, \leftrightarrow\, $ &$ h_2 E_{-\beta_2}-h_1
E_{\beta_2} $ \\
$ {\vec W}^{(36)} \, \leftrightarrow\, $ &$ h_3 E_{-\beta_2-\beta_3}+h_1 E_{\beta_2+\beta_3}  $
\\
$ {\vec W}^{(37)} \, \leftrightarrow\, $ &$ h_4 E_{-\beta_2-\beta_3-\beta_4}-h_1
E_{\beta_2+\beta_3+\beta_4} $ \\
$ {\vec W}^{(38)} \, \leftrightarrow\, $ &$ h_5 E_{-\beta_2-\beta_3-\beta_4-\beta_5}+h_1
E_{\beta_2+\beta_3+\beta_4+
\beta_5}  $    \\
$ {\vec W}^{(30)} \, \leftrightarrow\, $ &$ h_6 E_{-\beta_2-\beta_3-\beta_4-\beta_5-\beta_6}
-h_1 E_{\beta_2+\beta_3+
\beta_4+\beta_5+\beta_6}  $ \\
$ {\vec W}^{(45)} \, \leftrightarrow\, $ &$ -h_7 E_{-\beta_2-\beta_3-\beta_4-\beta_5-\beta_6
-\beta_7}+h_1 E_{\beta_2+
\beta_3+\beta_4+\beta_5+\beta_6+\beta_7} $    \\
$ {\vec W}^{(51)} \, \leftrightarrow\, $ &$h_1 E_{-\beta_1}+h_8
E_{\beta_1}  $  \\
$ {\vec W}^{(52)} \, \leftrightarrow\, $ &$ h_2 E_{-\beta_1-\beta_2}
+h_8 E_{\beta_1+\beta_2} $    \\
$ {\vec W}^{(53)} \, \leftrightarrow\, $ &$  h_3 E_{-\beta_1-\beta_2-\beta_3}-h_8
E_{\beta_1+\beta_2+\beta_3}  $  \\
$ {\vec W}^{(54)} \, \leftrightarrow\, $ &$ h_4 E_{-\beta_1-\beta_2-\beta_3-\beta_4}
+h_8 E_{\beta_1+\beta_2+\beta_3+
\beta_4} $   \\
$ {\vec W}^{(55)} \, \leftrightarrow\, $ &$   h_5 E_{-\beta_1-\beta_2-\beta_3-\beta_4-\beta_5}
-h_8 E_{\beta_1+\beta_2+
\beta_3+\beta_4+\beta_5} $  \\
$ {\vec W}^{(56)} \, \leftrightarrow\, $ &$  h_6 E_{-\beta_1-\beta_2-\beta_3-\beta_4
-\beta_5-\beta_6}+h_8 E_{\beta_1+
\beta_2+\beta_3+\beta_4+\beta_5+\beta_6}  $    \\
$ {\vec W}^{(29)} \, \leftrightarrow\, $ &$ -h_7 E_{-\beta_1-\beta_2-\beta_3-\beta_4
-\beta_5-\beta_6
-\beta_7}-h_8 E_{\beta_1+\beta_2+\beta_3+\beta_4+\beta_5+\beta_6+\beta_7}   $ \\
$ {\vec W}^{(43)} \, \leftrightarrow\, $ &$ -h_3 E_{-\beta_3}-h_2
E_{\beta_3}  $    \\
$ {\vec W}^{(42)} \, \leftrightarrow\, $ &$ -h_4 E_{-\beta_3-\beta_4}+h_2
E_{\beta_3+\beta_4}   $  \\
$ {\vec W}^{(39)} \, \leftrightarrow\, $ &$ -h_5
E_{-\beta_3-\beta_4-\beta_5}-h_2 E_{\beta_3+\beta_4+\beta_5}  $    \\
$ {\vec W}^{(31)} \, \leftrightarrow\, $ &$ -h_6
E_{-\beta_3-\beta_4-\beta_5-\beta_6}+h_2 E_{\beta_3+\beta_4+\beta_5+\beta_6}  $  \\
$ {\vec W}^{(46)} \, \leftrightarrow\, $ &$ h_7
E_{-\beta_3-\beta_4-\beta_5-\beta_6-\beta_7}-h_2 E_{\beta_3+\beta_4+\beta_5+\beta_6+\beta_7}   $    \\
$ {\vec W}^{(44)} \, \leftrightarrow\, $ &$h_4 E_{-\beta_4}+h_3
E_{\beta_4}  $ \\
$ {\vec W}^{(40)} \, \leftrightarrow\, $ &$ h_5 E_{-\beta_4-\beta_5}
-h_3 E_{\beta_4+\beta_5}  $    \\
$ {\vec W}^{(32)} \, \leftrightarrow\, $ &$  h_6
E_{-\beta_4-\beta_5-\beta_6}+ h_3 E_{\beta_4+\beta_5+\beta_6} $\\
$ {\vec W}^{(47)} \, \leftrightarrow\, $ &$  -h_7
E_{-\beta_4-\beta_5-\beta_6-\beta_7}- h_3 E_{\beta_4+\beta_5+\beta_6+\beta_7}  $    \\
$ {\vec W}^{(41)} \, \leftrightarrow\, $ &$ -h_5 E_{-\beta_5}-h_4
E_{\beta_5}  $ \\
$ {\vec W}^{(33)} \, \leftrightarrow\, $ &$   -h_6 E_{-\beta_5-\beta_6}+h_4
E_{\beta_5+\beta_6} $    \\
$ {\vec W}^{(48)} \, \leftrightarrow\, $ &$  h_7
E_{-\beta_5-\beta_6-\beta_7}-h_4
E_{\beta_5+\beta_6|\beta_7} $  \\
$ {\vec W}^{(34)} \, \leftrightarrow\, $ &$ h_6 E_{-\beta_6}+h_5
E_{\beta_6}   $    \\
$ {\vec W}^{(49)} \, \leftrightarrow\, $ &$ -h_7 E_{-\beta_6-\beta_7}-h_5
E_{\beta_6+\beta_7}  $ \\
$ {\vec W}^{(50)} \, \leftrightarrow\, $ &$  h_7 E_{-\beta_7}-h_6
E_{\beta_7} $ \\
%%%%%%%%%%%%%%%%%%%%%%%%%%%%%%%%%%%%%%%%%%%%%%%%%%%%%%%%%%%%
\hline
\end{tabular}
\end{center}
\end{table}
\subsection{The $\mathrm{CSO}\mathrm{(p,q,r)}$ algebras}
\label{csopqr}
The classification of table \ref{risulato}  besides the
obvious algebras $\mathrm{SO(p,q)}$ contains also the contracted algebras
$\mathrm{CSO}\mathrm{(p,q,r)}$. Since they are
relevant to the gaugings of maximal supergravity both in $D=4$ and
$D=5$ let me devote the present subsection to a brief description of their
structure. Indeed the only difference between the $D=4$ and $D=5$ case is that
in the former  we have $p+q+r=8$ why in the latter the condition
is $p+q+r=6$. So let the capital indices $I,J,K,\dots$ run on a
number $n$ of values that can either be $8$ or $6$.
%%%%%%%%%%%%%%
The generators of $\mathrm{SO(p,q)}$ (with $p+q=n$) in the vector representation are
\begin{equation}
(G^{IJ})^K_{~~L}=\delta^{[K}_J\eta^{L]I}
\end{equation}
where
\begin{equation}
\eta^{IJ}\equiv\,{\rm diag}(\overbrace{1,\dots,1}^p,\overbrace{-1,\dots,-1}^q)\,.
\end{equation}
They satisfy
\begin{equation}
[G^{IJ},G^{KL}]=f^{IJ,KL}_{MN}G^{MN}
\end{equation}
where
\begin{equation}
\label{structconst}
f^{IJ,KL}_{MN}=-2\delta^{[I}_{[M}\eta^{J][K}\delta^{L]}_{N]}\,.
\end{equation}
Their generalization, studied by Hull in the context of supergravity
\cite{hull},\cite{hull2} are the algebras $\mathrm{CSO}(p,q,\!r)$ with $p+q+r=n$,
defined by the
structure constants (\ref{structconst}) with
\begin{equation}
\eta^{IJ}\equiv\,{\rm diag}(\overbrace{1,\dots,1}^p,\overbrace{-1,\dots,-1}^q,
\overbrace{0,\dots,0}^r)\,.
\label{metriccpqr}
\end{equation}
Decomposing the indices
\begin{equation}
I=(\bar{I},\hat{I})~~~~\bar{I}=1,\dots,p+q,~\hat{I}=p+q+1,\dots,n\,,
\end{equation}
we have that $G^{\bar{I}\bar{J}}$ are the generators of
$\mathrm{SO(p,q)}\subset \mathrm{CSO}\mathrm{(p,q,r)}$ , while the $r(r-1)/2$ generators
$G^{\hat{I}\hat{J}}$ are central charges
\begin{equation}
[G^{\bar{I}\hat{J}},G^{\bar{K}\hat{L}}]={1\over 2}\eta^{\bar{I}\bar{K}}
G^{\hat{J}\hat{L}}\,.
\label{nonnull}
\end{equation}
They form an abelian subalgebra, and
\begin{equation}
\mathrm{SO(p,q)}\times \mathbb{R}^{r\left(r-1\right)\over 2}\subset
\mathrm{CSO(p,q,r)}\,.
\end{equation}
Notice that $\mathrm{\mathrm{CSO}(p,q,1)}=\mathrm{ISO(p,q)}$.
In the vector representation, the generators of the central charges are
identically null
\begin{equation}
(G^{\hat{I}\hat{J}})^K_{~L}=0\,,
\label{zerogen}
\end{equation}
while
\begin{equation}
(G^{\hat{I}\bar{J}})^K_{~L}={1\over 2}\delta^{\hat{I}}_L\eta^{\bar{J}K}\neq 0\,.
\end{equation}
It is worth noting that the Killing metric of $\mathrm{SO(p,q,r)}$ is
\begin{equation}
K^{IJ,KL}=f^{IJ,MN}_{PQ}f^{KL,PQ}_{MN}=-6\eta^{K[I}\eta^{J]L}\,.
\end{equation}
This notation is redundant, because the adjoint representation is
$n(n-1)/2$ dimensional. In the proper basis,
\begin{equation}
K^{IJ,KL}_{_{I<J,\,K<L}}=-3\eta^{IK}\eta^{JL}\,.
\end{equation}
This is a diagonal matrix of dimension $n(n-1)/2$, with components
$\eta^{II}\eta^{JJ}$. In general, the real sections of a given algebra
are characterized by the signature of the Killing metric\footnote{for non semisimple groups, by
signature we mean the number of positive, negative and null components of
the metric in its diagonal form}.
We see that, for the $\mathrm{CSO}\mathrm{(p,q,r)}$ algebras,
the signature of the Killing metric is equivalent to the signature of
the {\it vector} metric $\eta^{IJ}$. This explains why this tensor can give
an intrinsic characterization of such algebras.
%%%%%%%%%%%%%%%%%%%%%%%%%%%%%%%%%%%%%%
\section[Gauged supergravities in five dimensions and domain walls]{
Gauged supergravities in five dimensions and domain walls}
\label{gaugd5theo}
\setcounter{equation}{0}
%\section[$\mathcal{N}=2$ gauged supergravity in five--dimensions]{$\mathcal{N}=2$
%gauged supergravity in five--dimensions}
Recently the general form of \emph{gauged} $\mathcal{N}=2$ supergravity in five--dimensions
has been obtained. This occurred  through the contributions of two
groups of authors. In a first step G\"unaydin and Zagerman analysed
the problem of gauging in presence of an arbitrary number of vector
and tensor multiplets and in a series of papers \cite{Gunaydin:2000zx,Gunaydin:2000xk,Gunaydin:2001ph}
they established the key new features involved by the gauging procedure in this space-time dimension \footnote{
These main features are those described in detail in section \ref{gunasecti}
and can be summarized as follows:
\begin{enumerate}
  \item  N=2 vector fields outside the adjoint have to  be
converted to massive tensor fields
  \item Tensor fields have to sit in a symplectic
representation of the gauge group
  \item certain components of the $d_{\Lambda \Sigma \Gamma}$ tensor
  appearing in the Chern Simons coupling have to vanish when (gauged) tensor fields are present
  \item A certain group theoretical relation must exist between the
  above coefficients $d$ and the symplectic representation matrices
  acting on the tensor fields
\end{enumerate}
}
In a subsequent step Ceresole and Dall'Agata \cite{ceregatta}, utilizing the
general methods of the geometrical gauging
\cite{dewit1,Castellani:1986ka,D'Auria:1991fj,bertolo},  reconsidered
the problem and succeeded in including also the coupling to
hypermultiplets.
Since $\mathcal{N}=2$, that corresponds to $N_Q=8$ supercharges, is the minimal possible
number of supersymmetries in a five--dimensional space--time, it is clear
that this result is a relevant step in the quest for a minimal supersymmetrisation of
the Randall Sundrum scenario of the second type.
In the previous chapters I have already emphasized that this is the reason
why I choose precisely this theory as an example of \textbf{gauging}
in a \textbf{non--maximal, matter coupled} supergravity.
I already pointed out that in maximal supergravities  the number of
available gauge vectors is fixed a priori and the possible gauge
algebras fill a discrete set. In matter coupled supergravities, on the other hand,
the number of vector multiplets varies and one has much more
possibilities. If the number of supercharges $N_Q$ is larger than
eight the only available multiplets, beside the graviton multiplet,
are the vector or tensor multiplets; furthermore, given their number $n$,
the geometry of the scalar manifold is fixed and corresponds to a homogeneous space
$\mathcal{G}_n/\mathcal{H}_n$. At $N_Q=8$, instead, besides vector
or tensor multiplets (that can be dualised to vector multiplets) one
has also hypermultiplets so that the scalar manifold
$\mathcal{M}_{scalar}=\mathcal{M}_{vect. scal.}
\times \mathcal{M}_{hyp.scal.}$
is the tensor product of two submanifolds  containing the vector scalars
and the hyper scalars respectively.
As we have seen in sect.\ref{minid5geo}, although severely constrained,
the geometry of these two submanifolds is not completely
fixed by supersymmetry and can vary within an
ample class that contains both homogeneous and non homogeneous
spaces. As I also recalled in section \ref{minid5geo} there is a very
close structural relation between $N_Q=8$ supergravity in $D=4$
and in $D=5$ dimensions: the geometry of the hypermultiplet scalars
is the same in both theories, namely \emph{quaternionic geometry} (see
sect.\ref{hypgeosec}) while the vector scalars fill a \emph{special
K\"ahler} complex manifold in $D=4$ and a \emph{very special} real
manifold in $D=5$. Dimensional reduction on a circle maps $D=5$
theories into $D=4$ theories and provides a map from very special to
a subclass of special K\"ahler manifolds. Hence it is not
surprising that the \emph{gauging procedure} in $D=4$ and $D=5$
theories are extremely similar: yet there are some relevant differences
that had to be clarified before one could extend the
constructions of \cite{D'Auria:1991fj,bertolo} to one higher
space--time dimensions. These differences have essentially to do with
two specifically five--dimensional features:
\begin{enumerate}
  \item [a] Very special, differently from special K\"ahler manifolds
  are real and non--symplectic. So there is no notion of a moment-map
  for isometries
  \item [b] In presence of gauging vector and tensor multiplets
  become physically distinct and the vector fields that are in a non
  trivial non--adjoint representation of the gauge group have to be
  dualised to massive self--dual $2$--forms.
\end{enumerate}
In this section I will describe the general form of the
$\mathcal{N}=2$ gauging and compare it with the $\mathcal{N}=8$ gaugings in the
same dimensions. In another recent publication \cite{Andrianopoli:2000rs} these latter have
been obtained for all contracted and non contracted algebras
establishing in $D=5$ a complete classification of $\mathcal{N}=8$ gaugings fully
parallel to that derived in \cite{Cordaro:1998tx} for the $D=4$ case and
described in the previous section \ref{n8d4gaug}.
\par
To accomplish this programme  my first care is to discuss the general idea of the
moment map which constitutes an essential ingredient in the $\mathcal{N}=2$ case.
\subsection[The Moment Map]{The Moment Map}
\label{momentm1}
%%%%%%%%%%%%%%%%%%%%%%%%%%%%%%%%%%%%%%%%%%%%%%%%%%%%%%%%%%%%%%%%%
The moment map is a construction that applies to all manifolds
with a symplectic structure, in particular to K\"ahler, HyperK\"ahler
or quaternionic manifolds.
\par
I begin with the K\"ahler case, namely with the moment
map of holomorphic isometries which is the paradigma for all the other
cases. It is also the additional weapon one can use in gauging $D=4$ supergravity
while it is not available  for $D=5$ vector multiplets due to the real
structure of very special geometry.
The HyperK\"ahler and quaternionic cases correspond, instead,
to the moment map of triholomorphic isometries which equally applies to
$D=4$ and $D=5$ theories.
\subsubsection{The holomorphic moment map on K\"ahler manifolds}
I assume some basic knowledge of K\"ahler geometry which can
retrieved from any standard textbook.
Let  $g_{i {j^\star}}$ be the K\"ahler metric of a K\"ahler
manifold ${\cal M}$ and let us assume that  $g_{i {j^\star}}$ admits
a non trivial group of continuous isometries ${\cal G}$
generated by Killing vectors $k_\mathbf{I}^i$ ($\mathbf{I}=1, \ldots, {\rm dim}
\,{\cal G} )$ that define the infinitesimal variation of the complex
coordinates $z^i$ under the group action:
\begin{equation}
z^i \to z^i + \epsilon^\mathbf{I} k_\mathbf{I}^i (z)
\end{equation}
Let $k^i_{\mathbf{I}} (z)$ be a basis of holomorphic Killing vectors for
the metric $g_{i{j^\star}}$.  Holomorphicity means the following
differential constraint:
\begin{equation}
\partial_{j^*} k^i_{\mathbf{I}} (z)=0
\leftrightarrow \partial_j k^{i^*}_{\mathbf{I}} (\bar z)=0 \label{holly}
\end{equation}
while the generic Killing equation (suppressing the
gauge index $\mathbf{I}$):
\begin{equation}
\nabla_\mu k_\nu +\nabla_\mu k_\nu=0
\end{equation}
in holomorphic indices reads as follows:
\begin{equation}
\begin{array}{ccccccc}
\nabla_i k_{j} + \nabla_j k_{i} &=&0 & ; &
\nabla_{i^*} k_{j} + \nabla_j k_{i^*} &=& 0
\label{killo}
\end{array}
\end{equation}
where the covariant components are defined as
$k_{j }=g_{j i^*} k^{i^*}$ (and similarly for
$k_{i^*}$).
\par
The vectors $k_{\mathbf{I}}^i$ are generators of infinitesimal
holomorphic coordinate transformations $\delta z^i = \epsilon^\mathbf{I} k^i_{\mathbf{I}} (z)$
which leave the metric invariant. In the same way as the metric is
the derivative of a more fundamental
object, the Killing vectors in a K\"ahler manifold are the
derivatives of suitable prepotentials. Indeed the first of
eq.s (\ref{killo})  is automatically satisfied by holomorphic vectors
and the second equation reduces to the following one:
\begin{equation}
k^i_{\mathbf{I}}=i g^{i j^*} \partial_{j^*} {\cal P}_{\mathbf{I}},
\quad {\cal P}^*_{\mathbf{I}} = {\cal P}_{\mathbf{I}}\label{killo1}
\end{equation}
In other words if we can find a real function ${\cal P}^\mathbf{I}$ such
that the expression $i g^{i j^*} \partial_{j^*}
{\cal P}_{(\mathbf{I})}$ is holomorphic, then eq.(\ref{killo1}) defines a
Killing vector.
\par
The construction of the Killing prepotential can be stated in a more
precise geometrical fashion through the notion of {\it moment map}.
Let us review this construction.
\par
Consider a K\"ahlerian manifold ${\cal M}$ of real dimension $2n$.
Consider a compact Lie group ${\cal G}$ acting on
 ${\cal M}$  by means of Killing vector
fields $\overrightarrow{X}$ which are holomorphic
with respect to the  complex structure
${ J}$ of ${\cal M}$; then these vector
fields preserve also the K\"ahler 2-form
\begin{equation}
\begin{array}{ccc}
\matrix{
{\cal L}_{\scriptscriptstyle\overrightarrow{X}}g = 0 & \leftrightarrow &
\nabla_{(\mu}X_{\nu)}=0 \cr
{\cal L}_{\scriptscriptstyle\overrightarrow{X}}{  J}= 0 &\null &\null \cr }
  \Biggr \} & \Rightarrow &  0={\cal L}_{\scriptscriptstyle\overrightarrow{X}}
K = i_{\scriptscriptstyle\overrightarrow{X}}
dK+d(i_{\scriptscriptstyle\overrightarrow{X}}
K) = d(i_{\scriptscriptstyle\overrightarrow{X}}K) \cr
\end{array}
\label{holkillingvectors}
\end{equation}
Here ${\cal L}_{\scriptscriptstyle\overrightarrow{X}}$ and
$i_{\scriptscriptstyle\overrightarrow{X}}$
denote respectively the Lie derivative along
the vector field $\overrightarrow{X}$ and the contraction
(of forms) with it.
\par
If ${\cal M}$ is simply connected,
$d(i_{\overrightarrow{X}}K)=0$ implies the existence
of a function ${\cal P}_{\overrightarrow{X}}$ such
that
\begin{equation}
-\frac{1}{2\pi}d{\cal P}_{\overrightarrow{X}}=
i_{\scriptscriptstyle\overrightarrow{X}}K
\label{mmap}
\end{equation}
The function ${\cal P}_{\overrightarrow{X}}$ is defined up to a constant,
which can be arranged so as to make it equivariant:
\begin{equation}
\overrightarrow{X} {\cal P}_{\overrightarrow{Y}} =
{\cal P}_{[\overrightarrow{X},\overrightarrow{Y}]}
\label{equivarianza}
\end{equation}
${\cal P}_{\overrightarrow{X}}$ constitutes then a {\it moment map}.
This can be regarded as a map
\begin{equation}
{\cal P}: {\cal M} \, \longrightarrow \,
\mathbb{R} \otimes
{\mathbb{G} }^*
\end{equation}
where ${\mathbb{G}}^*$ denotes the dual of the Lie algebra
${\mathbb{G} }$ of the group ${\cal G}$.
Indeed let $x\in {\mathbb{G} }$ be the Lie algebra element
corresponding to the Killing vector $\overrightarrow{X}$; then, for a given
$m\in {\cal M}$
\begin{equation}
\mu (m)\,  : \, x \, \longrightarrow \,  {\cal P}_{\overrightarrow{X}}(m) \,
\in  \, \mathbb{R}
\end{equation}
is a linear functional on  ${\mathbb{G}}$.
If we expand
$\overrightarrow{X} = a^\mathbf{I} k_\mathbf{I}$ in a basis of Killing vectors
$k_\mathbf{I}$ such that
\begin{equation}
[k_\mathbf{I}, k_\mathbf{L}]= f_{\mathbf{I} \mathbf{L}}^{\ \ \mathbf{K}} k_\mathbf{K}
\label{blio}
\end{equation}
we have also
\begin{equation}
{\cal P}_{\overrightarrow{X}}\, = \, a^\mathbf{I} {\cal P}_\mathbf{I}
\end{equation}
In the following we  use the
shorthand notation ${\cal L}_\mathbf{I}, i_\mathbf{I}$ for the Lie derivative
and the contraction along the chosen basis of Killing vectors $ k_\mathbf{I}$.
\par
From a geometrical point of view the prepotential,
or moment map, ${\cal P}_\mathbf{I}$ is the Hamiltonian function providing the Poissonian
realization  of the Lie algebra on the K\"ahler manifold. This
is just another way of stating the already mentioned
{\it  equivariance}.
Indeed  the  very  existence  of the closed 2-form $K$ guarantees that
every K\"ahler space is a symplectic manifold and that we can define  a
Poisson bracket.
\par
Consider eqs.(\ref{killo1}). To every generator of the abstract  Lie algebra
${\mathbb{G}}$ we have associated a function  ${\cal P}_\mathbf{I}$ on
${\cal M}$; the Poisson bracket of
${\cal P}_\mathbf{I}$ with ${\cal P}_\mathbf{J}$ is defined as follows:
\begin{equation}
\{{\cal P}_\mathbf{I} , {\cal P}_\mathbf{J}\} \equiv 4\pi K
(\mathbf{I}, \mathbf{J})
\end{equation}
where $K(\mathbf{I}, \mathbf{J})
\equiv K (\vec k_\mathbf{I}, \vec k_\mathbf{J})$ is
the value of $K$ along the pair of Killing vectors.
\par
In reference \cite{D'Auria:1991fj} with D'Auria and Ferrara I proved the following
lemma:
\begin{lemma}
{\it{The following identity is true}}:
\begin{equation}
\{{\cal P}_\mathbf{I}, {\cal
P}_\mathbf{J}\}=f_{\mathbf{I}\mathbf{J}}^{\ \ \mathbf{L}}{\cal
P}_\mathbf{L} + C_{\mathbf{I} \mathbf{J}} \label{brack}
\end{equation}
{\it{where $C_{\mathbf{I} \mathbf{J}}$ is a constant fulfilling the
cocycle condition}}
\begin{equation}
f^{\ \ \mathbf{L}}_{\mathbf{I}\mathbf{M}} C_{\mathbf{L} \mathbf{J}} +
f^{\ \ \mathbf{L}}_{\mathbf{M}\mathbf{J}} C_{\mathbf{L} \mathbf{I}}+
f_{\mathbf{J}\mathbf{I}}^{\ \  \mathbf{L}} C_{\mathbf{L} \mathbf{M}}=0
\label{cocy}
\end{equation}
\end{lemma}
If the Lie algebra ${\mathbb{G}}$ has a trivial second cohomology group
$H^2({\mathbb{G}})=0$, then the cocycle $C_{\mathbf{I} \mathbf{J}}$ is a
coboundary; namely we have
\begin{equation}
C_{\mathbf{I} \mathbf{J}} = f^{\ \ \mathbf{L}}_{\mathbf{I} \mathbf{J}} C_\mathbf{L}
\end{equation}
where $C_\mathbf{L}$ are suitable constants. Hence, assuming
$H^2 (\mathbb{G})= 0$
we can reabsorb $C_\mathbf{L}$ in  the definition of ${\cal
P}_\mathbf{I}$:
\begin{equation}
{\cal P}_\mathbf{I} \rightarrow {\cal P}_\mathbf{I}+ C_\mathbf{I}
\end{equation}
and we obtain the stronger equation
\begin{equation}
\{{\cal P}_\mathbf{I}, {\cal P}_\mathbf{J}\} =
f_{\mathbf{I}\mathbf{J}}^{\ \  \mathbf{L}} {\cal P}_\mathbf{L}
\label{2.39}
\end{equation}
Note that $H^2({\mathbb{G}}) = 0$ is true for all semi-simple Lie
algebras.
Using eq.(\ref{brack}), eq.(\ref{2.39})
can be rewritten in components as follows:
\begin{equation}
{i\over 2} g_{ij^*}(k^i_\mathbf{I} k^{j^*}_\mathbf{J} -
k^i_\mathbf{J} k^{j^*}_\mathbf{I})=
{1\over 2} f_{\mathbf{I} \mathbf{J}}^{\  \  \mathbf{L}} {\cal
P}_\mathbf{L}
\label{2.40}
\end{equation}
Equation (\ref{2.40}) is identical with the equivariance condition
in eq.(\ref{equivarianza}).
\subsubsection{The triholomorphic moment map on quaternionic manifolds}
Next, closely following the original derivation of \cite{D'Auria:1991fj,mylecture}
I turn to a discussion of the triholomorphic isometries of the manifold $\mathcal{QM}$
associated with hypermultiplets.
Both in $D=4$ and in $D=5$ supergravity $\mathcal{QM}$ is  quaternionic
and we can gauge only those of its isometries
that are  triholomorphic  and that  either generate an abelian group $\mathcal{G}$
or are \emph{suitably realized}  as isometries also on the special manifold $\mathcal{SV}$
\footnote{I anticipate the meaning of {\it suitably realized} to be
discussed in later sections. By definitions the gauge vectors are in
the coadjoint representation of the gauge groups. The vectors
transform linearly under isometries as the sections
$X^{\widehat{\mathbf{I}}}$ defining very special geometry. It follows that
under the gauge algebra these latter must decompose in a
{\it coadjoint representation} plus, possibly, another representation
$R$. The vectors in the representation $R$ must be dualised to
massive self dual $2$--forms}. This means  that on $\mathcal{QM}$ we
have Killing vectors:
\begin{equation}
\vec k_\mathbf{I} = k^u_\mathbf{I} {\vec \partial\over \partial q^u}
\label{2.71}
\end{equation}
\noindent
satisfying the same Lie algebra  as the corresponding Killing
vectors on ${\cal VM}$. In other words
\begin{equation}
\hat{\vec{k}}_\mathbf{I} =
k^i_\mathbf{I} \vec \partial_i + k^{i^*}_\mathbf{I}
\vec\partial_{i^*} + k_\mathbf{I}^u \vec\partial_u
\label{2.72}
\end{equation}
\noindent
is a Killing vector of the block diagonal metric:
\begin{equation}
\hat g = \left (
\matrix { g^V_{ij} & \quad 0 \quad \cr \quad 0
\quad & h_{uv} \cr } \right )
\label{2.73}
\end{equation}
defined on the product manifold ${\cal VM}\otimes\mathcal{QM}$.
Let us first focus on the manifold $\mathcal{QM}$.
Triholomorphicity means that the Killing vector fields leave
the HyperK\"ahler structure invariant up to $\mathrm{SU(2)}$
rotations in the $\mathrm{SU(2)}$--bundle defined by eq.(\ref{su2bundle}).
Namely:
\begin{equation}
\begin{array}{ccccccc}
{\cal L}_\mathbf{I} K^x & = &\epsilon^{xyz}K^y
W^z_\mathbf{I} & ; &
{\cal L}_\mathbf{I}\omega^x&=& \nabla W^x_\mathbf{I}
\end{array}
\label{cambicchio}
\end{equation}
where $W^x_\mathbf{I}$ is an $\mathrm{SU(2)}$ compensator associated with the
Killing vector $k^u_\mathbf{I}$. The compensator $W^x_\mathbf{I}$ necessarily
fulfills  the cocycle condition:
\begin{equation}
{\cal L}_\mathbf{I} W^{x}_\mathbf{J} - {\cal L}_\mathbf{J} W^x_\mathbf{I} + \epsilon^{xyz}
W^y_\mathbf{I} W^z_\mathbf{J} = f_{\mathbf{I} \mathbf{J}}^{\cdot \cdot \mathbf{L}}
W^x_\mathbf{L}
\label{2.75}
\end{equation}
In the HyperK\"ahler case the $\mathrm{SU(2)}$--bundle is flat and the
compensator can be reabsorbed into the definition of the
HyperK\"ahler forms. In other words we can always find a
map
\begin{equation}
\mathcal{QM} \, \longrightarrow \, L^x_{\phantom{x}y} (q)
\, \in \, \mathrm{SO(3)}
\end{equation}
that trivializes the ${\cal SU}$--bundle globally. Redefining:
\begin{equation}
K^{x\prime} \, = \, L^x_{\phantom{x}y} (q) \, K^y
\label{enfantduparadis}
\end{equation}
the new HyperK\"ahler form  obeys the stronger equation:
\begin{equation}
{\cal L}_\mathbf{I} K^{x\prime} \, = \, 0
\label{noncambio}
\end{equation}
On the other hand, in the quaternionic case, the non--triviality of the
${\cal SU}$--bundle forbids to eliminate the $W$--compensator
completely. Due to the identification between HyperK\"ahler
forms and $\mathrm{SU(2)}$ curvatures eq.(\ref{cambicchio}) is rewritten
as:
\begin{equation}
\begin{array}{ccccccc}
{\cal L}_\mathbf{I} \Omega^x& = &\epsilon^{xyz}\Omega^y
W^z_\mathbf{I} & ; &
{\cal L}_\mathbf{I}\omega^x&=& \nabla W^x_\mathbf{I}
\end{array}
\label{cambiacchio}
\end{equation}
In both cases, anyhow, and in full analogy with the case of
K\"ahler manifolds, to each Killing vector
we can associate a triplet ${\cal
P}^x_\mathbf{I} (q)$ of 0-form prepotentials.
Indeed we can set:
\begin{equation}
{\bf i}_\mathbf{I}  K^x =
- \nabla {\cal P}^x_\mathbf{I} \equiv -(d {\cal
P}^x_\mathbf{I} + \epsilon^{xyz} \omega^y {\cal P}^z_\mathbf{I})
\label{2.76}
\end{equation}
where $\nabla$ denotes the $\mathrm{SU(2)}$ covariant exterior derivative.
\par
As in the K\"ahler case eq.(\ref{2.76}) defines a moment map:
\begin{equation}
{\cal P}: {\cal M} \, \longrightarrow \,
\mathbb{R}^3 \otimes
{\mathcal{G} }^*
\end{equation}
where ${\mathcal{G}}^*$ denotes the dual of the Lie algebra
${\mathcal{G} }$ of the group ${\cal G}$.
Indeed let $x\in {\mathcal{G} }$ be the Lie algebra element
corresponding to the Killing
vector $\overrightarrow{X}$; then, for a given
$m\in {\cal M}$
\begin{equation}
\mu (m)\,  : \, x \, \longrightarrow \,  {\cal P}_{\overrightarrow{X}}(m) \,
\in  \, \mathbb{R}^3
\end{equation}
is a linear functional on  ${\mathcal{G}}$. If we expand
$\overrightarrow{X} = a^\mathbf{I} k_\mathbf{I}$ on a basis of Killing vectors
$k_\mathbf{I}$ such that
\begin{equation}
[k_\mathbf{I}, k_\mathbf{L}]= f_{\mathbf{I} \mathbf{L}}^{\ \ \mathbf{K}} k_\mathbf{K}
\label{blioprime}
\end{equation}
and we also choose a basis ${\bf i}_x \, (x=1,2,3)$ for $\mathbb{R}^3$
we get:
\begin{equation}
{\cal P}_{\overrightarrow{X}}\, = \, a^\mathbf{I} {\cal P}_\mathbf{I}^x \, {\bf i}_x
\end{equation}
Furthermore we need a generalization of the equivariance defined
by eq.(\ref{equivarianza})
\begin{equation}
\overrightarrow{X} \circ {\cal P}_{\overrightarrow{Y}} \,=  \,
{\cal P}_{[\overrightarrow{X},\overrightarrow{Y}]}
\label{equivarianzina}
\end{equation}
In the HyperK\"ahler case, the left--hand side of
eq.(\ref{equivarianzina})
is defined as the usual action of a vector field on a $0$--form:
\begin{equation}
\overrightarrow{X} \circ {\cal P}_{\overrightarrow{Y}}\, =  \, {\bf i}_{\overrightarrow{X}} \, d
{\cal P}_{\overrightarrow{Y}}\, = \,
X^u \, {\frac{\partial}{\partial q^u}} \, {\cal P}_{\overrightarrow{Y}}\,
\end{equation}
The equivariance condition   implies
that we can introduce a triholomorphic Poisson bracket defined
as follows:
%%%%%%%%%%%%%%%%%%%%%%%%%
\begin{equation}
\{{\cal P}_\mathbf{I}, {\cal P}_\mathbf{J}\}^x \equiv 2 K^x (\mathbf{I},
\mathbf{J})
\label{hykapesce}
\end{equation}
leading to the triholomorphic Poissonian realization of the Lie
algebra:
\begin{equation}
\left \{ {\cal P}_\mathbf{I}, {\cal P}_\mathbf{J} \right \}^x \, = \,
f^{\mathbf{K}}_{\phantom{\mathbf{K}}\mathbf{I}\mathbf{J}} \, {\cal P}_\mathbf{K}^{x}
\label{hykapescespada}
\end{equation}
which in components reads:
\begin{equation}
K^x_{uv} \, k^u_\mathbf{I} \, k^v_\mathbf{J} \, = \, {\frac{1}{2}} \,
f^{\mathbf{K}}_{\phantom{\mathbf{K}}\mathbf{I}\mathbf{J}}\, {\cal P}_\mathbf{K}^{x}
\label{hykaide}
\end{equation}
%%%%%%%%%%%%%%%%%%%%%%%%%
In the quaternionic case, instead, the left--hand side of
eq.(\ref{equivarianzina})
is interpreted as follows:
\begin{equation}
\overrightarrow{X} \circ {\cal P}_{\overrightarrow{Y}}\, =  \, {\bf i}_{\overrightarrow{X}}\,  \nabla
{\cal P}_{\overrightarrow{Y}}\, = \,
X^u \, {\nabla_u} \, {\cal P}_{\overrightarrow{Y}}\,
\end{equation}
where $\nabla$ is the $\mathrm{SU(2)}$--covariant differential.
Correspondingly, the triholomorphic Poisson bracket is defined
as follows:
%%%%%%%%%%%%%%%%%%%%%%%%%
\begin{equation}
\{{\cal P}_\mathbf{I}, {\cal P}_\mathbf{J}\}^x \equiv 2 K^x (\mathbf{I},
\mathbf{J})  - { {\lambda}} \, \varepsilon^{xyz} \,
{\cal P}_\mathbf{I}^y  \, {\cal P}_\mathbf{J}^z
\label{quatpesce}
\end{equation}
and leads to the Poissonian realization of the Lie algebra
\begin{equation}
\left \{ {\cal P}_\mathbf{I}, {\cal P}_\mathbf{J} \right \}^x \, = \,
f^{\mathbf{K}}_{\phantom{\mathbf{K}}\mathbf{I}\mathbf{J}} \, {\cal P}_\mathbf{K}^{x}
\label{quatpescespada}
\end{equation}
which in components reads:
\begin{equation}
K^x_{uv} \, k^u_\mathbf{I} \, k^v_\mathbf{J} \, - \,
{ \frac{\lambda}{2}} \, \varepsilon^{xyz} \,
{\cal P}_\mathbf{I}^y  \, {\cal P}_\mathbf{J}^z\,= \,  {\frac{1}{2}} \,
f^{\mathbf{K}}_{\phantom{\mathbf{K}}\mathbf{I}\mathbf{J}}\, {\cal P}_\mathbf{K}^{x}
\label{quatide}
\end{equation}
Eq.(\ref{quatide}), which is the most convenient way of
expressing equivariance in a coordinate basis was originally written
in \cite{D'Auria:1991fj} and has played a fundamental
role in the construction of  supersymmetric actions  for gauged $\mathcal{N}=2$ supergravity
both in $D=4$ \cite{D'Auria:1991fj,bertolo} and in $D=5$
\cite{ceregatta}.
\subsection{$\mathcal{N}=2$ gaugings and the composite connections}
\label{gunasecti}
Equipped with the crucial geometric structure provided by the triholomorphic moment-map
let us come to the problem of gauging a general $\mathcal{N}=2$ matter coupled supergravity
as described by the bosonic lagrangian (\ref{genn2d5ung}). To single out
a viable gauge group we have to go through a few steps that have been derived by G\"unaydin and Zagerman
in \cite{Gunaydin:2000zx,Gunaydin:2000xk,Gunaydin:2001ph}.
\par
The first thing we have to consider is the isometry group $\mathcal{G}_{iso}$ of the
special manifold $\mathcal{SV}_n$. Later we have to see how it might be
represented on the quaternionic manifold $\mathcal{QM}$.
\par
By definition the vectors are in the representation $\mathbf{R}$ of $\mathcal{G}_{iso}$.
What we can gauge  is any  $n_V+1$--dimensional subgroup $G_g \subset
\mathcal{G}_{iso}$ such that certain conditions are satisfied. The
conditions are:
\begin{description}
\item[a)] The following branching must be true:
\begin{equation}
  \mathbf{R} \, \stackrel{G_g}{\Longrightarrow} \,
  \mathrm{Coadj}(G_g) \oplus \mathbf{D}_S
\label{brancio}
\end{equation}
where
\item [b)]
$\mathbf{D}_S$ denotes some reducible or irreducible
\textbf{symplectic representation} of the candidate gauge group $G_g$ different from the
coadjoint. By symplectic we mean the following.
Let us decompose the range of the index $\Lambda$ as in
eq.(\ref{Lamrang}) where
\begin{equation}
  \mathbf{I}=1,\dots, n_V+1 \equiv \mbox{dim}\, G_g
\label{dimGg}
\end{equation}
runs on the coadjoint representation of the $G_g$ Lie algebra whose
generators we denote $T_\mathbf{I}$  with commutation relations
\begin{equation}
  \left[ T_\mathbf{I}\, , \, T_\mathbf{J} \right] =
  f^\mathbf{K}_{\mathbf{IJ}}\,T_\mathbf{K}
\label{Ggalge}
\end{equation}
and
\begin{equation}
  \mathcal{M}=1,\dots ,n_T=\mbox{dim}\, \mathbf{D}_S
\label{D2repre}
\end{equation}
runs on a basis of the representation $\mathbf{D}_S$.
Let $\Lambda_{\mathbf{I}\mathcal{M}}^{\mathcal{N}}$ be the matrix
representing the generator $T_\mathbf{I}$ in $ \mathbf{D}_S$:
\begin{equation}
  T_\mathbf{I} \quad \rightarrow \quad \Lambda_\mathbf{I} \quad ;
  \quad \left[ \Lambda_I \, , \, \Lambda_\mathbf{J}\right] = f^K_{IJ}
  \, \Lambda_\mathbf{K}
\label{repregenesym}
\end{equation}
In order for the representation to be symplectic there must exist an
antisymmetric  $n_T \times n_T$ matrix
$\Omega^T=-\Omega$ that
squares to minus the identity $\Omega^2=-{\bf 1}$ and such that:
\begin{equation}
  \forall T_\mathbf{I} \in G_g \quad \Omega  \, \Lambda_\mathbf{I} +
  \Lambda_\mathbf{I}^T \, \Omega =0
\label{omegazer}
\end{equation}
Indeed this ensures that our algebra $G_g$ is a subalgebra of the
symplectic algebra $\mathrm{Sp}(n_T,\mathbb{R})$.
\item [c)]
The $d_{\Lambda\Sigma\Gamma}$ invariant tensor must
decompose under $G_g$ in the following way:
\begin{equation}
  d_{\Lambda\Sigma\Gamma}=\left\{ \begin{array}{rcl}
    d_{\mathbf{IJK}} & = & \mbox{Invariant tensor in the $\mathrm{Coadj}(G_g)$}  \\
    d_{\mathcal{MNP}} & = & 0 \\
    d_{\mathcal{M}\mathbf{IJ}} & = & 0 \\
    d_{\mathcal{MN}\mathbf{I}} &  = & -\frac{2}{3} \Omega_{\mathcal{MT}} \,
    \Lambda^\mathcal{T}_{\mathbf{I}\mathcal{N}}  \
  \end{array} \right.
\label{ergovit}
\end{equation}
\item [d)]  The group $G_g$ selected through the previous restriction
  must act as a triholomorphic isometry on the quaternionic manifold
  $\mathcal{QM}$ \footnote{This last requirement is that spelled out by Ceresole and Dall'Agata in \cite{ceregatta}
  who have extended to the D=5 case the methods and procedures of the geometrical gaugings originally introduced
  in \cite{dewit1,Castellani:1986ka,D'Auria:1991fj,bertolo}}.
\end{description}
The rationale for the above requirements is the following.
The reason for the requirement \textbf{a)}   is the same
as in four--dimensions. Since  the gauge vectors are by definition
in the coadjoint of the gauge algebra it is necessary that the
representation to which the vectors are pre--assigned should contain
the coadjoint of what we want to gauge. Note also that for
semisimple groups  adjoint and coadjoint representations are equivalent
but this is no longer true in the case of non semisimple gauge
algebras. An extreme possibility is provided by abelian algebras
\begin{equation}
  \mathcal{A}=\mathrm{U(1)}^{\ell} \otimes \mathbb{R}^m
\label{abelalg}
\end{equation}
where we were careful to distinguish compact from non compact
generators. In this case the coadjoint representation vanishes and
any set of $\ell +m$ vectors can be used to gauge an algebra $\mathcal{A}$
that has vanishing action on the very special manifold
$\mathcal{SV}_n$.
The rationale for the requirements \textbf{b)} and \textbf{c)}
is instead related to the consistent coupling of massive $2$--forms.
Gauge vectors that are in non trivial representations of a gauge
group different from the coadjoint representation are inconsistent
with their own gauge invariance. To cure this problem we have to
dualise them to massive self--dual $2$--forms, satisfying:
\begin{equation}
B^{\mathcal{M} |\mu\nu} = m
\epsilon^{\mu\nu\rho\sigma\lambda}{\cal{D}}_\rho B^{\mathcal{M}}_{\sigma\lambda}\,.
\label{selfconst}
\end{equation}
where the covariant derivative is :
\begin{equation}
  \mathcal{D}B^\mathcal{M} \equiv dB^\mathcal{M}
  + g  \Lambda^\mathcal{M}_{\mathbf{I}\mathcal{N}} A^\mathbf{I} \,
  \wedge \,
  B^\mathcal{N} \equiv \mathcal{H}^\mathcal{M}
\label{covderB}
\end{equation}
This is possible only if the part of
the Chern Simons term involving two $B$.s and one $A$
can be reabsorbed in  the kinetic term of the $2$--forms  that reads
as follows:
\begin{equation}
  \mathcal{L}^{kin}_{B} = \frac{1}{4} \sqrt{-g} \, \epsilon^{\mu \nu \rho \sigma \lambda
  } B^\mathcal{M}_{\mu \nu } \,\mathcal{D}_\rho  B^\mathcal{N}_{\sigma \lambda
  } \, \Omega_{\mathcal{MN}}
\label{Bkinterm}
\end{equation}
\par
The last requirement \textbf{(d)} deals with the possible presence of
hypermultiplets. In particular the action of $G_g$ on the quaternionic manifold
can be the identity action which is certainly triholomorphic.
In this case the hypermultiplets are simply neutral with respect to the gauge group.
Alternatively, we can consider an abelian algebra as in eq.(\ref{abelalg})
that has no action on the very special manifold but acts by  non
trivial triholomorphic isometries on the quaternionic manifold.
Both of these are extreme cases that allow more freedom of choice
for one of the two manifolds. The general case corresponds to a
choice of $G_g$ that acts non trivially both on $\mathcal{SV}_n$ and
$\mathcal{QM}_m$ and respects conditions \textbf{a)-d)}
\par
Assuming that the gauge algebra has been selected and satisfies the
above criteria, the gauging procedure becomes smooth and fully
parallel to the four--dimensional models we have already discussed.
The essential point is always the same, namely the \textbf{gauging} of the
\textbf{scalar vielbein} and of the \textbf{composite connections} acting on
the fermion fields. In the case of
maximal supersymmetry, where the scalar manifold is necessarily a
homogeneous space $\mathcal{G}/\mathcal{H}$, these two gaugings are
obtained in one stroke by gauging the Maurer--Cartan  $1$--forms,
as it was done in eq.(\ref{gaul1f}). In the non maximal case we have
to do it separately and specifically for the different factors
occurring in the scalar manifold. These latter are not necessarily
coset manifolds but have  a sufficiently \emph{special} geometric
structure to allow the generic construction of those ingredients that
are necessary for the gauging of the composite connections, most
relevant being the role of the triholomorphic moment-map.
\par
Let us begin with the gauging of the scalar vielbein. This is
equivalent to replacing the ordinary derivatives (or differentials) of
the scalar fields with covariant ones, as follows:
\begin{equation}
  \begin{array}{rcl}
 \mathcal{D}\phi^{i} &=& d \phi^{i} + g A^\mathbf{I} k_\mathbf{I}^{i} (\phi)\\
\mathcal{D}q^{u} &=& d q^{u} + g A^\mathbf{I} k_\mathbf{I}^{u} (q)\\
\end{array}
\label{covdiffe}
\end{equation}
where $g$ is the gauge coupling constant and $k^i_\mathbf{I}(\phi),k_\mathbf{I}^u(q)$
are the killing vectors expressing the action of the gauge algebra generators
$T_\mathbf{I}$ on the two scalar manifolds:
\begin{eqnarray}
\delta_I \phi^i & = &  k_\mathbf{I}^{i} (\phi) \nonumber\\
\delta_Iq^{u} & = & k_\mathbf{I}^{u} (q)
\label{deltafiq}
\end{eqnarray}
Next we have the gauging of the composite connections. There are three of
them corresponding to the three vector bundles of which the fermions are
sections:
\begin{enumerate}
\item The Levi--Civita connection $ \Gamma^i_{\phantom{i}j}$ on the
tangent bundle to the very special manifold $T\mathcal{SV}_n$. This
enters because the gauginos $\lambda^i_A$ carry a world index of the
very special manifold, namely are sections of $T\mathcal{SV}_n$
\item The $\mathrm{Sp}(2m,\mathbb{R})$ connection
$\Delta^{\alpha\beta}$. This enters because the hyperinos
$\zeta^\alpha$ are sections of the $\mathrm{Sp}(2m,\mathbb{R})$
bundle over the quaternionic manifold $\mathcal{QM}$. By definition
this latter has reduced holonomy, so that the structural group of the
tangent bundle $T\mathcal{QM}$ is $\mathrm{SU(2)}\times
\mathrm{Sp}(2m,\mathbb{R})$, as we know from section \ref{hypgeosec}.
\item The $\mathrm{SU_R(2)}$ connection of $R$--symmetry $\omega^{AB}$
  This connection enters the game because both the
gravitino $\psi_A$ and the gauginos $\lambda^i_A$ are sections of the
$SU(2)_R$ vector bundle in the fundamental doublet representation
(the index $A=1,2$ denotes this fact). On the other hand the
$R$--symmetry bundle is identified with the
$\mathcal{SU} \rightarrow \mathcal{QM}$ bundle over the quaternionic
manifold and this means that the connection $\omega^{AB}$ is the
connection of the $\mathcal{SU}$ bundle describe in section \ref{hypgeosec}.
\end{enumerate}
In terms of the Killing vectors and of the
the triholomorphic moment map $\mathcal{P}_\mathbf{I}^x(q) $
and just following the original recipe developed in
\cite{D'Auria:1991fj} and further clarified in \cite{bertolo}
the gauging of the connections is given:
\begin{equation}
\begin{array}{cccccc}
{ T\mathcal{VS}} & : & \mbox{tangent bundle} &
 \Gamma^{i}_{\phantom{i}j}& \to &{\hat \Gamma}^{i}_{\phantom{i}j} =
 \Gamma^{i}_{\phantom{i}j} +
 g\, A^\mathbf{I}\, \partial_j k^i_\mathbf{I} \cr
{\cal SU} & : & \mbox{$\mathrm{SU(2)}$ bundle} &
\omega^x &\to &{\hat \omega}^x = \omega^x + g_R\, A^\mathbf{I}\, {\cal
P}^x_\mathbf{I} \cr
{\cal SU}^{-1}\otimes{ T\mathcal{QM}} & : & \mbox{$\mathrm{Sp}(2m,\mathbb{R})$ bundle} &
\Delta^{\alpha\beta} &\to &{\hat  \Delta}^{\alpha\beta}=
\Delta^{\alpha\beta}  + g\, A^\mathbf{I}\,
 \partial_u k_\mathbf{I}^v \, {\cal U}^{u \vert  \alpha A}
 \, {\cal U}^\beta_{v \vert A} \cr
 \end{array}
\label{compogauging}
\end{equation}
where $g$ is the same gauge coupling constant as in
eq.(\ref{covdiffe}) while $g_R$ is an additional coupling constant
that allows to gauge or not to gauge the $R$-symmetry group $SU(2)$.
In the construction of the lagrangian and in checking the closure of
the supersymmetry algebra it turns out that $g$ and $g_R$ are
independent parameters \cite{ceregatta}.
\subsection{The Fermion shifts and gravitino mass-matrix}
Gauging the connections forces, through closure of the supersymmetry
algebra, the inclusion of new non--derivative terms in the \emph{susy}
rules of the fermions that are completely analogous to their
$4$--dimensional counterparts of eq.s (\ref{gravsusy}) and
(\ref{fermioshif}). Indeed as explained in sect.\ref{genaspect} the
gauging procedure of supergravity theories fits into a general and
uniform pattern for all space--time dimensions $D$ and for all number of supersymmetry
charges $N_Q$. The gravitino transformation rule (\ref{gensusrul})
becomes:
\begin{equation}
  \delta \psi_{A\mu} = \mathcal{D}_\mu \, \epsilon_A - \frac{1}{3} \,
  \mathcal{T}_{AB}^{\rho \sigma } \left( g_{\mu \rho } \, \gamma _\sigma
  - \frac{1}{8}\, \epsilon _{\mu \rho \sigma \lambda \nu } \, \gamma
  ^{\lambda \nu }   \right) \, \epsilon ^B\, + \,S_{AB} \, \epsilon ^B
\label{gaususrul}
\end{equation}
where the gravitino mass matrix is given by:
\begin{equation}
  S_{AB} = \mbox{i}\,g_R \, \frac{1}{6} \, X^\mathbf{I} (\phi) \,
  \mathcal{P}^x_\mathbf{I}(q) \, \left( \sigma^x \right)
  _A^{\phantom{A}C} \, \epsilon_{BC}
\label{ginmamatn2d5}
\end{equation}
while the transformation rules of the spin 1/2 fermions have been
determined by Ceresole and dall'Agata  (see \cite{ceregatta}) to have, apart
from some trivial choice of normalizations, an identical form to their
counterparts in $D=4$ $\mathcal{N}=2$ supergravity (see
\cite{D'Auria:1991fj,bertolo}).
Indeed one finds:
\begin{eqnarray}
\delta \zeta^\alpha  & = & \mbox{derivative terms} \, + \,  \Sigma^{\alpha \vert A} \,
\epsilon_A \nonumber\\
\delta \lambda^{i A}  & = & \mbox{derivative terms} \, + \,\Sigma^{iA \vert B} \,
\epsilon_B
\label{ceregashif}
\end{eqnarray}
where the fermion shifts take the following explicit form:
\begin{eqnarray}
\Sigma^{iA \vert B}&=& g \epsilon^{AB} W^i + g_R W^{iAB}\nonumber\\
W^i &=&\frac{1}{2}\,k_{\mathbf{I}}^i  X^\mathbf{I}\\
W^{iAB}&=&{\rm i}\frac{1}{2}(\sigma_x)_{C}^{\phantom{C}B} \epsilon^{CA} {\cal P}^x_{\mathbf{I}}
g^{ij} { f}_{j}^{\mathbf{I}}\nonumber\\
\Sigma^{\alpha \vert A} &=& g\, \epsilon ^{AB} \,\mathbf{N}_B^\alpha \nonumber\\
N_B^{\alpha}&=& \frac{1}{2} \,{\cal U}_{\alpha u}^B \,k^u_{\mathbf{I}}\,X^{\mathbf{I}}
\label{pesamatrice}
\end{eqnarray}
Indeed if one compares eq.s (\ref{ginmamatn2d5}),(\ref{pesamatrice}) with their
$4$--dimensional counterparts given in eq.s (8.23) of \cite{bertolo}
one sees that (apart from the overall normalization which can be
reabsorbed into the normalization of the corresponding fermionic
field)  the two sets of formulae are identical upon the
replacement of the complex section $L^\Lambda(z)$ of \emph{special K\"ahler
geometry} with the real section $X^\mathbf{I}(\phi)$ of \emph{very special
geometry}. The other noteworthy difference is that in $D=4$ the index
$\Lambda$ runs on the whole set of $n+1$ values, $n$ being the
dimension of the special K\"ahler manifold. In five dimensions,
instead, the index $ \mathbf{I}$ runs on the $n_V+1$ subset of values
corresponding to the gauged vectors while the total dimension of the
very special space is $n_V+n_T$. The remaining $n_T$ dimensions are,
as we know, associated with the massive self-dual $2$--forms.
\par
In five as in all other dimensions supersymmetry imposes a Ward
identity that is the straightforward generalization of
eq.(\ref{wardide}) namely:
\begin{equation}
  \alpha \, S^{AB} \, S_{BC} \, - \, \beta \, K_{ij} \Sigma^{i\vert
  A} \Sigma^{j \vert B} \, \epsilon_{BC} \, = \, - \,\delta^A_C \, \mathcal{V}
\label{d5wardide}
\end{equation}
where $K_{ij}$ is the kinetic matrix of the spin 1/2 fermions and
$\alpha$ and $\beta$ are just numerical coefficients that differ from
their analogues in $4$--dimensions only because of the differences in
Lorentz algebra and $\gamma$-matrix manipulations. Verifying such an
identity  whose explicit form is not written down in their paper,
the authors of \cite{ceregatta} have proved the supersymmetry of the
gauged action and calculated the final form of the potential that
reads as follows:
\begin{equation}
  \mathcal{V}=  \,g^2  \left[   X^\mathbf{I} \,X^\mathbf{J}
  \, \left( k_\mathbf{I}^i k_\mathbf{J}^j \, g_{ij} + k_\mathbf{I}^u
  \,
  k_\mathbf{J}^v \, h_{uv} \right) \right] - 4\,  g_R^2 \left[ \left(
  \frac{1}{3} \, X^\mathbf{I} \, X^\mathbf{J} \, -\,\ft 1 4 \,g^{ij}
  f_i^\mathbf{I} \, f_j^\mathbf{J}\right) \, P^x_\mathbf{I} \, P^x_\mathbf{J}\right]
\label{ceregatpot}
\end{equation}
in terms of the moment-map (\ref{2.76}) and of the section
$X^\mathbf{I}$ of very special geometry and its derivative
$f^\mathbf{I}_i =\partial _i X^\mathbf{I}$ (see eq.s(\ref{fidefi}).
\subsection{The scalar potential, supersymmetry
breaking and domain walls}
We can now summarize the results of the previous section writing the general
form of the  bosonic lagrangian for a general
\textbf{gauged} $\mathcal{N}=2,D=5$ supergravity. The ungauged action
(\ref{genn2d5ung}) is replaced by the following gauged one:
\begin{eqnarray}
  \mathcal{L}^{(gauged)}_{(D=5,\mathcal{N}=2)}&=&\sqrt{-g} \, \left(
  2
  \,R \, - \, \frac{1}{2}\,\mathcal{N}_{\mathbf{IJ}} (\phi)  F^\mathbf{I}_{\mu \nu
  } \, F^{\mathbf{J} \vert\mu \nu} \right.\nonumber\\
  &&\left. +  \, g_{ij}(\phi) \, \partial
  _\mu\phi^i \, \partial ^\mu \, \phi^j
  + \, h_{uv}(q) \, \partial
  _\mu q^u \, \partial ^\mu \, q^v - \mathcal{V}(\phi,q)\right ) \nonumber\\
  &&+  \left(\, \frac{1}{4} d_{\mathbf{IJK}} \,
   \, F^\mathbf{I}_{\mu \nu } \, F^\mathbf{J}_{\rho \sigma } \, A^\mathbf{K}_\tau
  + \frac{1}{2}\,\Omega_{\mathcal{MN}} \, B^\mathcal{M}_{\mu \nu } \,
  \mathcal{D}_\rho
  \, B^\mathcal{N}_{\sigma \tau } \,\right) \, \epsilon ^{\mu \nu \rho \sigma \tau}
\label{genn2d5gau}
\end{eqnarray}
where the potential $\mathcal{V}(\phi,q)$ is that given in
eq.(\ref{ceregatpot}).
\paragraph{General pattern of supersymmetry breaking in $D=5$}
Following the general discussion of eq.s (\ref{genbreakpat})
  in $\mathcal{N}$-extended $D=5$ supergravity a conventional vacuum configuration
$\phi_0$ that preserves $\mathcal{N}_0$
supersymmetries is characterized by the existence of $\mathcal{N}_0$
vectors $\rho^A_{(\ell)}$ ($\ell=1,\ldots,\mathcal{N}_0$) of
$\mathrm{USp}(\mathcal{N})$, such that
\begin{eqnarray}
\label{d5breakpat} S_{AB} \left( \phi_0 \right) \, \rho^A_{(\ell)} & =
& e^{\rm i\theta}\, \sqrt{\ft{|\mathcal{V}(\phi_0)|}{48}} \, \rho_{A(\ell)}
%\sqrt{-V(\phi_0)/3}\, \rho_{A(\ell)}
~,\nonumber\\ \Sigma_A^{\phantom{A}i} \left( \phi_0 \right) \,
\rho^A_{(\ell)} & = & 0~,
\end{eqnarray}
where $\theta$ is an irrelevant phase, $\mathcal{V}$ is the scalar
potential and $S_{AB}$ is the gravitino mass-matrix, uniformly defined
for all $\mathcal{N}$ by eq.(\ref{gaususrul}).
\subsubsection{Properties of the $\mathcal{N}=2$ potential and anti de Sitter vacua}
As it was already pointed out in chapter \ref{bestia1}, the embedding
of the Randall--Sundrum scenario inside a supersymmetric
five--dimensional field theory requires that the following schematic
situation should be realized:
\begin{itemize}
  \item The candidate theory admits at least two different  anti de
  Sitter vacua with the same vacuum energy, namely two extrema  of the scalar potential
  $\phi_0^{[1]}$ and $\phi_0^{[2]}$ with $0>\mathcal{V}\left(\phi_0^{[1]} \right)=
  \mathcal{V}\left(\phi_0^{[2]} \right) $
  \item The two $\mathrm{AdS}$ vacua  $\phi_0^[1]$ and $\phi_0^[2]$
  are stable, that is the spectrum of small fluctuations around these
  points satisfies the Breitenl\"ohner Friedman bound\footnote{ For a
  review of this bound see for instance \cite{castdauriafre} Volume
  I}
  \item There exists a smooth domain wall solution interpolating
  between these two vacua.
\end{itemize}
In view of these facts it is specifically interesting to survey the
conditions for the existence of anti de Sitter vacua.
According to our general discussion following eq.(\ref{trecass}) we
have anti de Sitter vacua if  $\mathcal{V}(q_0, \phi_0) < 0$ for
 $\mathcal{V}^{\prime} (q_0, \phi_0) = 0$.
 Thus it is  straightforward to see that the only contribution
which can allow for such solutions is the term
\begin{equation}
 \mathcal{V} =  -
 \frac{4}{3} \, g^2_R X^\mathbf{I} \, X^\mathbf{J}  \, P^x_\mathbf{I} \, P^x_\mathbf{J}
 + \mbox{positive contributions}
 \label{pertica}
 \end{equation}
coming from the $R$--symmetry gauging of the gravitinos. Indeed this
is the only negative contribution to the potential.
This implies that a simple Yang--Mills gauging, even in presence
of both tensor and hypermultiplets, does not allow any anti de Sitter
solution.
\par
We can briefly analyze various cases.
\begin{description}
  \item[a)] If we set $m=0$ there are no hypermultiplets and the
  quaternionic manifold disappears. Correspondingly,
  as already noted in \cite{D'Auria:1991fj,bertolo} for the
  $4$-dimensional case, the killing vector $k_\mathbf{I}^u$ is zero
  while the triholomorphic moment maps are $\mathrm{SU(2)}$ Lie algebra valued
constants $\xi^r_\mathbf{I}$ that, because of eq.(\ref{quatide}), must satisfy the
condition:
\begin{equation}
  g \, f^\mathbf{K}_{\mathbf{IJ}} \, \xi_\mathbf{K}^r = \ft 1 2
  \, g_R \, \e^{rst} \, \xi_\mathbf{I}^s \, \xi_\mathbf{J}^t
\label{identitaxi}
\end{equation}
Generically the $\xi_\mathbf{I}^r$ break $\mathrm{SU(2)} \to
\mathrm{U(1)}$. If the gauge group
$\mathcal{G}$ contains a subgroup $SU(2)$, this can be identified
with the R--symmetry group setting $\xi^r_\mathbf{I} =
\delta_\mathbf{I}^r$.
  \item[a1)] If at $m=0$ one makes the choice
$\xi_\mathbf{I} = (0, V_\mathbf{I}, 0)$, the condition (\ref{identitaxi})
reduces to
\begin{equation}
f^\mathbf{K}_{\mathbf{IJ}} V_\mathbf{K} = 0
\label{pirlon}
\end{equation}
As already noted in \cite{D'Auria:1991fj,bertolo} for the four--dimensional case
this is the Fayet--Iliopoulos
phenomenon which corresponds, in mathematical language to the
possibility of lifting the moment-maps to a non zero level for all
the generators belonging to the center of the gauge Lie algebra.
\item[b)] If we both set $m=0$, namely we include no hypermultiplet
but we also set $n_T=0$ namely we consider only vector fields in the
coadjoint representation of the gauge group (i.e. the symplectic
representation $\mathbf{D}_S$ of the massive two forms is deleted),
then one can easily prove that
\begin{equation}
  X^\mathbf{I} \, k^i_\mathbf{I}=0
\label{ortogcond}
\end{equation}
This implies that the scalar potential \ref{ceregatpot} reduces to:
\begin{eqnarray}
  \mathcal{V} & = &-\frac{1}{3},g_R^2  \left[ 4 \, W^2 \, - \, 3 g^{ij}
  \partial_i W \, \partial_j W \right ]\label{townsendo}\\
  W(\phi) & \equiv & V_\mathbf{I} X^\mathbf{I}(\phi)
\label{superpotential}
\end{eqnarray}
the constant coefficients $V_\mathbf{I}$ being those introduced above
and satisfying the consistency condition (\ref{pirlon}). The
interesting thing about the potential (\ref{townsendo}) is that it
follows in the general class of potentials of the form:
\begin{equation}
  \mathcal{V}  =  -  \alpha^2 \left[ (D-1) \, W^2 \, - \, (D-2) g^{ij}
  \partial_i W \, \partial_j W \right ]\label{townskend}
\end{equation}
where $W(\phi)$ is a real function named the \emph{superpotential},
$D$ denotes the space--time dimensions and $g^{ij}$ is the positive
definite kinetic metric of the scalar fields. In \cite{townsupot}
Townsend has shown that the structure (\ref{townskend}) is precisely
that required for vacuum stability. This is an encouraging starting
point for the search of domain wall solutions fitting the Randall
Sundrum scenario but unfortunately all attempts in this direction
have so far been rebuked. There is actually a negative result due to
Kallosh and Linde \cite{renandrei} that excludes such solutions
within this class of models and also within the larger class that
includes also tensor multiplets $n_T \neq 0$. This result does not
exclude, for the time being cases involving also the hypermultiplets,
where the situation is still not completely clarified.
\item[c)] If we set $n_V+n_T=0$ there are no vector multiplets and
we have simply hypermultiplets. Then $X^0=1$ and there is just one gauge vector:
the graviphoton whose action on the quaternionic manifold is
described by the triholomorphic Killing vector $k_\mathbf{0}^u$
The potential is still non--zero and becomes
\begin{equation}
\mathcal{V} = \,g^2 \, k_\mathbf{I}^u \, k_\mathbf{J}^v \, h_{uv}  -
g_R^2 \, \frac{4}{3} \,   P^x_\mathbf{0} \, P^x_\mathbf{0}
  \label{novectors}
\end{equation}
which in principle can admit anti de Sitter vacua.
\item [d)] Pure $5$--dimensional supergravity is retrieved as a
subcase of the above case setting also the number of hypermultiplets
to zero.
\end{description}
%%%%%%%%%%%%%%%%%%%%%%%%%%%%%%%%%
% Inserire CONTINUA
%%%%%%%%%%%%%%%%%%%%%%%%%%%%%%%%%
\subsection{Comparison with the $\mathcal{N}=8$ gaugings}
It is quite instructive to make a comparison of the gauged $N=2$
theory with the gaugings of the five dimensional $\mathcal{N}=8$ theory
described in section \ref{minid5geo}.
\par
The main issue is the choice of the gauge group. Here the isometry of
the scalar manifold is $E_{6(6)}$ and the $27$ vectors (prior to
gauging) sit in the fundamental representation of $E_{6(6)}$ which is
precisely $27$--dimensional. In full analogy to eq.(\ref{brancio})
the gauge algebra $\mathcal{G}$ must be chosen in such a way that:
\begin{equation}
\label{gaugprop}
{\bf 27}\stackrel{\mathcal{G}\subset E_{\left(6\right)6}}{\longrightarrow}
\mathrm{Coadj}\left(\mathcal{G}\right)\oplus \mathbf{D}_S
\end{equation}
where $\mathbf{D}_S$ is a symplectic representation of $\mathcal{G}$.
It turns out that this request is satisfied if and only if
$\mathcal{G}$ is a fifteen--dimensional
subgroup of $\mathrm{SL}(6,\mathbb{R})\subset E_{\left(6\right)6}$ whose adjoint
is identified with the $\bf 15$ representation of
$\mathrm{SL}(6,\mathbb{R})$. Indeed the $\bf 27$ of $\mathrm{E}_{\left(6\right)6}$
decomposes under
\begin{equation}
\mathrm{SL}(6,\mathbb{R})\times \mathrm{SL}(2,\mathbb{R})\subset E_{\left(6\right)6}
\end{equation}
as
\begin{equation}
\bf 27 \longrightarrow ({\bar{15}},1)\oplus (6,2)\,
\end{equation}
(for example, $\mathbb{L}_{\Lambda}^{~AB}\longrightarrow
 (\mathbb{L}^{IJAB},\mathbb{L}_{I\alpha}^{~~AB})$)
so that the property (\ref{gaugprop}) is satisfied, $(6,2)$ being the
requested symplectic representation $\mathbf{D}_S$.
The subgroups of $\mathrm{SL}(6,\mathbb{R})$ whose adjoint is the $\bf 15$ of
$\mathrm{SL}(6,\mathbb{R})$ are the $\mathrm{SO(p,q)}$ groups with $p+q=6$ and
their contractions
$\mathrm{CSO}\mathrm{(p,q,r)}$ (see section \ref{csopqr} for the
relevant definitions).
The possible gaugings are then restricted to these groups. The normalizer in
$E_{\left(6\right)6}$ of all these groups is the same as the normalizer of
$\mathrm{SL}(6,\mathbb{R}) $, namely
 $\mathrm{SL}(2,\mathbb{R})$. Therefore this latter is the
residual global symmetry for all possible gaugings.
The 27 vectors $A^{\Lambda}$ are then decomposed into the vectors $A_{IJ}$
in the $\bf (\bar{15},1)$, that
gauge $\mathcal{G}$, and the vectors  in the $\bf (6,2)$, which do not gauge anything
and are then forced  to be dualised into two--forms $B^{I\alpha}$. In
comparison with the $\mathcal{N}=2$ case we see that the pair of indices
${I\alpha}$ is the analogue of the symplectic index $\mathcal{M}$
while the antisymmetric pair of indices $IJ$ is the analogue of the
index $\mathbf{I}$ labeling the gauge group generators.
\par
The fifteen generators $G^{IJ}$ of $\mathcal{G}$ can be expressed as linear
combinations of the $35$ generators $G_r$ ($r\!\!=\!\!1,\dots 35$) of
$\mathrm{SL}(6,\mathbb{R})$: $G^{IJ}\!=\!G_r \,e^{rIJ}$
where $e^{rIJ}$ is the {\it embedding matrix}  which describes the
embedding of $\mathcal{G}$ into $\mathrm{SL}(6,\mathbb{R})$. This is
fully analogous to the embedding matrix used in the gaugings of
maximal $D=4$ supergravity (see eq.(\ref{alettomat}).
For all the admissible cases  in the fundamental $\bf 6$--dimensional  representation
the generators of the gauge group $\mathcal{G}$ take the form
\cite{gunwar}
\begin{equation}
(G^{IJ})^K_{~L}=\delta^{[I}_L\eta^{J]K}
\label{genfund}
\end{equation}
where $\eta^{JK}$ is a diagonal matrix with $p$ eigenvalues equal to $1$,
 $q$ eigenvalues equal to $(-1)$   and, only in the case of contracted groups,
  $r$  null eigenvalues. This signature completely characterizes
the gauge groups and correspondingly the
gauged theory.
From (\ref{genfund}) one can build the generators of
$\mathcal{G}\subset E_{\left(6\right)6}$ in the
$\bf 27$ representation of $E_{\left(6\right)6}$, namely
some suitable matrices $\,(G^{IJ})_{\Lambda}^{~\Sigma}$.
According to the general framework outlined in section
\ref{3.3.3}, in presence of gauging,
the composite $H$--connection of $\mathrm{USp(8)}$ and the scalar vielbein,
defined in (\ref{defungaugedconnection}) are replaced by their gauged analogues:
\begin{equation}
\mathbb{L}^{-1~\Lambda}_{AB}d\mathbb{L}_{\Lambda}^{~AB}+
g(\mathbb{L}^{-1})_{AB}^{~~\Lambda}(G^{IJ})_{\Lambda}^{~\Sigma}\mathbb{L}_{\Sigma}^{~CD}
A_{IJ}=\hat{\mathcal{Q}}_{AB}^{~~~CD}+\hat{\mathcal{P}}_{AB}^{~~~CD}\,,
\label{defgaugedconnection}
\end{equation}
where $g$ is the gauge coupling constant.
The covariant $\mathrm{USp(8)}$ derivative of a field $V_A$
is defined as
\begin{equation}
\nabla V_A=\mathcal{D} V_A+\hat{\mathcal{Q}}_A^{~B}\wedge V_B
\label{covQder}
\end{equation}
where $\mathcal{D}$ is the Lorentz--covariant exterior derivative.
The covariant derivative with respect to $\mathcal{G}$
of a field $V^I$ in the $\bf 6$ of $ \mathrm{SL(6, \mathbb{R})}$ is instead defined as
follows:
\begin{equation}
DV^I\equiv \nabla V^I+g(G^{KL})^I_{~J}A_{KL}\wedge V^J\,.
\label{covGder}
\end{equation}
The decomposition of the field content of maximal supergravity according
to the gauge and R--symmetry group representations is given in table \ref{gaugedfieldcontent}
\begin{table}[ht]
\caption{\label{gaugedfieldcontent} The field content of maximal $D=5$ supergravity}
\begin{center}
$$
\begin{array}{|c|c|c|c|c|}
\hline
\#& \hbox{Field} & \left(\mathrm{SU(2)}\times \mathrm{SU(2)}\right) \hbox{--spin~rep.} &
\mathrm{USp(8)}
\hbox{~rep.} & \mathcal{G}
\hbox{~rep.} \\
\hline
1& V^a & (1,1) & {\bf 1} & {\bf 1} \\
\hline
8& \psi^A & (1,1/2)\oplus (1/2,1) & {\bf 8} & {\bf 1} \\
\hline
15& A_{IJ} & (1/2,1/2) & {\bf 1} & {\bf 15} \\
\hline
12& B^{I\alpha} & (1,0)\oplus (0,1) & {\bf 1} & {\bf 6\oplus \overline{{6}}} \\
\hline
48& \chi^{ABC} & (1/2,0)\oplus (0,1/2) & {\bf 48} & {\bf 1} \\
\hline
42& \mathbb{L}_{\Lambda}^{~AB}\left(\phi\right) & (0,0) & {\bf 27} & {\bf\overline{{27}}} \\
\hline
\end{array}
$$
\end{center}
\end{table}
The solution of superspace Bianchi identities leading to a closed
supersymmetry algebra that implies consistent supersymmetric field equations has
been  recently obtained in \cite{Andrianopoli:2000rs} and shown to
exist for all $\mathrm{SO(p,q)}$ and $\mathrm{CSO(p,q,r)}$ algebras.
The first set of semisimple gaugings had been constructed long ago by
Gunaydin and  Warner \cite{gunwar}. The non semisimple gaugings
$\mathrm{CSO(p,q,r)}$ are instead new, since they appear to be
theories similar to type IIB supergravity where only the field
equations exist but it is not easy to write a conventional action.
What is important is that, irrespectively of the choice of the gauge
group within the allowed class we can write the fermion shifts and
the scalar potential.
For the gravitino mass--matrix we find:
\begin{equation}
  S_{AB} = - \frac{2}{45} \, g \, T_{AB}
\label{n8d5gravmass}
\end{equation}
while for the dilatino shifts we have:
\begin{eqnarray}
  \delta \xi _{ABC}  &=& \mbox{derivative terms} + \Sigma^D_{ABC} \, \epsilon
  _D \nonumber\\
\Sigma^D_{ABC} &=& g \, \frac{1}{\sqrt{2}} \, A^D_{ABC}
\label{d5n8shif}
\end{eqnarray}
Both structures are extracted from the gauge Maurer Cartan equations,
in particular from the $\mathrm{USp(8)}$ tensors:
\begin{eqnarray}
Y^{AB}_{~~~CDEF}&\equiv&
\mathbb{L}^{-1~~\Lambda}_{~CD}(G^{IJ})_{\Lambda}^{~\Sigma}
\mathbb{L}_{\Sigma}^{~AB}\mathbb{L}^{-1}_{EFIJ}\nonumber\\\\
Y^{\pm}_{ABCDEF}&\equiv&\frac{1}{2}(Y_{ABCDEF}\pm Y_{CDABEF})\nonumber\\\\
T^A_{~~BCD}&\equiv&Y^{AF}_{~~~BFCD}\,.
\label{YTtensors}
\end{eqnarray}
Indeed we have:
and the tensors $T_{AB}(\phi),\,A^D_{ABC}(\phi)$ are defined as
\footnote{$\left[\dots\right]\!\vert$
denotes the symplectic traceless antisymmetrization.}
\begin{equation}
T_{AB}=T^C_{~ACB},~~A^D_{ABC}=T^D_{\left[ABC\right]\vert}\,.
\end{equation}
Then, from the general Ward identity (\ref{d5wardide}) one finds the
scalar potential that reads:
\begin{equation}
\mathcal{V}=-4 \, g^2\left[{2\over 675}T_{AB}T^{AB}-{1\over 96}A_{ABCD}A^{ABCD}\right].
\label{d5potential}
\end{equation}
As the reader can see all this is completely analogous to the
construction of the $\mathcal{N}=8$ gaugings described in section
\ref{n8d4gaug} and parallels the construction of the matter coupled
$\mathcal{N}=2$ gauged theory in all respects.
\section{On the quest for supersymmetric brane worlds}
Unfortunately a careful analysis of the scalar potentials (\ref{d5potential})
and (\ref{ceregatpot}) in the case of non semisimple, non compact
gauge algebras is not available at the moment. This is not due to conceptual
problems but simply to the fact that such an analysis has not been
done yet. It is clearly a programme that will be accomplished
soon. Indeed  there are many reasons, in my opinion, to believe that this is the right
corner where to look for possible embeddings  of \emph{brane world
solutions} within a supersymmetric theory. This is mainly due to the
close relationship between partial supersymmetry breaking mechanisms
and non compact gaugings. Of this relation there are many examples in
the existing literature, in particular in $\mathcal{N}=2,D=4$
supergravity \cite{Fre:1997js,Girardello:1997hf,Zinovev:1992mw,Ferrara:1996gu}.
On the other hand it appears to me that the same properties of the
scalar potential and of the fermion shifts that allow for the
existence of non trivial Killing spinors are those that might  allow
for suitable non trivial \emph{brane--world like} solutions of the
bosonic field equations. An indirect argument to support
this viewpoint comes from the combination of two recent developments
related to Kaluza Klein consistent truncations.
The first result obtained in  \cite{duffoliuste} is the following.
Consider the compactification of type IIB
supergravity on $AdS_5 \times S^5$ and the infinite tower of Kaluza
Klein supermultiplets. If one truncates to the massless modes the
resulting theory is $D=5$ maximal supergravity with the compact gauging
$\mathrm{SO(6)}$. Its potential is encoded in eq.(\ref{d5potential})
with the appropriate choice of the embedding matrix (\ref{genfund}).
It was known that this theory does not support any Randall-Sundrum
brane world solution. However if one truncates to a larger theory
that includes also the massive multiplet corresponding to the
$S^5$ breathing mode then the situation is reversed and a
supersymmetric realization of the Randall-Sundrum brane world can be
found. The second result was derived in \cite{noin0101,noin0102} and
goes as follows. Considering the entire Kaluza Klein spectrum of
M--theory compactifications on $AdS_4 \times X^7$ where $X^7$ is some
compact $7$-manifold one finds that there are massive multiplets
with rational conformal dimensions that are linked to the massless
multiplets by a curious and very general pairing named
\emph{shadowing}. One can prove that the truncation to the massless
graviton multiplet plus its massive shadow is always a consistent
truncation. Furthermore the shadow multiplet always contains the
breathing mode considered in \cite{duffoliuste}. In the case of $\mathcal{N}=3$
compactifications it turns out that the shadow multiplet is a massive
gravitino multiplet so that the consistent truncation that includes
the shadow, being the union of an $\mathcal{N}=3$ massless graviton
multiplet with a massive gravitino multiplet looks like a
spontaneously broken version of an $\mathcal{N}=4$ theory. However it was
shown in \cite{noin0101} that no existing version of gauged $\mathcal{N}=4$
supergravity can encode the breaking pattern realized by the Kaluza
Klein spectrum. It is a really recent result to be
published in a forthcoming paper \cite{shadsug} that the appropriate
$\mathcal{N}=4$ theory realizing the Kaluza Klein breaking pattern has been
found. It is a very particular instance of non compact non semisimple
gauging. This concludes my heuristic argument. Combining these two
informations one has a strong hint that brane world solutions are
related to partial supersymmetry breaking mechanisms and that both
are embedded in supergravity theories where a non semisimple solvable
algebra has been gauged. In view of this it appears quite mandatory
that the structure of the potentials (\ref{d5potential}) and
(\ref{ceregatpot}) should be carefully analyzed in the case of non
semisimple gaugings. The best approach to this analysis seems to be
provided by the solvable parametrizations of the scalar coset
manifolds (see section \ref{maxsugra}) where the coset
representatives are always polynomials and where the correspondence
between scalar fields and compactified Ramond and Neveu Schwarz
$p$-forms is quite effective and punctual (see table \ref{dideals})
%%%%%%%%%%%%%%%%%%%%%%%%%%%%%%%%%%%%
\chapter[Solvable Lie Algebras in supergravity and superstrings]{Solvable
Lie Algebras in supergravity and superstrings}
\label{solvchap}
\section[Introduction: gaugings versus BPS black hole classification]{Introduction:
 gaugings versus BPS black hole classification}
In the previous chapters I have illustrated the geometric structures
that underlie  supergravity lagrangians, emphasizing that they are
essentially dictated by the number of supercharges $N_Q$ and by the
dimensionality $D$ of space--time. So doing I tried to illustrate the
interplay between the geometry of the scalar manifold  in \emph{ungauged supergravity}
and the possible choices of a \emph{gauging}, which is generically triggered
by the \emph{near brane geometry} of any $p$--brane
configuration.
\par
A fundamental problem that remains so far open is that of  giving a
$Dp$--brane interpretation to all the compact and non compact gaugings of
supergravities in diverse dimensions. Conversely one would like to
predict the gauged supergravity in $p+2$--dimensions of which the
the \emph{near brane geometry} of a $p$--brane is a classical solution.
\par
Although the solution of such a problem is unknown at the present
time I want to stress that there is a very similar problem which was
instead completely solved, at least in the case of the maximal number
of supercharges $N_Q=32$. I refer to the classification of all $BPS$
black--hole  solutions of $\mathcal{N}=8$ supergravity and to their
microscopic interpretation in terms of $Dp$--brane configurations.
Since the pioneering work on supersymmetric black holes of the middle
nineties
\cite{Kallosh:1992ii,Ferrara:1995ih,Ferrara:1996dd,Ferrara:1996um,Ferrara:1997tw}
and the statistical interpretation of the Bekenstein Hawking entropy
in terms of D--brane microstates found in 1995 by Strominger and Vafa \cite{Strominger:1996sh},
it became evident that a classification of  $BPS$ black
hole solutions in four dimensions and a derivation of their geometry
from microscopic $Dp$--brane configurations was an essential step
forward in understanding quantum gravity. The essential point in such
a programme is the need to master the $\mathrm{U}$--duality transformations
that map Ramond states into Neveu Schwarz ones and extend the $\mathrm{S
\times T}$ duality transformations respecting the two sectors of
superstrings but relating  type IIA to  type IIB $p$--branes.
At the level of supergravity all such transformations are part of the
isometry group $G_{iso}$ of the scalar manifold, while at the
microscopic superstring level they play well distinguished roles.
Hence a necessary bridge to relate microscopic superstring physics to
macroscopic classical solutions of supergravity is  given
by a some suitable treatment of the scalar manifold able to
separate Ramond from Neveu Schwarz directions.   For all
supergravities where $\mathcal{M}_{scalar}$ is a homogeneous coset manifold
the appropriate tool is provided by the so called Solvable Lie algebra
parametrization of the non compact coset $\mathrm{G/H}$.
In particular, as anticipated in chapter \ref{bestia2} the sequential toroidal
compactifications of either type IIA or type IIB superstrings can be
algebraically understood in terms of certain very specific chains of Solvable Lie
algebras. This approach, established in \cite{solvab1,solvab2,noi3},
was further extended by Bertolini and Trigiante
\cite{Bertolini:2000uz}
who succeed \cite{Bertolini:2000ya,Bertolini:2000ei,Bertolini:2000uz,Bertolini:1999je}
in deriving a general five parameter dependent solution for
$\mathcal{N}=8$ black holes preserving $\ft 1 2$ of the supersymmetry
and relating it in a well defined way to $Dp$--brane configurations
characterized by Ramond charges and angles between the branes. The
essential token in such a derivation was indeed the Solvable Lie algebra
technology leading to an algebraic characterization of  scalar  and   vector
fields which can be made so precise as to associate each component of
the compactified metric and  $p+1$ forms to the various roots and weights of the
$\mathrm{U}$--duality group.
\par
It is quite natural to think that the same technology  that links
microscopic $Dp$--configurations to black solutions should link
classical domain walls of gauged supergravities in $p+2$ dimensions  to their microscopic
description in terms of $D$--brane systems.  For this reason I devote
the last  chapter of my lectures  to a review of this
general algebraic framework relating the superstring origin of supergravity
(and hence of its $p$--brane solutions) to its scalar geometry.
\section[Solvable Lie algebras:  NS and RR scalar fields]{Solvable Lie algebras:
 NS and RR scalar fields}
Let us name $\mathcal{G}=\mathrm{U}_{(D,\mathcal{N})}$, $\mathcal{H}=
\mathrm{H}_{(D,\mathcal{N})}$ the
isometry and isotropy groups, respectively, that define the scalar coset manifold in
$D$--dimensional, $\mathcal{N}$--extended supergravity.
The  exciting developments of the second string revolution
have started from the conjecture \cite{huto} that
an appropriate restriction to integers $\mathrm{U}_{(D,\mathcal{N})}(\mathbb{Z})$ of
the Lie group $\mathrm{U}_{(D,\mathcal{N})}$
is an exact non perturbative symmetry of string theory. Eventually it
permutes the elementary, electric states of the perturbative string
spectrum with the non perturbative BPS saturated states like the
magnetic $p$--branes of various type. This U--duality
unifies S--duality (strong--weak duality) with T--duality
(large--small radius duality).
\par
As discussed in \cite{solvab1,solvab2}, utilizing a well established mathematical
framework \cite{helgason}, in all these cases the scalar coset manifold $\mathrm{U/H}$ can be
identified with the group manifold of a normed solvable Lie algebra:
\begin{equation}
  \mathrm{U/H} \sim \exp[{Solv}]
  \label{solvagniz}
\end{equation}
\par
The representation of the supergravity scalar manifold $\mathcal{M}_{scalar}= \mathrm{U/H}$
as the group manifold associated with a {\it  normed solvable Lie algebra}
introduces a one--to--one correspondence between the scalar fields $\phi^I$ of
supergravity and the generators $T_I$ of the solvable Lie algebra $Solv\, (\mathrm{U/H})$.
Indeed the coset representative $\mathbb{L}(\mathrm{U/H})$ of the homogeneous space $\mathrm{U/H}$ is
identified with:
\begin{equation}
\mathbb{L}(\phi) \, =\, \exp [ \phi^I \, T_I ]
\label{cosrep1}
\end{equation}
where $\{ T_I \}$ is a basis of $Solv\, (\mathrm{U/H})$.
\par
As a consequence of this fact the tangent bundle to the scalar manifold $T\mathcal{M}_{scalar}$
is identified with the solvable Lie algebra:
\begin{equation}
T\mathcal{M}_{scalar} \, \sim \,Solv \, (\mathrm{U/H})
\label{cosrep2}
\end{equation}
and any algebraic property of the solvable algebra has a corresponding physical interpretation in
terms of string theory massless field modes.
\par
Furthermore, the local differential geometry of the scalar manifold is described
 in terms of the solvable Lie algebra structure.
Given the euclidean scalar product on $Solv$:
\begin{eqnarray}
  <\, , \, > &:& Solv \otimes Solv \rightarrow \mathbb{R}
\label{solv1}\\
<X,Y> &=& <Y,X>\label{solv2}
\end{eqnarray}
the covariant derivative with respect to the Levi Civita connection is given by
the Nomizu operator \cite{alex}:
\begin{equation}
\forall X \in Solv : \IL_X : Solv \to Solv
\end{equation}
\begin{eqnarray}
  \forall X,Y,Z \in Solv & : &2 <Z,\IL_X Y> \nonumber\\
&=& <Z,[X,Y]> - <X,[Y,Z]> - <Y,[X,Z]>
\label{nomizu}
\end{eqnarray}
and the Riemann curvature 2--form is given by the commutator of two Nomizu
operators:
\begin{equation}
 <W,\{[\IL_X,\IL_Y]-\IL_{[X,Y]}\}Z> = R^W_{\ Z}(X,Y)
\label{nomizu2}
\end{equation}
In the case of maximally extended supergravities in $D=10-r$ dimensions the scalar
manifold has a universal structure:
\begin{equation}
 { \mathrm{U}_D\over \mathrm{H}_D}  = {\mathrm{E}_{r+1(r+1)} \over \mathrm{H}_{r+1}}
\label{maximal1}
\end{equation}
where the Lie algebra of the $\mathrm{U}_D$--group $\mathrm{E}_{r+1(r+1)} $ is the
maximally non compact real section of the exceptional $\mathrm{E}_{r+1}$ series
of the simple complex Lie Algebras
and $\mathrm{H}_{r+1}$ is its maximally compact subalgebra \cite{cre}.
As shown in \cite{solvab1,solvab2},
the manifolds $\mathrm{E}_{r+1(r+1)}/\mathrm{H}_{r+1}$
share the distinctive  property of being non--compact homogeneous spaces of maximal rank
$r+1$, so that the associated solvable Lie algebras,
 such that ${\mathrm{E}_{r+1(r+1)}}/{\mathrm{H}_{r+1}} \, = \, \exp \left [ Solv_{(r+1)} \right ]
$,  have the particularly simple structure:
\begin{equation}
Solv\, \left ( \mathrm{E}_{r+1}/\mathrm{H}_{r+1} \right )\, = \,
{\cal H}_{r+1} \, \oplus_{\alpha \in
\Phi^+(\mathrm{E}_{r+1})} \, \mathbb{E}^\alpha
\label{maxsolv1}
\end{equation}
where $\mathbb{E}^\alpha \, \subset \, \mathrm{E}_{r+1}$ is the
1--dimensional subalgebra associated
with the root $\alpha$
and $\Phi^+(\mathrm{E}_{r+1})$ is the positive part of the $\mathrm{E}_{r+1}$--root--system.
\par
The generators of the solvable Lie algebra  are in one to one
correspondence with the scalar fields of the theory.
Therefore they can be characterized as Neveu Schwarz or Ramond Ramond
depending on their origin in compactified string theory. From the
algebraic point of view the generators of the solvable algebra are of
three possible types:
\begin{enumerate}
\item {Cartan generators }
\item { Roots that belong to the adjoint representation of the
$D_r \equiv\mathrm{SO(r,r)} \subset \mathrm{E}_{r+1(r+1)}$ subalgebra (= the T--duality algebra) }
\item {Roots which are weights of an irreducible representation
 of the $D_r$ algebra.}
\end{enumerate}
The scalar fields associated with generators of type 1 and 2 in the above
list are Neveu--Schwarz fields while the fields of type 3 are
Ramond--Ramond fields.
\par
In the $r=6$ case, corresponding to $D=4$, there is one extra root,
besides those listed above, which is also of the Neveu--Schwarz type.
From the dimensional reduction viewpoint the origin of this extra
root is the following: it is associated with the axion $B_{\mu\nu}$
which only in 4--dimensions becomes equivalent to a scalar field.
This root (and its negative) together with the 7-th Cartan generator
of $\mathrm{O(1,1)}$ promotes the S--duality in $D=4$ from $\mathrm{O(1,1)}$, as it is in
all other dimensions, to $\mathrm{SL}(2,\mathbb{R})$.
%%%%%%%%%%%%%%%%%%%%%%%%%%%%%%%%%%%%%%%%%%%%%%%%%%%%%%%%%%%%%%%%%%%%%
\section[Non compact cosets and solvable Lie algebras: the general setup]{Non compact cosets and
solvable Lie algebras: the general setup}
%%%%%%%%%%%%%%%%%%%%%%%%%%%%%%%%%%%%%%%%%%%
\label{solgensetup}
The relation between coset manifolds and solvable Lie algebras
illustrated above in the case of maximal supergravities can be
generalized to all instances of non compact cosets as we have already
emphasized. Let us dwell a little more on the general idea of this
relation according  to which any homogeneous non-compact coset manifold may be expressed as
a group manifold generated by a suitable solvable Lie algebra. \cite{alex}
\par
Let us start by giving few preliminary definitions.
A {\it solvable } Lie algebra $Solv$ is a Lie algebra whose $n^{th}$ order
(for some $n\geq 1$) derivative algebra vanishes:
\begin{eqnarray}
{\cal D}^{(n)}Solv&=&0 \nonumber \\
{\cal D}Solv=[Solv,Solv]&;&\quad {\cal D}^{(k+1)}Solv=[{\cal D}^{(k)}Solv,
{\cal D}^{(k)}Solv]\nonumber
\end{eqnarray}
A {\it metric} Lie algebra $(\mathbb{G},\langle,\rangle)$ is a Lie algebra endowed with an
euclidean metric $\langle,\rangle$. An important theorem states that if a Riemannian
manifold
$(\mathcal{M},g)$ admits a transitive group of isometries ${\cal G}_s$ generated by
a solvable Lie algebra $Solv$ of the same dimension as $\mathcal{M}$, then:
\begin{eqnarray}
\mathcal{M}\sim {\cal G}_s&=&\exp(Solv)\nonumber\\
 g_{|e\in \mathcal{M}}&=&\langle,\rangle \nonumber
\end{eqnarray}
where $\langle,\rangle$ is an euclidean metric defined on $Solv$.
Therefore there is a one to one correspondence between Riemannian manifolds fulfilling
the hypothesis stated above and solvable metric Lie algebras $(Solv,\langle,\rangle)$.\\
Consider now an homogeneous coset manifold $\mathcal{M}=\mathrm{U} /\mathcal{H}$,
$ \mathrm{U}$ being a non compact real
 form of a semisimple Lie group and $\mathrm{H}$ its maximal compact
subgroup. If $\mathbb{U}$ is the Lie algebra generating $\mathrm{U}$, the so
called {\it Iwasawa
decomposition} ensures the existence of a solvable Lie subalgebra
$Solv\subset
\mathbb{U}$, acting transitively on $\mathcal{M}$, such that \cite{helgason}:
\begin{equation}
 \mathbb{U}= \mathbb{H}+ Solv \qquad \mbox{dim }Solv=\mbox{dim } \mathcal{M} \nonumber
\end{equation}
$\mathbb{H}$ being the maximal compact subalgebra of $\mathbb{U}$ generating
$\mathrm{H}$.
In virtue of the previously stated theorem, $\mathcal{M}$ may be expressed
as a solvable group manifold generated by $Solv$, namely
eq.(\ref{solvagniz})is true.\\
The algebra $Solv$ is constructed as follows \cite{helgason}.
Consider the Cartan decomposition
\begin{equation}
\mathbb{U} = \mathbb{H} \oplus \mathbb{K}
\end{equation}
where $\mathbb{K}$ is the subspace consisting of all the non compact
generators of $\mathbb{U}$.
Let us denote by $\mathcal{C}_K$ the  maximal abelian subspace
of $\mathbb{K}$ and by $\mathcal{C}$ the Cartan subalgebra of $\mathbb{U}$.
It can be shown \cite{helgason} that $\mathcal{C}_K = \mathcal{C} \cap \mathbb{K}$,
namely  it consists of all non compact elements of the Cartan subalgebra $\mathcal{C}$.
Furthermore let $h_{\alpha_i}$ denote the elements
of $\mathcal{C}_K$, $\{\alpha_i\}$ being
 a subset of the positive roots of $\mathbb{U}$ and $\Delta^+$
 the set of positive roots $\beta$ not orthogonal to
all the $\alpha_i$ (i.e. the corresponding ``shift'' operators
$\mathrm{E}_\beta$ do not commute with $\mathcal{C}_K$).
It can be shown that the solvable algebra $Solv$
defined by the Iwasawa decomposition
is constructed expressed in the following way:
\begin{equation}
  \label{iwa}
  Solv = \mathcal{C}_K \oplus \{\sum_{\alpha \in \Delta^+}\mathrm{E}_\alpha \cap \mathbb{U} \}
\end{equation}
where  the intersection with $\mathbb{U} $ means that $Solv$ is generated
by those suitable complex combinations of the ``shift'' operators
which belong to the real form of the isometry algebra $\mathbb{U}$.
\par
The {\it rank} of a homogeneous  coset manifold is defined as
the maximum number of commuting semisimple
elements of the non compact subspace $\mathbb{K}$. Therefore it
coincides with the dimension of $\mathcal{C}_K$,
i.e. with the number of non compact Cartan generators of $\mathbb{U}$.
A  coset manifold is {\it maximally non compact} if
$\mathcal{C} =\mathcal{C}_K \subset Solv$. As we have seen in the previous
section this kind of
manifolds correspond to the scalar manifolds of  {maximally extended
supergravities}. Indeed eq. (\ref{maxsolv1}) is just a particular
case of eq.(\ref{iwa}) where all the positive roots are retained and
the Cartan subalgebra is completely non--compact.
\section[Counting  of massless modes in the type IIA superstring]{Counting
of massless modes in sequential toroidal compactifications
of $D=10$ type IIA superstring}
In order to make the pairing between scalar field modes and solvable
Lie algebra generators explicit, it is convenient to organize the counting of bosonic zero modes
in a sequential way that goes down from $D=10$ to $D=4$ in 6 successive steps.
\par
The useful feature of this sequential viewpoint is that it has a direct algebraic
counterpart in the successive embeddings of the exceptional Lie Algebras $\mathrm{E}_{r+1}$
one into the next one:
{\footnotesize
\begin{equation}
  \matrix{\mathrm{E}_{7(7)}&\supset &  \mathrm{E}_{6(6)}&\supset & \mathrm{E}_{5(5)}&\supset
  & \mathrm{E}_{4(4)}&\supset
  & \mathrm{E}_{3(3)}&\supset & \mathrm{E}_{2(2)}&\supset & \mathrm{O}(1,1) \cr
D=4 & \leftarrow & D=5 & \leftarrow & D=6 & \leftarrow & D=7 & \leftarrow & D=8 &
\leftarrow & D=9 & \leftarrow & D=10 \cr}
\end{equation}}
If we consider the bosonic massless spectrum \cite{gsw} of  type II theory in $D=10$
in the Neveu--Schwarz sector we have the metric, the axion and the dilaton,
while in the Ramond--Ramond sector we have a 1--form and a 3--form:
\begin{equation}
 D=10 \quad : \quad  \cases{
 NS: \quad g_{\mu\nu}, B_{\mu\nu} , \Phi \cr
 RR: \quad  A_{\mu} , A_{\mu\nu\rho} \cr}
 \label{d10spec}
\end{equation}
corresponding to the following counting of degrees of freedom:
$\# $ d.o.f. $g_{\mu\nu} = 35$, $\# $ d.o.f. $B_{\mu\nu} = 28$, $\# $ d.o.f.
$A_{\mu} = 8$, $\# $ d.o.f. $A_{\mu\nu\rho} = 56$
so that the total number of degrees of freedom is $64$  both in the Neveu--Schwarz
and in the Ramond:
 \begin{eqnarray}
  \mbox{Total $\#$ of NS degrees of freedom}&=&{ 64}={ 35}+{ 28}+ { 1} \nonumber\\
  \mbox{Total $\#$ of RR degrees of freedom}&=&{ 64}={ 8}+{ 56}
  \label{64NSRR}
 \end{eqnarray}
 \par
It is worth noticing that the number of degrees of freedom of N--S and R--R sectors are equal, both for
bosons and fermions, to $128= (64)_{NS} + (64)_{RR}$. This is merely a consequence
of type II supersymmetry.
Indeed, the entire Ramond sector (both in type IIA and type IIB) can be thought as a spin $3/2$
multiplet of the second supersymmetry generator.
\par
Let us now organize the degrees of freedom as they appear after toroidal compactification
on a $r$--torus \cite{pope}:
\begin{equation}
\mathcal{M}_{10} = \mathcal{M}_{D-r} \, \otimes T_r
\end{equation}
Naming with Greek letters the world indices on the $D$--dimensional
space--time and with Latin letters the internal indices referring to
the torus dimensions we obtain the results displayed in Table \ref{tabu1} and number--wise we
obtain the counting of Table \ref{tabu2}:
\par
\vskip 0.3cm
\begin{table}[ht]
\begin{center}
\caption{Dimensional reduction of type IIA fields}
\label{tabu1}
 \begin{tabular}{|l|l|c|c|c|c|r|}\hline
\vline & \null & \vline & Neveu Schwarz & \vline & Ramond Ramond & \vline \\
 \hline
 \hline
\vline & Metric & \vline &  $g_{\mu\nu}$& \vline  & \null & \vline   \\ \hline
\vline & 3--forms & \vline &  \null & \vline  & $A_{\mu\nu \rho}$ & \vline \\ \hline
\vline & 2--forms & \vline   & $B_{\mu\nu}$ & \vline   & $A_{\mu\nu i}$ & \vline  \\ \hline
\vline & 1--forms & \vline   &  $g_{\mu i}, \quad B_{\mu i}$ & \vline  & $A_{\mu},
 \quad A_{\mu ij}$ & \vline \\ \hline
\vline & scalars  & \vline  & $\Phi, \quad g_{ij}, \quad B_{ij}$ &
 \vline  & $A_{i}, \quad A_{ijk}$ & \vline \\ \hline
 \end{tabular}
 \end{center}
\end{table}
 \par
 \vskip 0.3cm
 \par
\vskip 0.3cm
\begin{table}[ht]
\begin{center}
\caption{Counting of type IIA fields}
\label{tabu2}
 \begin{tabular}{|l|l|c|c|c|c|r|}\hline
\vline & \null & \vline & Neveu Schwarz & \vline & Ramond Ramond & \vline \\
 \hline
 \hline
\vline & Metric & \vline &  ${ 1}$& \vline  & \null & \vline   \\ \hline
\vline & $\#$ of 3--forms & \vline &  \null & \vline  & ${  1}$ & \vline \\ \hline
\vline &$\#$ of  2--forms & \vline   & ${  1}$ & \vline   & ${  r}$ & \vline  \\ \hline
\vline &$\#$ of 1--forms & \vline   &  $ {  2 r}$ & \vline  & $ {  1} +
\frac 1 2 \, r \, (r-1)$ & \vline \\ \hline
\vline & scalars  & \vline  & $1 \, +  \, \frac 1 2 \, r \, (r+1)$&
\vline  & $ r \, +\, \frac 1 6 \, r \, (r-1) \, (r-2)  $ & \vline \\
\vline &  \null   & \vline & $  +  \, \frac 1 2 \, r \, (r-1)  $ & \vline & \null   & \vline \\
\hline
\end{tabular}
\end{center}
\end{table}
\vskip 0.2cm
We can easily check that the total number
of degrees of freedom in both sectors is indeed $64$ after
dimensional reduction as it was before.
%%%%%%%%%%%%%%%%%%%%%%%%%%%%%%%%%%%%%%%%%
\section[$\mathrm{E}_{r+1}$ subalgebra chains and their string interpretation]{ $\mathrm{E}_{r+1}$ subalgebra chains and their string interpretation}
We can now inspect the algebraic properties of the solvable Lie algebras
$Solv_{r+1}$ defined by eq. (\ref{maxsolv1}) and illustrate the match between
these properties and the physical properties of the sequential compactification.
\par
Due to the specific structure (\ref{maxsolv1}) of a maximal rank solvable Lie algebra
every chain of {\it regular embeddings}:
\begin{equation}
\mathrm{E}_{r+1} \, \supset \,K^{0}_{r+1} \, \supset \, K^{1}_{r+1}\, \supset \, \dots \, \supset \,
 K^{i}_{r+1}\, \supset \, \dots
\label{aletto1}
\end{equation}
where $K^{i}_{r+1}$ are subalgebras of the same rank and with
the same Cartan subalgebra ${\cal H}_{r+1}$ as
$\mathrm{E}_{r+1}$ reflects into  a corresponding sequence of embeddings
of solvable Lie algebras and, henceforth, of  homogeneous non--compact scalar manifolds:
\begin{equation}
\mathrm{E}_{r+1}/\mathrm{H}_{r+1} \, \supset \,K^{0}_{r+1}/Q^{0}_{r+1} \,\supset \,  \dots  \,\supset \,
K^{i}_{r+1}/Q^{i}_{r+1}
\label{caten1}
\end{equation}
which must be endowed with a physical interpretation.
In particular we can consider embedding chains such that \cite{witten}:
\begin{equation}
K^{i}_{r+1}= K^{i}_{r} \oplus X^{i}_{1}
\label{spacco}
\end{equation}
where $K^{i}_{r}$ is a regular subalgebra of $rank= r$ and $X^{i}_{1}$
is a regular subalgebra of rank one.
Because of the relation between the rank and the number
of compactified dimensions such chains clearly
correspond to the sequential dimensional reduction of either type IIA (or B) or of M--theory.
Indeed the first of such regular embedding chains we can consider is:
\begin{equation}
K^{i}_{r+1}=\mathrm{E}_{r+1-i}\, \oplus_{j=1}^{i} \, \mathrm{O}(1,1)_j
\label{caten2}
\end{equation}
This chain simply tells us that the scalar manifold of
supergravity in dimension $D=10-r$ contains the
direct product of the supergravity scalar manifold  in dimension $D=10-r+1$
with the 1--dimensional moduli
space of a $1$--torus (i.e. the additional compactification radius one gets by making a further
step down in compactification).
\par
There are however additional embedding chains that originate from the different choices
of maximal
ordinary subalgebras admitted by the exceptional Lie algebra of the $\mathrm{E}_{r+1}$ series.
\par
All the $\mathrm{E}_{r+1}$ Lie algebras contain a subalgebra $D_{r}\oplus \mathrm{O}(1,1)$ so
that we can write the chain \cite{solvab1,solvab2}:
\begin{equation}
K^{i}_{r+1}=D_{r-i}\, \oplus_{j=1}^{i+1} \, \mathrm{O}(1,1)_j
\label{dueachain}
\end{equation}
As we discuss more extensively later on, the embedding chain (\ref{dueachain})
corresponds to the decomposition of the scalar manifolds into submanifolds spanned by either
 N-S or  R-R fields, keeping moreover track of the way they originate at each level of the
sequential dimensional reduction. Indeed the N--S fields correspond to generators of the
solvable Lie algebra that behave as integer (bosonic) representations of the
\begin{equation}
D_{r-i} \, \equiv \, \mathrm{SO}(r-i,r-i)
\label{subalD}
\end{equation}
while R--R fields correspond to generators of the solvable Lie algebra assigned to the spinorial
representation of the subalgebras (\ref{subalD}).
A third chain of subalgebras is the following one:
\begin{equation}
K^{i}_{r+1}=A_{r-1-i}\,\oplus  \, A_1 \, \oplus_{j=1}^{i+1} \, \mathrm{O}(1,1)_j
\label{duebchain}
\end{equation}
and a fourth one is
\begin{equation}
K^{i}_{r+1}=A_{r-i}\,  \oplus_{j=1}^{i+1} \, \mathrm{O}(1,1)_j
\label{elechain}
\end{equation}
The physical interpretation of the (\ref{duebchain}), illustrated in the next subsection, has its
origin in type IIB string theory. The same supergravity effective lagrangian can be viewed as
the result of compactifying either version of type II string theory. If we take the IIB
interpretation
the distinctive fact is that there is, already at the $10$--dimensional level a complex scalar
field $\Sigma$ spanning the non--compact coset manifold $\mathrm{SL}(2,\mathbb{R})_U/\mathrm{O}(2)$.
The $10$--dimensional U--duality
group  $\mathrm{SL}(2,\mathbb{R})_U$ must therefore be present in all lower dimensions and it
corresponds to the addend
$A_1$ of the chain (\ref{duebchain}).
\par
The fourth chain (\ref{elechain}) has its origin  in an M--theory interpretation or in a
 physical problem posed by the
$D=4$ theory.
\par
If we compactify the $D=11$ M--theory to $D=10-r$ dimensions using an $(r+1)$--torus $T_{r+1}$,
the flat metric on this is parametrized by the coset manifold $\mathrm{GL}(r+1) / \mathrm{O}(r+1)$.
The isometry group of the $(r+1)$--torus moduli space is therefore $\mathrm{GL}(r+1)$ and its
Lie Algebra is $A_r + \mathrm{O}(1,1)$, explaining the chain (\ref{elechain}).
Alternatively, we may consider the origin of the same chain from a $D=4$ viewpoint.
There the electric vector field strengths do not span an irreducible representation
of the U--duality group $\mathrm{E}_7$ but sit
together with their magnetic counterparts in the irreducible
fundamental ${\bf 56}$ representation.  An important question therefore is that of
establishing which subgroup $\mathrm{G}_{el}\subset \mathrm{E}_7$ has an electric action
on the field strengths. The answer is \cite{hull}:
\begin{equation}
\mathrm{G}_{el} \, = \, \mathrm{SL}(8, \mathbb{R} )
\label{elecgrou}
\end{equation}
since it is precisely with respect to this subgroup that the fundamental ${\bf 56}$ representation
of $\mathrm{E}_7$
splits into: ${\bf 56}= {\bf 28}\oplus {\bf 28}$. The Lie algebra of the electric subgroup is
$A_7 \, \subset \, \mathrm{E}_7$ and it contains an obvious subalgebra $A_6 \oplus \mathrm{O}(1,1)$.
The intersection
of this latter with the subalgebra chain (\ref{caten2}) produces the electric chain
(\ref{elechain}).
In other words, by means of equation (\ref{elechain}) we can trace back in each upper dimension
which symmetries will maintain an electric action also at the end point of the dimensional
reduction sequence, namely also in $D=4$.
\par
We have  spelled out the embedding chains of subalgebras that are physically significant from
a string theory viewpoint. The natural question to pose now  is  how to understand their
algebraic origin and how to encode them in an efficient description holding true sequentially in all
dimensions,
namely for all choices of the rank $r+1=7,6,5,4,3,2$. The answer is provided by reviewing the
explicit construction of the $\mathrm{E}_{r+1}$ root spaces in terms of $r+1$--dimensional
euclidean vectors
\cite{gilmore}.
%%%%%%%%%%%%%%%%%%%%%%%%%%%
\subsection[Dynkin diagrams of the  $\mathrm{E}_{r+1(r+1)}$ root spaces and structure
of the associated solvable algebras]{Dynkin diagrams of the  $\mathrm{E}_{r+1(r+1)}$ root spaces and structure
of the associated solvable algebras}
\label{dynsect1}
The root system  of type $\mathrm{E}_{r+1(r+1)}$  can be described
for all values of $1\le r \le 6$ in the following way. As any other
root system it is a finite subset of vectors $\Phi_{r+1}\, \subset\, \mathbb{R}^{r+1}$
such that $\forall \alpha ,\beta \, \in \Phi_{r+1}$ one has
$ \langle \alpha , \beta \rangle \, \equiv  2 (\alpha , \beta )/ (\alpha , \alpha) \,
\in \, \mathbb{Z} $ and such that $\Phi_{r+1}$ is invariant with respect to
the reflections generated by any of its elements. For an explicit
listing of the roots we refer the reader to \cite{solvab1,solvab2}. We just
recall that
the most efficient way to deal simultaneously with all the above root systems and
see the emergence of the above mentioned embedding chains is to embed them in the
largest, namely in the $\mathrm{E}_7$ root space. Hence the various root systems $\mathrm{E}_{r+1}$
will be represented by appropriate subsets of the full set of $\mathrm{E}_7$ roots. In this
fashion for all choices of $r$ the $\mathrm{E}_{r+1}$ are anyhow represented by 7--components
Euclidean vectors of length 2.
\par
Given a basis of seven simple roots $\alpha_1 , \dots \, \alpha_7$ whose
scalar products are those predicted by the $\mathrm{E}_7$ Dynkin diagram:
\begin{eqnarray}
\alpha_1 =\left \{-\frac {1}{2},-\frac {1}{2},-\frac {1}{2}, -\frac {1}{2}, -\frac {1}{2},
-\frac {1}{2}, \frac{1}{\sqrt{2}}\right \}\nonumber\\
\alpha_2 = \left \{ 0,0,0,0,1,1,0 \right \}\nonumber\\
\alpha_3 = \left \{ 0,0,0,1,-1,0,0 \right \}\nonumber\\
\alpha_4 = \left \{ 0,0,0,0,1,-1,0 \right \} \nonumber\\
\alpha_5 = \left \{ 0,0,1,-1,0,0,0 \right \} \nonumber\\
\alpha_6 = \left \{ 0,1,-1,0,0,0,0 \right \} \nonumber\\
\alpha_7 = \left \{ 1,-1,0,0,0,0,0 \right \} \nonumber\\
\label{e7simple}
\end{eqnarray}
the embedding of chain (\ref{caten2}) is easily described. By considering the subset of
$r$ simple roots
$\alpha_1 , \alpha_2 \, \dots \, \alpha_r$ we realize the Dynkin diagrams of type $\mathrm{E}_{r+1}$.
Correspondingly,
the subset of all roots pertaining to the root system $\Phi(\mathrm{E}_{r+1}) \, \subset \,
\Phi(\mathrm{E}_7)$ can be explicitly found.
At each step of the sequential embedding one  generator of the $r+1$--dimensional
Cartan subalgebra
${\cal H}_{r+1}$ becomes orthogonal to the roots of the subsystem
$\Phi(\mathrm{E}_{r})\subset\Phi(\mathrm{E}_{r+1})$,
while the remaining $r$ span the Cartan subalgebra of $\mathrm{E}_{r}$.
In order to visualize the other chains of subalgebras it is convenient to make two observations.
The first is to note that the simple roots selected in eq. (\ref{e7simple}) are of two types: six
of them have integer components and span the Dynkin diagram of a $D_6 \equiv \mathrm{SO}(6,6)$ subalgebra,
while the seventh simple root has half integer components and it is actually a spinor weight
with respect to this subalgebra. This observation leads to the embedding chain (\ref{dueachain}).
Indeed it suffices to discard one by one the last simple root to see the embedding of the
$D_{r-1}$ Lie algebra into $D_{r}\subset \mathrm{E}_{r+1}$. As discussed in the next section $D_{r}$
is the Lie algebra of the T--duality group in type IIA toroidally compactified string theory.
\par
The next  observation is that the $\mathrm{E}_7$ root system contains an exceptional pair of
roots $\beta =\pm \sqrt{2} \epsilon_7 \equiv \pm \sqrt{2} (0,0,0,0,0,0,1)$,
which does not belong to any of the other $\Phi (\mathrm{E}_r)$
root systems. Physically the origin of this exceptional pair is very clear. It is associated
with the axion field $B_{\mu\nu}$ which in $D=4$ and only in $D=4$ can be dualized to an
additional scalar field. This root has not been chosen to be a simple root in
eq.(\ref{e7simple})
since it can be regarded as a composite root in the $\alpha_i$ basis. However we have the
possibility
of discarding either $\alpha_2$ or $\alpha_1$ or  $\alpha_4$ in favour of $\beta$ obtaining a new
basis for the $7$-dimensional euclidean space $\mathbb{R}^7$. The three choices in this operation
lead to the three different Dynkin diagrams given in fig.s (\ref{stdual}) and (\ref{elecal}),
corresponding to
the Lie Algebras:
\begin{equation}
 A_5 \oplus A_2\, , \quad   D_6\oplus A_1  \, , \quad
  A_7
\label{splatto}
\end{equation}
\iffigs
\begin{figure}
\caption{\label{stdual} $S$ and $T$ duality subalgebras from Dynkin diagrams}
\epsfxsize = 10cm
\epsffile{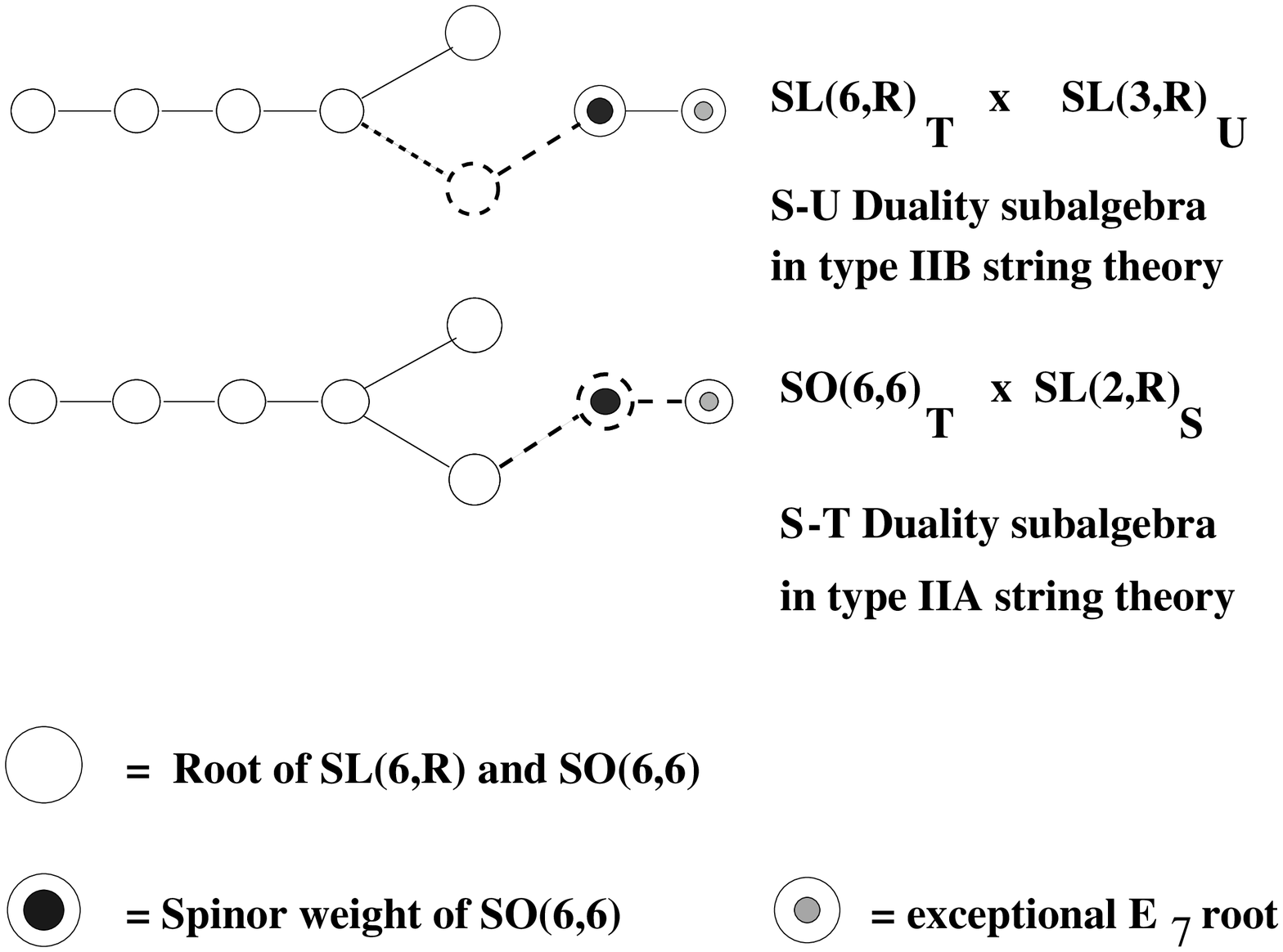}
\vskip -0.1cm
\unitlength=1mm
\end{figure}
\fi
\iffigs
\begin{figure}
\caption{\label{elecal} Dynkin diagrams and the Electric subalgebra}
\epsfxsize = 10cm
\epsffile{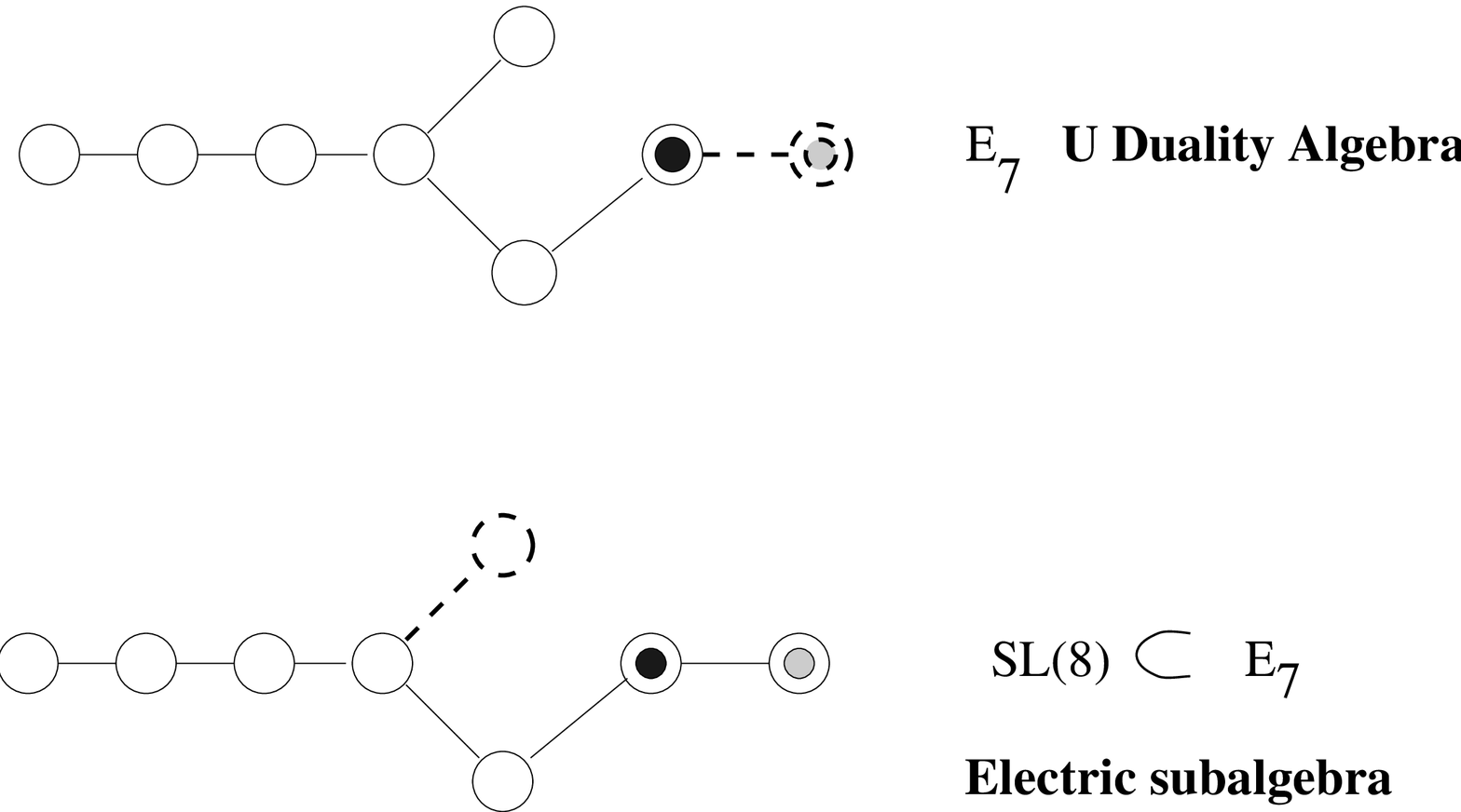}
\vskip -0.1cm
\unitlength=1mm
\end{figure}
\fi
From these embeddings occurring at the $\mathrm{E}_7$ level, namely in $D=4$,
one deduces the three embedding chains
(\ref{dueachain}),(\ref{duebchain}),(\ref{elechain}): it just suffices to peal
off the last $\alpha_{r+1}$ roots
one by one and also the $\beta$ root that occurs only in $D=4$.
One observes that the appearance of the
$\beta$ root is always responsible for an enhancement of the S--duality group.
In the type IIA case
this group is enhanced from $\mathrm{O(1,1)}$ to $\mathrm{SL}(2,\mathbb{R})$ while in the
type IIB case
it is enhanced from
the $\mathrm{SL}(2,\mathbb{R})_U$ already existing in $10$--dimensions to
$\mathrm{SL}(3,\mathbb{R})$.
Physically this occurs by
combining the original dilaton field with the compactification radius of
the latest compactified
dimension.
%%%%%%%%%%%%%%%%%%%%%%%%%%%%%%%%%%%%%%%%%%%%%%%%%%%%%%%%%%%%
\subsection{String theory interpretation of the sequential embeddings:
Type $IIA$, type $IIB$ and $M$ theory chains}
\label{dynsect2}
We now turn to a closer analysis of the physical meaning of
the embedding chains we have been illustrating.
\par
Let us begin with the chain of eq.(\ref{duebchain})that, as anticipated, is
related with the type IIB interpretation of supergravity theory.
The distinctive feature of this chain of embeddings is the presence
of an addend $A_1$ that is already present in 10 dimensions. Indeed
this $A_1$ is the Lie algebra of the $\mathrm{SL}(2,R)_\Sigma $ symmetry of type $IIB$
D=10 superstring. We can name this group the U--duality symmetry $\mathrm{U}_{10}$ in
$D=10$. We can use the chain (\ref{duebchain}) to trace it in lower dimensions.
Thus let us  consider the decomposition
\begin{eqnarray}
\mathrm{E}_{r+1(r+1)} & \rightarrow & N_r \otimes \mathrm{SL}(2,\mathbb{R}) \nonumber\\
N_r & = &   A_{r-1} \otimes \mathrm{O(1,1)}
\label{sl2r}
\end{eqnarray}
Obviously $N_r$ is not contained in the  $T$-duality group $\mathrm{O}(r,r)$ since the
$NS$ tensor field $B_{\mu \nu}$ (which  mixes with the metric under
$T$-duality) and the $RR$--field $B^c_{\mu \nu}$ form a doublet with
respect $\mathrm{SL}(2,\mathbb{R})_U$. In fact, $\mathrm{SL}(2,\mathbb{R})_U$ and  $\mathrm{O}(r,r)$ generate
the whole U--duality group $\mathrm{E}_{r+1(r+1)}$. The appropriate interpretation
of the normaliser of $\mathrm{SL}(2,R)_\Sigma$ in $\mathrm{E}_{r+1(r+1)}$  is
\begin{equation}
 N_r  = \mathrm{O(1,1)} \otimes \mathrm{SL}(r,\mathbb{R}) \equiv \mathrm{GL}(r,\mathbb{R})
 \label{agnosco}
\end{equation}
where $\mathrm{GL}(r,\mathbb{R})$ is the isometry group of the  classical moduli
space for the $T_r$ torus:
\begin{equation}
 \frac{\mathrm{GL}(r,\mathbb{R})}{\mathrm{O}(r)}.
\end{equation}
The decomposition of the U--duality group appropriate for the type $IIB$ theory is
\begin{equation}\mathrm{E}_{r+1} \rightarrow \mathrm{U}_{10} \otimes \mathrm{GL}(r,\mathbb{R}) =
\mathrm{SL}(2,\mathbb{R})_U \otimes \mathrm{O(1,1)} \otimes \mathrm{SL}(r,\mathbb{R}).
\label{sl2rii}
\end{equation}Note that since $\mathrm{GL}(r,\mathbb{R}) \supset \mathrm{O(1,1)}^r$,
this translates into $\mathrm{E}_{r+1} \supset
\mathrm{SL}(2,\mathbb{R})_U \otimes \mathrm{O(1,1)}^r$.
(In Type $IIA$, the corresponding chain would
be $\mathrm{E}_{r+1} \supset \mathrm{O(1,1)} \otimes \mathrm{O}(r,r) \supset
\mathrm{O(1,1)}^{r+1}$.)  Note that while
$\mathrm{SL}(2,\mathbb{R})$ mixes $RR$ and $NS$ states,
$\mathrm{GL}(r,\mathbb{R})$ does not. Hence we can write the following
decomposition for the solvable Lie algebra:
\begin{eqnarray}
Solv \left( \frac{\mathrm{E}_{r+1}}{\mathrm{H}_{r+1}} \right) &=&
Solv \left(\frac{\mathrm{GL}(r,\mathbb{R})}{\mathrm{O}(r)} \otimes
\frac{\mathrm{SL}(2,\mathbb{R})}{\mathrm{O}(2)} \right) +
\left(\frac{\bf r(r-1)}{\bf 2}, {\bf 2} \right) \oplus {\bf X}
\oplus {\bf Y}  \nonumber \\
\mbox{dim }Solv \left( \frac{\mathrm{E}_{r+1}}{\mathrm{H}_{r+1}} \right)&=&
\frac{d(3d-1)}{2} + 2 + x + y.
\label{solvii}
\end{eqnarray}
where $x=\mbox{dim }{\bf X} $ counts the scalars coming from the internal part of the $4$--form
$A^+_{{ \mu}{ \nu}{ \rho}{\sigma}}$ of type IIB string theory.
We have:
\begin{equation}
x =  \left \{
\matrix { 0 & r<4 \cr
\frac{r!}{4!(r-4)!} & r\geq 4 \cr}\right.
\label{xscal}
\end{equation}
and
\begin{equation}
y =\mbox{dim }{\bf Y} = \cases{
\matrix { 0 &  r  < 6   \cr 2 &  r = 6  \cr}\cr}.
\label{yscal}
\end{equation}
counts the scalars arising from dualising the two-index tensor
fields in $r=6$.
\par
For example, consider the $ D=6$ case. Here the type $IIB$  decomposition is:
\begin{equation}
\mathrm{E}_{5(5)}=\frac{\mathrm{O}(5,5)}{\mathrm{O}(5) \otimes \mathrm{O}(5)} \rightarrow \frac{\mathrm{GL}(4,\mathbb{R})}{\mathrm{O}(4)}
\otimes \frac{\mathrm{SL}(2,\mathbb{R})}{\mathrm{O}(2)}
\label{exii}
\end{equation}
whose compact counterpart is given by $\mathrm{O}(10) \rightarrow \mathrm{SU}(4) \otimes \mathrm{SU}(2) \otimes U(1)$,
corresponding to the decomposition: ${\bf 45} =
{\bf (15,1,1)}+ {\bf (1,3,1)} + {\bf (1,1,1)} +{\bf (6,2,2)} + {\bf (1,1,2)}$. It follows:
\begin{equation}Solv(\frac{\mathrm{E}_{5(5)}}{\mathrm{O}(5) \otimes \mathrm{O}(5)}) = Solv (\frac{\mathrm{GL}(4,\mathbb{R})}{\mathrm{O}(4)}
\otimes \frac{\mathrm{SL}(2,\mathbb{R})}{\mathrm{O}(2)}) +({\bf 6},{\bf 2})^+ + ({\bf 1},{\bf 1})^+.
\label{solvexii}
\end{equation}
where the factors on the right hand side parametrize the internal part of the metric $g_{ij}$,
the dilaton and the $RR$ scalar ($\phi$, $\phi^c$), ($B_{ij}$, $B^c_{ij}$) and $A^+_{ijkl}$
respectively.
\par
There is a connection between the decomposition (\ref{sl2r}) and the corresponding chains
in M--theory. The type IIB chain is given by eq.(\ref{duebchain}),
namely by
\begin{equation}
\mathrm{E}_{r+1(r+1)} \rightarrow \mathrm{SL}(2,\mathbb{R}) \otimes \mathrm{GL}(r,\mathbb{R})
\end{equation}
 while the $M$ theory is given by eq.(\ref{elechain}), namely by
 \begin{equation}
\mathrm{E}_{r+1} \rightarrow \mathrm{O(1,1)} \otimes \mathrm{SL}(r+1,\mathbb{R})
\end{equation}
coming from the moduli space of $T^{11-D} = T^{r+1}$.
We see that these decompositions involve the classical moduli spaces of $T^r$
 and of $T^{r+1}$ respectively.
Type $IIB$ and $M$ theory decompositions
become identical if we decompose further $\mathrm{SL}(r, \mathbb{R}) \rightarrow \mathrm{O(1,1)}
\times \mathrm{SL}(r-1,\mathbb{R})$ on the type $IIB$ side and
$\mathrm{SL}(r+1, \mathbb{R}) \rightarrow \mathrm{O(1,1)}
\otimes \mathrm{SL}(2,\mathbb{R}) \otimes \mathrm{SL}(r-1,\mathbb{R})$
on the $M$-theory side. Then we obtain for both theories
\begin{equation}
\mathrm{E}_{r+1} \rightarrow \mathrm{SL}(2,\mathbb{R}) \times \mathrm{O(1,1)}
\otimes \mathrm{O(1,1)} \otimes \mathrm{SL}(r-1,\mathbb{R}),
\label{sl2rall}
\end{equation}
and we see that the group $\mathrm{SL}(2,\mathbb{R})_U$ of type $IIB$ is identified with the
complex structure of the $2$-torus factor of the total
compactification torus $T^{11-D} \rightarrow T^2 \otimes T^{9-D}$.
\par
Note that according to  (\ref{splatto}) in 8 and 4 dimensions, ($r=2$ and $6$)
in the decomposition (\ref{sl2rall}) there is the following enhancement:
\begin{eqnarray}
&  \mathrm{SL}(2,\mathbb{R}) \times \mathrm{O(1,1)}  \rightarrow \mathrm{SL}(3,\mathbb{R}) \quad (\mbox{for} \, r=2,6) & \\
 &\left\{\matrix{\mathrm{O(1,1)} & \rightarrow & \mathrm{SL}(2,\mathbb{R})
 \quad (\mbox{for} \, r=2) \cr
\mathrm{SL}(5,\mathbb{R}) \times \mathrm{O(1,1)} & \rightarrow & \mathrm{SL}(6,
\mathbb{R}) \quad (\mbox{for} \, r=6) \cr}\right. &
\end{eqnarray}
Finally, by looking at fig.(\ref{wite5}) let us observe that
$\mathrm{E}_{7(7)}$ admits also a subgroup $\mathrm{SL}(2,\mathbb{R})_T$ $\otimes
(\mathrm{SO}(5,5)_S$ $\equiv \mathrm{E}_{5(5)})$ where the $\mathrm{SL}(2,\mathbb{R})$ factor is a
T--duality group, while the factor $(\mathrm{SO}(5,5)_S$ $\equiv \mathrm{E}_{5(5)})$
is an S--duality group which mixes RR and NS states.
\iffigs
\begin{figure}
\caption{}
\label{wite5}
\epsfxsize = 10cm
\epsffile{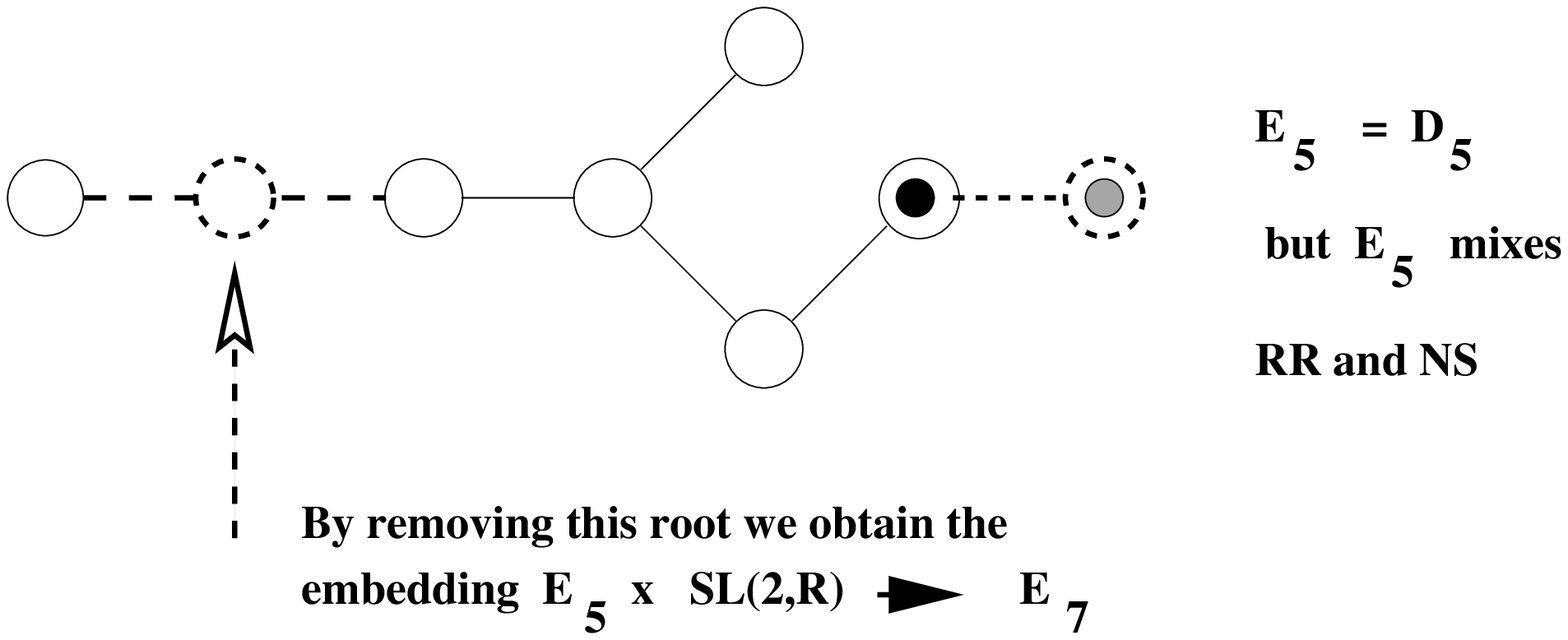}
\vskip -0.1cm
\unitlength=1mm
\end{figure}
\fi
%%%%%%%%%%%%%%%%%%%%%%%%%%%%%%%%%%%%%%%%%%%%%%%%%%%%%%%%%%%%%%%%%
%%%%%%%%%%%%%%%%%%%%%%%%%%%%%%%%%%%%%%%%%%%%%%%%%%%%%%%%%%%%%%%%%
\subsection[The maximal abelian ideals ${\cal A}_{r+1} \subset Solv_{r+1}$
of the solvable Lie algebra]{The maximal abelian ideals ${\cal A}_{r+1} \subset Solv_{r+1}$
of the solvable Lie algebra}
It is interesting to work out the maximal abelian ideals
${\cal A}_{r+1} \subset Solv_{r+1}$  of the
solvable Lie algebras generating the scalar manifolds
of maximal supergravity in dimension $D=10-r$.
The maximal abelian ideal of a solvable Lie
algebra is defined as the maximal subset of nilpotent generators  commuting among themselves.
\par
We derive ${\cal A}_{r+1}$ and we explore its relation with the
space of vector fields in one dimension above the dimension
we each time consider.  From such analysis we obtain a
filtration of the solvable Lie algebra which provides us
with a canonical polynomial parametrization of the supergravity
scalar coset manifold $\mathrm{U}_{r+1}/\mathrm{H}_{r+1}$
\par
%%%%%%%%%%%%%%%%%%%%%%%%%%%%%%%%%%%%%%%%%
\subsubsection{The maximal abelian ideal from an algebraic viewpoint}
Algebraically the maximal abelian ideal can be characterized by
looking at the decomposition of the U--duality algebra $\mathrm{E}_{r+1(r+1)}$ with
respect to the U--duality algebra  in one dimension above.
In other words we have to consider the decomposition of $\mathrm{E}_{r+1(r+1)}$ with
respect to the subalgebra $\mathrm{E}_{r (r )} \, \otimes \, \mathrm{O}(1,1)$. This
decomposition follows a general pattern which is given by the
next formula:
\begin{equation}
 \mbox{adj }  \mathrm{E}_{r+1(r+1)} \, = \, \mbox{adj }  \mathrm{E}_{r(r)}  \, \oplus
 \, \mbox{adj }  \mathrm{O}(1,1) \, \oplus ( \mathbb{D}^+_{r} \oplus \mathbb{D}^-_{r} )
 \label{genpat}
\end{equation}
where $\mathbb{D}^+_{r}$ is at the same time an irreducible representation of
the  U--duality algebra $\mathrm{E}_{r (r )}$ in $D+1$ dimensions and coincides with the
maximal abelian ideal
\begin{equation}
\mathbb{D}^+_{r}    \, \equiv \, {\cal A}_{r+1}   \, \subset \, Solv_{(r+1)}
\label{genmaxab}
\end{equation}
of the solvable  Lie algebra we are looking for. In eq. \eqn{genpat}
the subspace $\mathbb{D}^-_{r}$ is just a second identical copy of the representation
 $\mathbb{D}^{+}_{r}$ and it is made of negative rather than of positive weights
 of $\mathrm{E}_{r (r )}$. Furthermore   $\mathbb{D}^{+}_{r}$  and   $\mathbb{D}^-_{r}$
 correspond to the eigenspaces belonging respectively to the eigenvalues
 $\pm 1$ with respect to the adjoint action of the S--duality group
 $\mathrm{O}(1,1)$.
%%%%%%%%%%%%%%%%%%%%%%%%%%%%%%%%%%%%%%%%%%%%
\subsubsection{The maximal abelian ideal from a physical perspective:
the vector fields in one
dimension above and translational symmetries}
\label{abeideale}
Here, we would like to show that the dimension of the abelian
ideal in $D$ dimensions is equal to the number of vectors in dimensions $D+1$.
Denoting the number of compactified dimensions by $r$ (in string theory,
$r=10-D$), we will label the $U$-duality group in $D$ dimensions by $\mathrm{U}_D
= \mathrm{E}_{11-D} = \mathrm{E}_{r+1}$. The $T$-duality group is $\mathrm{O}(r,r)$, while the
$S$-duality group is $\mathrm{O}(1,1)$ in  dimensions higher than four, $SL(2,R)$
in $D=4$ (and it is inside $\mathrm{O}(8,8)$ in $D=3$).
\par
It follows from \eqn{genpat}
that the total dimension of the abelian ideal is given by
\begin{equation}{\rm dim} \, \mathcal{A}_{D} \,\equiv \,  {\rm dim} \, \mathcal{A}_{r+1}
 \,\equiv \,  {\rm dim} \,
\mathbb{D}_{r}
\label{abeliii}
\end{equation}where $\mathbb{D}_{r} $ is a representation of $\mathrm{U}_{D+1}$ pertaining to the vector fields.
 According to \eqn{genpat} we have (for $D \ge 4$):
\begin{equation}\mbox{adj } \mathrm{U}_D = \mbox{adj } \mathrm{U}_{D+1} \oplus {\bf 1} \oplus ({\bf 2}, \mathbb{D}_r).
\label{irrepu}
\end{equation}This is just an immediate consequence of the embedding chain \eqn{caten2}
which at the first level of iteration yields
$\mathrm{E}_{r+1} \rightarrow \mathrm{E}_r \times \mathrm{O}(1,1)$. For example, under
$\mathrm{E}_7 \rightarrow \mathrm{E}_6 \times \mathrm{O}(1,1)$ we have the branching rule:
${\rm adj} \, \mathrm{E}_7 = {\rm adj} \, \mathrm{E}_6 + {\bf 1} + ({\bf 2},{\bf 27})$ and
the abelian ideal is given by the ${\bf 27^+}$ representation of the $\mathrm{E}_{6(6)}$ group.
The $70$ scalars of the $D=4, N=8$ theory are
naturally decomposed as ${\bf 70} = {\bf 42} +{\bf 1} +{\bf 27^+}$.
To see the splitting of
the abelian ideal scalars into $NS$  and $RR$ sectors, one has to consider
the decomposition of $\mathrm{U}_{D+1}$ under the T--duality group $T_{D+1} = \mathrm{O}(r-1, r-1)$,
namely the second iteration of the embedding chain \eqn{caten2}: $\mathrm{E}_{r+1}
\rightarrow \mathrm{O}(1,1) \times \mathrm{O}(r-1,r-1)$. Then the vector
representation of $\mathrm{O}(r-1, r-1)$
gives  the $NS$ sector, while the  spinor representation yields  the $RR$
sector. The  example of $\mathrm{E}_7$ considered above is somewhat exceptional, since
we have ${\bf 27} \rightarrow ({\bf 10} + {\bf 1} +{\bf 16})$.
Here in addition to the expected ${\bf 10}$ and
${\bf 16}$ of $\mathrm{O}(5,5)$ we find an extra $NS$ scalar:  physically this is due
to the fact that in
four dimensions the two-index antisymmetric tensor field  $B_{\mu \nu}$
is dual to a scalar, algebraically this  generator is associated with the
exceptional root $ \sqrt {2} \epsilon_7$.
To summarize, the $NS$ and $RR$ sectors are separately  invariant under
$\mathrm{O}(r,r)$ in $D=10-r$ dimensions, while the abelian $NS$  and $RR$ sectors
are invariant under $\mathrm{O}(r-1, r-1)$. The standard parametrization of
the
$\mathrm{U}_D/\mathrm{H}_D$ and $\mathrm{U}_{D+1}/\mathrm{H}_{D+1}$ cosets gives a clear
illustration of this
fact:
\begin{equation}\frac{\mathrm{U}_D}{\mathrm{H}_D} \sim ( \frac{\mathrm{U}_{D+1}}{\mathrm{H}_{D+1}},
r_{D+1}, {\bf V}_r^{D+1}).
\label{cosetideal}
\end{equation}Here $r_{D+1}$ stands for the compactification radius, and ${\bf V}_r^{D+1} $
are the compactified vectors yielding the abelian ideal in $D$ dimensions.
\par
Note that:
\begin{equation}
  \mbox{adj}\, \mathrm{H}_D = \mbox{adj} \, \mathrm{H}_{D+1} \, +\,
\mbox{adj} \,\mbox{Irrep} \, \mathrm{U}_{D+1}
\end{equation}
so it appears  that the abelian ideal forms a representation not only of
$\mathrm{U}_{D+1}$ but also of the compact isotropy subgroup $\mathrm{H}_{D+1}$ of the scalar coset
manifold.
\par
In the  above $r=6$ example we find
${\rm adj} \, \mathrm{SU}(8) = {\rm adj} \, \mathrm{USp}(8) \oplus {\bf 27^-}$,
$ \Longrightarrow $ ${\bf 63}= {\bf 36}+ {\bf 27^-}$.
%%%%%%%%%%%%%%%%%%%%%%%%%%%%%%%%%%%%%%%%%%%%%%%%%%%%%%%%
%%%%%%%%%%%%%%%%%%%%%%%%%%%%%%%%%%%%%%%%%%%%%%%%%%%%%%%%%%%%%%%%%%%%%%
%%%%%%%%%%%%%%%%%%%%%%%%%%%%%%%%%%%%%%%%%%%%%%%%%%%%%%%%
%%%%%%%%%%%%%%%%%%%%%%%%%%%%%%%%%%%%%%%%%%%%%%%%%%%%%%%%%
\subsection[Roots and Weights and the fundamental representation of $\mathrm{E_{7(7)}}$]{Roots
and Weights and the fundamental representation of $\mathrm{E_{7(7)}}$}
\label{pesanti}
As an explicit illustration of the general ideas so far discussed
and in view of its further use in a relevant example of
\emph{gauging}, namely that of the $ \mathrm{N} = 8$ theory in four
dimensions, in this section we consider the explicit construction of
the  weights and roots  of $E_{7(7)}$ and their identification with
massless scalar fields produced in the toroidal compactification of the type IIA
superstring.
\par
We showed above that the $63$--dimensional
positive part $\Phi^+(E_7)$ of the $\mathrm{E_7}$ root space can be decomposed
as follows:
\begin{equation}
\Phi^+(E_7) = \mathbb{D}^+_1 \oplus \mathbb{D}^+_2  \oplus \mathbb{D}^+_3  \oplus \mathbb{D}^+_4
\oplus \mathbb{D}^+_5  \oplus \mathbb{D}^+_6
\label{filtro}
\end{equation}
where $\mathbb{D}^+_{r}$ are the maximal abelian ideals of the nested
U--duality algebras $\dots \subset E_{r(r)}\subset
E_{r+1(r+1)}\subset \dots$ in dimension $D=10-r$ ($\mathbb{D}^+_{r}$ being
the  ideal of $E_{r+1(r+1)}$). The dimensions of these abelian
ideals is:
\begin{equation}
\begin{array}{rclcrcl}
\mbox{dim}\,\mathbb{D}_1 &=& 1 &;& \mbox{dim}\,\mathbb{D}_2 &=&  3  \\
\mbox{dim}\,\mathbb{D}_3 &=& 6 &;& \mbox{dim}\,\mathbb{D}_4 &=&  10  \\
\mbox{dim}\,\mathbb{D}_5 &=& 16 &;& \mbox{dim}\,\mathbb{D}_6 &=& 27   \\
\end{array}
\label{didimens}
\end{equation}
The filtration (\ref{filtro}) provides a convenient way to
enumerate the $63$ positive roots which are
were associated in one--to--one way with the massless bosonic fields of
compactified string theory.
We name the roots as follows:
\begin{equation}
 {\vec \alpha }_{i,j} \, \in \, \mathbb{D}_i \quad ; \quad \cases { i=1,
 \dots , 6 \cr
 j=1,\dots , \mbox{dim} \, \mathbb{D}_i \cr }
 \label{filtroname}
\end{equation}
Each positive root can be decomposed along a basis of simple roots
$\alpha_\ell$ (i=1,\dots, 7):
\begin{equation}
{\vec \alpha }_{i,j} = n_{i,j}^\ell \, \alpha_\ell  \, \quad  n_{i,j}^\ell \in \mathbb{Z}
\end{equation}
The explicit correspondence between the roots and the fields of type
IIA supergravity in compact directions was derived in \cite{noi3} and is given in
table \ref{dideals}.
%\label{tabelle}
%%%%%%%%%%%%%%%%%%%%%%%%%%%%%%%%%%%%%%%%%%%%%%%%%%%%%%%%%%%%
% Abelian ideals in E7 filtration %%%%%%%%%%%%%%%%%%%%%%%%%%
%%%%%%%%%%%%%%%%%%%%%%%%%%%%%%%%%%%%%%%%%%%%%%%%%%%%%%%%%%%%
\begin{table}[ht]
\caption{{\bf
The abelian ideals $ \ID^{+}_{r}$ and the roots of $E_{7(7)}$}.
$A_{\widehat{\mu}}$, $A_{\widehat{\mu} \widehat{\nu} \widehat{\rho} }$
denote the Ramond--Ramond $1$--form and $3$--forms of
type IIA supergravity, while $g_{\widehat{\mu} \widehat{\nu} }$ and  $B_{\widehat{\mu}
\widehat{\nu} }$ are the metric and the
Neveu--Schwarz $B$--field, respectively. The toroidal compact directions have
been chosen to be $5,6,7,8,9,10$. The indices $\mu,\nu,..=0,1,2,3$ are instead
along the four non compact directions. The Dynkin labels are, by definition, the
components of the root in a basis of simple roots.}
\label{dideals}
\begin{center}
\begin{tabular}{||lcl|c|lcl||}
\hline
\hline
  Type IIA & Root & Dynkin & \null  &Type IIA & Root & Dynkin \\
  field    & name & labels & \null  &field    & name & labels \\
\hline
\null & \null & \null & $\ID^+_1$ & \null & \null & \null \\
$ A_{10}$ &  ${\vec \alpha}_{1,1}$ & $\{ 0,0,0,0,0,0,1\}$ & \null &
\null & \null & \null \\
\hline
\null & \null & \null & $\ID^+_2$ & \null & \null & \null \\
$B_{9,10}$ & ${\vec \alpha}_{2,1}$ &$\{ 0,0,0,0,0,1,0\}$ & \null &
$g_{9,10}$ & ${\vec \alpha}_{2,2}$& $\{ 0,0,0,0,1,0,0\}$ \\
$A_9$ & ${\vec \alpha}_{2,3}$ & $\{ 0,0,0,0,0,1,1\}$ & \null &
\null & \null & \null \\
\hline
\null & \null & \null & $\ID^+_3$ & \null & \null & \null \\
$B_{8,9}$ & ${\vec \alpha}_{3,1}$ & $\{ 0,0,0,1,1,1,0\}$ & \null &
$g_{8,9}$ & ${\vec \alpha}_{3,2}$ & $\{ 0,0,0,1,0,0,0\}$ \\
$B_{8,10} $ & ${\vec \alpha}_{3,3} $ & $\{ 0,0,0,1,0,1,0\} $ & \null &
$g_{8,10}$ & ${\vec \alpha}_{3,4} $ & $\{ 0,0,0,1,1,0,0\} $    \\
$A_8 $ & ${\vec \alpha}_{3,5} $ & $\{ 0,0,0,1,1,1,1\} $    & \null &
$A_{8,9,10} $ & ${\vec \alpha}_{3,6} $ & $\{ 0,0,0,1,0,1,1\} $ \\
\hline
\null & \null & \null & $\ID^+_4$ & \null & \null & \null \\
$B_{7,8} $ & ${\vec \alpha}_{4,1} $ & $ \{ 0,0,1,2,1,1,0\}  $ & \null &
$g_{7,8} $ & ${\vec \alpha}_{4,2} $ & $ \{ 0,0,1,0,0,0,0\}  $ \\
$B_{7,9} $ & ${\vec \alpha}_{4,3} $ & $ \{ 0,0,1,1,1,1,0\}  $ & \null &
$g_{7,9} $ & ${\vec \alpha}_{4,4} $ & $ \{ 0,0,1,1,0,0,0\}  $ \\
$B_{7,10} $ & ${\vec \alpha}_{4,5} $ & $ \{ 0,0,1,1,0,1,0\}  $ & \null &
$g_{7,10} $ & ${\vec \alpha}_{4,6} $ & $ \{ 0,0,1,1,1,0,0\}  $ \\
$A_{7,9,10}$ & ${\vec \alpha}_{4,7} $ & $ \{ 0,0,1,2,1,1,1\}  $ & \null &
$A_{7,8,10}$ & ${\vec \alpha}_{4,8} $ & $ \{ 0,0,1,1,1,1,1\}  $ \\
 $A_{7,8,9}$ & ${\vec \alpha}_{4,9} $ & $ \{ 0,0,1,1,0,1,1\}  $ & \null &
$A_{7} $ & ${\vec \alpha}_{4,10} $ & $ \{ 0,0,1,2,1,2,1\} $ \\
\hline
\null & \null & \null & $\ID^+_5$ & \null & \null & \null \\
$ B_{6,7}$ & ${\vec \alpha}_{5,1} $ & $ \{ 0,1,2,2,1,1,0\}  $ & \null &
$ g_{6,7}$ & ${\vec \alpha}_{5,2} $ & $ \{ 0,1,0,0,0,0,0\}  $ \\
$ B_{6,8}$ & ${\vec \alpha}_{5,3} $ & $ \{ 0,1,1,2,1,1,0\}  $ & \null &
$ g_{6,8}$ & ${\vec \alpha}_{5,4} $ & $ \{ 0,1,1,0,0,0,0\}  $ \\
$ B_{6,9}$ & ${\vec \alpha}_{5,5} $ & $ \{ 0,1,1,1,1,1,0\}  $ & \null &
$ g_{6,9}$ & ${\vec \alpha}_{5,6} $ & $ \{ 0,1,1,1,0,0,0\}  $ \\
$ B_{6,10}$ & ${\vec \alpha}_{5,7} $ & $ \{ 0,1,1,1,0,1,0\}  $ & \null &
$ g_{6,10}$ & ${\vec \alpha}_{5,8} $ & $ \{ 0,1,1,1,1,0,0\}  $ \\
$ A_{6,8,9}$ & ${\vec \alpha}_{5,9} $ & $ \{ 0,1,2,2,1,1,1\}  $ & \null &
$ A_{6,7,9}$ & ${\vec \alpha}_{5,10} $ & $ \{ 0,1,1,2,1,1,1\}  $ \\
$ A_{6,7,8}$ & ${\vec \alpha}_{5,11} $ & $ \{ 0,1,1,1,1,1,1\}  $ & \null &
$ A_{\mu\nu\rho}$ & ${\vec \alpha}_{5,12} $ & $ \{ 0,1,1,1,0,1,1\}  $ \\
$ A_{6,7,10}$ & ${\vec \alpha}_{5,13} $ & $ \{ 0,1,1,2,1,2,1\}  $ & \null &
$ A_{6,8,10}$ & ${\vec \alpha}_{5,14} $ & $ \{ 0,1,2,2,1,2,1\}  $ \\
$ A_{6,9,10}$ & ${\vec \alpha}_{5,15} $ & $ \{ 0,1,2,3,1,2,1\}  $ & \null &
$ A_6 $ & ${\vec \alpha}_{5,16} $ & $ \{ 0,1,2,3,2,2,1\}$ \\
\hline
\null & \null & \null & $\ID^+_6$ & \null & \null & \null \\
$ B_{5,6}$ & ${\vec \alpha}_{6,1} $ & $ \{ 1,2,2,2,1,1,0\}  $ & \null &
$ g_{5,6}$ & ${\vec \alpha}_{6,2} $ & $ \{ 1,0,0,0,0,0,0\}  $ \\
$ B_{5,7}$ & ${\vec \alpha}_{6,3} $ & $ \{ 1,1,2,2,1,1,0\}  $ & \null &
$ g_{5,7}$ & ${\vec \alpha}_{6,4} $ & $ \{ 1,1,0,0,0,0,0\}  $ \\
$ B_{5,8}$ & ${\vec \alpha}_{6,5} $ & $ \{ 1,1,1,2,1,1,0\}  $ & \null &
$ g_{5,8}$ & ${\vec \alpha}_{6,6} $ & $ \{ 1,1,1,0,0,0,0\}  $ \\
$ B_{5,9}$ & ${\vec \alpha}_{6,7} $ & $ \{ 1,1,1,1,1,1,0\}  $ & \null &
$ g_{5,9}$ & ${\vec \alpha}_{6,8} $ & $ \{ 1,1,1,1,0,0,0\}  $ \\
$ B_{5,10}$ & ${\vec \alpha}_{6,9} $ & $ \{ 1,1,1,1,0,1,0\}  $ & \null &
$ g_{5,10}$ & ${\vec \alpha}_{6,10} $ & $ \{ 1,1,1,1,1,0,0\}  $ \\
$ B_{\mu\nu}$ & ${\vec \alpha}_{6,11} $ & $ \{ 1,2,3,4,2,3,2\}  $ & \null &
$ A_5 $ & ${\vec \alpha}_{6,12} $ & $ \{ 1,2,3,4,2,3,1\}  $ \\
$ A_{\mu\nu 6}$ & ${\vec \alpha}_{6,13} $ & $ \{ 1,2,2,2,1,1,1\}  $ & \null &
$ A_{\mu\nu 7} $ & ${\vec \alpha}_{6,14} $ & $ \{ 1,1,2,2,1,1,1\}  $ \\
$ A_{\mu\nu 8} $ & ${\vec \alpha}_{6,15} $ & $ \{ 1,1,1,2,1,1,1\}  $ & \null &
$ A_{\mu\nu 9} $ & ${\vec \alpha}_{6,16} $ & $ \{ 1,1,1,1,1,1,1\}  $ \\
$ A_{\mu\nu 10} $ & ${\vec \alpha}_{6,17} $ & $ \{ 1,1,1,1,0,1,1\}  $ & \null &
$ A_{5,6,7} $ & $ {\vec \alpha}_{6,18} $ & $ \{ 1,1,1,2,1,2,1\}  $ \\
$ A_{5,6,8} $ & ${\vec \alpha}_{6,19} $ & $ \{ 1,1,2,2,1,2,1\}  $ & \null &
$ A_{5,6,9} $ & ${\vec \alpha}_{6,20} $ & $ \{ 1,1,2,3,1,2,1\}  $ \\
$ A_{5,6,10} $ & ${\vec \alpha}_{6,21} $ & $ \{ 1,1,2,3,2,2,1\}  $ & \null &
$ A_{5,7,8} $ & ${\vec \alpha}_{6,22} $ & $ \{ 1,2,2,2,1,2,1\}  $ \\
$ A_{5,7,9} $ & ${\vec \alpha}_{6,23} $ & $ \{ 1,2,2,3,1,2,1\}  $ & \null &
$ A_{5,7,10} $ & ${\vec \alpha}_{6,24} $ & $ \{ 1,2,2,3,2,2,1\}  $ \\
$ A_{5,8,9} $ & ${\vec \alpha}_{6,25} $ & $ \{ 1,2,3,3,1,2,1\}  $ & \null &
$ A_{5,8,10} $ & ${\vec \alpha}_{6,26} $ & $ \{ 1,2,3,3,2,2,1\}  $ \\
$ A_{5,9,10} $ & ${\vec \alpha}_{6,27} $ & $ \{ 1,2,3,4,2,2,1\}  $ & \null &
\null & \null & \null \\
\hline
\hline
\end{tabular}
\end{center}
\end{table}
%%%%%%%%%%%%%%%%%%%%%%%%%%%%%%%%%%%%%%%%%%%%%%%%%%%%%%%%%%%%%%%%
Each positive root can be decomposed along a basis of simple roots
$\alpha_\ell$ (i=1,\dots, 7):
\begin{equation}
{\vec \alpha }_{i,j} = n_{i,j}^\ell \, \alpha_\ell  \, \quad  n_{i,j}^\ell
\in \mathbb{Z}
\end{equation}
It turns out that as simple roots we can choose:
\begin{equation}
\begin{array}{rclcrclcrcl}
\alpha_1 & = & {\vec \alpha }_{6,2} & ; & \alpha_2 & = & {\vec \alpha }_{5,2}
& ; & \alpha_3 & = & {\vec \alpha }_{4,2} \\
\alpha_4 & = & {\vec \alpha }_{3,2} & ; & \alpha_5 & = & {\vec \alpha }_{2,2}
& ; & \alpha_6 & = & {\vec \alpha }_{2,1} \\
\alpha_7 & = & {\vec \alpha }_{1,1} & \null & \null & \null & \null
& \null & \null & \null & \null \\
\end{array}
\label{simplerut}
\end{equation}
Having fixed this basis, each root is intrinsically identified by its
Dynkin labels, namely by its integer valued components in the basis
\eqn{simplerut}. The next step for the construction of
fundamental representation  of $\mathrm{E_{7(7)}}$ embedded in the fundamental
$\mathrm{SpD}(56,\mathbb{R})$  is the knowledge of
the corresponding weight vectors ${\vec W}$.\par
A particularly relevant property of the maximally non--compact
real sections of a simple complex Lie algebra is that all of
its irreducible representations are real.  $\mathrm{E_{7(7)}}$ is the
maximally non compact real section of the complex Lie algebra $\mathrm{E_7}$, hence
all of its irreducible representations $\Gamma$  are real.
This implies that if an element of the  weight lattice ${\vec W} \, \in \, \Lambda_w$ is
a weight of a given irreducible representation
${\vec W}\in \Gamma$ then also its negative is a weight of the
same representation: $-{\vec W}\in \Gamma$. Indeed changing sign to
the weights corresponds to complex conjugation.
According to standard Lie algebra lore
every irreducible representation of a simple Lie algebra $\mathbb{G}$ is identified by a unique
{\it highest} weight ${\vec W}_{max}$. Furthermore all weights can be expressed as
integral non--negative linear  combinations of the {\it simple}
weights ${\vec W}_{\ell}\,(\ell=1,...,r=\mbox{rank}(\mathbb{G})) $.  The components of
 components of a weight in this basis are named its Dynkin labels.
The simple weights ${\vec W}_{i}$ of $\mathbb{G}$ are the generators of the
dual lattice to the root lattice and are defined by the condition:
\begin{equation}
\frac{2 ({\vec W}_{i}\, ,\, {\vec \alpha}_{j})}{({\vec \alpha}_{j}\, ,\,
{\vec \alpha}_{j})}=\delta_{ij}
\end{equation}
In the simply laced $\mathrm{E_{7(7)}}$ case, the previous equation simplifies as follows
\begin{equation}
({\vec W}_{i}\, ,\, {\vec \alpha}_{j})=\delta_{ij}
\label{simw}
\end{equation}
where ${\vec \alpha}_{j}$ are the the simple roots.
Using eq.(\ref{simplerut}), table \ref{dideals} and the Dynkin diagram of
$\mathrm{E_{7(7)}}$ (see fig.\ref{stande7}) from eq.(\ref{simw}) we can easily
obtain the explicit expression of the simple weights.
The Dynkin labels of the highest weight of an irreducible
representation $\Gamma$ give  the Dynkin labels of the
representation. Therefore the representation is usually denoted by
$\Gamma[n_1,...,n_{r}]$. All the weights ${\vec W}$ belonging
to the representation $\Gamma$ can be described by $r$ integer non--negative numbers
$q^\ell$ defined by the following equation:
\begin{equation}
{\vec W}_{max}-{\vec W}=\sum_{\ell=1}^{r}q^\ell{\vec \alpha}_{\ell}
\label{qi}
\end{equation}
where $\alpha_\ell$ are the simple roots.
\iffigs
\begin{figure}
\begin{center}
\caption{$E_7$ Dynkin diagram}
\label{stande7}
\epsfxsize = 10cm
\epsffile{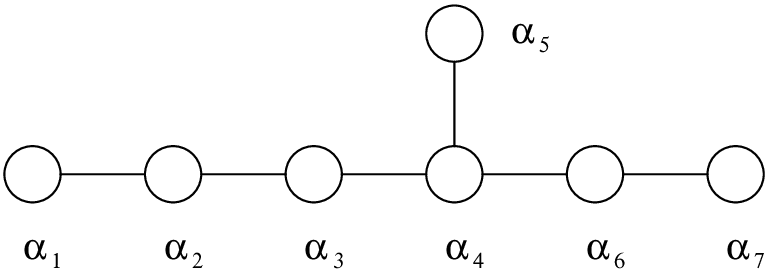}
\vskip -0.1cm
\unitlength=1mm
\end{center}
\end{figure}
\fi
According to this standard formalism the fundamental real
$\mathrm{SpD}(56,\mathbb{R})$ representation
of $\mathrm{E_{7(7)}}$ is $\Gamma[1,0,0,0,0,0,0]$
and the expression of its weights in terms of $q^\ell$ is given in table
\ref{e7weight}, the highest weight being ${\vec W}^{(51)}$.
\par
\begin{table}[ht]\caption{{\bf
Weights of the ${\bf 56}$ representation of $E_{7(7)} $:}}
\label{e7weight}
\begin{center}
\begin{tabular}{||cl|c|cl||}
\hline
\hline
    Weight & $q^\ell$  & \null & Weight & $q^\ell$ \\
      name & vector  & \null & name & vector \\
\hline
\null & \null & \null & \null & \null \\
$ {\vec W}^{(1)} \, =\, $ &$  \{ 2,3,4,5,3,3,1\} $  & \null &
$ {\vec W}^{(2)} \, =\, $ &$  \{ 2,2,2,2,1,1,1\} $   \\
$ {\vec W}^{(3)} \, =\, $ &$  \{ 1,2,2,2,1,1,1\} $  & \null &
$ {\vec W}^{(4)} \, =\, $ &$  \{ 1,1,2,2,1,1,1\} $    \\
$ {\vec W}^{(5)} \, =\, $ &$  \{ 1,1,1,2,1,1,1\} $  & \null &
$ {\vec W}^{(6)} \, =\, $ &$  \{ 1,1,1,1,1,1,1\} $    \\
$ {\vec W}^{(7)} \, =\, $ &$  \{ 2,3,3,3,1,2,1\} $  & \null &
$ {\vec W}^{(8)} \, =\, $ &$  \{ 2,2,3,3,1,2,1\} $    \\
$ {\vec W}^{(9)} \, =\, $ &$  \{ 2,2,2,3,1,2,1\} $  & \null &
$ {\vec W}^{(10)} \, =\, $ &$  \{ 2,2,2,2,1,2,1\} $   \\
$ {\vec W}^{(11)} \, =\, $ &$  \{ 1,2,2,2,1,2,1\} $  & \null &
$ {\vec W}^{(12)} \, =\, $ &$  \{ 1,1,2,2,1,2,1\} $    \\
$ {\vec W}^{(13)} \, =\, $ &$  \{ 1,1,1,2,1,2,1\} $  & \null &
$ {\vec W}^{(14)} \, =\, $ &$  \{ 1,2,2,3,1,2,1\} $    \\
$ {\vec W}^{(15)} \, =\, $ &$  \{ 1,2,3,3,1,2,1\} $  & \null &
$ {\vec W}^{(16)} \, =\, $ &$  \{ 1,1,2,3,1,2,1\} $    \\
$ {\vec W}^{(17)} \, =\, $ &$  \{ 2,2,2,2,1,1,0\} $  & \null &
$ {\vec W}^{(18)} \, =\, $ &$  \{ 1,2,2,2,1,1,0\} $    \\
$ {\vec W}^{(19)} \, =\, $ &$  \{ 1,1,2,2,1,1,0\} $  & \null &
$ {\vec W}^{(20)} \, =\, $ &$  \{ 1,1,1,2,1,1,0\} $    \\
$ {\vec W}^{(21)} \, =\, $ &$  \{ 1,1,1,1,1,1,0\} $  & \null &
$ {\vec W}^{(22)} \, =\, $ &$  \{ 1,1,1,1,1,0,0\} $    \\
$ {\vec W}^{(23)} \, =\, $ &$  \{ 3,4,5,6,3,4,2\} $  & \null &
$ {\vec W}^{(24)} \, =\, $ &$  \{ 2,4,5,6,3,4,2\} $    \\
$ {\vec W}^{(25)} \, =\, $ &$  \{ 2,3,5,6,3,4,2\} $  & \null &
$ {\vec W}^{(26)} \, =\, $ &$  \{ 2,3,4,6,3,4,2\} $    \\
$ {\vec W}^{(27)} \, =\, $ &$  \{ 2,3,4,5,3,4,2\} $  & \null &
$ {\vec W}^{(28)} \, =\, $ &$  \{ 2,3,4,5,3,3,2\} $    \\
$ {\vec W}^{(29)} \, =\, $ &$  \{ 1,1,1,1,0,1,1\} $  & \null &
$ {\vec W}^{(30)} \, =\, $ &$  \{ 1,2,3,4,2,3,1\} $    \\
$ {\vec W}^{(31)} \, =\, $ &$  \{ 2,2,3,4,2,3,1\} $  & \null &
$ {\vec W}^{(32)} \, =\, $ &$  \{ 2,3,3,4,2,3,1\} $    \\
$ {\vec W}^{(33)} \, =\, $ &$  \{ 2,3,4,4,2,3,1\} $  & \null &
$ {\vec W}^{(34)} \, =\, $ &$  \{ 2,3,4,5,2,3,1\} $    \\
$ {\vec W}^{(35)} \, =\, $ &$  \{ 1,1,2,3,2,2,1\} $  & \null &
$ {\vec W}^{(36)} \, =\, $ &$  \{ 1,2,2,3,2,2,1\} $   \\
$ {\vec W}^{(37)} \, =\, $ &$  \{ 1,2,3,3,2,2,1\} $  & \null &
$ {\vec W}^{(38)} \, =\, $ &$  \{ 1,2,3,4,2,2,1\} $    \\
$ {\vec W}^{(39)} \, =\, $ &$  \{ 2,2,3,4,2,2,1\} $  & \null &
$ {\vec W}^{(40)} \, =\, $ &$  \{ 2,3,3,4,2,2,1\} $   \\
$ {\vec W}^{(41)} \, =\, $ &$  \{ 2,3,4,4,2,2,1\} $  & \null &
$ {\vec W}^{(42)} \, =\, $ &$  \{ 2,2,3,3,2,2,1\} $    \\
$ {\vec W}^{(43)} \, =\, $ &$  \{ 2,2,2,3,2,2,1\} $  & \null &
$ {\vec W}^{(44)} \, =\, $ &$  \{ 2,3,3,3,2,2,1\} $    \\
$ {\vec W}^{(45)} \, =\, $ &$  \{ 1,2,3,4,2,3,2\} $  & \null &
$ {\vec W}^{(46)} \, =\, $ &$  \{ 2,2,3,4,2,3,2\} $    \\
$ {\vec W}^{(47)} \, =\, $ &$  \{ 2,3,3,4,2,3,2\} $  & \null &
$ {\vec W}^{(48)} \, =\, $ &$  \{ 2,3,4,4,2,3,2\} $    \\
$ {\vec W}^{(49)} \, =\, $ &$  \{ 2,3,4,5,2,3,2\} $  & \null &
$ {\vec W}^{(50)} \, =\, $ &$  \{ 2,3,4,5,2,4,2\} $    \\
$ {\vec W}^{(51)} \, =\, $ &$  \{ 0,0,0,0,0,0,0\} $  & \null &
$ {\vec W}^{(52)} \, =\, $ &$  \{ 1,0,0,0,0,0,0\} $    \\
$ {\vec W}^{(53)} \, =\, $ &$  \{ 1,1,0,0,0,0,0\} $  & \null &
$ {\vec W}^{(54)} \, =\, $ &$  \{ 1,1,1,0,0,0,0\} $   \\
$ {\vec W}^{(55)} \, =\, $ &$  \{ 1,1,1,1,0,0,0\} $  & \null &
$ {\vec W}^{(56)} \, =\, $ &$  \{ 1,1,1,1,0,1,0\} $   \\
%%%%%%%%%%%%%%%%%%%%%%%%%%%%%%%%%%%%%%%%%%%%%%%%%%%%%%%%%%%%
\hline
\end{tabular}
\end{center}
\end{table}
I can now explain the specific ordering of the weights I have
adopted.
\par
First of all I have separated the $56$ weights in two
groups of $28$ elements so that the first group:
\begin{equation}
{\vec \Lambda}^{(n)}={\vec W}^{(n)} \quad n=1,...,28
\label{elecweight}
\end{equation}
contains the weights of  the irreducible {\bf 28} dimensional representation of the
{\sl electric} subgroup {\bf $\mathrm{SL}(8,\mathbb{R}) \subset \mathrm{E_{7(7)}}$}.
The remaining group of $28$ weight vectors are the weights for the
transposed representation of the same group that I name ${\bf \bar{28}}$.
\par
Secondly the $28$ weights ${\vec \Lambda}$
have been arranged according to the decomposition with respect to the
T--{\it duality} subalgebra $\mathrm{SO(6,6)}\subset \mathrm{E_7(7)}$:  the first $16$
correspond to R--R vectors and are the weights of the spinor
representation of $\mathrm{SO(6,6)}$ while the last $12$ are associated with N--S fields
and correspond to the weights of the vector representation of $\mathrm{SO(6,6)}$.
\subsubsection[Matrices of the $E_{7(7)}$ algebra in the fundamental
representation]{Matrices of the $E_{7(7)}$ algebra in the fundamental
representation}
Equipped with the weight vectors we can  proceed to the explicit
construction of the fundamental  representation
of $\mathrm{E_{7(7)}}$. The basis vectors
are the $56$ weights, according to the enumeration of table
\ref{e7weight} and
what we look for are
\begin{enumerate}
  \item the $56\times 56$ matrices associated with the  $7$
Cartan generators $H_{{\vec \alpha}_i}$ ($i=1,\dots , 7$)
  \item the $56\times 56$ matrices associated with the $126$ step operators $E^{\vec \alpha}$
  \end{enumerate}
 Both are expressed in this basis   by:
\begin{eqnarray}
\left[ \mathrm{SpD}_{56}\left( H_{{\vec \alpha}_i} \right)\right]_{nm}&  \equiv &  \langle
{\vec W}^{(n)} \vert \,  H_{ {\vec \alpha}_i } \,\vert {\vec W}^{(m)}
\rangle \nonumber\\
\left[ \mathrm{SpD}_{56}\left( E^{ {\vec \alpha} } \right)\right]_{nm}&  \equiv &  \langle
{\vec W}^{(n)} \vert \,  E^{ {\vec \alpha} }  \,\vert {\vec W}^{(m)}
\rangle
\label{sp56defmat}
\end{eqnarray}
Let us begin with the Cartan generators. As a basis of the
Cartan subalgebra we use the generators $H_{\vec \alpha_i}$ defined by the
commutators:
\begin{equation}
\left[ E^{{\vec \alpha }_i}, E^{-{\vec \alpha }_i} \right] \, \equiv
\, H_{\vec \alpha_i}
\label{cartbadefi}
\end{equation}
In the $\mathrm{SpD(56)}$ representation the corresponding matrices are
diagonal and of the form:
\begin{equation}
\langle {\vec W}^{(p)} \vert\, H_{\vec \alpha_{i}}\, \vert
{\vec W}^{(q)}\rangle\, =\,\left({\vec W}^{(p)},{\vec \alpha_{i}}\right)
\delta_{p\, q}\quad ; \quad ( p,q\, =\, 1,...,56)
\label{cartane7}
\end{equation}
Next we construct the matrices associated with the step operators. The first
observation is that it suffices to consider the positive roots. Since
the representation is real the matrix associated with the
negative of a root is just the transposed of that associated with the
root itself:
\begin{equation}
E^{-\alpha} = \left[ E^\alpha \right]^T \, \leftrightarrow \, \langle
{\vec W}^{(n)} \vert \,  E^{- {\vec \alpha} }  \,\vert {\vec W}^{(m)} \rangle \, = \,
\langle {\vec W}^{(m)} \vert \,  E^{ {\vec \alpha} }  \,\vert {\vec W}^{(n)} \rangle
\label{transpopro}
\end{equation}
The method to obtain the matrices for all the
positive roots is that of constructing first the $56\times 56$
matrices for the step operators $E^{\vec \alpha_{\ell}}\, (\ell=1,...,7)$
associated with the simple roots and then generating all the others
through their commutators. The construction rules for the $\mathrm{SpD}(56)$
representation of  the six operators $E^{\alpha_{\ell}} \, (\ell\neq 5)$
are:
\begin{equation}
\ell \, \neq \, 5 \quad \,
\Biggl \{ \matrix {\langle {\vec W}^{(n)} \vert\, E^{\vec \alpha_{\ell}}\, \vert
{\vec W}^{(m)}\rangle & = & \delta_{{\vec W}^{(n)},
{\vec W}^{(m)}+{\vec \alpha}_\ell} &;& n,m=1,\dots, 28 \cr
\langle {\vec W}^{(n+28)} \vert\, E^{\vec \alpha_{\ell}}\, \vert
{\vec W}^{(m+28)}\rangle & = & -\delta_{{\vec W}^{(n+28)},
{\vec W}^{(m+28)}+{\vec \alpha}_\ell} & ; & n,m=1,\dots, 28 \cr }
\label{repineq5}
\end{equation}
The six simple roots ${\vec \alpha_{\ell}} $ with $ \ell \neq 5$
belong also to the the Dynkin diagram of the electric subgroup
$\mathrm{SL}(8,\mathbb{R})$ (see fig.\ref{elecal}).  Thus their shift
operators have a block diagonal action on the {\bf 28} and ${\bf \bar{28}}$
subspaces of the $\mathrm{SpD}(56)$ representation that are irreducible
under the electric subgroup. Indeed from eq.(\ref{repineq5}) we conclude
that:
\begin{equation}
\ell \, \neq \, 5 \quad \,\mathrm{SpD}_{56}\left(E^{{\vec \alpha}_\ell} \right)=
\left(\matrix { A[{{\vec \alpha}_\ell}]
& {\bf 0} \cr {\bf 0} & - A^T[{{\vec \alpha}_\ell}] \cr} \right)
\label{spdno5}
\end{equation}
the $28 \times 28$ block  $A[{{\vec \alpha}_\ell}]$ being defined
by the first line of eq.(\ref{repineq5}).\par
On the other hand the operator $E^{\vec \alpha_{5}}$, corresponding to the only
root of the $\mathrm{E_7}$ Dynkin diagram that is not also part of the $\mathrm{A_7}$
diagram is represented by a matrix whose non--vanishing $28\times 28$ blocks
are off--diagonal. We have
\begin{equation}
\mathrm{SpD}_{56}\left(E^{{\vec \alpha}_5} \right)=
\left(\matrix { {\bf 0} & B[{{\vec \alpha}_5}]
\cr  C[{{\vec \alpha}_5}] & {\bf 0} \cr} \right)
\label{spdyes5}
\end{equation}
where both $B[{{\vec \alpha}_5}]=B^T[{{\vec \alpha}_5}]$ and
$C[{{\vec \alpha}_5}]=C^T[{{\vec \alpha}_5}]$ are symmetric $28
\times 28$ matrices. More explicitly the  matrix
 $SpD_{56}\left(E^{{\vec \alpha}_5} \right)$
is given by:
\begin{eqnarray}
&& \langle {\vec W}^{(n)} \vert\, E^{\vec \alpha_{5}}\, \vert
{\vec W}^{(m+28)}\rangle \, =\,  \langle {\vec W}^{(m)} \vert\,
E^{\vec \alpha_{5}}\, \vert {\vec W}^{(n+28)}\rangle \nonumber \\
&& \langle {\vec W}^{(n+28)} \vert\, E^{\vec \alpha_{5}}\, \vert
{\vec W}^{(m)}\rangle \, =\,  \langle {\vec W}^{(m+28)} \vert\,
E^{\vec \alpha_{5}}\, \vert {\vec W}^{(n)}\rangle
\label{sim5}
\end{eqnarray}
with
\begin{equation}
\begin{array}{rcrcrcr}
  \langle {\vec W}^{(7)} \vert\, E^{\vec \alpha_{5}}\, \vert
{\vec W}^{(44)}\rangle & =& -1 & \null &
  \langle {\vec W}^{(8)} \vert\, E^{\vec \alpha_{5}}\, \vert
{\vec W}^{(42)}\rangle & = & 1  \\
   \langle {\vec W}^{(9)} \vert\, E^{\vec \alpha_{5}}\, \vert
{\vec W}^{(43)}\rangle & = & -1   & \null &
   \langle {\vec W}^{(14)} \vert\, E^{\vec \alpha_{5}}\, \vert
{\vec W}^{(36)}\rangle & = & 1 \\
   \langle {\vec W}^{(15)} \vert\, E^{\vec \alpha_{5}}\, \vert
{\vec W}^{(37)}\rangle & = & -1 & \null &
 \langle {\vec W}^{(16)} \vert\, E^{\vec \alpha_{5}}\, \vert
{\vec W}^{(35)}\rangle & = & -1  \\
   \langle {\vec W}^{(29)} \vert\, E^{\vec \alpha_{5}}\, \vert
{\vec W}^{(6)}\rangle  & = & -1 & \null &
   \langle {\vec W}^{(34)} \vert\, E^{\vec \alpha_{5}}\, \vert
{\vec W}^{(1)}\rangle  & = & -1 \\
  \langle {\vec W}^{(49)} \vert\, E^{\vec \alpha_{5}}\, \vert
{\vec W}^{(28)}\rangle & = & 1  &\null &
  \langle {\vec W}^{(50)} \vert\, E^{\vec \alpha_{5}}\, \vert
{\vec W}^{(27)}\rangle & = & -1 \\
   \langle {\vec W}^{(55)} \vert\, E^{\vec \alpha_{5}}\, \vert
{\vec W}^{(22)}\rangle & = &  -1 & \null &
   \langle {\vec W}^{(56)} \vert\, E^{\vec \alpha_{5}}\, \vert
{\vec W}^{(21)}\rangle & =& 1 \\
\end{array}
\label{sim5bis}
\end{equation}
In this way we have completed the construction of the $E^{{\vec \alpha}_\ell}$
operators associated with simple roots. For the matrices associated
with higher roots we just proceed iteratively in the following way.
As usual we can organize the roots by their height :
\begin{equation}
{\vec \alpha}=n^\ell \, {\vec \alpha}_\ell \quad \rightarrow
\quad \mbox{ht}\,{\vec \alpha} \, = \, \sum_{\ell=1}^{7} n^\ell
\label{altezza}
\end{equation}
and for the roots $\alpha_i + \alpha_j$ of height $\mbox{ht}=2$ we
set:
\begin{equation}
\mathrm{SpD}_{56} \left( E^{ \alpha _i + \alpha _j} \right) \equiv \left[
\mathrm{SpD}_{56}\left(E^{\alpha _i} \right) \, , \,
\mathrm{SpD}_{56}\left(E^{\alpha _i} \right) \right] \quad ; \quad i<j
\label{alto2}
\end{equation}
Next for the roots of $\mbox{ht}=3$ that can be written as $\alpha_i
+ \beta $ where $\alpha_i$ is simple and $\mbox{ht}\, \beta\, =\, 2$
we can write:
\begin{equation}
\mathrm{SpD}_{56} \left( E^{ \alpha _i + \beta} \right) \equiv \left[
\mathrm{SpD}_{56}\left(E^{\alpha _i} \right) \, , \,
\mathrm{SpD}_{56}\left(E^{\beta} \right) \right]
\label{alto3}
\end{equation}
Obtained the matrices for the roots of $\mbox{ht}=3$ one proceeds in
a similar way for those of the next height and so on up to exhaustion
of all the $63$ positive roots.
%%%%%%%%%%%%%%%%%%%%%%%%%%%%%%%%%%%%%%%%%%%%%%%%%%
%% Matteus %%%%%%%%%%%%%%%%%%%%%%%%%%%%%%%%%%%%%%%
%%%%%%%%%%%%%%%%%%%%%%%%%%%%%%%%%%%%%%%%%%%%%%%%%%
\section{Completing the type IIA versus type IIB algebraic correspondence}
%%%%%%%%%%%%%%%%%%%%%%%%%%%%%%%%%%%%%%%%%%%%%%%%%%%%%%%%%%%%%%%%%%%%%%%%%
Following the results of Bertolini and Trigiante \cite{Bertolini:2000uz}
we can now make the algebraic embedding of type IIA and type IIB states in
maximal supergravities completely systematic and handy.
%%%%%%%%%%%%%%%%%%%%%%%%%%%%%%%%%%%%%%%%%%%%%%%%%%%%%%%%%%%%%%%%%%
\par
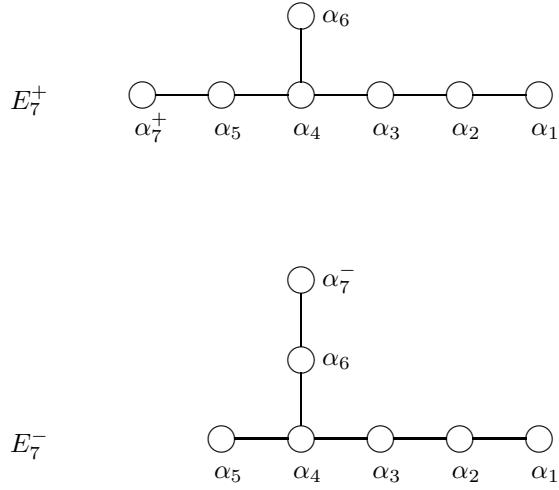
\begin{figure}
\centering
\begin{picture}(100,200)
%%%%%%%%%%%%%%%%%%%%%%%%%%
%%%%%%%%%%%%%%%%%%%%%%%%%%%%%%%%%
\put (-70,165){$E_7^+$} \put (-20,170){\circle {10}} \put
(-23,155){$\alpha_7^+$} \put (-15,170){\line (1,0){20}} \put
(10,170){\circle {10}} \put (7,155){$\alpha_5$} \put (15,170){\line
(1,0){20}} \put (40,170){\circle {10}} \put (37,155){$\alpha_4$} \put
(40,200){\circle {10}} \put (48,197.8){$\alpha_6$} \put
(40,175){\line (0,1){20}} \put (45,170){\line (1,0){20}} \put
(70,170){\circle {10}} \put (67,155){$\alpha_{3}$} \put
(75,170){\line (1,0){20}} \put (100,170){\circle {10}} \put
(97,155){$\alpha_{2}$} \put (105,170){\line (1,0){20}} \put
(130,170){\circle {10}} \put (127,155){$\alpha_1$}
%%%%%%%%%%%%%%%%%%%%%%%%%%
%%%%%%%%%%%%%%%%%%%%%%%%%%%%%%%%%
%%%%%%%%%%%%%%%%%%%%%%%%%%%%%%%%%
\put (-70,35){$E_7^-$}  \put
(10,40){\circle {10}} \put (7,25){$\alpha_5$} \put (15,40){\line
(1,0){20}} \put (40,40){\circle {10}} \put (37,25){$\alpha_4$} \put
(40,70){\circle {10}} \put (48,67.8){$\alpha_6$}
\put (40,75){\line(0,1){20}}\put
(40,100){\circle {10}} \put (48,97.8){$\alpha_7^-$}
 \put (40,45){\line
(0,1){20}} \put (45,40){\line (1,0){20}} \put (70,40){\circle {10}}
\put (67,25){$\alpha_{3}$} \put (75,40){\line (1,0){20}} \put
(100,40){\circle {10}} \put (97,25){$\alpha_{2}$} \put (105,40){\line
(1,0){20}} \put (130,40){\circle {10}} \put (127,25){$\alpha_1$}
%%%%%%%%%%%%%%%%%%%%%%%%%%
%%%%%%%%%%%%%%%%%%%%%%%%%%
\end{picture}
\vskip 0.1cm \caption{The two realizations of the $E_{7(7)}$ algebra corresponding to the
the type IIA and type IIB superstring respectively} \label{except}
\end{figure}

From an algebraic point of view,
the Dynkin diagram of  $\mathrm{E_{7(7)}} $ is
constructed by adding to the $D_6$ Dynkin diagram,  that consists of the
simple roots $(\alpha_i)_{i=1,\dots,6}$, the highest weight
$\alpha_7^\pm$ of one of the two spinorial representations ${\bf 32}^\pm$
of $\mathrm{SO(12)}$. The crucial observation in \cite{Bertolini:2000uz}
is that the choice of this chirality is what distinguishes (at the compactified level)
type IIA from the type IIB superstrings. Therefore it is convenient
the adopt an explicit realization of the $E_{7(7)}$ simple
roots which is slightly different from that used in
eq.(\ref{e7simple}).
Let us name $\mathrm{E_{7(7)}}^+$ the algebra obtained by
attaching $\alpha_7^+$ to  $\alpha_5$ and  $\mathrm{E_{7(7)}}^-$ the algebra
obtained by attaching $\alpha_7^-$  to $\alpha_6$.
We can say that in $D=4$ the $70$ scalars of the $\mathcal{N}=8$ theory
parametrize the $Solv(\mathrm{{E}_{7(7)}}^+)$ coset when we compactify
type IIA $D=10$ supergravity on a six torus $T^6$  and that they parametrize
the $Solv(\mathrm{{E}_{7(7)}}^-)$ coset
when  it is the type IIB theory that is compactified on $T^6$.
Writing the roots in an orthonormal
basis $\epsilon_n$ we have:
\begin{eqnarray}
\alpha_1\,&=&\,\epsilon_1-\epsilon_2\,\,;\,\,\alpha_2\,=\,\epsilon_2-\epsilon_3\,\,;
\,\,\alpha_3\,=\,\epsilon_3-\epsilon_4\nonumber\\
\alpha_4\,&=&\,\epsilon_4-\epsilon_5\,\,;\,\,\alpha_5\,=\,\epsilon_5-\epsilon_6\,\,;
\,\,\alpha_6\,=\,\epsilon_5+\epsilon_6\nonumber\\ \alpha_7^\pm\,&=&\,-
\frac{1}{2}\left(\epsilon_1+\epsilon_2+\epsilon_3+\epsilon_4+
\epsilon_5\mp\epsilon_6\right)+\frac{\sqrt{2}}{2}\epsilon_7
\label{dynke7pm}
\end{eqnarray}
and the corresponding Dynkin diagrams are displayed in fig. \ref{except}
Depending on the two interpretations we can give two
different $\mathrm{SLA}$ descriptions of the scalar manifold which are consistent
with the geometric characterization of $T$--duality as described below.
\par
According to our previous general discussion,
the relevant SLA is generated by the non--compact Cartan generators,
which in the case of maximal supergravities means all of them,
and the shift operators corresponding to roots with positive restriction on the
non--compact Cartan generators. For the common NS--NS sector of type IIA and type IIB, a
suitable basis of non--compact Cartan generators is parametrized
by the radii of the torus and by the ten dimensional dilaton:
\begin{eqnarray}
{\cal C}_K ({\rm IIB})&=&\sum_{i=1}^6-\sigma_i
H_{\epsilon_i+\frac{\epsilon_7} {\sqrt{2}}} +\phi
H_{\sqrt{2}\epsilon_7}\nonumber\\ {\cal C}_K ({\rm
IIA})&=&\sum_{i=1}^5-\sigma_i H_{\epsilon_i+\frac{\epsilon_7}
{\sqrt{2}}} -\sigma_6 H_{-\epsilon_6+\frac{\epsilon_7}{\sqrt{2}}}+
\phi H_{\sqrt{2}\epsilon_7}\nonumber\\ \sigma_i&=& {\rm
ln}(\rho_{i+3})
\label{strange}
\end{eqnarray}
where $\rho_k$ ($k=4,\dots ,9$) are the radii of the internal torus
in the  $x^k$ directions.
In the expressions (\ref{strange})  the overall coefficient of
$H_{\sqrt{2}\epsilon_7}$  is the four dimensional dilaton:
\begin{eqnarray}
\phi_4&=& \phi-\frac{1}{2}\sum_{i=1}^6\sigma_i=\phi-\frac{1}{4}
{\rm ln}\left({\rm det}(G_{ij})\right)
\end{eqnarray}
The  non--orthonormal basis in (\ref{strange}) is defined by
the decomposition of  the $\mathrm{U}$--duality group in $D$ dimensions with
respect to the $\mathrm{U}$--duality group in $D+1$  which I have
already illustrated in eq.(\ref{caten2}). Explicitly we have
\begin{eqnarray}
\mathrm{{E}_{r+1(r+1)}}&\rightarrow & \mathrm{O(1,1)}_r+\mathrm{{
E}_{r(r)}}\,\,\,(r=10-d)
\end{eqnarray}
where $\mathrm{{E}_{r(r)}}$ is obtained by deleting the extreme root in
the Dynkin diagram  of $\mathrm{{E}_{r+1(r+1)}}$ (on the branch of
$\alpha_1,\,\alpha_2,\dots$) and substituting  it with the Cartan
generator $\mathrm{O(1,1)}_r$ orthogonal to the rest of the Dynkin diagram.
These $\mathrm{O(1,1)}_r$ define the basis in (\ref{strange}) and are naturally
parametrized by  $-\sigma_{7-r}$.
\par
The remaining NS--NS fields  are
parameterized by the positive roots of $\mathrm{SL}(2,\mathbb{R})\times
\mathrm{SO(6,6)}$. According to the definition of the ordering relation among
the roots with respect to the orthonormal basis $(H_{\epsilon_i})$,
which determines who contributes to the SLA and who does not, one can
find different but equivalent SLAs, usually related by some automorphism of
the $D_6$ subalgebra. It is natural to associate  the fields $G_{ij}$  ($i\neq
j$) coming from the
metric with the roots $\pm (\epsilon_i-\epsilon_j)$ and  the fields $B_{ij}$ coming
from the torsion with
the roots $\pm (\epsilon_i+\epsilon_j)$.  Indeed in the fundamental ${\bf 12}$--dimensional
representation of $\mathrm{SO(6,6)}$ in which the generators have the form:
\begin{eqnarray}
\left(M_{\Lambda\Sigma}\right)_\Delta^{\phantom{\delta}\Gamma}&=&\eta_{\Lambda\Delta}
\delta_\Sigma^\Gamma-\eta_{\Sigma\Delta}\delta_\Lambda^\Gamma\nonumber\\
\eta_{\Lambda\Delta}&=& {\rm diag}(++++++------)\nonumber\\
H_{\epsilon_i}&=& M_{ii+6}
\end{eqnarray}
the shift operators corresponding to the former set of roots have a symmetric
$6\times 6$ off--diagonal block, while those corresponding to the
latter set of roots have an antisymmetric off--diagonal block. Using a
lexicographic ordering with respect to  the basis $(H_{\epsilon_i})$
the roots contributing to the SLA and the corresponding scalar fields
are listed in table \ref{table2},\ref{table2b}.
\par
 As already pointed out, the R--R fields correspond to the
shift operators associated with the roots that are spinor weights in the
${\bf 32}^\pm$ of $\mathrm{SO(6,6)}$. These roots are:
\begin{eqnarray}
{\bf 32}^+ :\,\,\alpha^+ &=&-\frac{1}{2}\left(\pm \epsilon_1\pm
\epsilon_2\pm \epsilon_3\pm \epsilon_4\pm \epsilon_5\pm
\epsilon_6\right)+\frac{\sqrt{2}}{2}\epsilon_7\nonumber\\ &&
(\mbox{odd number of ``$+$'' signs within brackets})\nonumber\\ {\bf
32}^- :\,\,\alpha^- &=&-\frac{1}{2}\left(\pm \epsilon_1\pm
\epsilon_2\pm \epsilon_3\pm \epsilon_4\pm \epsilon_5\pm
\epsilon_6\right)+\frac{\sqrt{2}}{2}\epsilon_7\nonumber\\
&&(\mbox{even number of ``$+$'' signs within brackets})
\end{eqnarray}
Indeed the {\it chirality} operator $\gamma$ is easily computed in
terms of the product of the Cartan  generators
$(H_{\epsilon_i})_{i=1,\dots,6}$ in the spinorial representation
($(S(H_{\epsilon_i}))_{i=1,\dots,6}$):
\begin{eqnarray}
\{\gamma_\Lambda,\gamma_\Sigma\}&=&2 \eta_{\Lambda\Sigma}\nonumber\\
S(H_{\epsilon_i}) &\equiv &
\gamma_i\gamma_{i+6}\,,\,\,(i,1,\dots,6)\nonumber\\ \gamma &=&
\gamma_1\gamma_2\cdots \gamma_{12}=-
S(H_{\epsilon_1})S(H_{\epsilon_2})\cdots S(H_{\epsilon_6})
\end{eqnarray}
and we can easily verify that $\gamma$ is positive on the $\alpha^+$ and
negative on the $\alpha^-$. A precise correspondence between the
spinorial roots and scalar fields from type IIA and type IIB theories
is again given in table \ref{table2},\ref{table2b}.
\subsection[$ S\times T$ duality made precise]{$ S\times T$ duality made precise}
In sections \ref{dynsect1} and \ref{dynsect2} I have already
explained how  certain subalgebras  sequentially embedded in
$E_{r+1(r+1)}$ can be interpreted as the  subalgebras of strong/weak
and target space dualities at each step of the sequential toroidal
compactification. It still remained to be seen how the target space
dualities that connect the type IIA to the type IIB theory are
algebraically realized. Following the set up of Bertolini and
Trigiante  outlined above this can be done in a very neat
way.
\par
Their crucial observation is that we can characterize the effect  of an $\mathrm{S}$
or $\mathrm{T}$--duality on a supergravity  $p$--brane solution  as the action
of an element of the automorphism group $\mathrm{Aut}(\mathrm{S\times T})$
on the SLA that generates the scalar manifold.
\par
 Automorphisms of a semisimple Lie algebra $\mathbb{G}$ are isomorphisms
of the algebra on  itself and can be {\it inner} if their action can
be expressed as a conjugation of the algebra by means of a group
element generated by the algebra itself, or {\it outer} if they do not
admit such a representation (see for instance \cite{jacob})
 A generic automorphism may be reduced, through
the composition with a suitable (nilpotent) inner automorphism,
to an isometric mapping which leaves the Cartan subalgebra stable.
Let us focus on the latter kind of transformations, which we denote by $\psi_\tau$.
It is proved in mathematical textbooks that the restriction of the group $\{\psi_\tau\}$
to the Cartan subalgebra is isomorphic to the automorphism group
of the root space $\Delta$. This latter consists of the transformations $\tau$ on the weight
lattice that leave the Cartan--Killing matrix invariant ({\it rotations}).
It can also be shown that the inner automorphisms $\psi_\tau$ correspond to
$\tau$.s that belong to the Weyl group $Weyl(\mathbb{G})$ while in the case of
an outer automorphism
$\psi_\tau$, the element $\tau$ can be reduced,
 modulo Weyl transformations, to symmetries of the Dynkin diagram (permutations
 of the simple roots).\par
Conversely, given a rotation $\tau$ acting on the root basis $\Delta$,
one may associate to it an
automorphism $\tilde{\psi}_\tau$ of the whole Lie algebra
$\mathbb{G}$ whose action on its canonical basis reads as follows:
\begin{eqnarray}
\tilde{\psi}_\tau(H_\beta)&=&H_{\tau(\beta)}\,\,;\,\,\tilde{\psi}_\tau(E_\alpha)
 \propto E_{\tau(\alpha)}\nonumber\\
\alpha,\beta \,&&\mbox{roots}
\label{simplerecipe}
\end{eqnarray}
A general $\psi_\tau$ has the form $\psi_\tau=\tilde{\psi}_\tau \cdot \omega$,
where $\omega$ is an automorphism leaving the Cartan subalgebra
$ \mathcal{H} \subset \mathbb{G}$ pointwise fixed.
These automorphisms are all inner \cite{jacob}.
 \par
As an example we can focus on the $D=4$ case where the $\mathrm{S \times T}$
duality algebra is $\mathbb{G}=\mathrm{SL}(2,\mathbb{R})\times
\mathrm{SO(6,6)}$, yet it will be clear from our discussion how the
argument can be applied to higher dimensions and hence to
$\mathbb{G}=\mathrm{O(1,1)}\times
\mathrm{SO(r,r)}$. For the choice $\mathbb{G}=\mathrm{SL}(2,\mathbb{R})\times
\mathrm{SO(6,6)}$
the rotation corresponding to an outer
automorphism $\psi_\tau$ can be reduced (modulo Weyl transformations)
to the only symmetry transformation of the $D_6$ Dynkin diagram, i.e.
\begin{equation}
  \alpha_5\leftrightarrow
\alpha_6 \quad \quad \mbox{or equivalently} \quad \quad \epsilon_6\rightarrow -\epsilon_6
\label{D6auto}
\end{equation}
In particular it can be shown \cite{jacob} that rotations on the root space
amounting to a \textbf{change of sign of an odd number} of $\epsilon_i$ ($i=1,\dots, 6$)
 define \textbf{outer automorphisms}. Since the automorphisms preserve algebraic
structures, they will map  solvable subalgebras into solvable
subalgebras. Of course we do not expect  all the automorphisms of $\mathrm{S\times T}$
to be automorphisms of $\mathrm{E_{7(7)}}$ since, for instance, the Dynkin diagram of
the latter does not have any symmetry. Indeed it is easy to check that
outer automorphisms of $\mathrm{SO(6,6)}$ map $\mathrm{E_{7(7)}}^\pm$ into
$\mathrm{{E}_{7(7)}}^\mp$ (this follows from the fact that  changing sign to an
odd number of $\epsilon_i$ maps $\alpha^\pm$ into $\alpha^\mp$).
\par
These observation lead to a very neat algebraic characterization of a
\emph{large radius $\leftrightarrow $  small radius}
\textbf{T--duality transformation}  along a
compact direction $x^k$ ($k=4,\dots,9$).  It is simply the action of an \textbf{outer
automorphism} $\psi_\tau$ corresponding to
\begin{equation}
  \tau:\,\,\epsilon_{k-3}\rightarrow -\epsilon_{k-3}
\label{Tdualalg}
\end{equation}
Similarly a \emph{strong coupling) $\leftrightarrow $ weak coupling}
\textbf{S--duality}  is an automorphism
  $\psi_\tau$ such that
\begin{equation}
  \tau:\,\,\epsilon_7\rightarrow -\epsilon_7
\label{Sdualalg}
\end{equation}
For example let us consider   $\mathrm{T}$--duality transformation along the
direction  $x^9$ and its effect on $\rho_9$ ($\sigma_6$) and
the dilaton $\phi$ starting from type IIB fields. According to the
algebraic recipe given  above we have:
\begin{eqnarray}
\left(\phi-\frac{\sigma_6}{2}\right) H_{\sqrt{2}\epsilon_{7}}-\sigma_6
H_{\epsilon_6}&=&  \left(\phi^\prime-\frac{\sigma^\prime_6}{2}\right)
H_{\sqrt{2}\epsilon_{7}}- \sigma^\prime_6 H_{-\epsilon_6}\nonumber\\
\sigma^\prime_6 =-\sigma_6\,&\Rightarrow &
\rho_9^\prime=1/\rho_9\nonumber\\ \phi^\prime = \phi-\sigma_6
&=&\phi-{\rm ln}(\rho_9)
\label{tdualphi}
\end{eqnarray}
where the primed fields are the corresponding type IIA fields and the
last equation  is the known transformation rule for the dilaton under
a $\mathrm{T}$--duality along a compact  direction (in the units
$\alpha^\prime=1$).  As far as the other fields are concerned, the
action of this automorphism is to map the roots
$\epsilon_i\pm\epsilon_6$ into $\epsilon_i\mp\epsilon_6$. If we extend
the rotation $\tau:\,\,\epsilon_6\rightarrow -\epsilon_6$
to the whole Lie algebra using this simple recipe (\ref{simplerecipe}) the fields $G_{i9}$ and
$B_{i9}$ are  mapped (modulo proportionality constants $c_{1,2}$
to be fixed) into $B^\prime_{i9}$ and $G^\prime_{i9}$ respectively:
\begin{eqnarray}
G_{i9}E_{\epsilon_i-\epsilon_9}+B_{i9}E_{\epsilon_i+\epsilon_9}=
G_{i9}E_{\epsilon_i+(-\epsilon_9)}+B_{i9}E_{\epsilon_i-(-\epsilon_9)}=
c_1 B^\prime_{i9}E_{\epsilon_i+(-\epsilon_9)}+c_2 G^\prime_{i9}E_{\epsilon_i-(-\epsilon_9)}
\nonumber \\
\end{eqnarray}
The transformation on the R--R fields, applying a similar rationale,
can be read off table \ref{table2},\ref{table2b}.
\par
\paragraph{Vector fields}
\par It is clearly essential, in discussing the effect of  $\mathrm{S \times T}$ dualities on
$p$--branes to have a clear algebraic characterization of their action also on
vector fields. The same remark obviously applies at the level of
gaugings. This is not difficult, since the vector fields are
associated with weights while the scalars are associated with roots.
So given the action on $\mathrm{S \times T}$ on the root lattice we
easily lift it to the weight lattice. Let me recall the results of
section \ref{pesanti} where the fundamental $Sp(56)_D$ representation of
$E_{7(7)}$ has been constructed in the symplectic real basis
\footnote{By $Sp(56)_D$ we denote the ${\bf 56}$ symplectic
representation of $E_{7(7)}$ in which the Cartan generators are
diagonal.}.  There  the representation  was described in terms of 56 weights
$W^{(\lambda)}$ ($\lambda=1,\dots,56$), whose difference from the {\it
highest weight} $W^{(51)}$ are suitable combinations of the simple
roots with positive integer coefficients (the first 28 weights
correspond to {\it magnetic} charges, the last 28 to {\it electric}
charges). In the Bertolini and Trigiante's conventions adopted in the present section
depending on whether we consider $\mathrm{{E}_{7(7)}}^{\pm}$ (type IIA/IIB) we have two
set of weights ${W^{(\lambda)}}^{\mp}$. These weights provide a suitable
basis also for the two  representations ${\bf 28}$ and ${\bf
\bar{28}}$ in which the ${\bf 56}$ decomposes  with respect to
$\mathrm{SU(8)}$: the  ${\bf 28}$ is generated by $W^{(\alpha)}$ with
$\alpha=1,\dots,28$ and the  ${\bf \bar{28}}$ by
$W^{(\alpha+28)}=-W^{(\alpha)}$.    The weights $W^{(\lambda)}$  can
be naturally put in correspondence with the vector fields obtained
from the dimensional reduction of the  type IIA or IIB theory
respectively. Both the first 28 magnetic charges and the last 28
electric charges decompose into a first set of 16 R--R charges (which
contribute to a $ {\bf 32}^\pm $ of $\mathrm{SO(6,6)}$) and a second set of 12
NS--NS charges. Representing these weights (as well as the roots for
the scalar fields in table \ref{table2},\ref{table2b}) in the basis of
$(\epsilon_i)_{i=1,\dots,7}$ the correspondence weights
$\leftrightarrow $ vectors (or roots $\leftrightarrow $ scalars)
become natural and consistent with our characterization of $\mathrm{S\times T}$
duality. Indeed, as far as the R--R fields are concerned, the natural
correspondence is between the inner indices of the dimensionally
reduced form (which gives rise either to a scalar or to a vector) and
the number and positions of the ``$+$'' signs multiplying the
$(\epsilon_i)_{i=1,\dots,6}$ in the corresponding weight.\footnote{For
example the vector $A_{\mu ijkl}$ corresponds to the weight $(1/2)
(..+_i..+_j..+_k..+_l..)$.}  In tables \ref{table2},\ref{table2b} and
\ref{table3},\ref{table3b} this correspondence has
been ``nailed'' down for a particular  $S\times T$--duality gauge (so
that the fields (weights) of IIB and  IIA are related by a
$\mathrm{T}$--duality along the compact direction $x^9$ (automorphism
$\epsilon_6\rightarrow -\epsilon_6$)), making it possible to infer the
transformation rules  of the fields under a generic $\mathrm{S\times T}$
transformation.
%%%%%%%%%%%%%%%%%%%%%%%%%%%%%%%%%%%%%%
\begin{table} [ht]
%\vskip -25pt
\begin{center}
{\tiny
\begin{tabular}{|c|c|c|c|c|}
\hline
{\small IIA} & {\small IIB} & {\small $\alpha_{m,n}^\mp$ ($IIB/IIA$) } &
 {\small $\epsilon_i$--components}
& {\small $\alpha_i$--components} \\
\hline
\hline
$A_{9}$ & $\rho$ & $\alpha_{1,1}$ & $\frac{1}{2}(-1,-1,-1,-1,-1, {\mp} 1,\sqrt{2})$ &
$(0,0,0,0,0,0,1)$\\
\hline
\hline
$B_{8\,9}$ & $B_{8\,9}$ & $\alpha_{2,1}$ & $(0,0,0,0,1,1,0)$ & $(0,0,0,0,0,1,0)$\\
\hline
$G_{8\,9}$ & $G_{8\,9}$ & $\alpha_{2,2}$ & $(0,0,0,0,1,-1,0)$ & $(0,0,0,0,1,0,0)$\\
\hline
$A_{8}$ & $A_{8\, 9}$ & $\alpha_{2,3}$ & $\frac{1}{2}(-1,-1,-1,-1,1, {\pm} 1,\sqrt{2})$ &
$(0,0,0,0,0,1,1)$\\
\hline
\hline
$B_{7\,8}$ & $B_{7\,8}$ & $\alpha_{3,1}$ & $(0,0,0,1,1,0,0)$ & $(0,0,0,1,1,1,0)$\\
\hline
$G_{7\,8}$ & $G_{7\,8}$ & $\alpha_{3,2}$ & $(0,0,0,1,-1,0,0)$ & $(0,0,0,1,0,0,0)$\\
\hline
$B_{7\,9}$ & $B_{7\,9}$ & $\alpha_{3,3}$ & $(0,0,0,1,0,1,0)$ & $(0,0,0,1,0,1,0)$\\
\hline
$G_{7\,9}$ & $G_{7\,9}$ & $\alpha_{3,4}$ & $(0,0,0,1,0,-1,0)$ & $(0,0,0,1,1,0,0)$\\
\hline
$A_{7\,8\,9}$ & $A_{7\, 8}$ & $\alpha_{3,4}$ & $\frac{1}{2}(-1,-1,-1,1,1, {\mp} 1,\sqrt{2})$ &
$(0,0,0,1,1,1,1)$\\
\hline
$A_{7}$ & $A_{7\, 9}$ & $\alpha_{3,6}$ & $\frac{1}{2}(-1,-1,-1,1,-1,{\pm} 1,\sqrt{2})$ &
$(0,0,0,1,0,1,1)$\\
\hline
\hline
$B_{6\,7}$ & $B_{6\,7}$ & $\alpha_{4,1}$ & $(0,0,1,1,0,0,0)$ & $(0,0,1,2,1,1,0)$\\
\hline
$G_{6\,7}$ & $G_{6\,7}$ & $\alpha_{4,2}$ & $(0,0,1,-1,0,0,0)$ & $(0,0,1,0,0,0,0)$\\
\hline
$B_{6\,8}$ & $B_{6\,8}$ & $\alpha_{4,3}$ & $(0,0,1,0,1,0,0)$ & $(0,0,1,1,1,1,0)$\\
\hline
$G_{6\,8}$ & $G_{6\,8}$ & $\alpha_{4,4}$ & $(0,0,1,0,-1,0,0)$ & $(0,0,1,1,0,0,0)$\\
\hline
$B_{6\,9}$ & $B_{6\,9}$ & $\alpha_{4,4}$ & $(0,0,1,0,0,1,0)$ & $(0,0,1,1,0,1,0)$\\
\hline
$G_{6\,9}$ & $G_{6\,9}$ & $\alpha_{4,6}$ & $(0,0,1,0,0,-1,0)$ & $(0,0,1,1,1,0,0)$\\
\hline
$A_{6\,7\,9}$ & $A_{6\, 7}$ & $\alpha_{4,7}$ & $\frac{1}{2}(-1,-1,1,1,-1, {\mp} 1,\sqrt{2})$ &
$(0,0,1,2,1,1,1)$\\
\hline
$A_{6\,8\,9}$ & $A_{6\, 8}$ & $\alpha_{4,8}$ & $\frac{1}{2}(-1,-1,1,-1,1,{\mp} 1,\sqrt{2})$ &
$(0,0,1,1,1,1,1)$\\
\hline
$A_{6}$ & $A_{6\, 9}$ & $\alpha_{4,9}$ & $\frac{1}{2}(-1,-1,1,-1,-1,{\pm} 1,\sqrt{2})$ &
$(0,0,1,1,0,1,1)$\\
\hline
$A_{6\,7\,8}$ & $A_{6\, 7\,8\,9}$ & $\alpha_{4,10}$ & $\frac{1}{2}(-1,-1,1,1,1,
 {\pm} 1,\sqrt{2})$ &
$(0,0,1,2,1,2,1)$\\
\hline
\hline
$B_{5\,6}$ & $B_{5\,6}$ & $\alpha_{5,1}$ & $(0,1,1,0,0,0,0)$ & $(0,1,2,2,1,1,0)$\\
\hline
$G_{5\,6}$ & $G_{5\,6}$ & $\alpha_{5,2}$ & $(0,1,-1,0,0,0,0)$ & $(0,1,0,0,0,0,0)$\\
\hline
$B_{5\,7}$ & $B_{5\,7}$ & $\alpha_{5,3}$ & $(0,1,0,1,0,0,0)$ & $(0,1,1,2,1,1,0)$\\
\hline
$G_{5\,7}$ & $G_{5\,7}$ & $\alpha_{5,4}$ & $(0,1,0,-1,0,0,0)$ & $(0,0,1,0,0,0,0)$\\
\hline
$B_{5\,8}$ & $B_{5\,8}$ & $\alpha_{5,5}$ & $(0,1,0,0,1,0,0)$ & $(0,1,1,1,1,1,0)$\\
\hline
$G_{5\,8}$ & $G_{5\,8}$ & $\alpha_{5,6}$ & $(0,1,0,0,-1,0,0)$ & $(0,0,0,1,0,0,0)$\\
\hline
$B_{5\,9}$ & $B_{5\,9}$ & $\alpha_{5,7}$ & $(0,1,0,0,0,1,0)$ & $(0,1,1,1,0,1,0)$\\
\hline
$G_{5\,9}$ & $G_{5\,9}$ & $\alpha_{5,8}$ & $(0,1,0,0,0,-1,0)$ & $(0,0,0,0,1,0,0)$\\
\hline
$A_{5\,6\,9}$ & $A_{5\, 6}$ & $\alpha_{5,9}$ & $\frac{1}{2}(-1,1,1,-1,-1,{\mp} 1,\sqrt{2})$ &
$(0,1,2,2,1,1,1)$\\
\hline
$A_{5\,7\,9}$ & $A_{5\, 7}$ & $\alpha_{5,10}$ & $\frac{1}{2}(-1,1,-1,1,-1, {\mp} 1,\sqrt{2})$ &
$(0,1,1,2,1,1,1)$\\
\hline
$A_{5\,8\,9}$ & $A_{5\, 8}$ & $\alpha_{5,11}$ & $\frac{1}{2}(-1,1,-1,-1,1, {\mp} 1,\sqrt{2})$ &
$(0,1,1,1,1,1,1)$\\
\hline
$A_{5}$ & $A_{5\, 9}$ & $\alpha_{5,12}$ & $\frac{1}{2}(-1,1,-1,-1,-1,{\pm} 1,\sqrt{2})$ &
$(0,1,1,1,0,1,1)$\\
\hline
$A_{5\,7\,8}$ & $A_{5\, 7\,8\,9}$ & $\alpha_{5,13}$ & $\frac{1}{2}(-1,1,-1,1,1, {\pm} 1,
\sqrt{2})$ &
$(0,1,1,2,1,2,1)$\\
\hline
$A_{5\,6\,8}$ & $A_{5\,6\,8\, 9}$ & $\alpha_{5,14}$ & $\frac{1}{2}(-1,1,1,-1,1,
{\pm} 1,\sqrt{2})$ &
$(0,1,2,2,1,2,1)$\\
\hline
$A_{5\,6\,7}$ & $A_{5\,6\,7\, 9}$ & $\alpha_{5,15}$ & $\frac{1}{2}(-1,1,1,1,-1,
{\pm} 1,\sqrt{2})$ &
$(0,1,2,3,1,2,1)$\\
\hline
$A_{\mu\nu\rho}$ & $A_{5\,6\,7\, 8}$ & $\alpha_{5,16}$ &
$\frac{1}{2}(-1,1,1,1,1,{\mp} 1,\sqrt{2})$ &
$(0,1,2,3,1,2,1)$\\
\hline
\hline
\end{tabular}}
\end{center}
\caption{\small The correspondence between the positive
roots $\alpha_{m,n}^\pm$ of the $U$--duality algebra ${\cal E}_{7(7)}^\pm$
and the scalar fields parameterizing the moduli space for either IIA or IIB
compactifications on $T^6$. The notation $\alpha_{m,n}$ for the positive roots was
explained in section \ref{abeideale}.}
\label{table2}
\end{table}
%%%%%%%%%%%%%%%%%%%%%%%%%%%%%%%%%%%%%%%%%%%%%%%%%%%%%%%%%%%%%%%%%%%%%%%%%%%%%%%%%%%%%%
%
%%%%%%%%%%%%%%%%%%%%%%%%%%%%%%%%%%%%%%%%%%%%%%%%%%%%%%%%%%%%%%%%%%%%%%%%%%%%%%%%%%%%%
\begin{table} [ht]
%\vskip -25pt
\begin{center}
{\tiny
\begin{tabular}{|c|c|c|c|c|}
\hline
{\small IIA} & {\small IIB} & {\small $\alpha_{m,n}^\mp$ ($IIB/IIA$) } &
 {\small $\epsilon_i$--components}
& {\small $\alpha_i$--components} \\
\hline
\hline
$B_{4\,5}$ & $B_{4\,5}$ & $\alpha_{6,1}$ & $(1,1,0,0,0,0,0)$ & $(1,2,2,2,1,1,0)$\\
\hline
$G_{4\,5}$ & $G_{4\,5}$ & $\alpha_{6,2}$ & $(1,-1,0,0,0,0,0)$ & $(1,0,0,0,0,0,0)$\\
\hline
$B_{4\,6}$ & $B_{4\,6}$ & $\alpha_{6,3}$ & $(1,0,1,0,0,0,0)$ & $(1,1,2,2,1,1,0)$\\
\hline
$G_{4\,6}$ & $G_{4\,6}$ & $\alpha_{6,4}$ & $(1,0,-1,0,0,0,0)$ & $(1,1,0,0,0,0,0)$\\
\hline
$B_{4\,7}$ & $B_{4\,7}$ & $\alpha_{6,5}$ & $(1,0,0,1,0,0,0)$ & $(1,1,1,2,1,1,0)$\\
\hline
$G_{4\,7}$ & $G_{4\,7}$ & $\alpha_{6,6}$ & $(1,0,0,-1,0,0,0)$ & $(1,1,1,0,0,0,0)$\\
\hline
$B_{4\,8}$ & $B_{4\,8}$ & $\alpha_{6,7}$ & $(1,0,0,0,1,0,0)$ & $(1,1,1,1,1,1,0)$\\
\hline
$G_{4\,8}$ & $G_{4\,8}$ & $\alpha_{6,8}$ & $(1,0,0,0,-1,0,0)$ & $(1,1,1,1,0,0,0)$\\
\hline
$B_{4\,9}$ & $B_{4\,9}$ & $\alpha_{6,9}$ & $(1,0,0,0,0,1,0)$ & $(1,1,1,1,0,1,0)$\\
\hline
$G_{4\,9}$ & $G_{4\,9}$ & $\alpha_{6,10}$ & $(1,0,0,0,0,-1,0)$ & $(1,1,1,1,1,0,0)$\\
\hline
$B_{\mu\nu}$ & $B_{\mu\nu}$ & $\alpha_{6,11}$ & $(0,0,0,0,0,0,\sqrt{2})$ &
$(1,2,3,4,2,3,2)$\\
\hline
$A_{\mu\nu\,9}$ & $A_{\mu\nu}$ & $\alpha_{6,12}$ & $\frac{1}{2}(1,1,1,1,1,{\pm} 1,\sqrt{2})$ &
$(1,2,3,4,2,3,1)$\\
\hline
$A_{4\,5\,9}$ & $A_{4\,5}$ & $\alpha_{6,13}$ & $\frac{1}{2}(1,1,-1,-1,-1,{\mp} 1,\sqrt{2})$ &
$(1,2,2,2,1,1,1)$\\
\hline
$A_{4\,6\,9}$ & $A_{4\,6}$ & $\alpha_{6,14}$ & $\frac{1}{2}(1,-1,1,-1,-1,{\mp} 1,\sqrt{2})$ &
$(1,1,2,2,1,1,1)$\\
\hline
$A_{4\,7\,9}$ & $A_{4\,7}$ & $\alpha_{6,15}$ & $\frac{1}{2}(1,-1,-1,1,-1,{\mp} 1,\sqrt{2})$ &
$(1,1,1,2,1,1,1)$\\
\hline
$A_{4\,8\,9}$ & $A_{4\,8}$ & $\alpha_{6,16}$ & $\frac{1}{2}(1,-1,-1,-1,1,{\mp} 1,\sqrt{2})$ &
$(1,1,1,1,1,1,1)$\\
\hline
$A_{4}$ & $A_{4\,9}$ & $\alpha_{6,17}$ & $\frac{1}{2}(1,-1,-1,-1,-1,{\pm} 1,\sqrt{2})$ &
$(1,1,1,1,0,1,1)$\\
\hline
$A_{4\,7\,8}$ & $A_{4\,7\,8\,9}$ & $\alpha_{6,18}$ &
$\frac{1}{2}(1,-1,-1,1,1,{\pm} 1,\sqrt{2})$ & $(1,1,1,2,1,2,1)$\\
\hline
$A_{4\,6\,8}$ & $A_{4\,6\,8\,9}$ & $\alpha_{6,19}$ &
$\frac{1}{2}(1,-1,1,-1,1,{\pm} 1,\sqrt{2})$ & $(1,1,2,2,1,2,1)$\\
\hline
$A_{4\,6\,7}$ & $A_{4\,6\,7\,9}$ & $\alpha_{6,20}$ &
$\frac{1}{2}(1,-1,1,1,-1,{\pm} 1,\sqrt{2})$ & $(1,1,2,3,1,2,1)$\\
\hline
$A_{\mu\nu\,5}$ & $A_{4\,6\,7\,8}$ & $\alpha_{6,21}$ &
$\frac{1}{2}(1,-1,1,1,1,{\mp} 1,\sqrt{2})$ & $(1,1,2,3,2,2,1)$\\
\hline
$A_{4\,5\,8}$ & $A_{4\,5\,8\,9}$ & $\alpha_{6,22}$ &
$\frac{1}{2}(1,1,-1,-1,1,{\pm} 1,\sqrt{2})$ & $(1,2,2,2,1,2,1)$\\
\hline
$A_{4\,5\,7}$ & $A_{4\,5\,7\,9}$ & $\alpha_{6,23}$ &
$\frac{1}{2}(1,1,-1,1,-1,{\pm} 1,\sqrt{2})$ & $(1,2,2,3,1,2,1)$\\
\hline
$A_{\mu\nu\,6}$ & $A_{4\,5\,7\,8}$ & $\alpha_{6,24}$ &
$\frac{1}{2}(1,1,-1,1,1,{\mp} 1,\sqrt{2})$ & $(1,2,2,3,2,2,1)$\\
\hline
$A_{4\,5\,6}$ & $A_{4\,5\,6\,9}$ & $\alpha_{6,25}$ &
$\frac{1}{2}(1,1,1,-1,-1,{\pm} 1,\sqrt{2})$ & $(1,2,3,3,1,2,1)$\\
\hline
$A_{\mu\nu\,7}$ & $A_{4\,5\,6\,8}$ & $\alpha_{6,26}$ &
$\frac{1}{2}(1,1,1,-1,1,{\mp} 1,\sqrt{2})$ & $(1,2,3,3,2,2,1)$\\
\hline
$A_{\mu\nu\,8}$ & $A_{4\,5\,6\,7}$ & $\alpha_{6,27}$ &
$\frac{1}{2}(1,1,1,1,-1,{\mp} 1,\sqrt{2})$ & $(1,2,3,4,2,2,1)$\\
\hline
\end{tabular}}
\end{center}
\caption{\small The correspondence between the positive
roots $\alpha_{m,n}^\pm$ of the $\mathrm{U}$--duality algebra $\mathrm{{E}_{7(7)}}^\pm$
and the scalar fields {\it continued...}.}
\label{table2b}
\end{table}
%%%%%%%%%%%%%%%%%%%%%%%%%%%%%%%%%%%%%%%%%%%%%%%%%
\begin{table} [ht]
%\vskip -45pt
\begin{center}
{\tiny
\begin{tabular}{|c|c|c|c|}
\hline
{\small IIA} & {\small IIB} & {\small ${W^{(\lambda)}}^{\pm}$ ($IIB/IIA$)} &
{\small $\epsilon_i$--components: $IIB/IIA$} \\
\hline
\hline
$A_{\mu}$ & $A_{\mu 9}$ & $W^{(1)} $ & $\frac{1}{2}\left(-1,-1,-1,-1,-1,{\pm} 1,0 \right) $\\
\hline
$A_{\mu 5678}$ & $A_{\mu 56789}$ & $W^{(2)} $ & $\frac{1}{2}\left(-1,1,1,1,1,{\pm} 1,0 \right) $\\
\hline
$A_{\mu 4678}$ & $A_{\mu 46789}$ & $W^{(3)} $ & $\frac{1}{2}\left(1,-1,1,1,1,{\pm} 1,0 \right) $\\
\hline
$A_{\mu 4578}$ & $A_{\mu 45789}$ & $W^{(4)} $ & $\frac{1}{2}\left(1,1,-1,1,1,{\pm} 1,0 \right) $\\
\hline
$A_{\mu 4568}$ & $A_{\mu 45689}$ & $W^{(5)} $ & $\frac{1}{2}\left( 1,1,1,-1,1,{\pm} 1,0\right) $\\
\hline
$A_{\mu 4567}$ & $A_{\mu 45679}$ & $W^{(6)} $ & $\frac{1}{2}\left(1,1,1,1,-1,{\pm} 1,0 \right) $\\
\hline
$A_{\mu 6789}$ & $A_{\mu 678}$ & $W^{(7)} $ & $\frac{1}{2}\left(-1,-1,1,1,1,{\mp} 1,0 \right) $\\
\hline
$A_{\mu 5789}$ & $A_{\mu 578}$ & $W^{(8)} $ & $\frac{1}{2}\left(-1,1,-1,1,1, {\mp} 1,0 \right) $\\
\hline
$A_{\mu 5689}$ & $A_{\mu 568}$ & $W^{(9)} $ & $\frac{1}{2}\left( -1,1,1,-1,1,{\mp} 1,0 \right) $\\
 \hline
$A_{\mu 5679}$ & $A_{\mu 567}$ & $W^{(10)} $ & $\frac{1}{2}\left(-1,1,1,1,-1,{\mp} 1,0 \right) $\\
\hline
$A_{\mu 4679}$ & $A_{\mu 467}$ & $W^{(11)} $ & $\frac{1}{2}\left(1,-1,1,1,-1,{\mp} 1,0 \right) $\\
\hline
$A_{\mu 4579}$ & $A_{\mu 457}$ & $W^{(12)} $ & $\frac{1}{2}\left(1,1,-1,1,-1,{\mp} 1,0 \right) $\\
 \hline
$A_{\mu 4569}$ & $A_{\mu 456}$ & $W^{(13)} $ & $\frac{1}{2}\left( 1,1,1,-1,-1,{\mp} 1,0\right) $\\
 \hline
$A_{\mu 4689}$ & $A_{\mu 468}$ & $W^{(14)} $ & $\frac{1}{2}\left(1,-1,1,-1,1,{\mp} 1,0 \right) $\\
 \hline
$A_{\mu 4789}$ & $A_{\mu 478}$ & $W^{(15)} $ & $\frac{1}{2}\left(1,-1,-1,1,1,{\mp} 1,0 \right) $\\
 \hline
$A_{\mu 4589}$ & $A_{\mu 458}$ & $W^{(16)} $ & $\frac{1}{2}\left(1,1,-1,-1,1,{\mp} 1,0 \right) $\\
 \hline
$B_{\mu 4}$ & $B_{\mu 4}$ & $W^{(17)} $ & $\left(-1,0,0,0,0,0,{\frac{1}{{\sqrt{2}}}}\right)$\\
 \hline
$B_{\mu 5}$ & $B_{\mu 5}$ & $W^{(18)} $ & $\left(0,-1,0,0,0,0,{\frac{1}{{\sqrt{2}}}}\right)$\\
 \hline
$B_{\mu 6}$ & $B_{\mu 6}$ & $W^{(19)} $ & $\left(0,0,-1,0,0,0,{\frac{1}{{\sqrt{2}}}}\right)$\\
 \hline
$B_{\mu 7}$ & $B_{\mu 7}$ & $W^{(20)} $ & $\left(0,0,0,-1,0,0,{\frac{1}{{\sqrt{2}}}}\right)$\\
 \hline
$B_{\mu 8}$ & $B_{\mu 8}$ & $W^{(21)} $ & $\left(0,0,0,0,-1,0,{\frac{1}{{\sqrt{2}}}}\right)$\\
 \hline
$G_{\mu 9}$ & $G_{\mu 9}$ & $W^{(22)} $ & $\left(0,0,0,0,0,1,{\frac{1}{{\sqrt{2}}}}\right)$\\
 \hline
$G_{\mu 4}$ & $G_{\mu 4}$ & $W^{(23)} $ & $\left(-1,0,0,0,0,0,-{\frac{1}{{\sqrt{2}}}}\right)$\\
 \hline
$G_{\mu 5}$ & $G_{\mu 5}$ & $W^{(24)} $ & $\left(0,-1,0,0,0,0,-{\frac{1}{{\sqrt{2}}}}\right)$\\
 \hline
$G_{\mu 6}$ & $G_{\mu 6}$ & $W^{(25)} $ & $\left(0,0,-1,0,0,0,-{\frac{1}{{\sqrt{2}}}}\right)$\\
 \hline
$G_{\mu 7}$ & $G_{\mu 7}$ & $W^{(26)} $ & $\left(0,0,0,-1,0,0,-{\frac{1}{{\sqrt{2}}}}\right)$\\
 \hline
$G_{\mu 8}$ & $G_{\mu 8}$ & $W^{(27)} $ & $\left(0,0,0,0,-1,0,-{\frac{1}{{\sqrt{2}}}}\right)$\\
 \hline
$B_{\mu 9}$ & $B_{\mu 9}$ & $W^{(28)} $ & $\left(0,0,0,0,0,1,-{\frac{1}{{\sqrt{2}}}}\right)$\\
 \hline
\end{tabular}}
\end{center}
\caption{\small Correspondence between the weights ${W^{(\lambda)}}^{\mp}$ of
the ${\bf 56}$ of $\mathrm{{E}_{7(7)}}^{\pm}$ and the vectors deriving from the dimensional
 reduction of
type IIA and type IIB fields.}
\label{table3}
\end{table}
%%%%%%%%%%%%%%%%%%%%%%%%%%%%%%%%%%%%%%%%%%%%%%%%%%%%%%%%%%%%%%%%%%%%%%%%%%%%%%%%%%%%%%%%%%%%%%
%
%%%%%%%%%%%%%%%%%%%%%%%%%%%%%%%%%%%%%%%%%%%%%%%%%%%%%%%%%%%%%%%%%%%%%%%%%%%%%%%%%%%%%%%%%%%%%
\begin{table} [ht]
%\vskip -45pt
\begin{center}
{\tiny
\begin{tabular}{|c|c|c|c|}
\hline
{\small IIA} & {\small IIB} & {\small ${W^{(\lambda)}}^{\pm}$ ($IIB/IIA$)} &
 {\small $\epsilon_i$--components: $IIB/IIA$} \\
\hline
\hline
$A_{\mu 456789}$ & $A_{\mu 45678}$ & $W^{(29)} $ & $-\frac{1}{2}\left(-1,-1,-1,-1,-1,
{\pm} 1\right) $\\ \hline
$A_{\mu 49}$ & $A_{\mu 4}$ & $W^{(30)} $ & $-\frac{1}{2}\left(-1,1,1,1,1,{\pm} 1,0 \right) $\\
\hline
$A_{\mu 59}$ & $A_{\mu 5}$ & $W^{(31)} $ & $-\frac{1}{2}\left(1,-1,1,1,1,{\pm} 1,0 \right) $\\
\hline
$A_{\mu 69}$ & $A_{\mu 6}$ & $W^{(32)} $ & $-\frac{1}{2}\left(1,1,-1,1,1,{\pm} 1,0 \right) $\\
 \hline
$A_{\mu 79}$ & $A_{\mu 7}$ & $W^{(33)} $ & $-\frac{1}{2}\left( 1,1,1,-1,1,{\pm} 1,0\right) $\\
 \hline
$A_{\mu 89}$ & $A_{\mu 8}$ & $W^{(34)} $ & $-\frac{1}{2}\left(1,1,1,1,-1,{\pm} 1,0 \right) $\\
\hline
$A_{\mu 45}$ & $A_{\mu 459}$ & $W^{(35)} $ & $-\frac{1}{2}\left(-1,-1,1,1,1,{\mp} 1,0 \right) $\\
 \hline
$A_{\mu 46}$ & $A_{\mu 469}$ & $W^{(36)} $ & $-\frac{1}{2}\left(-1,1,-1,1,1,{\mp} 1,0 \right) $\\
 \hline
$A_{\mu 47}$ & $A_{\mu 479}$ & $W^{(37)} $ & $-\frac{1}{2}\left( -1,1,1,-1,1,{\mp} 1,0 \right) $\\
 \hline
$A_{\mu 48}$ & $A_{\mu 489}$ & $W^{(38)} $ & $-\frac{1}{2}\left(-1,1,1,1,-1,{\mp} 1,0 \right) $\\
 \hline
$A_{\mu 58}$ & $A_{\mu 589}$ & $W^{(39)} $ & $-\frac{1}{2}\left(1,-1,1,1,-1,{\mp} 1,0 \right) $\\
 \hline
$A_{\mu 68}$ & $A_{\mu 689}$ & $W^{(40)} $ & $-\frac{1}{2}\left(1,1,-1,1,-1,{\mp} 1,0 \right) $\\
 \hline
$A_{\mu 78}$ & $A_{\mu 789}$ & $W^{(41)} $ & $-\frac{1}{2}\left( 1,1,1,-1,-1,{\mp} 1,0\right) $\\
 \hline
$A_{\mu 57}$ & $A_{\mu 579}$ & $W^{(42)} $ & $-\frac{1}{2}\left(1,-1,1,-1,1,{\mp} 1,0 \right) $\\
 \hline
$A_{\mu 56}$ & $A_{\mu 569}$ & $W^{(43)} $ & $-\frac{1}{2}\left(1,-1,-1,1,1,{\mp} 1,0 \right) $\\
\hline
$A_{\mu 67}$ & $A_{\mu 679}$ & $W^{(44)} $ & $-\frac{1}{2}\left(1,1,-1,-1,1,{\mp} 1,0 \right) $\\
\hline
$B_{\mu 4}$ & $B_{\mu 4}$ & $W^{(45)} $ & $\left(1,0,0,0,0,0,-{\frac{1}{{\sqrt{2}}}}\right)$\\
\hline
$B_{\mu 5}$ & $B_{\mu 5}$ & $W^{(46)} $ & $\left(0,1,0,0,0,0,-{\frac{1}{{\sqrt{2}}}}\right)$\\
\hline
$B_{\mu 6}$ & $B_{\mu 6}$ & $W^{(47)} $ & $\left(0,0,1,0,0,0,-{\frac{1}{{\sqrt{2}}}}\right)$\\
\hline
$B_{\mu 7}$ & $B_{\mu 7}$ & $W^{(48)} $ & $\left(0,0,0,1,0,0,-{\frac{1}{{\sqrt{2}}}}\right)$\\
\hline
$B_{\mu 8}$ & $B_{\mu 8}$ & $W^{(49)} $ & $\left(0,0,0,0,1,0,-{\frac{1}{{\sqrt{2}}}}\right)$\\
\hline
$G_{\mu 9}$ & $G_{\mu 9}$ & $W^{(50)} $ & $\left(0,0,0,0,0,-1,-{\frac{1}{{\sqrt{2}}}}\right)$\\
\hline
$G_{\mu 4}$ & $G_{\mu 4}$ & $W^{(51)} $ & $\left(1,0,0,0,0,0,{\frac{1}{{\sqrt{2}}}}\right)$\\
\hline
$G_{\mu 5}$ & $G_{\mu 5}$ & $W^{(52)} $ & $\left(0,1,0,0,0,0,{\frac{1}{{\sqrt{2}}}}\right)$\\
\hline
$G_{\mu 6}$ & $G_{\mu 6}$ & $W^{(53)} $ & $\left(0,0,1,0,0,0,{\frac{1}{{\sqrt{2}}}}\right)$\\
\hline
$G_{\mu 7}$ & $G_{\mu 7}$ & $W^{(54)} $ & $\left(0,0,0,1,0,0,{\frac{1}{{\sqrt{2}}}}\right)$\\
\hline
$G_{\mu 8}$ & $G_{\mu 8}$ & $W^{(55)} $ & $\left(0,0,0,0,1,0,{\frac{1}{{\sqrt{2}}}}\right)$\\
\hline
$G_{\mu 9}$ & $B_{\mu 9}$ & $W^{(56)} $ & $\left(0,0,0,0,0,-1,{\frac{1}{{\sqrt{2}}}}\right)$\\
\hline
\end{tabular}}
\end{center}
\caption{\small Correspondence between the weights ${W^{(\lambda)}}^{\mp}$ of
the ${\bf 56}$ of $\mathrm{E_{7(7)}}^{\pm}$ and the vectors
{\it continued...}.}
\label{table3b}
\end{table}
\section{Concluding Remarks}
The algebraic machinery  I have reviewed in this chapter is a
powerful tool that has not been fully exploited yet. It has proved
very useful in understanding the general structure of $BPS$ black
holes and in pinpointing their \emph{microscopic/macroscopic
correspondence} but its application to the general problem of
the same \emph{microscopic/macroscopic correspondence} at the level
of  gaugings, domain walls and other $p$--brane solutions  is just a
programme for the future. Nonetheless I have no doubt that it is a
very fruitful direction to be pursued. In this chapter I have mainly
focused on the case of maximal supergravities with particular
attention to the $D=4$ case, but as I explained in section
\ref{solgensetup} the Solvable Lie algebra approach exists for all
supergravities where the scalar manifold is a homogeneous coset
$\mathrm{G/H}$. As we know this is the most frequent case and
applies, in particular, to the $\mathcal{N}=4$ theory in $D=5$.
Before gauging we just have the graviton multiplet that contains one real scalar
and $n$ vector, each of which contains $5$ scalars. The scalar manifold
in this case is:
\begin{equation}
  \mathcal{M}_{scalar}= SO(1,1) \, \times \,
  \frac{\mathrm{SO(5,n)}}{\mathrm{SO(5) \times \mathrm{SO(n)}}}
\label{scaln4d5}
\end{equation}
For $n=5$ eq.(\ref{scaln4d5}) describes the submanifold of NS-NS
scalars for type IIA supergravity compactified on a $T^5$ torus. The
remaining $16$ scalars are Ramond fields. Consider next the case of a
$D3$--brane placed placed at the singular point of a transverse space
$\mathbb{R}^2 \times \mathbb{C}^2 /\Gamma$ where $\Gamma \subset
SU(2)$ is a discrete subgroup of $SU(2)$. If $\Gamma$ were the
identity the $D3$--brane would be a regular one leading to $\mathrm{AdS}_5
\times \mathrm{S}^5$ as near horizon geometry. The corresponding near
brane supergravity is the maximal compact gauging of the
$\mathcal{N}=8$ where the gauge algebra is $\mathrm{SO(6)}$. When
$\Gamma$ is non trivial the non trivial holonomy of the transverse space
reduces the number of preserved supercharges from $32$ down to $16$
and the near brane supergravity is some gauged version of the
$\mathcal{N}=4$ theory with a number of vector/tensor multiplets that
is:
\begin{equation}
  n=5+ {\#} \, \mbox{of twisted multiplets}
\label{twistnum}
\end{equation}
It is a challenging problem to obtain a precise relation between such
a gauging and the microscopic description of the parent $D3$--brane
system. It is natural to me to think that the solvable Lie algebra machinery
described in the previous pages should provide the appropriate tool to bridge
this gap. This would be particularly rewarding in the case of
fractional $D3$--branes where there is a non trivial $B$--flux at the
singular point \cite{Bertolini:2000dk,Billo:2000yb}.
\par
The above discussion is just an explicit example of the many applications
of the SLA machinery to the \emph{microscopic/macroscopic correspondence} that can be
conceived. The general programme is to exploit the algebraic
characterization of $\mathrm{S \times T}$ dualities in a systematic
way.
\section*{Acknowledgements}
It is my great pleasure to thank here Riccardo D'Auria, Anna
Ceresole, Antoine Van Proeyen and Marco Bill\'o from whose scientific advice and
reading of the earlier versions of this manuscript I have  much
profited. I also had important and clarifying discussions with Mario
Trigiante, Matteo Bertolini, Igor Pesando and Alberto Lerda to whom I am very much
grateful. I want to express my gratitude to Eugene Cremmer and Kelly Stelle
for their  excellent organization of the
$RTN$--school at the \emph{Institut Henry Poincar\'e} of Paris and
for their kind invitation (together with the other organizers) to
give this series of lectures. I have much benefited in writing the
final version of these notes from all the comments and constructive
criticisms raised by the audience during the lectures and I would
like to thank all the participants for this. Finally my constant
gratitude goes to Toine Van Proeyen whose skillful, friendly and
stimulating managing of the $RTN$ network is the essential basis for
the success of all network activities.
%%%%%%%%%%%%%%%%%%%%%%%%%%%%%%%%%%%%%%%%%%%%%%%%%%%%%%%%%%%%%%%%%%
\appendix
\chapter[Conventions]{Conventions}
The conventions used in these lecture notes are those used throughout
the whole development of the  \emph{rheonomy approach} to
supergravity, as exposed in the book \cite{castdauriafre}, and
subsequently utilized  in the original papers on the derivation of
coordinate free special geometry
\cite{Castellani:1990zd,Castellani:1990tp}, on the $\mathcal{N}=2$ gauging
\cite{D'Auria:1991fj} and formulation of the most general $\mathcal{N}=2$ lagrangian
\cite{bertolo}. The same conventions were also used in the whole
series of papers dealing with the central charges in extended
supergravity
\cite{Ceresole:1996ca,Andrianopoli:1997wf,Andrianopoli:1997pn,Andrianopoli:1998pg,
Billo:1999ip,D'Auria:1999fa}, in the series of papers on the AdS/CFT
correspondence via harmonic analysis on $\mathrm{G/H}$ manifolds
\cite{Ghbrane,m111,sergiotorino,sanssergio,adscftcheckers,poliv52,noin0101,noin0102}
and in the papers on the gauging of maximal supergravities in $D=4$
and $D=5$ \cite{Cordaro:1998tx,Andrianopoli:2000rs} or in the papers
on partial supersymmetry breaking
\cite{Ferrara:1996gu,Fre:1997js,Girardello:1997hf}.
In order to have a uniform language and a possibility to compare
theories with different number of supersymmetries and in different
dimensions I have made an effort to translate also the results of
\cite{ceregatta} and the formalism of very special geometry
\cite{GST1,GST2,deWit:1992nm,deWit:1993wf,deWit:1995tf} to the
same set of consistent notations used in the other papers mentioned
above and adopted in these lecture notes. They are as follows.
\section{Listing of the conventions}
I list the conventions by items
\subsection{Space--time signature}
The space--time metric in all dimensions is the mostly minus metric:
\begin{equation}
  \eta=\mbox{diag}\left( + , \underbrace{-, \dots , -}_{D-1 \,\,
  times}\right)
\label{mostminu}
\end{equation}
\subsection{Gamma matrix algebra}
The conventions for gamma matrices are always:
\begin{eqnarray}
\left( \gamma^0 \right )^\dagger & = & \gamma^0 \nonumber\\
\left( \gamma^i \right )^\dagger & = & - \gamma^i \quad
(i=1,\dots,D-1)
\label{gamma}
\end{eqnarray}
In particular for the $D=4$ theories we have:
\subsubsection{Gamma matrices properties in $D=4$}
\begin{eqnarray}
\gamma_5\equiv\ft{i}{4!}\epsilon_{mnpq}\gamma^m\gamma^n\gamma^p\gamma^q\qquad
\{\gamma_5,\gamma^m\}=0\nonumber\\
\gamma_5^2=1\qquad\gamma_5^\dagger=\gamma_5\qquad\gamma_5^T=\gamma_5\nonumber\\
C(\gamma^m)^TC^{-1}=-\gamma^m\qquad C^2=-1\qquad
C^\dagger=-C\qquad [C,\gamma_5]=0
\end{eqnarray}
and:
\begin{equation}\label{gamma2}
  \left.\begin{array}{c}
  C^T=-C\\
  (C\gamma_m)^T=C\gamma_m\\
  (C\gamma^{mn})^T=C\gamma^{mn}\\
  (C\gamma_5)^T=-C\gamma_5\\
  (C\gamma_5\gamma_m)^T=-C\gamma_5\gamma_m\\
  (C\gamma_5\gamma^{mn})^T=C\gamma_5\gamma^{mn}
  \end{array}\right\}
\end{equation}
Some useful identities:
\begin{equation}
  \left\{\begin{array}{ccc}
  \gamma^p\gamma^{mn}&=&\eta^{pm}\gamma^n-\eta^{pn}\gamma^m+\gamma^{pmn}\\
  \gamma^{mn}\gamma^p&=&\eta^{pn}\gamma^m-\eta^{pm}\gamma^n+\gamma^{pmn}\\
  \end{array}\right.
\end{equation}
\begin{equation}
  \gamma^m\gamma^{pqr}=\eta^{mp}\gamma^{qr}+\eta^{mq}\gamma^{rp}
  +\eta^{mr}\gamma^{pq}+\gamma^{mpqr}
\end{equation}
\begin{eqnarray}
  \gamma^{mn}\gamma^{pq}&=&\eta^{mq}\eta^{np}-\eta^{mp}\eta^{nq}
  \nonumber\\
  &&+\left(\gamma^{mq}\eta^{np}-\gamma^{mp}\eta^{nq}
  -\gamma^{nq}\eta^{mp}+\gamma^{np}\eta^{mq}\right)\nonumber\\
  &&+\gamma^{mnpq}
\end{eqnarray}
\begin{equation}
  \gamma_5\gamma_{pqr}=i\epsilon_{pqrs}\gamma^s
\end{equation}
\begin{equation}
  \gamma^{mn}\epsilon_{mnpq}=2i\gamma_5\gamma_{pq}
\end{equation}
\subsection{Index conventions}
\subsubsection{Space--time indices}
In most instances I use a differential form language so that the
curved space--time indices are not mentioned. When they are mentioned
they appear as low case Greek letters from the middle of the alphabet
\begin{equation}
  \lambda,\mu,\nu,\dots = \mbox{space--time curved indices} \, = \,
  0,1,\dots,D-1
\label{spacetimind}
\end{equation}
Large use is instead made of the flat (=tangent) space--time indices
that are denoted with low case Latin letters form the beginning of
the alphabet:
\begin{equation}
a,b,c,\dots  = \mbox{\emph{flat} space--time indices } \, = \, 0,1,2,\dots,D-1
\label{flatinde}
\end{equation}
\subsubsection{R-symmetry indices}
A convention that has been universally adopted in the whole
development of the rheonomy approach and hence in all the papers,
books and reviews mentioned above concerns the R-symmetry indices,
namely those labeling the supersymmetry charges and transforming in
the fundamental of $\mathrm{SU}(\mathcal{N})$ in four dimensions, in
the fundamental of $\mathrm{USp}( \mathcal{N}) $ in five dimensions
and so on. For all $D$.s and for all $N$.s these indices are denoted
as capital Latin letters from the beginning of the alphabet:
\begin{equation}
  A,B,C,D,\dots = \mbox{R symmetry indices}
\label{Rsymmetry}
\end{equation}
\subsubsection{Vector Field indices}
Following my general discussion in section \ref{dualsym} on duality
transformations and the whole set up of supergravity gaugings I stress the
importance of having a well separated index convention for the
enumeration of the vector fields or in general for the $p+1$--forms.
For this purpose, in every dimension $D$ and for all $\mathcal{N}$--extended
supersymmetries we use the Capital Greek indices:
\begin{equation}
  \Lambda, \Sigma, \Gamma, \Delta, \dots = \mbox{indices enumerating vector fields or
  $p+1$--forms}
\label{vectinde}
\end{equation}
Group--theoretically these indices are assigned to a linear
representation $\mathbb{D}$ of the isometry group of the scalar manifold. In the
case of $p+1$--forms that are not self dual the dimension of the linear representation
$\mathbb{D}$ equals the number $n$ of $p+1$--forms, so that the range of the indices
$\Lambda,\Sigma, \dots$ exhausts such a dimension. Instead, in the case of
self--dual $p+1$ forms   we have $\mbox{dim} \mathbb{D}=2n$. Here
the following convention has been uniformly adopted in all the papers, books and reviews
quoted above: an object $V$
transforming in $\mathbb{D}$ is labeled as follows:
\begin{equation}
  V = \left( \begin{array}{c}
    V^\Lambda \\
    V_\Sigma \
  \end{array}\right) \quad \quad ; \quad \quad
  \Lambda,\Sigma=1,\dots,n
\label{Vvector}
\end{equation}
The upper $n$ indices refer to the electric fields while the lower
$n$ indices refer to the magnetic duals.
\par
This general convention consistently implies the conventions adopted
for special K\"ahler geometry and very special geometry. Since  in
both cases the section $X^\Lambda $  entering the definition of these
geometries transforms as the electric field strengths it naturally
carries the same Capital Greek indices.
\subsection{Notations for special and very special geometry}
As just mentioned we name:
\begin{equation}
  V = \left( \begin{array}{c}
    X^\Lambda(z) \\
    F_\Sigma(z) \
  \end{array}\right) \quad \quad ; \quad \quad
  \Lambda,\Sigma=1,\dots,n_V = 1+n
\label{holosection}
\end{equation}
the holomorphic section of the flat symplectic bundle governing
the definition of special K\"ahler geometry $n$ being the number of
vector multiplets. The complex coordinates of the special K\"ahler
manifold (= the scalar fields) are denoted $z^i$ and the Latin low
case letters from the middle of the alphabet are used to denote the
\emph{world indices} on the complex scalar manifold:
\begin{eqnarray}
  i,j,k,\ell,\dots &=& \mbox{holomorphic world indices on the scalar
  manifold}\\
  i^\star,j^\star,k^\star,\ell^\star,\dots &=& \mbox{antiholomorphic world indices on the scalar
  manifold}\\
\label{holoantiholo}
\end{eqnarray}
Similarly for very special geometry we name
\begin{equation}
  V = \left( \begin{array}{c}
    X^\Lambda(\phi) \\
    F_\Sigma(\phi)=\frac{\partial }{\partial X^\Sigma} \ln \mathrm{N}(X)  \
  \end{array}\right) \quad \quad ; \quad \quad
  \Lambda,\Sigma=1,\dots,n_V = 1+n
\label{holosect}
\end{equation}
the real section of the flat bundle governing this geometry. As
explained in the text at the level of gauging we need to distinguish
those of the vector fields that are true vectors and those that are
dualised to massive two--forms. This is done as in eq.(\ref{Lamrang}), namely
\begin{eqnarray}
  \Lambda &=& 1,\dots,n+1 \, = \left \{\underbrace{\, 0\, , \, I }_{ \mathbf{I}}\,,
   , \, \mathcal{M}\right\} \nonumber\\
   \mathbf{I,J,K},\dots, &=& \mbox{indices enumerating gauged vectors
   } \, = \, 1,\dots, \mbox{dim} \, \mathcal{G}_{gauge}
   \nonumber\\
   \mathcal{M,N,},\dots & = & \mbox{indices enumerating tensor
   multiplets} \, = \, 1,\dots,n_T
\label{Lamrangus}
\end{eqnarray}
\subsection{Notations for quaternionic geometry}
The three complex structures closing the quaternionic algebra are
labeled with low case Latin indices from the almost end of the alphabet:
\begin{equation}
  x,y,z,\dots = \mbox{vector indices of $\mathrm{SU_R(2)}\sim
  \mathrm{SO_R(3)}$}\, = \, 1,2,3
\label{su2vectind}
\end{equation}
On the other hand low case Greek letters from the beginning of the
alphabet are used to denote the symplectic indices transforming in
the fundamental of $Sp(2m,\mathbb{R})$ where $m$ is the number of
hypermultiplets:
\begin{equation}
  \alpha,\beta,\gamma,\dots \, = \, \mbox{$Sp(2m,\mathbb{R})$
  indices}\, = \, 1,2,\dots,2m.
\label{sympinde}
\end{equation}
The low case Latin  letters from the very end of the alphabet are
instead used to denote \emph{world indices} on the quaternionic
manifold:
\begin{equation}
  u,v,w,\dots \, = \, \mbox{world indices on the quaternionic manifold}\,
  1,2,\dots,4m
\label{quateinde}
\end{equation}
Other conventions are:
\begin{eqnarray}
K^x & = & \mbox{HyperK\"ahler forms} \nonumber\\
\Omega^x & = & \mbox{curvatures of the $\mathrm{SU_R(2)}$
connection}\nonumber\\
\mathcal{P}_{\overrightarrow{X}}^x & = & \mbox{triholomorphic moment map
of the Killing vector $\overrightarrow{X}$ } \nonumber\\
\label{triholonot}
\end{eqnarray}
\subsection{Conventions in differential geometry and General Relativity}
In dealing with forms I use the conventions used throughout all the
papers, books and reviews mentioned at the beginning of this
appendix, namely:
\begin{description}
  \item[a] The exterior derivative $d$ always acts from the left to
  the right on differential forms:
\begin{equation}
  d \omega^{[p]} = \partial _{[\mu_1} \,
  \omega_{\mu_2,\dots,\mu_{p+1}]} \, dx^{\mu _1} \, \wedge \, \dots
  \,dx^{\mu_{p+1}}
\label{exteder}
\end{equation}
  \item[b] The components of a $p$-form are always written to the
  left of the differentials (which is important in the case of
  fermionic forms) and they are normalized to strength $1$. For
  instance, for vector fields:
\begin{eqnarray}
  F &=& dA= F_{\mu\nu} \, dx^\mu \wedge dx^\nu \nonumber\\
  F_{\mu\nu} & = & \ft {1}{2} \, \left( \partial _\mu  A_\nu  - \partial
  _\nu  A_\mu \right)
\label{FverDA}
\end{eqnarray}
Note the factor $\ft 1 2$ difference with the normalization of $F_{\mu \nu
}$ which is customary in most field theory textbooks.
 \item[c] In General Relativity the curvature $2$--form is defined as
\begin{equation}
  R^{ab} = d\omega^{ab} -\omega^{ac} \, \wedge \, \omega^{db} \,
  \eta_{cd}
\label{curv2form}
\end{equation}
The Riemann tensor (and consequently the scalar curvature) is defined
as the component in the vielbein basis of the curvature $2$--form
(\ref{curv2form}):
\begin{equation}
  R^{ab} = R^{ab}_{cd} \, V^c \, \wedge \, V^d
\label{Riemanconv}
\end{equation}
Note once again the factor $\ft 12$ difference (and also the sign)
with respect to the traditional normalizations of the Riemann tensors
used throughout most of the GR textbooks.
\end{description}

%%%%%%%%%%%%%%%%%%%%%%%%%%%%%%%%%%
% FILE TRSTbib %%%%%%%%%%%%%%%%%%%
%%%%%%%%%%%%%%%%%%%%%%%%%%%%%%%%%%
\References

\end{document}